# Projective Loop Quantum Gravity
# I. State Space

**Suzanne Lanéry[1,2] and Thomas Thiemann[1]**

[1] Institute for Quantum Gravity, Friedrich-Alexander University Erlangen-Nürnberg, Germany

[2] Mathematics and Theoretical Physics Laboratory, François-Rabelais University of Tours, France

**Abstract**

Instead of formulating the state space of a quantum field theory over one big Hilbert space, it has been proposed by Kijowski [14] to describe quantum states as projective families of density matrices over a collection of smaller, simpler Hilbert spaces. Beside the physical motivations for this approach, it could help designing a quantum state space holding the states we need. In [24] the description of a theory of Abelian connections within this framework was developed, an important insight being to use building blocks labeled by combinations of edges and surfaces. The present work generalizes this construction to an arbitrary gauge group $G$ (in particular, $G$ is neither assumed to be Abelian nor compact). This involves refining the definition of the label set, as well as deriving explicit formulas to relate the Hilbert spaces attached to different labels.

If the gauge group happens to be compact, we also have at our disposal the well-established Ashtekar-Lewandowski Hilbert space, which is defined as an inductive limit using building blocks labeled by edges only. We then show that the quantum state space presented here can be thought as a natural extension of the space of density matrices over this Hilbert space. In addition, it is manifest from the classical counterparts of both formalisms that the projective approach allows for a more balanced treatment of the holonomy and flux variables, so it might pave the way for the development of more satisfactory coherent states.

## Contents

# 1 Introduction

The context of the present work is a formalism, originally introduced by Jerzy Kijowski [14], in which the state space of a quantum field theory is presented in the form of a projective limit: the key idea is, instead of describing quantum states as density matrices over a (very) large Hilbert space, to describe them as families of density matrices over a collection of 'small' Hilbert spaces. The labels indexing these Hilbert spaces are to be thought as selecting finitely many degrees of freedom out of the considered infinite dimensional theory. Whenever the degrees of freedom retained by some label $\eta$ are also covered by a finer label $\eta'$, the small Hilbert space $\mathcal{H}_\eta$ associated to $\eta$ should be identified as a tensor product factor in $\mathcal{H}_{\eta'}$: this allows to formulate, in terms of the corresponding partial traces, the projective consistency conditions that the families of density matrices representing our quantum states need to fulfill. This formalism has been further developed by Andrzej Okołów [23, 24, 25], achieving in particular its application to models involving real-valued connections (in [24, section 5] and [25]). Our goal here is to expand this line of research to the construction of suitable projective quantum state spaces for theories of connections having a possibly non-Abelian gauge group. We will rely strongly on a previous series of articles [17, 18], in which the projective approach has been investigated at a fairly general level.

Our motivation for adopting this approach lies in its potential to deliver bigger state spaces [18, subsection 2.2], as it sidesteps the need to specialize into a single representation of the algebra of observables. This may help to ensure that the quantum theory will actually contain the particular states we are looking for. More specifically, we will be interested in applications to Loop Quantum Gravity (LQG, [1, 31]), and to the construction of semi-classical and related states in this context. There seems indeed to exist serious obstructions [30, 16, 10] to find such states within the Ashtekar-Lewandowski Hilbert space used in LQG [2], arising from the intrinsic asymmetry in the role played by the configuration and momentum variables (ie. the holonomies and fluxes, see eg. the discussion in [5]). This asymmetry can be traced back to the fact that the formalism is build on a vacuum which is an eigenstate of the flux observables (thus having maximal uncertainties in the holonomies). The states are then obtained as discrete excitations around this vacuum. The trouble is that, no matter how many discrete excitations are piled up on top of the vacuum, this will never be sufficient to mask this initial bias.

Note that the situation here differs crucially from similar constructions routinely used when doing quantum field theory on a flat Minkowski background: the vacuum state there is a *coherent* state with respect to a certain set of canonically conjugate variables (as provided by mode decomposition), so it does not favor half of the variables to the detriment of the others. This is however not an option for LQG, because the Ashtekar-Lewandowski vacuum is the only diffeomorphism invariant state at our disposal [21]: using a genuinely non-diffeomorphism-invariant state as vacuum would (in much the same way as just argued) lead to a quantum theory breaking the diffeomorphism invariance, which would be many orders of magnitude more disastrous than the above concerns.

The expectation that the projective approach could allow for a more balanced treatment of the configuration and momentum variables is supported by the respective classical precursors of both formalisms. While the inductive limit construction underlying the Ashtekar-Lewandowski Hilbert space can be seen as emerging from a projective limit of *configuration* spaces [22, 31], we emphasized in [18, section 3] that projective quantum state spaces are naturally obtained as the quantization of a projective limit of *phase* spaces.

Following the general procedure laid out in [18, subsection 3.1], we will, once equipped with an appropriate label set (theorem 2.16), proceed to set up a factorizing system of configuration spaces [17, def. 2.15] (theorem 3.7): this is indeed a fairly generic way of writing down a projective limit of phase spaces, provided that a family of real polarizations can be chosen consistently across all partial phase spaces. We can then quantize each of these partial theories in the corresponding position representation and the consistency of the polarizations ensures that the resulting small Hilbert spaces can eventually be arranged into a projective system of quantum state spaces. Equivalently, the tensor product factorizations needed for the quantum projective system can be read out directly from the classical factorizing system [18, prop. 3.3].

One might be worried that relying on the position representation to perform the quantization of the individual partial theories would reintroduce an unwanted singularization of the configuration variables. This is however not the case, because they only play a special role as far as the quantization of the *finite dimensional* small phase spaces is concerned: this is therefore comparable to the choice of a representation in quantum *mechanics* (in opposition to quantum field theory), which is known to be rather innocuous [32]. Indeed, the small phase spaces we will be working with will be cotangent bundles on Lie groups, whose position representation (namely, the $L_2$ space on the considered group) can be shown to be unitarily equivalent to a suitably defined momentum representation [9, 5]. Although we will thus be using the same building blocks as in the usual approach, the critical difference comes from the alternative way of gluing them together to compose the state space of the full theory: design choices in this regard are precisely the most likely to have irreversible consequences on the final quantum theory.

On the other hand, working in the position representation is convenient to study the relation between these two approaches. Any factorizing system of configuration spaces yields, by forgetting about the more precise information it provides regarding the links between the partial theories, an associated projective system of configuration spaces. On the quantum side, we then have an embedding, mapping the states over the corresponding inductive limit Hilbert space into the quantum projective state space [18, prop. 3.5]. We will make use of this device to understand how the state space we are proposing here extends the established one (theorem 3.20, props. 3.21 and 3.22).

As a side remark, for the construction we will be presenting here, the gauge group does not need to be compact, nor do we have to impose any particular restriction beyond the assumption of a finite-dimensional Lie group (should it be of any use, even countable discrete groups fall into this category, as the 0-dimensional case). By contrast, the Ashtekar-Lewandowski vacuum (and therefore the Hilbert space built on it) only exists if the gauge group is compact (note that the just highlighted results concerning the relation between the projective and inductive approaches of course only apply when the latter can at all be defined). The possibility of setting up a quantum state space in the case of a non-compact gauge group, although not in the focus of our interest, might find application in the treatment of the complex Ashtekar variables, requiring $\mathcal{SL}(2, \mathbb{C})$ as gauge group (see eg. the discussion in [23] and references therein).

In the following $\Sigma$ will denote a finite-dimensional, analytic manifold [20], $d$ its dimension ($d \geqslant 2$), $G$ a finite-dimensional Lie group and $\mathfrak{g}$ its Lie algebra.

# 2 Label set

The construction routinely employed in LQG relies on an inductive limit of 'small' Hilbert spaces, or, equivalently, on a projective limit of finite dimensional configuration spaces, with building blocks labeled by *graphs*. Each graph corresponds to a selection of position variables, namely the holonomies along its edges.

In principle, we could associate to any such graph a corresponding phase space, as the cotangent bundle on its configuration space. However, if we now consider a big graph $\gamma'$ and a subgraph $\gamma$ of $\gamma'$, there is no preferred way of defining a projection from the phase space $\mathcal{M}_{\gamma'}$ thus associated to $\gamma'$ into the phase space $\mathcal{M}_\gamma$ associated to $\gamma$. In order to define unambiguously such a projection, we would indeed have to specify how the impulsion variables described by $\mathcal{M}_\gamma$ should be transported to $\mathcal{M}_{\gamma'}$ (having in mind that a downward projection between the phase spaces is dual to an upward injection between the algebras of observables). As pointed out in [17, prop. 2.10], a projection between the phase spaces encapsulates, at least locally, the same information as a factorization of $\mathcal{M}_{\gamma'}$ into a Cartesian product of $\mathcal{M}_\gamma$ times a 'complementary' phase space $\mathcal{M}_{\gamma' \to \gamma}$. In addition, if the projection we are considering is compatible with the splitting of the phase spaces into position and momentum variables, this factorization should go down to a factorization $\mathcal{C}_{\gamma'} \approx \mathcal{C}_{\gamma' \to \gamma} \times \mathcal{C}_\gamma$ of the underlying configuration spaces.

In other words, if we specify not only which configuration variables are to be retained by $\gamma$, but also which momentum variables, we have a preferred choice of complementary configuration variables: they are simply characterized by their vanishing Poisson brackets with the retained momentum variables. By contrast, if we are only provided with a projection between configuration spaces, we cannot single out a choice of complementary variables within $\mathcal{C}_{\gamma'}$ that would span $\mathcal{C}_{\gamma' \to \gamma}$. These considerations suggest that the desired projective structure should rely on labels that are made not only of edges but also of surfaces, whose role will be, for each label $\eta$, to select which fluxes are to be the momentum variables associated to $\eta$. Besides, such mixed labels clearly sounds promising in view of giving the holonomies and fluxes a more symmetric status.

The need to include surfaces in the labels was already recognized by Okołów in [24, 25]. The label set he was using is however not immediately applicable to the non-Abelian case, which requires, as we will see, to impose more restrictive conditions on the relative disposition of the edges and surfaces. The reason why complications emerge in the non-Abelian case is the following. As mentioned above, a projection from the phase spaces associated to a finer label $\eta'$ into the one associated to a coarser label $\eta$ is dual to an embedding of the algebra of observables selected by $\eta$ into the algebra of observables selected by $\eta'$. Moreover, this embedding is linear and preserves the Poisson brackets [17, prop. 2.2], ie. it is an injective algebra morphism. But this requires that the vector space generated by the observables associated to $\eta$, within the algebra of $\eta'$, should be closed under Poisson brackets, and that the algebra structure thus induced by $\eta'$ should match the one seen from $\eta$.

This is a rather harmless requirement in the Abelian case, for there the flux operators commute with each other and the Poisson bracket of a flux variable with an holonomy variable is just a constant (possibly 0 depending on the intersection of the corresponding edge and surface), so the set of observables associated to a collection of edges and surfaces will be automatically closed under Poisson brackets. The aim of the present section will therefore be to determine which collections of edges and surfaces are admissible when the gauge group is arbitrary, and to check that the label set they are forming, although much reduced, is still directed.

Actually there do exist possibilities to write the state space of a theory of connection in projective form while using labels made of edges only: two such models have been for example proposed in [23] and, at the classical level, in [29]. In subsection 3.2, we will discuss in more details how a non ambiguous choice of complementary variables is achieved in these proposals without explicitly referring to the momentum variables in the definition of the labels, and why the thus obtained projective structures would altogether not fit our purpose.

## 2.1 Definitions

To fix our notations and definitions, we begin by writing down what we mean precisely by edges and surfaces, and we recall a few elementary properties, that we will use again and again in the following [31, section II.6].

As a technical side note, the class of edges we are considering here is pretty restrictive (namely, they are fully analytic edges embedded in a single analytic path), in contrast to the class of semi-analytic edges commonly used in LQG [31, section IV.20]. This is purely for convenience, and as far as the construction of the quantum state space is concerned, it has absolutely no incidence (for this restricted class of edges is a cofinal part of the more usual one, so that in particular the corresponding inductive limits of Hilbert spaces are identical, see subsection 3.2 and in particular the proof of theorem 3.20 for more details).

**Definition 2.1** An analytic, encharted edge in $\Sigma$ is an analytic diffeomorphism $\breve{e} : U \to V$, where $U$ is an open neighborhood of $[0,1] \times \{0\}^{d-1}$ in $\mathbb{R}^d$, and $V$ is an open subset of $\Sigma$. We call $\breve{\mathcal{L}}_{\text{edges}}$ the set of all encharted edges, and for $\breve{e} \in \breve{\mathcal{L}}_{\text{edges}}$ we define its starting point $b(\breve{e}) := \breve{e}(0,0)$, its ending point $f(\breve{e}) := \breve{e}(1,0)$ and its range $r(\breve{e}) := \breve{e}\left\langle [0,1] \times \{0\}^{d-1} \right\rangle$.

We say that $\breve{e}_1, \breve{e}_2 \in \breve{\mathcal{L}}_{\text{edges}}$ are equivalent, and we write $\breve{e}_1 \sim \breve{e}_2$, iff:

$$r(\breve{e}_1) = r(\breve{e}_1) \quad \& \quad b(\breve{e}_1) = b(\breve{e}_2).$$

This defines an equivalence relation on $\breve{\mathcal{L}}_{\text{edges}}$. Its set of equivalence classes will be denoted by $\mathcal{L}_{\text{edges}}$. An element $e \in \mathcal{L}_{\text{edges}}$ is called an edge, and we can define its starting point $b(e)$, its ending point $f(e)$ and its range $r(e)$, since these are the same for any representative of $e$.

**Proposition 2.2** Let $e \in \mathcal{L}_{\text{edges}}$ and let $p \neq p'$ be two distinct points in $r(e)$. Then, there exists a unique edge $e_{[p,p']} \in \mathcal{L}_{\text{edges}}$ such that:

$$r\left(e_{[p,p']}\right) \subset r(e), \quad b\left(e_{[p,p']}\right) = p \quad \& \quad f\left(e_{[p,p']}\right) = p'. \tag{2.2.1}$$

We denote by $e^{-1}$ the reversed edge $e^{-1} := e_{[f(e),b(e)]}$. We also define a strict, total order on the points of $r(e)$ by:

$$\forall p \in r(e), \quad b(e) <_{(e)} p \Leftrightarrow b(e) \neq p$$

$$\& \quad \forall p, p' \in r(e) \setminus \{b(e)\}\,, \quad p <_{(e)} p' \;\Leftrightarrow\; r\left(e_{[b(e),p]}\right) \subsetneq r\left(e_{[b(e),p']}\right) \tag{2.2.2}$$

For any $p_1 \neq p_4 \in r(e)$ and any $p_2 \neq p_3 \in r\left(e_{[p_1,p_4]}\right)$, we have:

1. $\left(e_{[p_1,p_4]}\right)_{[p_2,p_3]} = e_{[p_2,p_3]}$, so in particular $\left(e^{-1}\right)^{-1} = e$, $\left(e^{-1}\right)_{[p_1,p_4]} = e_{[p_1,p_4]}$ and $\left(e_{[p_1,p_4]}\right)^{-1} = e_{[p_4,p_1]}$;

2. $r\left(e_{[p_1,p_4]}\right) = \begin{cases} \{p \in r(e) \mid p_1 \leqslant_{(e)} p \leqslant_{(e)} p_4\} & \text{if } p_1 <_{(e)} p_4 \\ \{p \in r(e) \mid p_4 \leqslant_{(e)} p \leqslant_{(e)} p_1\} & \text{if } p_1 >_{(e)} p_4 \end{cases}$ ;

3. $p_2 <_{(e_{[p_1,p_4]})} p_3 \;\Leftrightarrow\; \begin{cases} p_2 <_{(e)} p_3 & \text{if } p_1 <_{(e)} p_4 \\ p_2 >_{(e)} p_3 & \text{if } p_1 >_{(e)} p_4 \end{cases}$ .

**Proof** *Existence and uniqueness.* Let $p \neq p' \in r(e)$ and let $\breve{e} : U \to V$ be a representative of $e$. Let $t \neq t' \in [0,1]$ such that $\breve{e}(t,0) = p$ and $\breve{e}(t',0) = p'$. The map $\varphi_1$, defined by:

$$\begin{array}{rccl} \varphi_1 & : & U & \to & \mathbb{R}^d \approx \mathbb{R} \times \mathbb{R}^{d-1} \\ & & (\tau, x) & \mapsto & (t + \tau\,(t' - t),\, x) \end{array},$$

is an analytic diffeomorphism onto its image $W_1 := \varphi_1 \langle U \rangle$, with $\varphi_1(0,0) = (t,0)$, $\varphi_1(1,0) = (t',0)$ and $\varphi_1 \left\langle [0,1] \times \{0\}^{d-1} \right\rangle = [t,t'] \times \{0\}^{d-1}$. Next, $U$ is an open neighborhood of $[0,1] \times \{0\}^{d-1}$ in $\mathbb{R}^d$, thus also of $[t,t'] \times \{0\}^{d-1}$. Hence, $U_1 := \varphi_1^{-1} \langle U \rangle$ is an open neighborhood of $[0,1] \times \{0\}^{d-1}$ in $\mathbb{R}^d$. Defining $W_1' = \varphi_1 \langle U_1 \rangle$ and $V_1 := \breve{e} \langle W_1' \rangle$, $\breve{e}_1 := \breve{e}|_{W_1' \to V_1} \circ \varphi_1|_{U_1 \to W_1'}$ is an encharted edge, with $b(\breve{e}_1) = p$, $f(\breve{e}_1) = p'$ and $r(\breve{e}_1) = \breve{e} \left\langle [t,t'] \times \{0\}^{d-1} \right\rangle \subset r(\breve{e})$.

Let $\breve{e}_2 : U_2 \to V_2$ be an encharted edge such that $b(\breve{e}_2) = p$, $f(\breve{e}_2) = p'$ and $r(\breve{e}_2) \subset r(\breve{e})$. Since $V$ is an open neighborhood of $r(\breve{e})$ in $\Sigma$, $W_2 := (\breve{e}_2)^{-1} \langle V \rangle$ is an open neighborhood of $[0,1] \times \{0\}^{d-1}$ in $\mathbb{R}^d$ and $\varphi_2 = \left(\breve{e}^{-1}\right) \circ \left(\breve{e}_2|_{W_2}\right) : W_2 \to U$ is an analytic diffeomorphism onto its image. Moreover, we have $\varphi_2(0,0) = (t,0)$, $\varphi_2(1,0) = (t',0)$ and $\varphi_2 \left\langle [0,1] \times \{0\}^{d-1} \right\rangle \subset [0,1] \times \{0\}^{d-1}$. So, by the intermediate value theorem, $\varphi_2 \left\langle [0,1] \times \{0\}^{d-1} \right\rangle = [t,t'] \times \{0\}^{d-1}$, and therefore $r(\breve{e}_1) = r(\breve{e}_2)$. Since we also have $b(\breve{e}_2) = p = b(\breve{e}_1)$, $\breve{e}_1$ and $\breve{e}_2$ are two representative of the same edge $e_{[p,p']}$.

Prop. 2.2.1 then follows immediately from eq. (2.2.1).

*Order on $r(e)$.* That eq. (2.2.2) unambiguously defines is a strict order on $r(e)$ (ie. an irreflexive and transitive relation) can be checked directly. Moreover, if $\breve{e}$ is a representative of $e$ and $t, t' \in [0,1]$ are such that $\breve{e}(t,0) = p$ and $\breve{e}(t',0) = p'$, we have, from the previous point:

$$p <_{(e)} p' \;\Leftrightarrow\; t < t'.$$

In particular, $<_{(e)}$ is therefore a total order.

Let $p_1 \neq p_4 \in r(e)$ and $p_2 \neq p_3 \in r\left(e_{[p_1,p_4]}\right)$. Using the explicit expression above for a representative of $e_{[p_1,p_4]}$, there exist $t_1 \neq t_4 \in [0,1]$ and $t_2 \neq t_3 \in [t_1,t_4]$ such that $\forall i \leqslant 4$, $\breve{e}(t_i,0) = p_i$ and we have:

$$r\left(e_{[p_1,p_4]}\right) = \breve{e}\left\langle [t_1, t_4] \times \{0\}^{d-1} \right\rangle \quad \text{and} \quad p_2 <_{(e_{[p_1,p_4]})} p_3 \;\Leftrightarrow\; \left( \frac{t_2 - t_1}{t_4 - t_1} < \frac{t_3 - t_1}{t_4 - t_1} \right),$$

which yields props. 2.2.2 and 2.2.3. □

**Proposition 2.3** We say that $e_1, \ldots, e_n \in \mathcal{L}_{\text{edges}}$ are composable iff there exist an edge $e \in \mathcal{L}_{\text{edges}}$ and points $p_0, p_1, \ldots, p_n$ in $r(e)$ such that:

$$b(e) = p_0 <_{(e)} p_1 <_{(e)} \ldots <_{(e)} p_{n-1} <_{(e)} p_n = f(e) \;\;\&\;\; \forall i \in \{1, \ldots, n\}\,,\; e_i = e_{[p_{i-1},p_i]}\,. \quad (2.3.1)$$

Then $e$ is uniquely determined by $e_1, \ldots, e_n$ and we write $e = e_n \circ \ldots \circ e_1$. Moreover, the following properties holds:

1. $e_n^{-1}, \ldots, e_1^{-1}$ are composable and $e^{-1} = e_1^{-1} \circ \ldots \circ e_n^{-1}$;
2. $\forall i \leqslant j \in \{1, \ldots, n\}$, $e_i, \ldots, e_j$ are composable and $e_{[b(e_i),f(e_j)]} = e_j \circ \ldots \circ e_i$;
3. if, for all $i \in \{1, \ldots, n\}$, there exist composable edges $e_{i,1}, \ldots, e_{i,m_i} \in \mathcal{L}_{\text{edges}}$ such that $e_i = e_{i,m_i} \circ \ldots \circ e_{i,1}$, then $e_{1,1}, \ldots, e_{1,m_1}, \ldots, \ldots, e_{n,1}, \ldots, e_{n,m_n}$ are composable and:

$$e = e_{n,m_n} \circ \ldots \circ e_{n,1} \circ \ldots \circ \ldots \circ e_{1,m_1} \circ \ldots \circ e_{1,1}\,.$$

**Proof** Let $e_1, \ldots, e_n \in \mathcal{L}_{\text{edges}}$ and let $e \in \mathcal{L}_{\text{edges}}$ and $p_0, p_1, \ldots, p_n \in r(e)$ be as in eq. (2.3.1). Let $\breve{e}$ be a representative of $e$ and $t_0, t_1, \ldots, t_n \in [0, 1]$ such that $\breve{e}(t_i, 0) = p_i$ for all $i \leqslant n$. Then, using auxiliary results from the proof of prop. 2.2, eq. (2.3.1) can be rewritten as:

$$0 = t_0 < t_1 < \ldots < t_{n-1} < t_n = 1$$

$$\text{and} \quad \forall i \in \{1, \ldots, n\}\,,\; \breve{e}\left\langle [t_{i-1}, t_i] \times \{0\}^{d-1} \right\rangle = r(e_i) \;\;\&\;\; \breve{e}(t_{i-1}, 0) = b(e_i)\,.$$

Thus, $r(e) = \bigcup_{i=1}^{n} r(e_i)$ and $b(e) = b(e_1)$, and therefore $e$ is uniquely determined by $e_1, \ldots, e_n$.

Then, props. 2.3.1 to 2.3.3 can be checked using props. 2.2.1 to 2.2.3. □

**Definition 2.4** A graph is a *finite* set of edges $\gamma \subset \mathcal{L}_{\text{edges}}$ such that:

$$\forall e \neq e' \in \gamma,\; r(e) \cap r(e') \subset \{b(e),\, f(e)\} \cap \{b(e'),\, f(e')\}\,.$$

We denote the set of graphs by $\mathcal{L}_{\text{graphs}}$ and we equip it with the preorder (reflexive and transitive relation):

$$\forall \gamma, \gamma' \in \mathcal{L}_{\text{graphs}},$$
$$\gamma \preccurlyeq \gamma' \;\Leftrightarrow\; \left( \forall e \in \gamma,\, \exists\, e_1, \ldots, e_n \in \gamma',\, \exists\, \epsilon_1, \ldots, \epsilon_n \in \{\pm 1\} \;/\; e = e_n^{\epsilon_n} \circ \ldots \circ e_1^{\epsilon_1} \right). \quad (2.4.1)$$

(The transitivity of $\preccurlyeq$ follows from props. 2.3.1 and 2.3.3.)

As a warming up for the more difficult proof of directedness that we will carry out in subsection 2.2 (where we will be dealing with labels that are made of edges and surfaces), we recall here why the set of analytic graphs $\mathcal{L}_{\text{graphs}}$ is directed [22, 31]. Note that it is only in lemma 2.6 (and in its analogue for the intersection of an edge with a surface, viz. lemma 2.10) that the analyticity actually plays a role. Hence, any class of edges (and surfaces) that could provide such an intersection property would do as well for the whole construction [4].

**Proposition 2.5** Let $\widetilde{\gamma}$ be a finite set of edges. Then, there exists $\gamma \in \mathcal{L}_{\text{graphs}}$ such that $\forall e \in \widetilde{\gamma},\ \{e\} \preccurlyeq \gamma$.

In particular, $\mathcal{L}_{\text{graphs}}, \preccurlyeq$ is a directed preordered set.

**Lemma 2.6** Let $e, e' \in \mathcal{L}_{\text{edges}}$ such that:

$$\forall p \in r(e) \setminus \{b(e)\}\,,\ \exists\, p' \in r(e) \ \ / \ \ b(e) <_{(e)} p' <_{(e)} p \quad \& \quad p' \in r(e')\,. \tag{2.6.1}$$

Then, there exists $p \in r(e) \setminus \{b(e)\}$ such that $r\left(e_{[b(e),p]}\right) \subset r(e')$.

**Proof** Let $\breve{e} : U \to V$, resp. $\breve{e'} : U' \to V'$, be a representative of $e$, resp. $e'$. Eq. (2.6.1) can be rewritten:

$$\forall t \in ]0,1],\ \exists\, t' \in ]0,t[ \ \ / \ \ \breve{e}(t',0) \in r(e')\,. \tag{2.6.2}$$

$U$ being an open neighborhood of $[0,1] \times \{0\}^{d-1}$ in $\mathbb{R}^d$ there exists $\epsilon \in ]0,1]$ such that $]-\epsilon, \epsilon[ \times \{0\}^{d-1} \subset U$. Now, the map $t' \mapsto \breve{e}(t',0)$ is continuous from $]-\epsilon,\epsilon[$ into $\Sigma$ and $r(e')$ is compact, so $b(e) = \breve{e}(0,0) \in r(e') \subset V'$. Hence, since $V'$ is an open subset of $\Sigma$, there exists $\epsilon' \in ]0,\epsilon]$ such that $\breve{e}\left\langle ]-\epsilon', \epsilon'[ \times \{0\}^{d-1} \right\rangle \subset V'$.

Thus, we can define a map $\psi : ]-\epsilon', \epsilon'[ \to \mathbb{R}^{d-1}$ by:

$$\begin{array}{rcl} \psi \ : \ ]-\epsilon', \epsilon'[ & \to & \mathbb{R}^{d-1} \\ t & \mapsto & s \circ \left(\breve{e'}\right)^{-1} \circ \breve{e}(t,0) \end{array},$$

where $s : \mathbb{R}^d \approx \mathbb{R} \times \mathbb{R}^{d-1} \to \mathbb{R}^{d-1}$ is the projection map on the second Cartesian factor. $\psi$ is analytic as a composition of analytic maps and, from eq. (2.6.2), 0 is an accumulation point of $\psi^{-1}\langle 0 \rangle$, hence $\psi \equiv 0$.

Next, we define the map $\psi' : ]-\epsilon', \epsilon'[ \to \mathbb{R}$ by:

$$\begin{array}{rcl} \psi' \ : \ ]-\epsilon', \epsilon'[ & \to & \mathbb{R} \\ t & \mapsto & p \circ \left(\breve{e'}\right)^{-1} \circ \breve{e}(t,0) \end{array},$$

where $p : \mathbb{R}^d \approx \mathbb{R} \times \mathbb{R}^{d-1} \to \mathbb{R}$ is the projection map on the first Cartesian factor. $\psi'$ is a continuous, injective map (combining $\psi \equiv 0$ with the bijectivity of $\breve{e}$ and $\breve{e'}$), $\psi'(0) \in [0,1]$, and from eq. (2.6.2) there exists $\epsilon'' \in ]0, \epsilon'[$ such that $\psi'(\epsilon'') \in [0,1]$, hence by the intermediate value theorem $\psi'\langle [0,\epsilon''] \rangle \subset [0,1]$. In other words, defining $p := \breve{e}(0,\epsilon'') \in r(e) \setminus \{b(e)\}$, we have $r\left(e_{[b(e),p]}\right) \subset r(e')$. □

**Proof of prop. 2.5** *Intersection of 2 edges.* Let $e_1, e_2 \in \mathcal{L}_{\text{edges}}$. We define:

$$C(e_1,\, e_2) := \{e_3 \in \mathcal{L}_{\text{edges}} \mid r(e_3) \subset r(e_1) \cap r(e_2)\}\,,$$

and:

$$c(e_1,\, e_2) := \left\{p \in r(e_1) \cap r(e_2) \,\middle|\, \forall e_3 \in C(e_1,\, e_2),\ \ p \in r(e_3) \Rightarrow p \in \{b(e_3),\, f(e_3)\}\right\}.$$

Then, for any $p \in r(e_1) \setminus \{b(e_1)\}$ we have, by applying lemma 2.6 to $e = e_{1,[p,b(e_1)]}$, $e' = e_2$ and using prop. 2.2:

$$\exists\, p' <_{(e_1)} p \;\;/\;\; \left(\forall p'' \in r(e_1),\;\; p' <_{(e_1)} p'' <_{(e_1)} p \Rightarrow p'' \notin r(e_2)\right) \;\text{ or }\; r\left(e_{1,[p',p]}\right) \subset r(e_2)\,,$$

and therefore:

$$\exists\, p' <_{(e_1)} p \;\;/\;\; \forall p'' \in r(e_1),\;\; p' <_{(e_1)} p'' <_{(e_1)} p \Rightarrow p'' \notin c(e_1,\, e_2)\,.$$

Similarly, for any $p \in r(e_1) \setminus \{f(e_1)\}$, applying lemma 2.6 to $e = e_{1,[p,f(e_1)]}$, $e' = e_2$ yields:

$$\exists\, p' >_{(e_1)} p \;\;/\;\; \forall p'' \in r(e_1),\;\; p <_{(e_1)} p'' <_{(e_1)} p' \Rightarrow p'' \notin c(e_1,\, e_2)\,.$$

Hence, choosing a representative of $e_1$ and using the explicit form of $<_{(e_1)}$ from the proof of prop. 2.2, we can, for any $p \in r(e_1)$, construct an open neighborhood $V_p$ of $p$ in $r(e_1)$ such that $c(e_1, e_2) \cap V_p \subset \{p\}$. Since $r(e_1)$ is compact, we thus have that $c(e_1, e_2)$ is finite.

Let $p \in r(e_1) \cap r(e_2) \setminus c(e_1,\, e_2)$. Using prop. 2.2.2 together with the definition of $c(e_1,\, e_2)$, there exists $p' \in r(e_1)$ such that:

$$p' <_{(e_1)} p \quad\text{and}\quad r\left(e_{1,[p',p]}\right) \subset r(e_2)\,.$$

Thus, using again the explicit form of $<_{(e_1)}$ in terms of some representative of $e_1$, we can define $p_{\text{inf}} \in r(e_1)$ by:

$$p_{\text{inf}} = \inf_{<_{(e_1)}} \left\{p' \in r(e_1) \;\middle|\; p' <_{(e_1)} p \quad \& \quad r\left(e_{1,[p',p]}\right) \subset r(e_2)\right\},$$

and $p_{\text{inf}} \in r(e_2)$ (for $r(e_1) \cap r(e_2)$ is closed in $r(e_1)$). Moreover, for any $p'' \in r(e_1)$ with $p_{\text{inf}} <_{(e_1)} p'' <_{(e_1)} p$, there exists $p' \in r(e_1)$ such that:

$$p' <_{(e_1)} p'' <_{(e_1)} p \quad\text{and}\quad r\left(e_{1,[p',p]}\right) \subset r(e_2)\,,$$

therefore $p'' \in r(e_2)$. Hence, $r\left(e_{1,[p_{\text{inf}},p]}\right) \subset r(e_2)$. On the other hand, if there exists $e_3 \in C(e_1,\, e_2)$ with $p_{\text{inf}} \in r(e_3)$, then $e_3 = e_{1,[b(e_3),f(e_3)]}$, thus there exists $p'' \in \{b(e_3),\, f(e_3)\}$ such that:

$$p'' \leqslant_{(e_1)} p_{\text{inf}} <_{(e_1)} p \quad\text{and}\quad r\left(e_{1,[p'',p]}\right) \subset r(e_2)\,,$$

so $p_{\text{inf}} = p''$. Therefore, $p_{\text{inf}} \in c(e_1,\, e_2)$. Similarly, we can construct $p_{\text{sup}} \in c(e_1,\, e_2)$ such that $p <_{(e_1)} p_{\text{sup}}$ and $r\left(e_{1,[p,p_{\text{sup}}]}\right) \subset r(e_2)$. To summarize, we have proved:

$$\forall p \in r(e_1) \cap r(e_2) \setminus c(e_1,\, e_2),$$
$$\exists\, p_{\text{inf}},\, p_{\text{sup}} \in c(e_1,\, e_2) \,/\, p_{\text{inf}} <_{(e_1)} p <_{(e_1)} p_{\text{sup}} \quad \& \quad r\left(e_{1,[p_{\text{inf}},p_{\text{sup}}]}\right) \subset r(e_2)\,. \tag{2.6.3}$$

*Intersection of 3 edges.* Let $e_1, e_2, e_3 \in \mathcal{L}_{\text{edges}}$ and consider $p \in r(e_1) \cap r(e_2) \cap r(e_3)$ with $p \notin c(e_1, e_3) \cup c(e_2, e_3)$. Then, for $i = 1, 2$, there exist $p'_i, p''_i \in r(e_3)$ such that:

$$p'_i <_{(e_3)} p <_{(e_3)} p''_i \quad\text{and}\quad r\left(e_{3,[p'_i,p''_i]}\right) \subset r(e_i)\,.$$

Hence, defining $p' := \max_{<_{(e_3)}} \left(p'_1,\, p'_2\right)$ and $p'' := \min_{<_{(e_3)}} \left(p''_1,\, p''_2\right)$, we have:

$$p' <_{(e_3)} p <_{(e_3)} p'' \quad\text{and}\quad e_{3,[p',p'']} \in C(e_1,\, e_2)\,.$$

Thus, $p \notin c(e_1,\, e_2)$. In other words, we get:

$$c(e_1,\, e_2) \cap r(e_3) \subset c(e_1, e_3) \cup c(e_2, e_3)\,.$$

*Directedness of $\mathcal{L}_{graphs}$.* Let $\widetilde{\gamma}$ be a finite subset of $\mathcal{L}_{\text{edges}}$, and define:

$$c(\widetilde{\gamma}) := \bigcup_{e_1, e_2 \in \widetilde{\gamma}} c(e_1,\, e_2)\,.$$

Let $e_1 \in \widetilde{\gamma}$. From the previous point, we have:

$$c(e_1,\, \widetilde{\gamma}) := c(\widetilde{\gamma}) \cap r(e_1) = \bigcup_{e_2 \in \widetilde{\gamma}} c(e_1,\, e_2)\,,$$

and since all $c(e_1,\, e_2)$ are finite, so is $c(e_1,\, \widetilde{\gamma})$. Moreover, $\{b(e_1),\, f(e_1)\} = c(e_1,\, e_1) \subset c(e_1,\, \widetilde{\gamma})$, so there exist $n_{e_1} \geqslant 1$ and $p_0^{e_1}, \dots, p_{n_{e_1}}^{e_1} \in r(e_1)$ such that:

$$b(e_1) = p_0^{e_1} <_{(e_1)} p_1^{e_1} <_{(e_1)} \dots <_{(e_1)} p_{n_{e_1}}^{e_1} = f(e_1) \quad \text{and} \quad c(e_1,\, \widetilde{\gamma}) = \left\{p_0^{e_1}, \dots, p_{n_{e_1}}^{e_1}\right\}.$$

Hence, $e_1 = e_{1,\, n_{e_1}} \circ \dots \circ e_{1,\, 1}$, where $e_{1,i} := e_{1,\, [p_{i-1}^{e_1}, p_i^{e_1}]}$.

Let $e_1,\, e_2 \in \widetilde{\gamma}$ and $i \in \{1, \dots,\, n_{e_1}\}$. Suppose that there exists $p \in r(e_{1,i}) \setminus \{p_{i-1}^{e_1},\, p_i^{e_1}\}$ such that $p \in r(e_2)$. By definition of $e_{1,i}$, we then have $p \in r(e_1) \cap r(e_2) \setminus c(e_1,\, e_2)$, so from eq. (2.6.3),

$$\exists\, k \leqslant i-1,\, \exists\, k' \geqslant i \;\; / \;\; r\left(e_{1,\, [p_k^{e_1}, p_{k'}^{e_1}]}\right) \subset r(e_2)\,.$$

Thus, we have in particular $r(e_{1,i}) \subset r(e_2)$. Moreover, $r(e_{1,i}) \cap c(e_2,\, \widetilde{\gamma}) = r(e_{1,i}) \cap c(\widetilde{\gamma}) = r(e_{1,i}) \cap c(e_1,\, \widetilde{\gamma}) = \{p_{i-1}^{e_1},\, p_i^{e_1}\}$. Hence, there exist $j \in \{1, \dots,\, n_{e_2}\}$ such that:

$$\left(p_{i-1}^{e_1} = p_{j-1}^{e_2} \quad \& \quad p_i^{e_1} = p_j^{e_2}\right) \quad \text{or} \quad \left(p_{i-1}^{e_1} = p_j^{e_2} \quad \& \quad p_i^{e_1} = p_{j-1}^{e_2}\right).$$

In other words, we have proved:

$$\begin{aligned} &\forall e_1,\, e_2 \in \widetilde{\gamma}, \;\; \forall i \in \{1, \dots,\, n_{e_1}\}\,, \\ &\qquad \left(r(e_{1,i}) \cap r(e_2) \subset \{p_{i-1}^{e_1},\, p_i^{e_1}\}\right) \;\; \text{or} \;\; \left(\exists\, j \in \{1, \dots,\, n_{e_2}\}\,,\, \exists\, \epsilon = \pm 1 \,\big/\, e_{1,i} = e_{2,j}^{\epsilon}\right). \end{aligned}$$

Moreover, we have from prop. 2.2.2:

$$\forall e_1 \in \widetilde{\gamma}, \;\; \forall i \neq j \in \{1, \dots,\, n_{e_1}\}\,, \;\; r(e_{1,i}) \cap r(e_{1,j}) \subset \{p_{i-1}^{e_1},\, p_i^{e_1}\}\,,$$

so we get:

$$\begin{aligned} &\forall e_1,\, e_2 \in \widetilde{\gamma}, \;\; \forall i \in \{1, \dots,\, n_{e_1}\}\,, \;\; \forall j \in \{1, \dots,\, n_{e_2}\}\,, \\ &\qquad \left(r(e_{1,i}) \cap r(e_{2,j}) \subset \{p_{i-1}^{e_1},\, p_i^{e_1}\}\right) \;\; \text{or} \;\; \left(\exists\, \epsilon = \pm 1 \,\big/\, e_{1,i} = e_{2,j}^{\epsilon}\right). \end{aligned} \tag{2.6.4}$$

Finally, we define the finite subset $\gamma' := \left\{e_{1,i} \,\middle|\, e_1 \in \widetilde{\gamma},\, i \in \{1, \dots,\, n_{e_1}\}\right\} \subset \mathcal{L}_{\text{edges}}$ and we can construct $\gamma \subset \gamma'$, such that:

$$\forall e \in \gamma', \;\; \exists!\, \epsilon = \pm 1 \,\big/\, e^{\epsilon} \in \gamma\,.$$

From eq. (2.6.4), we then have:

$$\forall e, e' \in \gamma, \;\; \left(r(e) \cap r(e') \subset \{b(e),\, f(e)\}\right) \;\; \text{or} \;\; \left(e = e'\right).$$

Therefore $\gamma \in \mathcal{L}_{\text{graphs}}$ and, by construction, $\forall e_1 \in \widetilde{\gamma}$, $\{e_1\} \preccurlyeq \gamma$.

In particular, for any $\gamma, \gamma' \in \mathcal{L}_{\text{graphs}}$, there exists $\gamma'' \in \mathcal{L}_{\text{graphs}}$, such that $\forall e \in \gamma \cup \gamma'$, $\{e\} \preccurlyeq \gamma''$, hence $\gamma, \gamma' \preccurlyeq \gamma''$. □

We now come to the surfaces. Our notion of surfaces is quite limiting here, since we restrict ourselves to fully analytic, 'round' surfaces. Like our class of edges, this is mostly a matter of convenience, and it would be relatively harmless to relax our definition (eg. we could cut an arbitrary compact piece out of an analytic plane, instead of only considering disk-shaped surfaces). Anyway, the flux operators will not be labeled directly by surfaces, but rather by finite intersections and differences thereof (in a sense that will be made precise in prop. 3.3), and those run through a considerably larger class of geometrical objects.

**Definition 2.7** An analytic, encharted surface in $\Sigma$ is an analytic diffeomorphism $\breve{S} : U \to V$, where $U$ is an open neighborhood of $\{0\} \times B^{(d-1)}$ in $\mathbb{R}^d$ ($B^{(d-1)}$ being the closed unit ball of $\mathbb{R}^{d-1}$), and $V$ is an open subset of $\Sigma$. We call $\breve{\mathcal{L}}_{\mathrm{surfcs}}$ the set of all encharted surfaces, and for $\breve{S} \in \breve{\mathcal{L}}_{\mathrm{surfcs}}$ we define its range $r\left(\breve{S}\right) := \breve{S}\left\langle \{0\} \times B^{(d-1)} \right\rangle$.

We say that $\breve{S}_1 : U_1 \to V_1, \breve{S}_2 : U_2 \to V_2 \in \breve{\mathcal{L}}_{\mathrm{surfcs}}$ are equivalent, and we write $\breve{S}_1 \sim \breve{S}_2$, iff:

$$r\left(\breve{S}_1\right) = r\left(\breve{S}_2\right) \quad \& \quad \breve{S}_1 \left\langle U_1' \cap \left(\mathbb{R}_+ \times \mathbb{R}^{d-1}\right)\right\rangle = \breve{S}_2 \left\langle U_2' \cap \left(\mathbb{R}_+ \times \mathbb{R}^{d-1}\right)\right\rangle ,$$

where $U_1'$ (resp. $U_2'$) is an open neighborhood of $\{0\} \times B^{(d-1)}$ in $U_1$ (resp. in $U_2$), and $\mathbb{R}_+$ is the set of non-negative reals. This defines an equivalence relation on $\breve{\mathcal{L}}_{\mathrm{surfcs}}$. Its set of equivalence classes will be denoted by $\mathcal{L}_{\mathrm{surfcs}}$. An element $S \in \mathcal{L}_{\mathrm{surfcs}}$ is called a surface, and we can define its range $r(S)$, since it is the same for any representative of $S$.

To determine the symplectic structure (or, equivalently, to specify, on the quantum side, how the flux operators act in the position representation), one has to discuss the relative positioning of edges with respect to surfaces. It will make the construction in subsection 3.1 appreciably simpler to consider 'one-sided' fluxes, that only interact with the edges reaching the surface from one side. Also, we will impose that all edges having a non-trivial interaction with a given surface should *start* from that surface (reorienting them if need be), so that flux operators always act at the beginning of edges. Thus, we classify the edges adapted to a surface as being above, below or indifferent to it (instead of the slightly different classification as outside/inside/up/down [31, section II.6.4]).

Since the surfaces we are considering are closed, one might be worried that an edge touching some surface precisely on its rand would have an unclear positioning: this is however not the case, for our surfaces have been defined in def. 2.7 as being embedded within an open analytic plane, which extends beyond the rand and allows to distinguish between above and below in a neighborhood of the surface. In particular, an edge intersecting the rand of a surface can only be indifferent to that surface if it runs along the same analytic plane.

Examples of well-positioned edges are shown in fig. 2.1. Note that all figures in the present article will be drawn in the case $d = 2$ (where both edges and surfaces are one-dimensional), as this is sufficient to illustrate most aspects of the construction (we will comment on subtleties arising in the physically more relevant case $d = 3$ when appropriate).

Given some surface, it is well-known that any edge can be subdivided into parts adapted to that surface [31, section II.6.4]. As announced earlier, this is the second place where the requirement for analyticity plays a critical role: it ensures that an edge cannot cross the plane of the surface more than finitely many times (fig. 2.2). Thus, we can cut this edge at each intersection point, and, splitting again each section in two parts, we can reorient these parts so that they start from the surface.

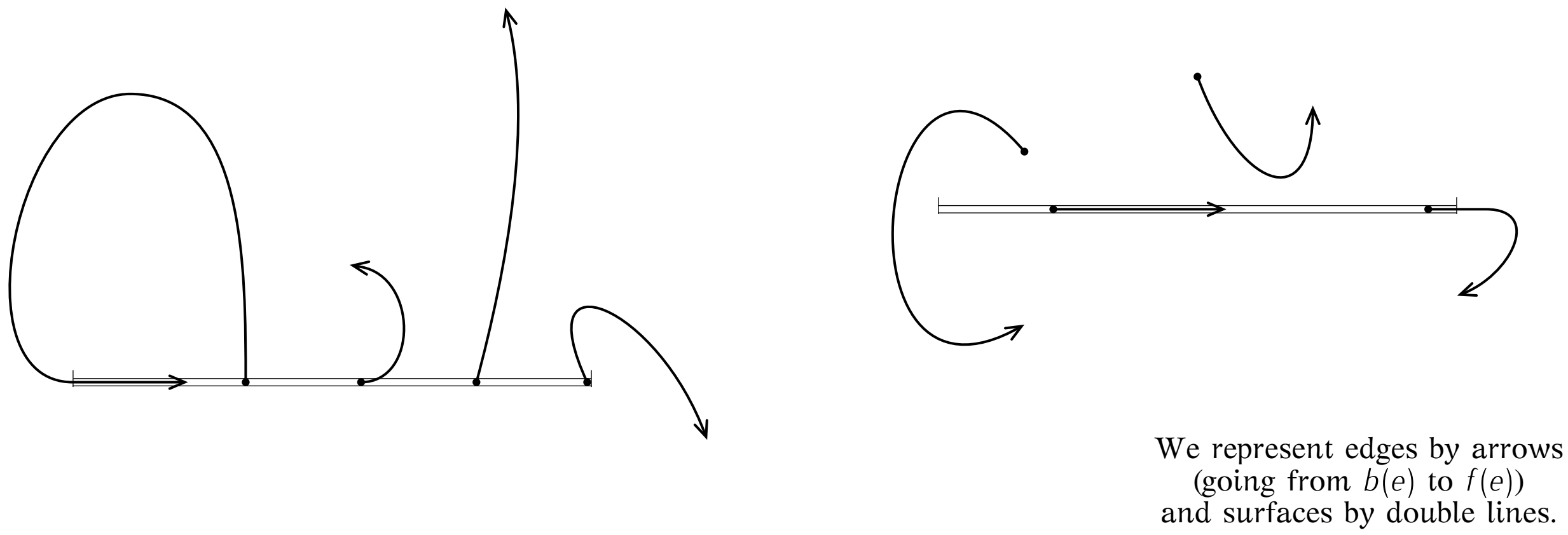

Figure 2.1 – Examples of edges above (on the left, assuming the surface is oriented upward) and indifferent to a surface (on the right)

**Proposition 2.8** Let $e \in \mathcal{L}_{\text{edges}}$ and $S \in \mathcal{L}_{\text{surfcs}}$. We say that:

1. $e$ is indifferent to $S$, and we write $e \curvearrowright S$, if there exist a representative $\check{S} : U \to V$ of $S$ and $e_1, \ldots, e_n \in \mathcal{L}_{\text{edges}}$ such that:

$$e = e_n \circ \ldots \circ e_1 \quad \& \quad \forall i \in \{1, \ldots, n\}\,, \;\; r(e_i) \cap r(S) = \varnothing \text{ or } r(e_i) \subset \check{S}\left\langle U_o \right\rangle,$$

where $U_o := U \cap \left(\{0\} \times \mathbb{R}^{d-1}\right)$;

2. $e$ is above $S$, and we write $e \uparrow S$, if there exist a representative $\check{S} : U \to V$ of $S$ and $e_1, e_2 \in \mathcal{L}_{\text{edges}}$ such that:

$$e = e_2 \circ e_1, \;\; e_2 \curvearrowright S, \;\; r(e_1) \cap r(S) = \{b(e)\} \quad \& \quad r(e_1) \setminus \{b(e)\} \subset \check{S}\left\langle U_+ \setminus U_o \right\rangle,$$

where $U_+ := U \cap \left(\mathbb{R}_+ \times \mathbb{R}^{d-1}\right)$;

3. $e$ is below $S$, and we write $e \downarrow S$, if there exist a representative $\check{S} : U \to V$ of $S$ and $e_1, e_2 \in \mathcal{L}_{\text{edges}}$ such that:

$$e = e_2 \circ e_1, \;\; e_2 \curvearrowright S, \;\; r(e_1) \cap r(S) = \{b(e)\} \quad \& \quad r(e_1) \setminus \{b(e)\} \subset \check{S}\left\langle U_- \setminus U_o \right\rangle,$$

where $U_- := U \cap \left(\mathbb{R}_- \times \mathbb{R}^{d-1}\right)$ and $\mathbb{R}_-$ is the set of non-positive reals.

We have the following properties:

4. these 3 cases are mutually disjoint;

5. if $e \curvearrowright S$, then $e^{-1} \curvearrowright S$;

6. if $e_1, e_2 \in \mathcal{L}_{\text{edges}}$ are such that $e = e_2 \circ e_1$, then, for any $\diamond \in \left\{\curvearrowright, \uparrow, \downarrow\right\}$:

$$e \diamond S \quad \Leftrightarrow \quad e_1 \diamond S \;\; \& \;\; e_2 \curvearrowright S\,.$$

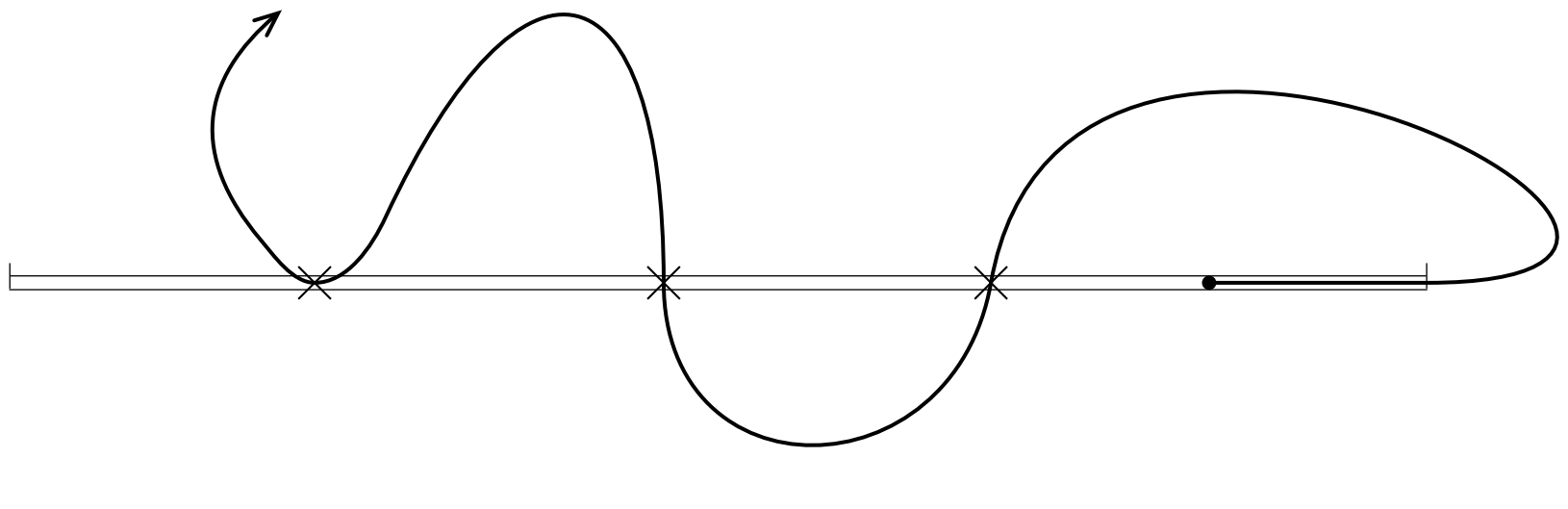

The points in $c(e,\, S)$ are marked by crosses.

Figure 2.2 – Adapting an edge to a given surface

**Proof** Assertion 2.8.4 follows from the definition of the equivalence relation in def. 2.7, assertions 2.8.5 and 2.8.6 from the properties of subedges (prop. 2.2) and edge compositions (prop. 2.3). $\square$

**Proposition 2.9** For any $e \in \mathcal{L}_{\text{edges}}$ and any $S \in \mathcal{L}_{\text{surfcs}}$, there exist $e_1, \dots, e_n \in \mathcal{L}_{\text{edges}}$ and $\epsilon_1, \dots, \epsilon_n \in \{\pm 1\}$ such that:

$$e = e_n^{\epsilon_n} \circ \dots \circ e_1^{\epsilon_1} \quad \& \quad \forall i \in \{1, \dots, n\}\,, \quad e_i \diamond_i S \quad \text{with} \quad \diamond_i \in \left\{\curvearrowright, \uparrow, \downarrow\right\}.$$

**Lemma 2.10** Let $e \in \mathcal{L}_{\text{edges}}$ and $S \in \mathcal{L}_{\text{surfcs}}$. Then, there exists $p \in r(e) \setminus \{b(e)\}$ such that $e_{[b(e),p]} \diamond S$ with $\diamond \in \left\{\curvearrowright, \uparrow, \downarrow\right\}$.

**Proof** If $b(e) \notin r(S)$, then there exists an open neighborhood of $b(e)$ in $r(e)$ that does not intersects $r(S)$, for $r(S)$ is compact. Hence, choosing some representative of $\breve{e}$ and using the explicit expression for the range of a subedge (from the proof of prop. 2.2), there exists $p \in r(e) \setminus \{b(e)\}$ such that $r\left(e_{[b(e),p]}\right) \cap r(S) = \varnothing$, so $e_{[b(e),p]} \curvearrowright S$.

We now assume $b(e) \in r(S)$ and we pick out representatives $\breve{e} : U \to V$ of $e$ and $\breve{S} : U' \to V'$ of $S$. $U$ is an open neighborhood of $[0,1] \times \{0\}^{d-1}$ in $\mathbb{R}^d$, $V'$ an open neighborhood of $r(S)$ in $\Sigma$, and $\breve{e}(0,0) \in r(S)$. Hence, as in the proof of lemma 2.6, there exists $\epsilon \in ]0,1]$ such that $\breve{e}\left\langle ]-\epsilon, \epsilon[ \times \{0\}^{d-1} \right\rangle \subset V'$ and we can define an analytic map $\psi : ]-\epsilon, \epsilon[ \to \mathbb{R}$ by:

$$\begin{array}{rcl} \psi \;:\; ]-\epsilon, \epsilon[ & \to & \mathbb{R} \\ t & \mapsto & p \circ \left(\breve{S}\right)^{-1} \circ \breve{e}(t,0) \end{array},$$

where $p : \mathbb{R}^d \approx \mathbb{R} \times \mathbb{R}^{d-1} \to \mathbb{R}$ is the projection map on the first Cartesian factor.

If there exists $\epsilon' \in \, ]0,\, \epsilon[$ such that $\forall t \in \, ]0,\, \epsilon'[\,, \psi(t) > 0$ (resp. $< 0$), we can take $p := \breve{e}(\epsilon'', 0)$, with some $\epsilon'' \in \, ]0,\, \epsilon'[$, and we have $e_{[b(e),p]} \uparrow S$ (resp. $\downarrow S$). Hence, there only remains to consider:

$$\forall \epsilon' \in \, ]0,\, \epsilon[\,,\ \exists\, t_1, t_2 \in \, ]0,\, \epsilon'[ \;/\; \psi(t_1) \geqslant 0 \quad \& \quad \psi(t_2) \leqslant 0\,.$$

The intermediate value theorem yields:

$$\forall \epsilon' \in \, ]0,\, \epsilon[\,,\ \exists\, t \in \, ]0,\, \epsilon'[ \;/\; \psi(t) = 0\,.$$

Therefore, 0 is an accumulation point of $\psi^{-1}\langle 0 \rangle$, so $\psi \equiv 0$. Defining $p = \breve{e}(\epsilon', 0)$ for some $\epsilon' \in \, ]0,\, \epsilon[$, we thus get $e_{[b(e),p]} \curvearrowright S$. $\square$

**Proof of prop. 2.9** *Transversal crossings* $c(e, S)$. Let $e \in \mathcal{L}_{\text{edges}}$ and $S \in \mathcal{L}_{\text{surfcs}}$. We define:

$$c(e, S) := \left\{ p \in r(e) \,\middle|\, \exists p' \in r(e) \setminus \{p\} \;/\; e_{[p,p']} \uparrow S \text{ or } e_{[p,p']} \downarrow S \right\}. \tag{2.10.1}$$

Let $p \in r(e) \setminus \{b(e)\}$. Applying lemma 2.10 to $e_{[p,b(e)]}$ and $S$, there exists $p' <_{(e)} p$ such that $e_{[p,p']} \diamond_p S$ with $\diamond_p \in \left\{ \curvearrowright, \uparrow, \downarrow \right\}$. Let $q \in r(e)$ such that $p' <_{(e)} q <_{(e)} p$ and assume that there exists $q' \in r(e) \setminus \{q\}$ such that $e_{[q,q']} \diamond_q S$ with $\diamond_q \in \left\{ \curvearrowright, \uparrow, \downarrow \right\}$. Since $q \in r(e_{[p,p']}) \setminus \{p, p'\}$, there exists $q'' \in \left( r(e_{[q,q']}) \setminus \{q, q'\} \right) \cap \left( r(e_{[p,p']}) \setminus \{p, p'\} \right)$. From prop. 2.8.6, $e_{[q,q'']} \diamond_q S$. On the other hand, we have either $e_{[p,p']} = e_{[q'',p']} \circ e_{[q,q'']} \circ e_{[p,q]}$ (if $q'' <_{(e)} q$) or $e_{[p,p']} = e_{[q,p']} \circ e_{[q'',q]} \circ e_{[p,q'']}$ (if $q'' >_{(e)} q$), so using twice prop. 2.8.6 (together with prop. 2.8.5), we get $e_{[q,q'']} \curvearrowright S$. Therefore, prop. 2.8.4 yields $\diamond_q = \curvearrowright$.

Thus, for any $p \in r(e) \setminus \{b(e)\}$, there exists $p' <_{(e)} p$ such that:

$$\forall q \in r(e), \;\; p' <_{(e)} q <_{(e)} p \;\;\Rightarrow\;\; q \notin c(e, S).$$

Similarly, for any $p \in r(e) \setminus \{f(e)\}$, there exists $p' >_{(e)} p$ such that:

$$\forall q \in r(e), \;\; p <_{(e)} q <_{(e)} p' \;\;\Rightarrow\;\; q \notin c(e, S).$$

As in the proof of prop. 2.5, this ensures that $c(e, S)$ is finite.

*Subedges with no transversal crossing are indifferent.* Let $p \neq p' \in r(e)$ such that $r(e') \cap c(e, S) = \varnothing$, with $e' := e_{[p,p']}$. Applying lemma 2.10 to $e'$ and recalling the definition of $c(e, S)$, there exists $p'' \in r(e') \setminus \{p\}$ such that $e_{[p,p'']} \curvearrowright S$. This allows to define $p_{\text{sup}} \in r(e') \setminus \{p\}$ by:

$$p_{\text{sup}} := \sup_{<_{(e')}} \left\{ p'' \in r(e') \setminus \{p\} \;\middle|\; e_{[p,p'']} \curvearrowright S \right\}.$$

Applying lemma 2.10 to $e_{[p_{\text{sup}},p]}$, there exists $p''$ such that:

$$p \leqslant_{(e')} p'' <_{(e')} p_{\text{sup}} \text{ and } e_{[p_{\text{sup}},p'']} \curvearrowright S.$$

On the other hand, by definition of $p_{\text{sup}}$, there exists $p'''$ such that:

$$p'' <_{(e')} p''' <_{(e')} p_{\text{sup}} \text{ and } e_{[p,p''']} \curvearrowright S.$$

Combining props. 2.8.5 and 2.8.6, we thus get $e_{[p,p_{\text{sup}}]} \curvearrowright S$.

Let $p'' \in r(e') \setminus \{p, p'\}$ such that $e_{[p,p'']} \curvearrowright S$. Applying lemma 2.10 to $e_{[p'',p']}$, there exists $p'''$ such that:

$$p'' <_{(e')} p''' \leqslant_{(e')} p' \text{ and } e_{[p'',p''']} \curvearrowright S,$$

hence $e_{[p,p''']} \curvearrowright S$ (using again prop. 2.8.6), and therefore $p'' <_{(e')} p_{\text{sup}}$. So we have $p_{\text{sup}} = p'$. Together with the previous point, this implies:

$$\forall p \neq p' \in r(e), \;\; r\left(e_{[p,p']}\right) \cap c(e, S) = \varnothing \;\Rightarrow\; e_{[p,p']} \curvearrowright S. \tag{2.10.2}$$

*Well-positioned subedges.* Let $p \neq p' \in r(e)$ such that $r\left(e_{[p,p']}\right) \cap c(e, S) \subset \{p\}$. Applying lemma 2.10 to $e_{[p,p']}$, there exists $p'' \in r\left(e_{[p,p']}\right) \setminus \{p\}$ such that $e_{[p,p'']} \diamond S$ with $\diamond \in \left\{ \curvearrowright, \uparrow, \downarrow \right\}$. If $p'' = p'$, then $e_{[p,p']} \diamond S$. Otherwise, we have $r\left(e_{[p'',p']}\right) \cap c(e, S) = \varnothing$, so from eq. (2.10.2), $e_{[p'',p']} \curvearrowright S$, and therefore $e_{[p,p']} \diamond S$ (using $e_{[p,p']} = e_{[p'',p']} \circ e_{[p,p'']}$ together with prop. 2.8.6). Thus, we have proved:

$$\forall p \neq p' \in r(e), \quad r\left(e_{[p,p']}\right) \cap c(e, S) \subset \{p\} \Rightarrow e_{[p,p']} \diamond S \text{ with } \diamond \in \left\{\curvearrowright, \uparrow, \downarrow\right\}. \tag{2.10.3}$$

*Decomposition of e adapted to S.* Since $c(e, S)$ is finite there exists $n \geqslant 1$, $\kappa \in \{0, 1\}$ and $p_0, p_1, \dots, p_n \in r(e)$ such that:

$$b(e) = p_0 <_{(e)} p_1 <_{(e)} \dots <_{(e)} p_{n-1} <_{(e)} p_n = f(e),$$

and:

$$c(e, S) = \{p_{2k+\kappa} \mid k \in \mathbb{N}, 2k + \kappa \leqslant n\}. \tag{2.10.4}$$

For $i \in \{1, \dots, n\}$, we define:

$$\epsilon_i = \begin{cases} +1 & \text{if } i + \kappa \text{ is odd} \\ -1 & \text{if } i + \kappa \text{ is even} \end{cases},$$

and $e_i = e^{\epsilon_i}_{[p_{i-1},p_i]}$. From eqs. (2.10.3) and (2.10.4), there exists $\diamond_i \in \left\{\curvearrowright, \uparrow, \downarrow\right\}$ such that $e_i \diamond_i S$. Moreover, we have $e = e_n^{\epsilon_n} \circ \dots \circ e_2^{\epsilon_2} \circ e_1^{\epsilon_1}$. □

As argued at the beginning of the present section, a satisfactory projective limit of phase spaces for conjugate holonomy and flux variables requires labels containing not only edges but also surfaces. The difficulty is that we cannot prevent the surfaces in a label to intersect wildly, for this would void the hopes for directedness: if a surface $S_1$ belongs to some label, and a surface $S_2$ belongs to some other label, there has to be, in a directed label set, a label containing both $S_1$ and $S_2$ at the same time. On the other hand, the set of variables described by a label should be closed under Poisson brackets: as already stressed, the algebra of observables associated to some label $\eta$ will be mounted by pullback into the algebra of any finer label $\eta'$ in a way that preserves the Poisson brackets [17, prop. 2.2], so the brackets between two variables should be correct all the way from the very first label in which these two variables appear. Since we know that fluxes associated to intersecting surfaces do not commute (at least as soon as the gauge group $G$ is non-Abelian), the additional variables arising as their Poisson brackets should therefore be included as soon as these surfaces are considered.

Thus, whenever the surfaces in a label $\eta$ intersect, the flux variables supported on their intersection are naturally among the observables selected by $\eta$. Accordingly, the momentum variables assigned to this label are not attached to the individual surfaces in $\eta$, but rather to so called 'faces', which enumerate all possible non trivial ways of positioning an edge with respect to these surfaces. It might be that different collections of surfaces actually result in the same set of momentum variables, which motivates the equivalence relation introduced in def. 2.12. Also, the ordering of the corresponding equivalence classes is prescribed by the comparison of their associated algebras of momentum variables (as will become clear in prop. 3.8).

**Proposition 2.11** Let $\widetilde{\lambda}$ be a *finite* set of surfaces. For $\diamond : \widetilde{\lambda} \to \left\{\curvearrowright, \uparrow, \downarrow\right\}$, $S \mapsto \diamond_S$, we define:

$$F_{\diamond}(\widetilde{\lambda}) := \left\{ e \in \mathcal{L}_{\text{edges}} \,\middle|\, \forall S \in \widetilde{\lambda},\ e \diamond_S S \right\}.$$

In particular (abusing notations by writing $\curvearrowright$ for the constant map $S \mapsto \curvearrowright$), we have the set of all edges that are indifferent to every surface in $\widetilde{\lambda}$:

$$F_{\curvearrowright}(\widetilde{\lambda}) = \left\{ e \in \mathcal{L}_{\text{edges}} \,\middle|\, \forall S \in \widetilde{\lambda},\ e \curvearrowright S \right\}.$$

The set of faces in $\widetilde{\lambda}$ is defined as:

$$\mathcal{F}(\widetilde{\lambda}) := \left\{ F_{\diamond}(\widetilde{\lambda}) \,\middle|\, \diamond\colon \widetilde{\lambda} \to \left\{ \curvearrowright, \uparrow, \downarrow \right\} \ \ / \ \ F_{\diamond}(\widetilde{\lambda}) \neq \varnothing \ \ \& \ \ \diamond \not\equiv \curvearrowright \right\},$$

(where $\diamond \not\equiv \curvearrowright$ stands for $\{S \mid \diamond_S \neq \curvearrowright\} \neq \varnothing$). In addition, we define:

$$F_{\text{any}}(\widetilde{\lambda}) := \bigcup_{F \in \mathcal{F}(\widetilde{\lambda})} F,$$

and, for $F$, $F' \subset \mathcal{L}_{\text{edges}}$:

$$F' \circ F := \{ e_2 \circ e_1 \mid e_1 \in F,\ e_2 \in F',\ \text{ and } e_1, e_2 \text{ are composable} \}.$$

We have the following properties:

1. the elements of $\mathcal{F}(\widetilde{\lambda})$ are disjoints;
2. for any $F \in \mathcal{F}(\widetilde{\lambda}) \cup \left\{ F_{\curvearrowright}(\widetilde{\lambda}) \right\}$, $\ F_{\curvearrowright}(\widetilde{\lambda}) \circ F = F$;
3. $F_{\curvearrowright}(\widetilde{\lambda}) = \left\{ e \in \mathcal{L}_{\text{edges}} \,\middle|\, \forall p \neq p' \in r(e),\ \ e_{[p,p']} \notin F_{\text{any}}(\widetilde{\lambda}) \right\}$;
4. for any $e \in \mathcal{L}_{\text{edges}}$ there exist $e_1, \ldots, e_n \in \mathcal{L}_{\text{edges}}$ and $\epsilon_1, \ldots, \epsilon_n \in \{\pm 1\}$ such that:
$$e = e_n^{\epsilon_n} \circ \ldots \circ e_1^{\epsilon_1} \ \text{ and } \ \forall i \in \{1, \ldots, n\},\ e_i \in F_{\text{any}}(\widetilde{\lambda}) \cup F_{\curvearrowright}(\widetilde{\lambda}).$$

**Proof** Assertions 2.11.1 and 2.11.2 follow from prop. 2.8.4 and 2.8.6 respectively.

*Assertion 2.11.3.* From Prop. 2.8, we have:

$$F_{\curvearrowright}(\widetilde{\lambda}) \subset \left\{ e \in \mathcal{L}_{\text{edges}} \,\middle|\, \forall p \neq p' \in r(e),\ \ e_{[p,p']} \notin F_{\text{any}}(\widetilde{\lambda}) \right\}.$$

We now want to prove the reverse inclusion.

Let $e \in \mathcal{L}_{\text{edges}}$ such that for any $p \neq p' \in r(e)$, $e_{[p,p']} \notin F_{\text{any}}(\widetilde{\lambda})$. Let $S_o \in \widetilde{\lambda}$ and $p \neq p_o \in r(e)$ such that $e_{[p,p_o]} \diamond_o S_o$ with $\diamond_o \in \left\{ \curvearrowright, \uparrow, \downarrow \right\}$. Choose an ordering of the finitely many remaining surfaces $\widetilde{\lambda} \setminus S = \{S_1, \ldots, S_m\}$, and define $p_i$ and $\diamond_i$ for $i \in \{1, \ldots, n\}$ such that:

$$p_i \in r\left(e_{[p,p_{i-1}]}\right) \setminus \{p\} \text{ and } e_{[p,p_i]} \diamond_i S_i,$$

by applying inductively lemma 2.10 to $e_{[p,p_{i-1}]}$ and $S_i$. From prop. 2.8.6, $e_{[p,p_m]} \in F_{\diamond}(\widetilde{\lambda})$ where $\diamond\colon \widetilde{\lambda} \to \left\{ \curvearrowright, \uparrow, \downarrow \right\}$ is defined by $\forall i \in \{0, \ldots, m\}$, $\diamond_{S_i} := \diamond_i$. Since $e_{[p,p_m]} \notin F_{\text{any}}(\widetilde{\lambda})$, $\diamond \equiv \curvearrowright$, therefore $\diamond_o = \curvearrowright$.

Hence, $c(e, S_o) = \varnothing$ (where $c(e, S_o)$ has been defined in eq. (2.10.1)). So, using eq. (2.10.2) with $p = b(e)$ and $p' = f(e)$, $e \looparrowright S_o$. As this holds for any $S_o \in \widetilde{\lambda}$, $e \in F_{\looparrowright}(\widetilde{\lambda})$.

*Assertion 2.11.4.* Let $e \in \mathcal{L}_{\text{edges}}$. We define:

$$c(e, \widetilde{\lambda}) := \bigcup_{S \in \widetilde{\lambda}} c(e, S).$$

$c(e, \widetilde{\lambda})$ is finite, for $\widetilde{\lambda}$ is finite and each $c(e, S)$ is finite. Moreover, eq. (2.10.3) becomes:

$$\forall p \neq p' \in r(e), \quad r\left(e_{[p,p']}\right) \cap c(e, \widetilde{\lambda}) \subset \{p\} \;\Rightarrow\; e_{[p,p']} \in F_{\text{any}}(\widetilde{\lambda}) \cup F_{\looparrowright}(\widetilde{\lambda}).$$

Thus we can form a decomposition of $e$ adapted to $\widetilde{\lambda}$ exactly like in the last step of the proof of prop. 2.9. □

**Definition 2.12** We define on the set of finite subsets of $\mathcal{L}_{\text{surfcs}}$ an equivalence relation by:

$$\widetilde{\lambda} \sim \widetilde{\lambda}' \Leftrightarrow \mathcal{F}(\widetilde{\lambda}) = \mathcal{F}(\widetilde{\lambda}').$$

Its set of equivalence classes will be denoted by $\mathcal{L}_{\text{profls}}$. An element $\lambda \in \mathcal{L}_{\text{profls}}$ is called a profile, and we can define its set of faces $\mathcal{F}(\lambda)$ and the corresponding set of indifferent edges $F_{\looparrowright}(\lambda)$, since these are the same for any representative of $\lambda$ (thanks to prop. 2.11.3).

**Proposition 2.13** We equip $\mathcal{L}_{\text{profls}}$ with the binary relation:

$$\forall \lambda, \lambda' \in \mathcal{L}_{\text{profls}},$$

$$\lambda \preccurlyeq \lambda' \Leftrightarrow \left(\forall F \in \mathcal{F}(\lambda), \exists F_1, \ldots, F_m \in \mathcal{F}(\lambda') \,\middle/\, F = F_{\looparrowright}(\lambda) \circ \bigcup_{i=1}^{m} F_i\right).$$

For any two finite sets of surfaces $\widetilde{\lambda}, \widetilde{\lambda}'$, we have:

$$\left[\widetilde{\lambda}\right]_{\text{profl}} \preccurlyeq \left[\widetilde{\lambda} \cup \widetilde{\lambda}'\right]_{\text{profl}},$$

where $[\,\cdot\,]_{\text{profl}}$ denotes the equivalence class in $\mathcal{L}_{\text{profls}}$.

In particular, $\mathcal{L}_{\text{profls}}, \preccurlyeq$ is a directed preordered set.

**Proof** Let $\widetilde{\lambda}, \widetilde{\lambda}'$ be two finite sets of surfaces and $F \in \mathcal{F}(\widetilde{\lambda})$. There exists $\diamond \colon \widetilde{\lambda} \to \left\{\looparrowright, \uparrow, \downarrow\right\}$, with $\diamond \not\equiv \looparrowright$, such that $F = F_{\diamond}(\widetilde{\lambda})$. We define:

$$\mathcal{F}(\widetilde{\lambda}, \widetilde{\lambda}', \diamond) := \left\{F_{\diamond'}(\widetilde{\lambda} \cup \widetilde{\lambda}') \,\middle|\, F_{\diamond'}(\widetilde{\lambda} \cup \widetilde{\lambda}') \in \mathcal{F}(\widetilde{\lambda} \cup \widetilde{\lambda}') \quad \& \quad \diamond'|_{\widetilde{\lambda}} = \diamond\right\}.$$

Let $e \in F$ ($F \neq \varnothing$ by definition of $\mathcal{F}(\widetilde{\lambda})$). By applying inductively lemma 2.10 to the surfaces in $\widetilde{\lambda}'$ and using prop. 2.8.6 (as in the proof of prop. 2.11.3), there exists $p \in r(e) \setminus \{b(e)\}$ and an extension $\diamond' \colon \widetilde{\lambda} \cup \widetilde{\lambda}' \to \left\{\looparrowright, \uparrow, \downarrow\right\}$ of $\diamond$ such that $e_{[b(e),p]} \in F_{\diamond'}(\widetilde{\lambda} \cup \widetilde{\lambda}')$. Let $p' \in r\left(e_{[b(e),p]}\right) \setminus \{b(e), p\}$. From prop. 2.8.6, we have $e_{[b(e),p']} \in F_{\diamond'}(\widetilde{\lambda} \cup \widetilde{\lambda}')$ and $e_{[p',f(e)]} \in F_{\looparrowright}(\widetilde{\lambda})$. Therefore, $e \in F_{\looparrowright}(\widetilde{\lambda}) \circ F_{\diamond'}(\widetilde{\lambda} \cup \widetilde{\lambda}')$.

In particular, $F_{\diamond'}(\widetilde{\lambda} \cup \widetilde{\lambda}') \neq \varnothing$ and, since $\diamond \not\equiv \looparrowright$, we also have $\diamond' \not\equiv \looparrowright$. Thus, $F_{\diamond'}(\widetilde{\lambda} \cup \widetilde{\lambda}') \in \mathcal{F}(\widetilde{\lambda}, \widetilde{\lambda}', \diamond)$. So $\mathcal{F}(\widetilde{\lambda}, \widetilde{\lambda}', \diamond) \neq \varnothing$ and there exists $F_1, \ldots, F_m \in \mathcal{F}(\widetilde{\lambda} \cup \widetilde{\lambda}')$ such that:

$$\mathcal{F}(\widetilde{\lambda}, \widetilde{\lambda}', \diamond) = \{F_1, \ldots, F_m\}.$$

and we just proved that $F \subset F_{\gimel}(\widetilde{\lambda}) \circ \bigcup_{i=1}^{m} F_i$.

Now, let $e_1$, $e_2$ be composable edges such that $e_1 \in F_i$ for some $i \in \{1, \dots, m\}$ and $e_2 \in F_{\gimel}(\widetilde{\lambda})$. By definition of $\mathcal{F}(\widetilde{\lambda}, \widetilde{\lambda}', \diamond)$, $e_1 \in F$, hence, by 2.11.2, $e_2 \circ e_1 \in F$. Therefore, $F_{\gimel}(\widetilde{\lambda}) \circ \bigcup_{i=1}^{m} F_i \subset F$. So, we have $\big[\widetilde{\lambda}\,\big]_{\text{profl}} \preccurlyeq \big[\widetilde{\lambda} \cup \widetilde{\lambda}'\,\big]_{\text{profl}}$.

To prove that $\mathcal{L}_{\text{profls}}, \preccurlyeq$ is a directed preordered set, only the transitivity of $\preccurlyeq$ remains to be checked. Let $\lambda, \lambda', \lambda'' \in \mathcal{L}_{\text{profls}}$ with $\lambda \preccurlyeq \lambda'$ and $\lambda' \preccurlyeq \lambda''$. Using the definition of $\preccurlyeq$ on $\mathcal{L}_{\text{profls}}$ together with prop. 2.11.3, we have:

$$F_{\gimel}(\lambda'') \subset F_{\gimel}(\lambda') \subset F_{\gimel}(\lambda).$$

Then, for any $F \in \mathcal{F}(\lambda'')$, we can use prop. 2.11.2 to write:

$$F_{\gimel}(\lambda) \circ F_{\gimel}(\lambda'') \circ F = F_{\gimel}(\lambda) \circ F = F_{\gimel}(\lambda) \circ F_{\gimel}(\lambda) \circ F,$$

so we get:

$$F_{\gimel}(\lambda) \circ F_{\gimel}(\lambda') \circ F = F_{\gimel}(\lambda) \circ F, \tag{2.13.1}$$

and therefore $\lambda \preccurlyeq \lambda''$. □

Finally, we are ready to describe what our labels should be, keeping in mind that each label is meant to be associated with a small, finite dimensional phase space, on which the observables it selects can be represented. As underlined many times, this phase space should be big enough for their Poisson algebra to be correctly reproduced. Yet it should not be too big either, otherwise the projection between the phase spaces corresponding to two labels $\eta \preccurlyeq \eta'$ would not be uniquely characterized by the sole prescription of how its pullback should mount the observables from $\eta$ to $\eta'$.

These considerations reveal that edges and faces should comes in conjugate pairs. In particular, if the label contains intersecting surfaces, it should also contain edges that will probe the intersection from every side, so that the additional momentum variables promoted above are supplied with suitable conjugate configuration variables (fig. 2.3).

**Definition 2.14** We define the label set $\mathcal{L}$ by:

$$\mathcal{L} := \Big\{(\gamma, \lambda) \in \mathcal{L}_{\text{graphs}} \times \mathcal{L}_{\text{profls}} \;\Big|\; \exists \chi : \gamma \to \mathcal{F}(\lambda) \text{ bijective } \Big/ \forall e \in \gamma,\, e \in \chi(e)\Big\}.$$

For $\eta = (\gamma, \lambda)$ we define its underlying graph $\gamma(\eta) := \gamma$ and profile $\lambda(\eta) := \lambda$, its set of faces $\mathcal{F}(\eta) := \mathcal{F}(\lambda)$ and its set of indifferent edges $F_{\gimel}(\eta) := F_{\gimel}(\lambda)$, as well as the unique bijective map $\chi_\eta : \gamma(\eta) \to \mathcal{F}(\eta)$ such that $\forall e \in \gamma(\eta)$, $e \in \chi_\eta(e)$ (uniqueness follows from the fact that the faces in $\mathcal{F}(\eta)$ are disjoints, see prop. 2.11.1).

We equip $\mathcal{L}$ with the product preorder, defined by:

$$\forall \eta, \eta' \in \mathcal{L},\; \eta \preccurlyeq \eta' \;\Leftrightarrow\; \big(\gamma(\eta) \preccurlyeq \gamma(\eta') \quad \& \quad \lambda(\eta) \preccurlyeq \lambda(\eta')\big).$$

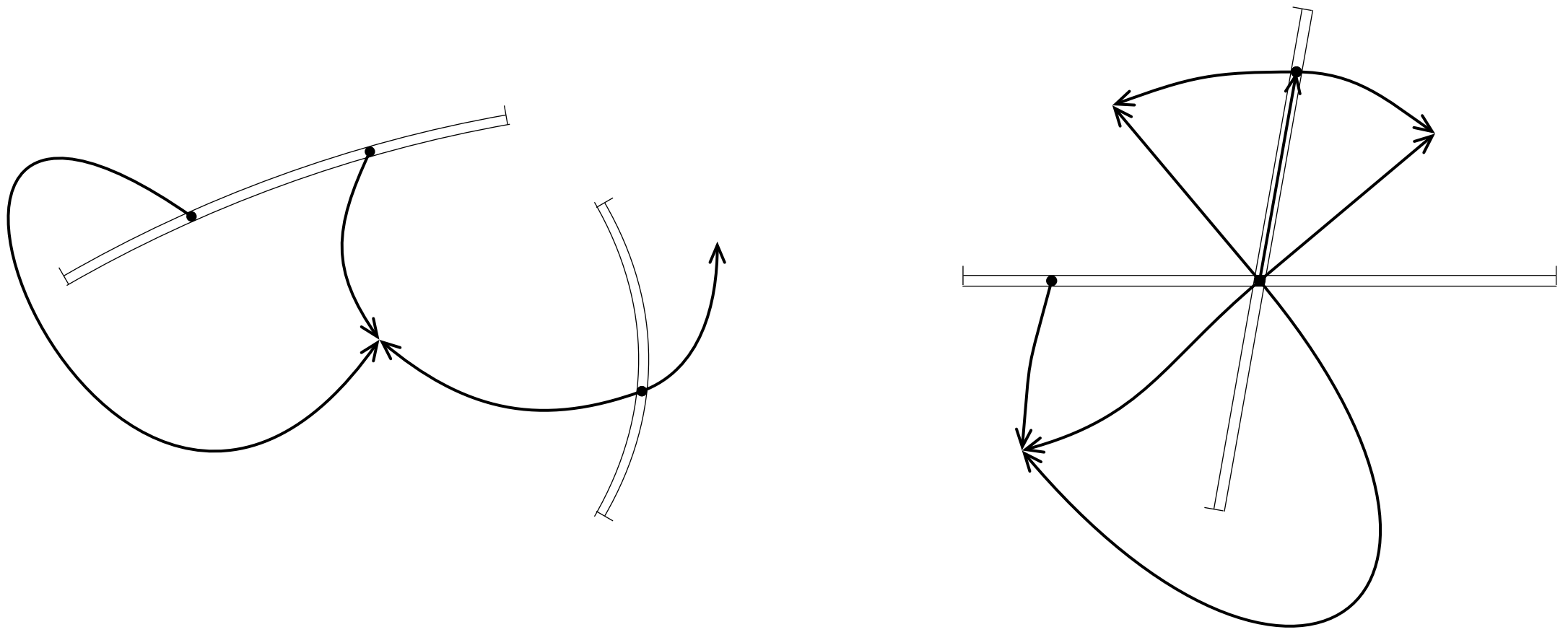

Figure 2.3 – Examples of valid labels

## 2.2 Directedness

It is of critical importance for the intended construction of a projective state space that the label set $\mathcal{L}$ should be directed (eg. the pivotal 'three-spaces consistency' condition gets truly useful in combination with the directedness of the label set). Since both $\mathcal{L}_{\text{graphs}}$ and $\mathcal{L}_{\text{profls}}$ are directed on their own, so is $\mathcal{L}_{\text{graphs}} \times \mathcal{L}_{\text{profls}}$, thus it is sufficient to show that $\mathcal{L}$ is a cofinal part of $\mathcal{L}_{\text{graphs}} \times \mathcal{L}_{\text{profls}}$. In other words, given some arbitrary graph $\gamma$ and profile $\lambda$, we want to construct a finer graph $\gamma'$ and a finer profile $\lambda'$ that are adapted to each other in the sense of def. 2.14.

For this, we will proceed in successive steps. First we will subdivide the edges of $\gamma$ to adapt them to $\lambda$ in the sense of prop. 2.9. Next we will add a bunch of small surfaces to ensure there is never more than one edge belonging to a given face, and we will add a few small edges to populate the faces that does not yet contain one edge. Finally we will add a few more small surfaces so that every edge has its fellow face. Note that the order of these steps is important, for we have to ensure that what has been achieved at a given point will be preserved by the subsequent steps. Also, we should take care that the whole procedure only requires a finite sequence of operations: graphs have been defined as *finite* sets of edges, while profiles arise from *finite* sets of surfaces, thus adding infinitely many edges or surfaces, or subdividing an edge into infinitely many parts, would not lead to a valid label.

**Definition 2.15** For any $\gamma \in \mathcal{L}_{\text{graphs}}$ and any $\lambda \in \mathcal{L}_{\text{profls}}$, we define:

1. $M^{(1)}_{(\gamma,\lambda)} := \left\{\chi : \gamma \to \mathcal{F}(\lambda) \cup \left\{F_{\daleth}(\lambda)\right\} \,\middle|\, \forall e \in \gamma,\, e \in \chi(e)\right\}$ ;
2. $M^{(2)}_{(\gamma,\lambda)} := \left\{\chi \in M^{(1)}_{(\gamma,\lambda)} \,\middle|\, \forall F \in \mathcal{F}(\lambda),\, \forall e, e' \in \chi^{-1}\langle F\rangle,\, e = e'\right\}$ ;
3. $M^{(3)}_{(\gamma,\lambda)} := \left\{\chi \in M^{(2)}_{(\gamma,\lambda)} \,\middle|\, \forall F \in \mathcal{F}(\lambda),\, \chi^{-1}\langle F\rangle \neq \varnothing\right\}$ ;
4. $M^{(4)}_{(\gamma,\lambda)} := \left\{\chi \in M^{(3)}_{(\gamma,\lambda)} \,\middle|\, \chi^{-1}\left\langle F_{\daleth}(\lambda)\right\rangle = \varnothing\right\} = \left\{\chi : \gamma \to \mathcal{F}(\lambda) \,\middle|\, \chi \text{ bijective } \;\&\; \forall e \in \gamma,\, e \in \chi(e)\right\}$.

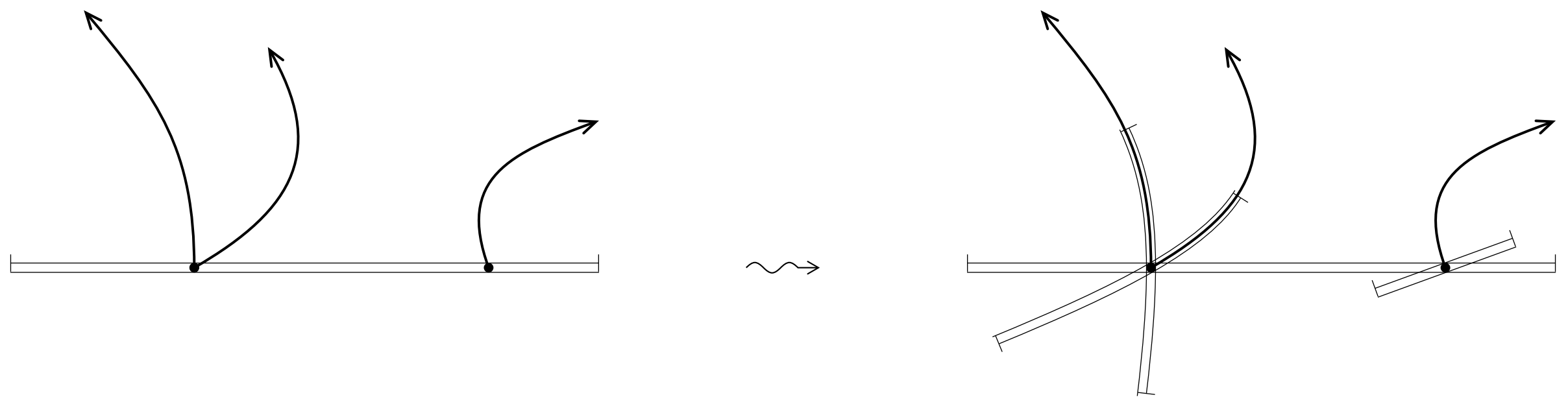

Figure 2.4 – Adding surfaces to separate the edges into different faces

**Theorem 2.16** $\mathcal{L}, \preccurlyeq$ is a directed preordered set.

**Lemma 2.17** Let $\gamma \in \mathcal{L}_{\text{graphs}}$ and $\lambda \in \mathcal{L}_{\text{profls}}$. Then, there exists $\gamma' \in \mathcal{L}_{\text{graphs}}$, such that $\gamma \preccurlyeq \gamma'$ and $M^{(1)}_{(\gamma',\lambda)} \neq \varnothing$.

**Proof** This follows from prop. 2.11.4 and the definition of $\preccurlyeq$ on $\mathcal{L}_{\text{graphs}}$ (def. 2.4). □

The next step is to deal with the faces that contain more than one edge of the graph. The key idea here is to add some small surfaces with respect to which these edges have distinct positionings: thus, they will not belong to the same face any more. If they have different starting points, we can simply add, for each of them, a small surface going through its starting point, with respect to witch it is, say, above (as we do for the edge on the right in fig. 2.4). If many edges start from the same point of the face, we will add, for each $e$ among these, a small surface arranged so that $e$ is indifferent to it, while all other edges starting from that point are either above or below it: this is another way of ensuring that two different edges will have a distinct positioning with respect to at least one of the added surfaces (see the two edges on the left in fig. 2.4).

Given an edge $e$ and a bunch of other edges $\{e'\}$ starting from the same point, we therefore want to construct a surface that contains an initial subedge of $e$ and intersect each $e'$ transversally at the common starting point $b(e) = b(e')$: in two dimensions, where "surfaces" are one-dimensional, we can cut such a surface out of the very analytic curve that provides $e$; in higher dimension, the family of surfaces that contains $e$ is parametrized by continuous parameters, so we just need to pick out one that passes in between the finitely many edges starting from $b(e)$. However, if the analytic extension of $e$ beyond $b(e)$ is among the edges $\{e'\}$, it will automatically be in the analytic plane of any surface containing an initial subedge of $e$, and thus indifferent to that surface, as illustrated on fig. 2.5. Note that it may seems at first a very unlucky special case, that precisely the analytic extension of $e$ also belongs to the graph, but we should keep in mind that this very situation is produced in great numbers when we subdivides the edges of $\gamma$ to adapt them to the given $\lambda$ (fig. 2.2). To deal with this case we need to also include an additional small surface with respect to which $e$ is of the "above" type, because its analytic continuation will then be of the "below" type: this will ensure that these two edges do not belong to the same face at the end. To avoid laborious case distinctions in the proof, we add surfaces quite liberally, and, for any initial face $F$ containing more than one edge, and any edge $e$ belonging to $F$, we will systematically add two small surfaces, one along $e$ and the other transverse to $e$. Proceeding this way we probably end up with much more faces that would have been strictly necessary to separate the edges into distinct faces: we generate a lot of new faces that do not contain any edge at all. On the other hand, we will have to deal with such faces in a latter step anyway, so it does not cost us more to add more of those in the present step.

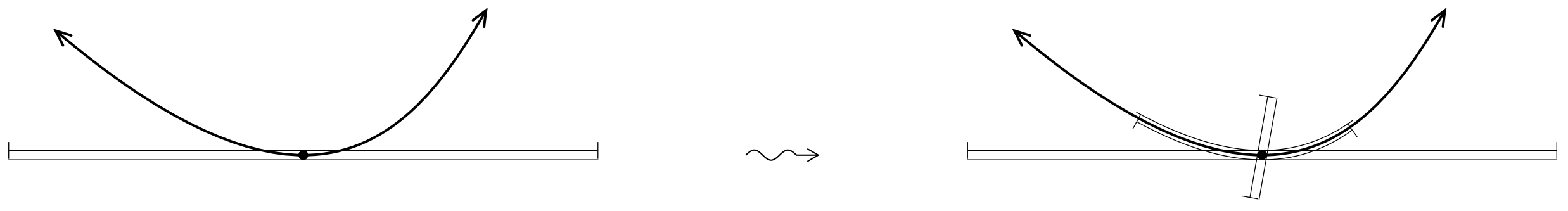

Figure 2.5 – Only a transversal surface can separate an edge from its analytic continuation

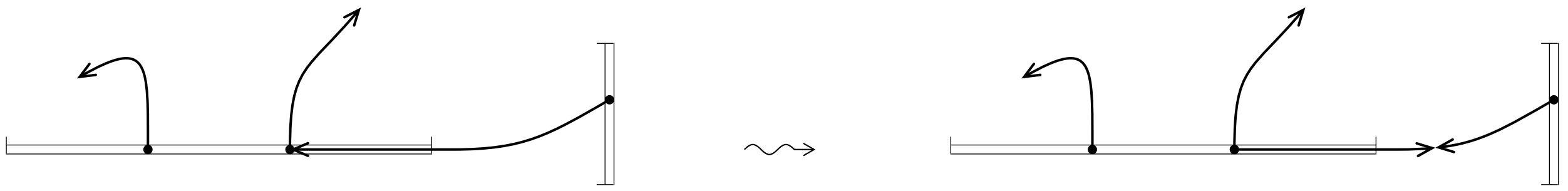

Figure 2.6 – Dealing with an edge that ends precisely where surfaces will have to be added

It is by contrast crucial that we preserve what has been achieved in the *previous* step, namely that all edges of the graph are adapted to the new profile. When adding a surface for an edge $e$, we can make it as small as we want around $b(e)$ without prejudice to the requirements above. In particular, we can ensure that it does not intersect any edge of the graph that does not go through $b(e)$. Moreover, for any edge $e'$ that start at $b(e)$, the analyticity of edges and surfaces will ensure that, provided the surface is chosen small enough around $b(e)$, $e'$ will either be above or below it or indifferent to it. The only edges $e'$ that might be problematic here are therefore those that contains $b(e)$ but not as starting point. Since $e$ and $e'$ are edges of a graph, $b(e)$ should then be $f(e')$, and we deal preventively with this potential source of difficulties by subdividing $e'$ and reorienting the second part so that it now *starts* from $b(e)$ (see fig. 2.6). Note that, by doing so, we admittedly increase the number of edges, but not the number of those for which we will then need to add small surfaces, for the first part of $e'$ takes the place of $e'$ in $\chi(e')$, while its second part is in $F_{\urcorner}(\lambda)$. Thus this does not cause infinite recursion, and the total number of surfaces added during the present step is finite, as it should.

**Lemma 2.18** Let $\gamma \in \mathcal{L}_{\text{graphs}}$ and $\lambda \in \mathcal{L}_{\text{profls}}$ such that $M^{(1)}_{(\gamma,\lambda)} \neq \varnothing$. Then, there exists $\gamma' \in \mathcal{L}_{\text{graphs}}$ and $\lambda' \in \mathcal{L}_{\text{profls}}$, such that $\gamma \preccurlyeq \gamma'$, $\lambda \preccurlyeq \lambda'$ and $M^{(2)}_{(\gamma',\lambda')} \neq \varnothing$.

**Proof** *Construction of $\gamma'$.* Let $\chi \in M^{(1)}_{(\gamma,\lambda)}$. We define:

$$\gamma_{(1,\chi)} := \left\{ e \in \gamma \;\middle|\; \exists\, e' \neq e \,/\, \chi(e) = \chi(e') \in \mathfrak{F}(\lambda) \right\},$$

and:

$$B\left(\gamma_{(1,\chi)}\right) := \left\{ b(e) \mid e \in \gamma_{(1,\chi)} \right\}.$$

Let $e \in \gamma$. Since $\gamma$ is a graph, $r(e) \cap B\left(\gamma_{(1,\chi)}\right) \subset \{b(e), f(e)\}$. If $f(e) \notin r(e) \cap B\left(\gamma_{(1,\chi)}\right)$, we define $\gamma'_e := \{e\}$. Otherwise, we choose some point $p \in r(e) \setminus \{b(e), f(e)\}$ and we define $\gamma'_e := \left\{e_{[b(e),p]},\, e_{[f(e),p]}\right\}$.

$\gamma' := \bigcup_{e\in\gamma}\gamma'_e$ is a graph such that $\gamma \preccurlyeq \gamma'$ and, from prop. 2.8, there exists $\chi' \in M^{(1)}_{(\gamma',\lambda)}$, satisfying $B\left(\gamma'_{(1,\chi')}\right) = B\left(\gamma_{(1,\chi)}\right)$. Moreover, we now have for any $e \in \gamma'$, $r(e) \cap B\left(\gamma'_{(1,\chi')}\right) \subset \{b(e)\}$.

*Construction of $\lambda'$.* Let $e \in \gamma'_{(1,\chi')}$ and define:

$$\gamma'_{(e)} := \{e' \in \gamma' \mid b(e) \in r(e')\} = \{e' \in \gamma' \mid b(e) = b(e')\}.$$

We choose a representative $\check{e} : U \to V$ of $e$ and a real $\epsilon > 0$ such that $B^{(d)}_\epsilon \subset U$ (where $B^{(d)}_\epsilon$ is the closed ball of radius $\epsilon$ and center 0 in $\mathbb{R}^d$). We define:

$$\begin{array}{rcl} \check{S}_{\check{e},\epsilon} : U_\epsilon & \to & V \\ x & \mapsto & \check{e}(\epsilon\, x) \end{array} \quad \text{with} \quad U_\epsilon := \left\{x \in \mathbb{R}^d \;\middle|\; \epsilon\, x \in U\right\}.$$

Since $\{0\} \times B^{(d-1)} \subset B^{(d)} \subset U_\epsilon$, $\check{S}_{\check{e},\epsilon}$ is an analytic, encharted surface in $\Sigma$. We denote by $S_{\check{e},\epsilon}$ the corresponding surface (def. 2.7). We also define, for any $\theta \in S^{(d-2)}$ (with $S^{(d-2)}$ the unit sphere in $\mathbb{R}^{d-1}$):

$$\begin{array}{rcl} R_\theta : \mathbb{R} \times \mathbb{R}^{d-1} & \to & \mathbb{R} \times \mathbb{R}^{d-1} \\ (t, y) & \mapsto & (-\theta.y,\; y + (t - \theta.y)\,\theta) \end{array}.$$

$R_\theta$ is an analytic diffeomorphism $\mathbb{R}^d \to \mathbb{R}^d$ and:

$$R_\theta\left\langle\{0\} \times B^{(d-1)}\right\rangle \subset R_\theta\left\langle B^{(d)}\right\rangle = B^{(d)} \subset U_\epsilon.$$

Hence, $\check{S}_{\check{e},\epsilon,\theta}$ defined by $\check{S}_{\check{e},\epsilon,\theta} := \check{S}_{\check{e},\epsilon} \circ R_\theta : R_\theta^{-1}\langle U_\epsilon\rangle \to V$ is an analytic, encharted surface in $\Sigma$. We denote by $S_{\check{e},\epsilon,\theta}$ the corresponding surface.

Let $e' \in \gamma'_{(e)} \setminus \{e\}$. We define:

$$K_{(e,e')} = \left\{\theta \in S^{(d-2)} \;\middle|\; \exists p \in r(e') \setminus \{b(e')\} \;/\; e'_{[b(e'),p]} \curvearrowright S_{\check{e},\epsilon,\theta}\right\}.$$

Since $b(e') = b(e) = \check{e}(0) \in r\left(S_{\check{e},\epsilon,\theta}\right)$ for any $\theta \in S^{(d-2)}$, we have from prop. 2.8 (together with the definition of the equivalence relation in def. 2.7):

$$\forall\theta \in K_{(e,e')},\, \exists p' \in r(e') \setminus \{b(e')\} \;/\; r\left(e'_{[b(e'),p']}\right) \subset \check{e}\left\langle U \cap R_\theta\left\langle\{0\} \times \mathbb{R}^{d-1}\right\rangle\right\rangle,$$

where we have used:

$$\check{S}_{\check{e},\epsilon,\theta}\left\langle R_\theta^{-1}\langle U_\epsilon\rangle \cap \left(\{0\} \times \mathbb{R}^{d-1}\right)\right\rangle = \check{e}\left\langle U \cap R_\theta\left\langle\{0\} \times \mathbb{R}^{d-1}\right\rangle\right\rangle.$$

Suppose now that there exist $d-1$ vectors $\theta_1, \ldots, \theta_{d-1} \in S^{(d-2)}$, linearly independent in $\mathbb{R}^{d-1}$, such that $\forall i \leqslant d-1$, $\theta_i \in K_{(e,e')}$. Then, there exists $p'' \in r(e') \setminus \{b(e')\}$ such that:

$$r\left(e'_{[b(e'),p'']}\right) \subset \check{e}\left\langle U \cap \bigcap_{i=1}^{d-1} R_{\theta_i}\left\langle\{0\} \times \mathbb{R}^{d-1}\right\rangle\right\rangle = \check{e}\left\langle U \cap \left(\mathbb{R} \times \{0\}\right)\right\rangle.$$

Since $b(e') = \check{e}(0)$ and $r(e) \cap r(e') \subset \{b(e), f(e)\}$ (for $e \neq e'$ and both belong to the graph $\gamma'$), we have, by the intermediate value theorem, $r\left(e'_{[b(e'),p'']}\right) = \check{e}\left\langle [-\alpha, 0] \times \{0\} \right\rangle$ for some $\alpha > 0$. Thus, $e'_{[b(e'),p'']} \downarrow S_{\check{e},\epsilon}$.

Accordingly, we define:

$$\gamma'_{(e,0)} := \{e\} \cup \left\{ e' \in \gamma'_{(e)} \;\middle|\; \exists\, p'' \in r(e') \setminus \{b(e')\} \;/\; e'_{[b(e'),p'']} \downarrow S_{\check{e},\epsilon} \right\},$$

and $\gamma'_{(e,1)} := \gamma'_{(e)} \setminus \gamma'_{(e,0)}$. Then, for any $e' \in \gamma'_{(e,1)}$, $K_{(e,e')}$ has measure zero in $S^{(d-2)}$ (for example with respect to the standard measure on $S^{(d-2)}$ if $d \geqslant 3$, and with respect to the counting measure if $d = 2$). Hence, there exists $\theta_1 \in S^{(d-2)}$ such that $\forall e' \in \gamma'_{(e,1)}$, $\theta_1 \notin K_{(e,e')}$.

Now, from lemma 2.10 and prop. 2.8.6, there exist, for any $e' \in \gamma'_{(e)}$, $p_{e'} \in r(e') \setminus \{b(e')\}$ and $\diamond_{e',e,0}, \diamond_{e',e,1} \in \left\{ \curvearrowright, \uparrow, \downarrow \right\}$ such that:

$$e'_{[b(e'),p_{e'}]} \diamond_{e',e,0} S_{\check{e},\epsilon} \;\text{ and }\; e'_{[b(e'),p_{e'}]} \diamond_{e',e,1} S_{\check{e},\epsilon,\theta_1}.$$

By construction, we have:

$$\forall e' \in \gamma'_{(e,0)} \setminus \{e\}, \; \diamond_{e',e,0} = \downarrow \quad\text{and}\quad \forall e' \in \gamma'_{(e,1)}, \; \diamond_{e',e,1} \neq \curvearrowright.$$

On the other hand, we can check from the definition of $S_{\check{e},\epsilon}$ and $S_{\check{e},\epsilon,\theta_1}$ that:

$$\diamond_{e,e,0} = \uparrow \quad\text{and}\quad \diamond_{e,e,1} = \curvearrowright.$$

Thus, we get $\forall e' \in \gamma'_{(e)} \setminus \{e\}$, $\left(\diamond_{e',e,0}, \diamond_{e',e,1}\right) \neq \left(\diamond_{e,e,0}, \diamond_{e,e,1}\right)$.

Next, using def. 2.7 and prop. 2.8, there exists $p'_{e'} \in r\left(e'_{[b(e'),p_{e'}]}\right) \setminus \{b(e'), p_{e'}\}$ such that:

$$r\left(e'_{[b(e'),p'_{e'}]}\right) \setminus \{b(e')\} \subset \check{S}_{\check{e},\epsilon}\left\langle U_\epsilon \cap D_{\diamond_{e',e,0}} \cap R_{\theta_1}\left\langle D_{\diamond_{e',e,1}} \right\rangle \right\rangle, \tag{2.18.1}$$

where $D_{\curvearrowright} := \{0\} \times \mathbb{R}^{d-1}$, $D_{\uparrow} := (\mathbb{R}_+ \setminus \{0\}) \times \mathbb{R}^{d-1}$ and $D_{\downarrow} := (\mathbb{R}_- \setminus \{0\}) \times \mathbb{R}^{d-1}$. Since:

$$\bigcup_{e' \in \gamma' \setminus \gamma'_{(e)}} r(e') \cup \bigcup_{e' \in \gamma'_{(e)}} r\left(e'_{[p'_{e'}, f(e')]}\right)$$

is a compact set of $\Sigma$ that does not contain $b(e)$, there exists an open neighborhood $W$ of $b(e)$ in $V$ such that:

$$\forall e' \in \gamma' \setminus \gamma'_{(e)}, \; r(e') \cap W = \varnothing \;\text{ and }\; \forall e' \in \gamma'_{(e)}, \; r\left(e'_{[p'_{e'}, f(e')]}\right) \cap W = \varnothing.$$

$\check{e}^{-1}\langle W \rangle$ is an open neighborhood of 0 in $\mathbb{R}^d$, hence there exists $\epsilon' \in ]0, \epsilon]$ such that $B^{(d)}_{\epsilon'} \subset \check{e}^{-1}\langle W \rangle$. We define:

$$S_{e,0} := S_{\check{e},\epsilon'} \;\text{ and }\; S_{e,1} := S_{\check{e},\epsilon',\theta_1}.$$

For $k \in \{0, 1\}$, we have $r\left(S_{e,k}\right) \subset W$, therefore:

$$\forall e' \in \gamma' \setminus \gamma'_{(e)}, \; e' \curvearrowright S_{e,k} \;\text{ and }\; \forall e' \in \gamma'_{(e)}, \; e'_{[p'_{e'}, f(e')]} \curvearrowright S_{e,k}.$$

In addition, we get from eq. (2.18.1):

$$\forall e' \in \gamma'_{(e)}, \; e'_{[b(e'),p'_{e'}]} \diamond_{e',e,k} S_{\check{e},k},$$

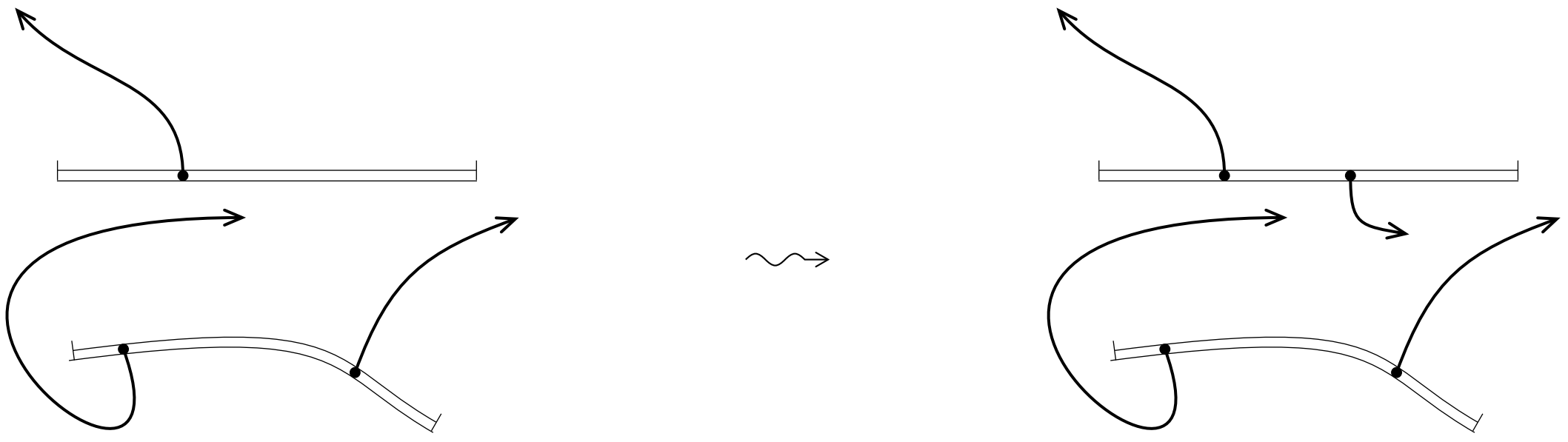

Figure 2.7 – Adding a small edge to populate a face that was empty

thus, using prop. 2.8.6:

$$\forall e' \in \gamma'_{(e)},\ e' \diamond_{e',e,k} S_{\check{e},k}\,.$$

To summarize, we have proven that, for any $e \in \gamma'_{(1,\chi')}$ there exist a finite set of surfaces $\widetilde{\lambda}'_e := \{S_{e,0},\ S_{e,1}\}$ and a map $\chi'_e : \gamma' \to \mathfrak{F}\big(\widetilde{\lambda}'_e\big) \cup F_{\neg}\big(\widetilde{\lambda}'_e\big)$ such that:

$$\forall e' \in \gamma',\ e' \in \chi'_e(e') \quad \& \quad \forall e' \in \gamma' \setminus \{e\}\,,\ \chi'_e(e') \neq \chi'_e(e)\,.$$

Finally, we choose $\widetilde{\lambda} \subset \mathcal{L}_{\text{surfcs}}$ such that $\lambda = \big[\widetilde{\lambda}\big]_{\text{profl}}$ and we define a profile $\lambda'$ by:

$$\lambda' := \left[\widetilde{\lambda} \cup \bigcup_{e \in \gamma'_{(1,\chi')}} \widetilde{\lambda}_e\right]_{\text{profl}}\,.$$

Then, $\lambda \preccurlyeq \lambda'$ (prop. 2.13) and the map $\chi'' : \gamma' \to \mathfrak{F}(\lambda') \cup \big\{F_{\neg}(\lambda')\big\}$, given by:

$$\forall e' \in \gamma',\ \chi''(e') = \chi'(e') \cap \bigcap_{e \in \gamma'_{(1,\chi')}} \chi'_e(e'),$$

belongs to $M^{(2)}_{(\gamma',\lambda')}$. □

We will now turn to populating faces that do not contain any edge yet. The basic idea is to pick some edge in the concerned face and to add it to our graph (remember that faces have been defined as non empty set of edges in prop. 2.11). Since the edge we add is chosen in a face of $\lambda$, the result of the first step is preserved, and since we add a single edge per face, and only for those faces that do not already contain an edge of the graph, there is no problem with the second step either. The only precaution required here is therefore to make sure that the added edges, together with the already present ones, form a graph, ie. intersects only at their extremities. Now, if we have picked some edge $e$ in a face $F$, any initial subedge of edge of $e$ will also be in $F$, and will do just as well for our purpose. This way, if $e$ intersects another edge $e'$ (either an edge of the graph, or one of the to-be-added edges), and if this intersection takes place *away* from $b(e)$, we can simply make $e$ shorter to avoid it (fig. 2.7 shows an edge being added to a face that were initially empty: we make the added edge small enough so that it stays away from any preexisting edges). Moreover, it cannot be that $e$ has a common initial subedge with another edge $e'$ for this would mean that $e'$ belongs to $F$, in contradiction with the preserved validity of step two (eg. the fact that we do not get more that one edge per face, as underlined above).

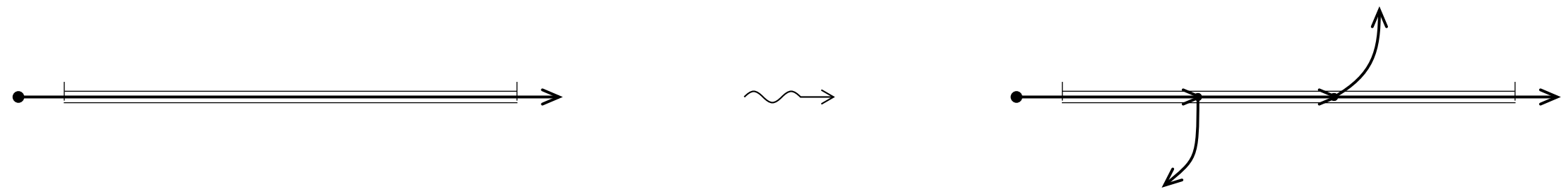

Figure 2.8 – An edge that runs along a surface (or an intersection of surfaces) may need to be subdivided when populating the corresponding faces

The only kind of intersection that cannot be fixed by shortening $e$ is therefore the situation in which $e \cap e' = \{b(e)\}$. As illustrated in fig. 2.8, it might not be possible to prevent this by a careful choice of $e$, because $b(e)$ has to belong to every surface involved in the face $F$ (this is indeed a necessary condition for $e$ to belongs to $F$): it might force $b(e)$ to be in the interior of some edge $e'$. Not that this concern is *not* an artifact of the dimension two: even when the surfaces have dimension $d-1>1$, there are faces that arises at the intersection of surfaces, and this intersection could be one-dimensional, with an edge running along it. Thus, the present step might require, beside the addition of new edges, also the subdivision of preexisting edges of the graph. Such a subdivision does not, however, threaten the two previous steps, nor does it lead to infinite recursion in the present step, for the first part of the subdivided edge $e'$ will belong to $\chi(e')$, and every subsequent parts will be of the $F_{\daleth}(\lambda)$ type.

**Lemma 2.19** Let $\gamma \in \mathcal{L}_{\text{graphs}}$ and $\lambda \in \mathcal{L}_{\text{profls}}$ such that $M^{(2)}_{(\gamma,\lambda)} \neq \varnothing$. Then, there exists $\gamma' \in \mathcal{L}_{\text{graphs}}$, such that $\gamma \preccurlyeq \gamma'$ and $M^{(3)}_{(\gamma',\lambda)} \neq \varnothing$.

**Proof** *Auxiliary result: Intersection of edges belonging to different faces.* Let $e \in \mathcal{L}_{\text{edges}}$ and $F \in \mathfrak{F}(\lambda)$ such that $e \in F$. Suppose that there exists $p \neq q \in r(e)$ and $F' \in \mathfrak{F}(\lambda) \cup \left\{F_{\daleth}(\lambda)\right\}$ such that the subedge $e_{[p,q]} \in F'$. Then, from prop. 2.8.6, there exists $q' \in r\left(e_{[p,q]}\right) \setminus \{p,q\}$ such that $e_{[p,q']} \in F'$. Since $q' \notin \{b(e), f(e)\}$, we are in one of the following situations:

$$\begin{array}{lll} e = e_{[q',f(e)]} \circ e_{[p,q']} & \text{if} & b(e) = p <_{(e)} q' <_{(e)} f(e) \\ e = e_{[q',f(e)]} \circ e_{[p,q']} \circ e_{[b(e),p]} & \text{if} & b(e) <_{(e)} p <_{(e)} q' <_{(e)} f(e) \\ e = \left(e_{[p,q']}\right)^{-1} \circ e_{[b(e),q']} & \text{if} & b(e) <_{(e)} q' <_{(e)} p = f(e) \\ e = e_{[p,f(e)]} \circ \left(e_{[p,q']}\right)^{-1} \circ e_{[b(e),q']} & \text{if} & b(e) <_{(e)} q' <_{(e)} p <_{(e)} f(e) \end{array}.$$

Hence, from props. 2.8.5 and 2.8.6, we have either $F' = F$ (if $p = b(e)$) or $F' = F_{\daleth}(\lambda)$ (otherwise).

Let $e_1, e_2 \in \mathcal{L}_{\text{edges}}$ and $F_1, F_2 \in \mathfrak{F}(\lambda)$ such that $e_1 \in F_1$, $e_2 \in F_2$ and:

$$\forall p \in r(e_1) \setminus \{b(e)\},\ \exists p' \in r(e_1) \,/\, b(e_1) <_{(e_1)} p' <_{(e_1)} p \quad \& \quad p' \in r(e_2).$$

Then, from lemma 2.6, there exists $q \in r(e_1) \setminus \{b(e_1)\}$ such that $r\left(e_{1,[b(e_1),q]}\right) \subset r(e_2)$, hence $e_{2,[b(e_1),q]} = e_{1,[b(e_1),q]}$. Since $e_{1,[b(e_1),q]} \in F_1 \neq F_{\daleth}(\lambda)$ (by definition of $\mathfrak{F}(\lambda)$), the previous argument applied to the subedge $e_{2,[b(e_1),q]}$ of $e_2$, together with prop. 2.11.1, implies that $F_1 = F_2$.

Thus, we have proven that, for any $e_1, e_2 \in \mathcal{L}_{\text{edges}}$ and any $F_1 \neq F_2 \in \mathfrak{F}(\lambda)$ such that $e_1 \in F_1$ and $e_2 \in F_2$, there exists $p \in r(e_1) \setminus \{b(e_1)\}$ such that $r\left(e_{1,[b(e_1),p]}\right) \cap r(e_2) \subset \{b(e_1), p\}$. Hence, there exists $p' \in r\left(e_{1,[b(e_1),p]}\right) \setminus \{b(e_1)\}$ such that $r\left(e_{1,[b(e_1),p']}\right) \cap r(e_2) \subset \{b(e_1)\}$.

*Construction of $\gamma'$.* Let $\chi \in M^{(2)}_{(\gamma,\lambda)}$ and define:

$$\mathcal{F}_{(2,\chi)}(\lambda) := \left\{F \in \mathcal{F}(\lambda) \;\middle|\; \chi^{-1}\langle F\rangle = \varnothing\right\}.$$

For any $F \in \mathcal{F}_{(2,\chi)}(\lambda)$, we choose an edge $e_F \in F$, and we define:

$$\gamma_{(2,\chi)} := \{e_F \mid F \in \mathcal{F}_{(2,\chi)}(\lambda)\}.$$

Let $e \in \gamma_{(2,\chi)}$ and $F \in \mathcal{F}_{(2,\chi)}(\lambda)$ such that $e \in F$. For any $\widetilde{e} \in \gamma$, $e \in \widetilde{F}$ with $\widetilde{F} = \chi(\widetilde{e}) \in \mathcal{F}(\lambda) \setminus \mathcal{F}_{(2,\chi)}(\lambda)$. And for any $\widetilde{e} \in \gamma_{(2,\chi)} \setminus \{e\}$, there exists $\widetilde{F} \in \mathcal{F}_{(2,\chi)}(\lambda) \setminus \{F\}$ such that $\widetilde{e} \in \widetilde{F}$. Hence, for any $\widetilde{e} \in \gamma \cup \gamma_{(2,\chi)} \setminus \{e\}$, there exists $p_{e,\widetilde{e}} \in r(e) \setminus \{b(e)\}$ such that $r\left(e_{[b(e),p_{e,\widetilde{e}}]}\right) \cap r(\widetilde{e}) \subset \{b(e)\}$. Since $\gamma \cup \gamma_{(2,\chi)} \setminus \{e\}$ is a finite set, there exists $p_e$ such that:

$$\forall \widetilde{e} \in \gamma \cup \gamma_{(2,\chi)} \setminus \{e\},\; r\left(e_{[b(e),p_e]}\right) \cap r(\widetilde{e}) \subset \{b(e)\}.$$

We define $\gamma'_e := \left\{e_{[b(e),p_e]}\right\}$ and $\chi'_e : \gamma'_e \to \mathcal{F}(\lambda) \cup \left\{F_{\urcorner}(\lambda)\right\}$, $e_{[b(e),p_e]} \mapsto F$. Then, for any $e' \in \gamma'_e$, we have $r(e') \subset r(e)$, $e' \in \chi'_e(e')$ and:

$$\forall \widetilde{e} \in \gamma \cup \gamma_{(2,\chi)} \setminus \{e\},\; r(e') \cap r(\widetilde{e}) \subset \{b(e'), f(e')\}.$$

Let $e \in \gamma$. The set $\{b(\widetilde{e}) \mid \widetilde{e} \in \gamma_{(2,\chi)} \;\;\&\;\; b(\widetilde{e}) \in r(e)\}$ is finite, hence there exist $n_e \geqslant 1$ and composable edges $e_1, \ldots, e_{n_e} \in \mathcal{L}_{\text{edges}}$ such that $e = e_{n_e} \circ \ldots \circ e_1$ and:

$$\{b(\widetilde{e}) \mid \widetilde{e} \in \gamma_{(2,\chi)} \;\;\&\;\; b(\widetilde{e}) \in r(e)\} \subset \bigcup_{i=1}^{n_e} \{b(e_i), f(e_i)\}.$$

We define $\gamma'_e := \{e_1, \ldots, e_{n_e}\}$. We also define the map $\chi'_e : \gamma'_e \to \mathcal{F}(\lambda) \cup \left\{F_{\urcorner}(\lambda)\right\}$ by:

$$\chi'_e(e_1) := \chi(e) \;\;\&\;\; \forall k \in \{2, \ldots, n_e\},\; \chi'_e(e_k) := F_{\urcorner}(\lambda).$$

Then, for any $e' \in \gamma'_e$, we have $r(e') \subset r(e)$, $e' \in \chi'_e(e')$ (combining $e \in \chi(e)$ with props. 2.8.6 and 2.3.3) and:

$$\forall \widetilde{e} \in \gamma \setminus \{e\},\; r(e') \cap r(\widetilde{e}) \subset r(e') \cap \{b(e), f(e)\} \subset \{b(e'), f(e')\},$$

where we have used that $\gamma \in \mathcal{L}_{\text{graphs}}$. For any $\widetilde{e} \in \gamma_{(2,\chi)}$, we have by definition of $\gamma'_e$ and $\gamma'_{\widetilde{e}}$:

$$\forall e' \in \gamma'_e,\, \forall \widetilde{e}' \in \gamma'_{\widetilde{e}},\; r(e') \cap r(\widetilde{e}') \subset r(e') \cap \bigcup_{i=1}^{n_e} \{b(e_i), f(e_i)\} \subset \{b(e'), f(e')\}.$$

Additionaly, we have from props. 2.3 and 2.2:

$$\forall e' \neq e'' \in \gamma'_e,\; r(e') \cap r(e'') \subset \{b(e'), f(e')\}.$$

Finally, we construct a finite set of edges $\gamma'$ as $\gamma' := \bigcup_{e \in \gamma \cup \gamma_{(2,\chi)}} \gamma'_e$, and we define a map $\chi' : \gamma' \to \mathcal{F}(\lambda) \cup \left\{F_{\urcorner}(\lambda)\right\}$ by:

$$\forall e \in \gamma \cup \gamma_{(2,\chi)},\, \forall e' \in \gamma'_e,\; \chi'(e') := \chi'_e(e')$$

($\chi'$ is well-defined since the $\gamma'_e$ for different $e$ are disjoints by construction). By putting together what we have proven above, we get:

$$\forall e' \neq \widetilde{e}' \in \gamma',\; r(e') \cap r(\widetilde{e}') \subset \{b(e'), f(e')\},$$

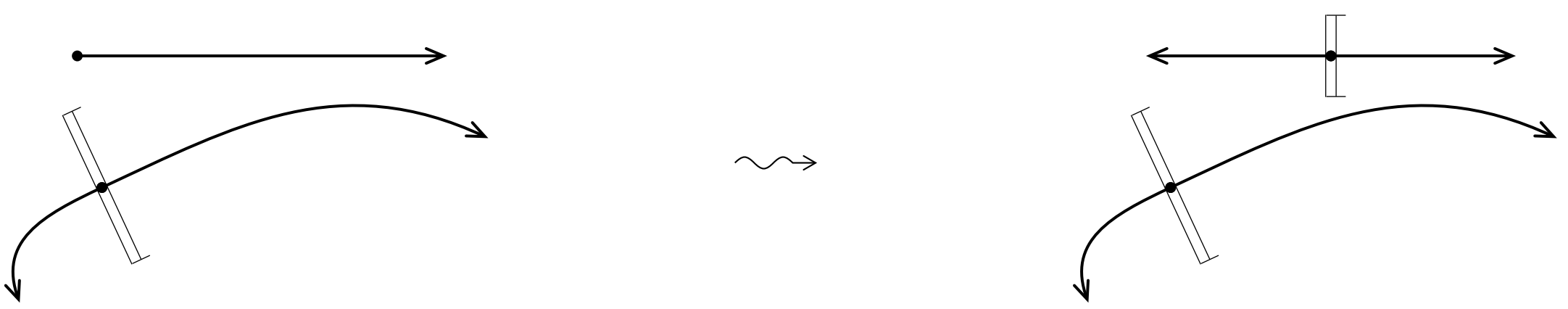

Figure 2.9 – Adding a small surface through the middle of a still unpaired edge

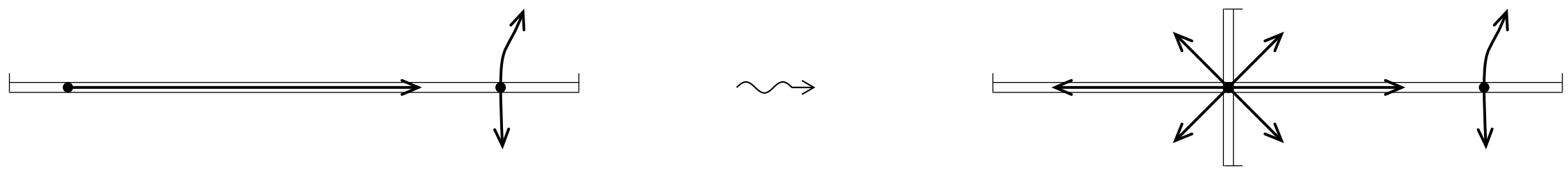

Figure 2.10 – Accidental extra faces need to be populated

thus $\gamma' \in \mathcal{L}_{\text{graphs}}$, and, by definition of $\gamma'_e$ for $e \in \gamma$, $\gamma \preccurlyeq \gamma'$. We also have $\forall e' \in \gamma'$, $e' \in \chi'(e')$ and:

$$\forall F \in \chi\langle\gamma\rangle\,,\ \chi'^{-1}\langle F\rangle = \left\{e_1 \mid e \in \chi^{-1}\langle F\rangle\right\} \quad \& \quad \forall F \in \mathcal{F}_{(2,\chi)}(\lambda),\ \chi'^{-1}\langle F\rangle = \gamma'_{e_F}\,.$$

Hence, using $\mathcal{F}(\lambda) = \mathcal{F}_{(2,\chi)}(\lambda) \sqcup \left(\chi\langle\gamma\rangle \setminus \left\{F_{\urcorner}(\lambda)\right\}\right)$ and $\chi \in M^{(2)}_{(\gamma,\lambda)}$, we obtain that $\chi' \in M^{(3)}_{(\gamma,\lambda)}$. □

Finally, we want to consider those edges that do not yet belong to any face. If $e$ is such an edge, we will let a small surface cut it through the middle, subdivide $e$ accordingly, and reorient the parts so that they start from the added surface (fig. 2.9): thus one part will be above the surface, and the other below. Since this surface goes through an interior point $p_e$ of $e$, we can ensure, by making it small enough, that it does not cross any other edges of the graph (by definition of a graph, $p_e$ can not belong to any other edge). Thus, the results of the first two steps are preserved.

However, if $e$ lies inside some preexisting surface, the achievement of the third step might have to be restored at this point: beside the two faces now populated by the two parts of $e$, there might be additional new faces corresponding to intersections of the added small surface with the preexisting ones. If this occurs, we will need to add a few edges to populate those faces (fig. 2.10), but we can make sure that all added edges, together with the two parts of $e$, intersects at most at their starting points. Also, by making the added edges shorter if required, we can prevent them to intersect any other edge of the graph. Thus, there is no need for further subdivision of the edges (in contrast to the situation depicted in fig. 2.8 that could arise in the previous step), and we have achieved the goal announced at the beginning of the present subsection.

**Lemma 2.20** Let $\gamma \in \mathcal{L}_{\text{graphs}}$ and $\lambda \in \mathcal{L}_{\text{profls}}$ such that $M^{(3)}_{(\gamma,\lambda)} \neq \varnothing$. Then, there exists $\gamma' \in \mathcal{L}_{\text{graphs}}$ and $\lambda' \in \mathcal{L}_{\text{profls}}$, such that $\gamma \preccurlyeq \gamma'$, $\lambda \preccurlyeq \lambda'$ and $M^{(4)}_{(\gamma',\lambda')} \neq \varnothing$.

**Proof** Let $\widetilde{\lambda} \subset \mathcal{L}_{\text{surfcs}}$ such that $\lambda = \left[\widetilde{\lambda}\right]_{\text{profl}}$ and $\chi \in M^{(3)}_{(\gamma,\lambda)}$. We define:

$$\gamma_{(3,\chi)} := \left\{e \in \gamma \mid \chi(e) = F_{\urcorner}(\lambda)\right\} \quad \& \quad \gamma'_0 := \gamma \setminus \gamma_{(3,\chi)}\,.$$

Thus, $\chi\left\langle\gamma'_0\right\rangle\subset\mathcal{F}(\lambda)$ and, by definition of $M^{(3)}_{(\gamma,\lambda)}$, $\chi_0:=\chi|_{\gamma'_0\to\mathcal{F}(\lambda)}$ is bijective.

For each $e\in\gamma_{(3,\chi)}$, we choose a representative $\check{e}:U_e\to V_e$ of $e$ and define $p_e:=\check{e}(0.5,0)$. Since, for any $e\in\gamma_{(3,\chi)}$, $\bigcup_{e'\in\gamma\setminus\{e\}}r(e')$ is a compact set that does not contain $p_e$ (for $p_e\in r(e)\setminus\{b(e),f(e)\}$ and $\gamma\in\mathcal{L}_{\text{graphs}}$), there exists an open neighborhood $W_e$ of $p_e$ in $V_e$ such that:

$$\forall e'\in\gamma\setminus\{e\}\,,\ r(e')\cap W_e=\varnothing\,.$$

Next, $\{p_e\mid e\in\gamma_{(3,\chi)}\}$ is finite and $\Sigma$ is Hausdorff, hence there exists a family $\left(W'_e\right)_{e\in\gamma_{(3,\chi)}}$ of *disjoint* open subsets of $\Sigma$ such that, for any $e\in\gamma_{(3,\chi)}$, $W'_e$ is an open neighborhood of $p_e$ in $W_e$. Thus, we have:

$$\forall e\in\gamma_{(3,\chi)},\ W'_e\cap\left(\bigcup_{e'\in\gamma\setminus\{e\}}r(e')\cup\bigcup_{e'\in\gamma_{(3,\chi)}\setminus\{e\}}W'_{e'}\right)=\varnothing\,. \tag{2.20.1}$$

Let $e\in\gamma_{(3,\chi)}$. There exists $\epsilon>0$ such that $\{0.5\}\times B^{(d-1)}_\epsilon\subset\check{e}^{-1}\left\langle W'_e\right\rangle$ and we define:

$$\begin{array}{rccl}\check{S}_e & : & U'_e & \to & V_e\\ & & t,y & \mapsto & \check{e}(t+0.5,\,\epsilon\,y)\end{array}\quad\text{with } U'_e:=\{(t,y)\mid(t+0.5,\,\epsilon\,y)\in U_e\}\,.$$

Since $\{0\}\times B^{(d-1)}\subset U'_e$, $\check{S}_e$ is an analytic, encharted surface in $\Sigma$. We denote by $S_e$ the corresponding surface. We can check from the definition of $\check{S}_e$ that $r(S_e)\subset W'_e$ and:

$$e_\uparrow:=e_{[p_e,f(e)]}\in F_{e,\uparrow}\quad\&\quad e_\downarrow:=e_{[p_e,b(e)]}\in F_{e,\downarrow}\,,$$

where $F_{e,\uparrow}:=\left\{e'\in F_{\urcorner}(\lambda)\,\middle|\,e'\uparrow S_e\right\}$ and $F_{e,\downarrow}:=\left\{e'\in F_{\urcorner}(\lambda)\,\middle|\,e'\downarrow S_e\right\}$ (using $e\in\chi(e)=F_{\urcorner}(\lambda)$ together with props. 2.8.5 and 2.8.6). Defining $\gamma'_{e,0}:=\{e_\uparrow,\,e_\downarrow\}$, we moreover have:

$$\forall e'\in\gamma'_{e,0},\ r(e')\cap\{b(e),\,f(e)\}\subset\{f(e')\}\,, \tag{2.20.2}$$

$$\text{and}\quad\forall e'\neq\widetilde{e}'\in\gamma'_{e,0},\, r(e')\cap r(\widetilde{e}')\subset\{b(e')\}=\{b(\widetilde{e}')\}\,. \tag{2.20.3}$$

Next, we define:

$$\mathcal{F}_{(e,\uparrow)}(\lambda):=\left\{F\in\mathcal{F}(\lambda)\,\middle|\,\exists\,e'\in F\,/\,e'\uparrow S_e\right\}\quad\&\quad\mathcal{F}_{(e,\downarrow)}(\lambda):=\left\{F\in\mathcal{F}(\lambda)\,\middle|\,\exists\,e'\in F\,/\,e'\downarrow S_e\right\}.$$

For each $F\in\mathcal{F}_{(e,\uparrow)}(\lambda)$ (resp. $F\in\mathcal{F}_{(e,\downarrow)}(\lambda)$), we define:

$$F_{e,F,\uparrow}:=\left\{e'\in F\,\middle|\,e'\uparrow S_e\right\}\quad(\text{resp. } F_{e,F,\downarrow}:=\left\{e'\in F\,\middle|\,e'\downarrow S_e\right\}),$$

and we choose an edge $e_{e,F,\uparrow}\in F_{e,F,\uparrow}$ (resp. $e_{e,F,\downarrow}\in F_{e,F,\downarrow}$). We also define:

$$\gamma'_{e,2}:=\{e_{e,F,\uparrow}\mid F\in\mathcal{F}_{(e,\uparrow)}(\lambda)\}\cup\{e_{e,F,\downarrow}\mid F\in\mathcal{F}_{(e,\downarrow)}(\lambda)\}\,.$$

The sets $\{F_{e,\uparrow},\,F_{e,\downarrow}\}$, $\{F_{e,F,\uparrow}\mid F\in\mathcal{F}_{(e,\uparrow)}(\lambda)\}$ and $\{F_{e,F,\downarrow}\mid F\in\mathcal{F}_{(e,\downarrow)}(\lambda)\}$ are disjoints subsets of $\mathcal{F}\left(\widetilde{\lambda}\cup\{S_e\}\right)$. Hence, defining:

$$\mathcal{F}_{(e)}\left(\widetilde{\lambda}\cup\{S_e\}\right):=\{F_{e,\uparrow},\,F_{e,\downarrow}\}\cup\{F_{e,F,\uparrow}\mid F\in\mathcal{F}_{(e,\uparrow)}(\lambda)\}\cup\{F_{e,F,\downarrow}\mid F\in\mathcal{F}_{(e,\downarrow)}(\lambda)\}\,,$$

$\mathfrak{F}_{(e)}\left(\widetilde{\lambda} \cup \{S_e\}\right) \subset \mathfrak{F}\left(\widetilde{\lambda} \cup \{S_e\}\right)$ and there exists a *bijective* map $\chi_{e,2} : \gamma'_{e,0} \cup \gamma'_{e,2} \to \mathfrak{F}_{(e)}\left(\widetilde{\lambda} \cup \{S_e\}\right)$ such that, for any $e' \in \gamma'_{e,0} \cup \gamma'_{e,2}$, $e' \in \chi_{e,2}(e')$.

Let $e' \in \gamma'_{e,2}$. There exists $F' := \chi_{e,2}(e') \in \mathfrak{F}\left(\widetilde{\lambda} \cup \{S_e\}\right)$ such that $e' \in F'$. Moreover, for any $\widetilde{e}' \in \gamma'_{e,0} \cup \gamma'_{e,2} \setminus \{e'\}$, there exists $\widetilde{F}' := \chi_{e,2}(\widetilde{e}') \in \mathfrak{F}\left(\widetilde{\lambda} \cup \{S_e\}\right) \setminus \{F'\}$ such that $\widetilde{e}' \in \widetilde{F}'$. Thus, using the auxiliary result from the proof of lemma 2.19, there exists, for any $\widetilde{e}' \in \gamma'_{e,0} \cup \gamma'_{e,2} \setminus \{e'\}$, a point $q_{e',\widetilde{e}'} \in r(e') \setminus \{b(e')\}$ such that $r\left(e'_{[b(e'),q_{e',\widetilde{e}'}]}\right) \cap r(\widetilde{e}') \subset \{b(e')\}$. Hence, there exists $q_{e'} \in r(e') \setminus \{b(e')\}$ such that:

$$\forall \widetilde{e}' \in \gamma'_{e,0} \cup \gamma'_{e,2} \setminus \{e'\},\ r\left(e'_{[b(e'),q_{e'}]}\right) \cap r(\widetilde{e}') \subset \{b(e')\}.$$

Since $W'_e$ is an open subset of $\Sigma$ containing $b(e')$ (for $b(e') \in r(S_e) \subset W'_e$), there exists $q'_{e'} \in r\left(e'_{[b(e'),q_{e'}]}\right) \setminus \{b(e')\}$ such that $r\left(e'_{[b(e'),q'_{e'}]}\right) \subset W'_e$. Also, we have $r\left(e'_{[b(e'),q'_{e'}]}\right) \subset r(e')$ and, from prop. 2.8.6, $e'_{[b(e'),q'_{e'}]} \in F'$.

Defining $\gamma'_{e,1} := \left\{e'_{[b(e'),q'_{e'}]} \,\middle|\, e' \in \gamma'_{e,2}\right\}$ and $\gamma'_e := \gamma'_{e,0} \cup \gamma'_{e,1}$, there exists therefore a bijective map $\chi_e : \gamma'_e \to \mathfrak{F}_{(e)}\left(\widetilde{\lambda} \cup \{S_e\}\right)$ such that, for any $e' \in \gamma'_e$, $e' \in \chi_e(e')$.
In addition, we have:

$$\forall e' \in \gamma'_{e,0},\ r(e') \subset r(e) \quad \& \quad \forall e' \in \gamma'_{e,1},\ r(e') \subset W'_e, \tag{2.20.4}$$

$$\text{and} \quad \forall e' \in \gamma'_{e,1},\ \forall \widetilde{e}' \in \gamma'_e \setminus \{e'\},\ r(e') \cap r(\widetilde{e}') \subset \{b(e')\}. \tag{2.20.5}$$

But, since for any $e' \in \gamma'_{e,1}$, $b(e') \in r(S_e)$, and for any $\widetilde{e}' \in \gamma'_{e,0}$, $r(\widetilde{e}') \cap r(S_e) \subset \{b(\widetilde{e}')\}$, eq. (2.20.5) together with eq. (2.20.3) implies:

$$\forall e' \neq \widetilde{e}' \in \gamma'_e,\ r(e') \cap r(\widetilde{e}') \subset \{b(e')\}. \tag{2.20.6}$$

Now, we define:

$$\gamma' := \gamma'_0 \cup \bigcup_{e \in \gamma_{(3,\chi)}} \left(\gamma'_e\right) \quad \& \quad \widetilde{\lambda}' := \widetilde{\lambda} \cup \bigcup_{e \in \gamma_{(3,\chi)}} \{S_e\}.$$

The fact that $\gamma$ is a graph, together with eqs. (2.20.1), (2.20.2), (2.20.4) and (2.20.6), ensures that $\gamma'$ is again a graph. Moreover, by definition of $\gamma'_{e,0}$ for $e \in \gamma_{(3,\chi)}$, $\gamma \preccurlyeq \gamma'$, and, from prop. 2.13, $\lambda \preccurlyeq \lambda'$ where $\lambda' := \left[\widetilde{\lambda}'\right]_{\text{profl}}$. Next, we define:

$$\mathfrak{F}_{(0)}(\widetilde{\lambda}') := \left\{F \cap F_{\daleth}(\widetilde{\lambda}' \setminus \widetilde{\lambda}) \,\middle|\, F \in \mathfrak{F}(\widetilde{\lambda})\right\},$$

$$\text{and, for any } e \in \gamma_{(3,\chi)},\ \mathfrak{F}_{(e)}(\widetilde{\lambda}') := \left\{F \cap F_{\daleth}\big(\widetilde{\lambda}' \setminus (\widetilde{\lambda} \cup \{S_e\})\big) \,\middle|\, F \in \mathfrak{F}_{(e)}(\widetilde{\lambda} \cup \{S_e\})\right\}.$$

Since, for any $e \in \gamma_{(3,\chi)}$, $r(S_e) \subset W'_e$, we have:

$$\left\{e' \in \mathcal{L}_{\text{edges}} \,\middle|\, \exists\, \diamond_e \in \left\{\uparrow, \downarrow\right\} \,/\, e' \diamond_e S_e\right\} \subset \{e' \in \mathcal{L}_{\text{edges}} \mid b(e') \in r(S_e)\}$$

$$\subset \{e' \in \mathcal{L}_{\text{edges}} \mid b(e') \in W'_e\} . \tag{2.20.7}$$

Therefore, for any $e \neq \tilde{e} \in \gamma_{(3,\chi)}$, we get, using $W'_e \cap W'_{\tilde{e}} = \varnothing$:

$$\left\{e' \in \mathcal{L}_{\text{edges}} \,\middle|\, \exists \diamond_e, \diamond_{\tilde{e}} \in \left\{\uparrow, \downarrow\right\} / \, e' \diamond_e S_e \quad \& \quad e' \diamond_{\tilde{e}} S_{\tilde{e}}\right\} = \varnothing .$$

This, together with the definition of $\mathcal{F}_{(e)}(\widetilde{\lambda} \cup \{S_e\})$ for $e \in \gamma_{(3,\chi)}$, implies:

$$\mathcal{F}(\widetilde{\lambda}') \quad \subset \bigcup_{e \in \{0\} \sqcup \gamma_{(3,\chi)}} \mathcal{F}_{(e)}(\widetilde{\lambda}') .$$

In addition, we can define for any $e \in \{0\} \sqcup \gamma_{(3,\chi)}$ a bijective map $\chi'_e : \gamma'_e \to \mathcal{F}_{(e)}(\widetilde{\lambda}')$ by:

$$\forall e' \in \gamma'_0,\, \chi'_0(e') := \chi_0(e') \cap F_{\urcorner}(\widetilde{\lambda}' \setminus \widetilde{\lambda}) \quad \& \quad \forall e' \in \gamma'_e,\, \chi'_e(e') := \chi_e(e') \cap F_{\urcorner}\big(\widetilde{\lambda}' \setminus (\widetilde{\lambda} \cup \{S_e\})\big) .$$

For any $e \in \{0\} \sqcup \gamma_{(3,\chi)}$, $\chi'_e$ satisfies:

$$\forall e' \in \gamma'_e,\, e' \in \chi'_e(e') ,$$

as follows from the corresponding property of $\chi_e$ together with eqs. (2.20.1), (2.20.4) and (2.20.7). In particular, we thus have:

$$\forall e \in \{0\} \sqcup \gamma_{(3,\chi)},\, \forall F \in \mathcal{F}_{(e)}(\widetilde{\lambda}'),\, F \neq \varnothing ,$$

so, we obtain:

$$\mathcal{F}(\widetilde{\lambda}') \quad = \bigcup_{e \in \{0\} \sqcup \gamma_{(3,\chi)}} \mathcal{F}_{(e)}(\widetilde{\lambda}') .$$

Since the domains of the bijective maps $\chi'_e$ for $e \in \{0\} \sqcup \gamma_{(3,\chi)}$ are disjoints, as well as their images, they can be combined into a well-defined bijective map $\chi' : \gamma' \to \mathcal{F}(\lambda')$. Hence, $M^{(4)}_{(\gamma',\lambda')} \neq \varnothing$. □

**Proof of theorem 2.16** Let $(\gamma, \lambda)$, $(\gamma', \lambda') \in \mathcal{L}$. Since $\mathcal{L}_{\text{graphs}}, \preccurlyeq$ and $\mathcal{L}_{\text{profls}}, \preccurlyeq$ are directed sets (props. 2.5 and 2.13), there exists $(\gamma_1, \lambda_1) \in \mathcal{L}_{\text{graphs}} \times \mathcal{L}_{\text{profils}}$ such that $\gamma, \gamma' \preccurlyeq \gamma_1$ and $\lambda, \lambda' \preccurlyeq \lambda_1$. By chaining lemmas 2.17 to 2.20 and using the transitivity of $\preccurlyeq$ on $\mathcal{L}_{\text{graphs}}$ and $\mathcal{L}_{\text{profls}}$, there exists $(\gamma'', \lambda'') \in \mathcal{L}_{\text{graphs}} \times \mathcal{L}_{\text{profils}}$ such that $\gamma_1 \preccurlyeq \gamma''$, $\lambda_1 \preccurlyeq \lambda''$ and $M^{(4)}_{(\gamma'',\lambda'')} \neq \varnothing$. Thus, $(\gamma'', \lambda'') \in \mathcal{L}$ and $(\gamma, \lambda)$, $(\gamma', \lambda') \preccurlyeq (\gamma'', \lambda'')$. □

# 3 Quantum state space

The labels introduced in the previous section are meant to identify corresponding small algebras of observables, and the ordering has been chosen such that, whenever $\eta \preccurlyeq \eta'$, there is a natural injection of the algebra labeled by $\eta$ into the one labeled by $\eta'$. By carefully adjusting under which conditions a collection of edges and surfaces can be turned into a label, we have ensured that these algebras of observables can be represented on small phase spaces $\mathcal{M}_\eta$ and $\mathcal{M}_{\eta'}$ respectively, and that the identification between observables on $\mathcal{M}_\eta$ and $\mathcal{M}_{\eta'}$ unambiguously prescribes a suitable projection from $\mathcal{M}_{\eta'}$ into $\mathcal{M}_\eta$.

Moreover, this projection is compatible with the symplectic structures [17, def. 2.1], and is actually of the form that were considered in [18, theorem 3.2] (as will be shown in prop. 3.8). Thus, we expect from this theorem that the obtained projective system of symplectic manifolds goes down to a factorizing system on the underlying configuration spaces [17, def. 2.15]. Indeed, we will prove that it is the case by giving the explicit expressions for the factorization maps (with the added benefit that no further restriction need to be imposed on the finite-dimensional Lie group $G$, while [18, theorem 3.2] have been derived in the case of simply-connected groups). Once we have such a factorizing system of configuration spaces, it can be straightforwardly quantized into a projective quantum state space (along the lines of [24] and [18, subsection 3.1]).

For setting up the quantum state space in subsection 3.1, the group $G$ will neither be required to be Abelian, nor compact. However, if $G$ actually happens to be compact, a different construction is also possible: in this case, we have at our disposal a family of normalizable measures on the configuration spaces, and this family of measure is compatible with the factorizations, so that instead of assembling the small Hilbert spaces (obtained by quantization of the small phase spaces using a position polarization) into a projective structure, we can assemble them into an inductive limit Hilbert space. As exposed in [18, prop. 3.5], there is then a natural injection mapping the density matrices on the inductive limit Hilbert space into the projective state space. Interestingly, this inductive limit can in our case be identified with the Ashtekar-Lewandowski Hilbert space [22]. This may at first seem surprising, since the former is made of building blocks labeled by edges and surfaces, while the latter use labels which are just graphs. The trick here is that the injections defining this inductive limit in fact do not depend on the disposition of the surfaces in the labels: the injection that mount the Hilbert space associated to a label $\eta$ into the one associated to a finer label $\eta'$ turns out to only depend on the underlying graphs $\gamma(\eta)$ and $\gamma(\eta')$, so that the inductive system labeled by elements of $\mathcal{L}$ actually collapses into an inductive system simply labeled by graphs. This is in fact the very observation that was spelled out at the beginning of section 2: projections between configuration spaces (which are the ingredients of an inductive limit construction) are less specific than factorizations, and it was precisely in order to distinguish between different possible factorizations corresponding to the same projection map that we had to introduce surfaces in the labels.

## 3.1 Factorizations

We start by attaching to each label $\eta$ a configuration space $\mathcal{C}_\eta$, as well as a momentum space $\mathcal{P}_\eta$ (later on, we will rely on left translations to identify the cotangent bundle $T^*(G)$ with $G \times \mathfrak{g}^*$, and thus $T^*(\mathcal{C}_\eta)$ with $\mathcal{C}_\eta \times \mathcal{P}_\eta$): $\mathcal{C}_\eta$ is nothing but the configuration space routinely associated in LQG to the graph $\gamma(\eta)$, while momentum variables are assigned to the faces of the label, as announced in subsection 2.1.

Also, we introduce a few notations to discuss how the holonomy along an edge, or the flux through a face, can be related to the variables in $\mathcal{C}_\eta$ and $\mathcal{P}_\eta$ (provided $\eta$ is fine enough to describe the desired observable). Since we want to deduce the correct projective structure from the interpretation of the labels in terms of observables, it is particularly important that the relation between these observables and the variables in the small phase spaces should be unambiguous: this will ensure that the factorization maps are well-defined, and will be essential for proving the so-called three-spaces consistency [17, eq. (2.11.1) and fig. 2.2] (combined with the directedness of $\mathcal{L}$, this consistency condition indeed expresses the concern for an univocal meaning of the variables attached to a label).

For a clean labeling of the flux observables we formulate in prop. 3.3 a notion of 'free-standing' faces. In prop. 2.11 the different faces corresponding to a certain collection of surfaces $\{S\}$ have been defined as particular sets of edges, and each such face $F$ can only contain edges that are adapted to every surface in $\{S\}$. In particular, an edge can be prevented from belonging to $F$ simply because it crosses transversally some surface of the collection, even this surface is in reality unrelated to the face $F$. Thus, a given flux operator is described in different profiles by different set of edges, and its characterization by a more intrinsic set is only obtained after compensating this effect.

**Definition 3.1** For any $\eta \in \mathcal{L}$, we define its associated configuration space:

$$\mathcal{C}_\eta := \{h : \gamma(\eta) \to G\} \approx G^{\#\gamma(\eta)},$$

where $\#\gamma(\eta) < \infty$ denotes the number of edges in $\gamma(\eta)$. Since $G$ is a finite-dimensional Lie group, $\mathcal{C}_\eta$ is a finite-dimensional smooth manifold.

Similarly, we define the corresponding momentum space:

$$\mathcal{P}_\eta := \{P : \mathcal{F}(\eta) \to \mathfrak{g}^*\} \approx (\mathfrak{g}^*)^{\#\mathcal{F}(\eta)},$$

which is a finite-dimensional real vector space.

**Proposition 3.2** Let $e \in \mathcal{L}_{\text{edges}}$. We define:

$$\mathcal{L}_{\text{graphs}/e} := \{\gamma \in \mathcal{L}_{\text{graphs}} \mid \{e\} \preccurlyeq \gamma\}.$$

For any $\gamma \in \mathcal{L}_{\text{graphs}/e}$, there exists a *unique* map $a_{\gamma\to e} : \{1, \dots, n_{\gamma\to e}\} \to \gamma$ (with $n_{\gamma\to e} \geqslant 1$) such that:

$$e = a_{\gamma\to e}(n_{\gamma\to e})^{\epsilon_{\gamma\to e}(n_{\gamma\to e})} \circ \dots \circ a_{\gamma\to e}(1)^{\epsilon_{\gamma\to e}(1)}, \tag{3.2.1}$$

where, for any $k \in \{1, \dots, n_{\gamma\to e}\}$:

$$\epsilon_{\gamma\to e}(k) := \begin{cases} +1 & \text{if } b\left(a_{\gamma\to e}(k)\right) <_{(e)} f\left(a_{\gamma\to e}(k)\right) \\ -1 & \text{if } b\left(a_{\gamma\to e}(k)\right) >_{(e)} f\left(a_{\gamma\to e}(k)\right) \end{cases}. \tag{3.2.2}$$

Moreover, $a_{\gamma\to e}$ then induces a bijection $\{1, \dots, n_{\gamma\to e}\} \to H_{\gamma\to e}$, where:

$$H_{\gamma\to e} := \{e' \in \gamma \mid r(e') \subset r(e)\}. \tag{3.2.3}$$

**Proof** Let $\gamma \in \mathcal{L}_{\text{graphs}/e}$. Since $\{e\} \preccurlyeq \gamma$, there exist $n_{\gamma\to e} \geqslant 1$ and a map $a_{\gamma\to e} : \{1, \dots, n_{\gamma\to e}\} \to \gamma$ such that:

$$e = a_{\gamma\to e}(n_{\gamma\to e})^{\epsilon_{\gamma\to e}(n_{\gamma\to e})} \circ \dots \circ a_{\gamma\to e}(1)^{\epsilon_{\gamma\to e}(1)},$$

where, for any $k \in \{1, \dots, n_{\gamma\to e}\}$, $\epsilon_{\gamma\to e}(k)$ is defined as in eq. (3.2.2). By definition of the composition of edges (prop. 2.3), $a_{\gamma\to e}$ is injective and for any $k \in \{1, \dots, n_{\gamma\to e}\}$, $a_{\gamma\to e}(k) \in H_{\gamma\to e}$.

Next, let $e' \in H_{\gamma\to e}$, and let $p \in r(e') \setminus \{b(e'), f(e')\}$. Since $p \in r(e)$, there exits $k \in \{1, \dots, n_{\gamma\to e}\}$ such that $p \in a_{\gamma\to e}(k)$, so, $\gamma$ being a graph, $e' = a_{\gamma\to e}(k)$. Thus, $a_{\gamma\to e}$ induces a bijection $\{1,\ldots,n_{\gamma\to e}\}\to H_{\gamma\to e}$. In particular, $n_{\gamma\to e}=\#H_{\gamma\to e}$.

Finally, defining the map $\mu : H_{\gamma\to e}\to r(e)$ by:

$$\forall e'\in H_{\gamma\to e},\ \mu(e') := \begin{cases} b(e') & \text{if } b(e') <_{(e)} f(e') \\ f(e') & \text{if } b(e') >_{(e)} f(e') \end{cases},$$

prop. 2.3 implies that $\mu\circ a_{\gamma\to e}$ is strictly increasing (using $<$ on $\{1,\ldots,n_{\gamma\to e}\}$ and $<_{(e)}$ on $r(e)$), hence the uniqueness. □

**Proposition 3.3** For any $F\subset\mathcal{L}_{\text{edges}}$, we define:

$$F^{\perp} := \left\{e\in\mathcal{L}_{\text{edges}}\ \middle|\ \forall p\neq p'\in r(e),\ e_{[p,p']}\notin F\right\},$$

and we will denote by $\mathcal{L}_{\text{faces}}$ the set:

$$\mathcal{L}_{\text{faces}} := \bigcup_{\lambda\in\mathcal{L}_{\text{profls}}}\left\{F^{\perp}\circ F\ \middle|\ F\in\mathcal{F}(\lambda)\right\}.$$

In addition, we define for any $\overline{F}\in\mathcal{L}_{\text{faces}}$:

$$\mathcal{L}_{\text{profls}/\overline{F}} := \left\{\lambda'\in\mathcal{L}_{\text{profls}}\ \middle|\ \exists\,\lambda\preccurlyeq\lambda',\ \exists\,F\in\mathcal{F}(\lambda)\,\big/\,\overline{F}=F^{\perp}\circ F\right\}.$$

Then, for any $\overline{F}\in\mathcal{L}_{\text{faces}}$ and any $\lambda'\in\mathcal{L}_{\text{profls}/\overline{F}}$, we have:

$$\overline{F}=\overline{F}^{\perp}\circ\bigcup_{F'\in H_{\lambda'\to\overline{F}}}F', \tag{3.3.1}$$

where $H_{\lambda'\to\overline{F}} := \left\{F'\in\mathcal{F}(\lambda')\ \middle|\ F'\subset\overline{F}\right\}$.

**Proof** Let $\lambda\in\mathcal{L}_{\text{profls}}$, $F\in\mathcal{F}(\lambda)$, $\overline{F}:=F^{\perp}\circ F$ and $\lambda'\in\mathcal{L}_{\text{profls}}$ such that $\lambda'\succcurlyeq\lambda$. From props. 2.8.5 and 2.8.6, we have:

$$F^{\perp}=F^{\perp}\circ F^{\perp}\quad\&\quad\overline{F}^{\perp}=\left(F^{\perp}\circ F\right)^{\perp}=F^{\perp}.$$

Hence, for any $F'\in H_{\lambda'\to\overline{F}}$:

$$\overline{F}^{\perp}\circ F'\subset\overline{F}^{\perp}\circ\overline{F}=F^{\perp}\circ F^{\perp}\circ F=\overline{F},$$

therefore $\overline{F}^{\perp}\circ\bigcup_{F'\in H_{\lambda'\to\overline{F}}}F'\subset\overline{F}$.

Now, by definition of the preorder $\preccurlyeq$ on $\mathcal{L}_{\text{profls}}$ (prop. 2.13), there exist $F'_1,\ldots,F'_m\in\mathcal{F}(\lambda')$ ($m\geqslant 1$) such that:

$$F=F_{\daleth}(\lambda)\circ\bigcup_{i=1}^{m}F'_i.$$

For any $i\in\{1,\ldots,m\}$, $F_{\daleth}(\lambda)\circ F'_i\subset F$ implies $F'_i\subset F\subset\overline{F}$ (using again props. 2.8.5 and 2.8.6), so $F'_i\in H_{\lambda'\to\overline{F}}$. And from prop. 2.11.3, $F_{\daleth}(\lambda)\subset F^{\perp}$, therefore:

$$\overline{F}\subset\overline{F}^{\perp}\circ\bigcup_{i=1}^{m}F'_i\subset\overline{F}^{\perp}\circ\bigcup_{F'\in H_{\lambda'\to\overline{F}}}F'.$$

□

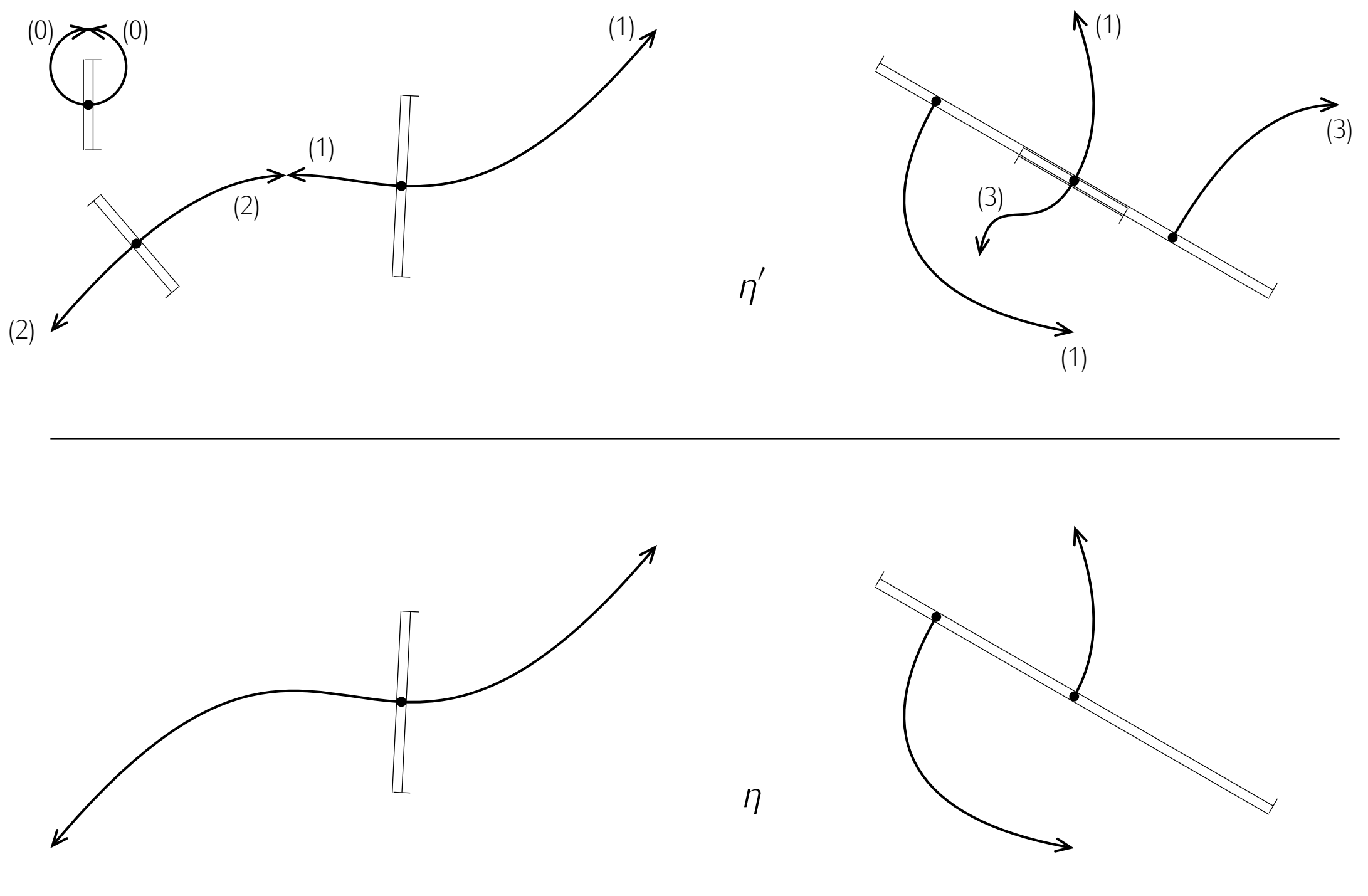

Figure 3.1 – Classification of the edges in $\eta' \succcurlyeq \eta$: a tag $^{(\kappa)}$ denotes an edge belonging to $H^{(\kappa)}_{\eta'\to\eta}$ (prop. 3.4)

Props. 3.2 and 3.3 make it possible to unambiguously attach a physical interpretation to the variables in $\mathcal{C}_\eta$ and $\mathcal{P}_\eta$, and are therefore at the root of the relation between the variables assigned to a label $\eta$ and the ones assigned to a finer label $\eta'$. Although we will directly give an analytic expression for the factorization map $\varphi_{\eta'\to\eta} : \mathcal{C}_{\eta'} \to \mathcal{C}_{\eta'\to\eta} \times \mathcal{C}_\eta$ and we will not explicitly make use of [18, theorem 3.2], the proof we gave for this result provides the right hints regarding why such a factorization map does exist, and how it should be defined so that it leads to the desired projection between the phase spaces.

The key idea is that the momentum variables assigned to the label $\eta$ can be mounted into the phase space associated to $\eta'$, and therefore correspond to certain vector fields on $\mathcal{C}_{\eta'}$. Each orbit under the finite transformations generated by these vector fields can then be naturally identified with $\mathcal{C}_\eta$ (for the relation between the configuration variables in $\mathcal{C}_\eta$ and $\mathcal{C}_{\eta'}$ yields a projection from $\mathcal{C}_{\eta'}$ into $\mathcal{C}_\eta$ which intertwines the action of these transformations). The complementary space $\mathcal{C}_{\eta'\to\eta}$ can thus be taken as the corresponding quotient space, which itself can be identified with the preimage of some point in $\mathcal{C}_\eta$ (eg. the function mapping every edge in $\gamma(\eta)$ to the identity element in $G$) under the projection $\mathcal{C}_{\eta'} \to \mathcal{C}_\eta$. Note that this is simply the non-linear version of the procedure described in [24, section 3.4].

This prescription can equivalently be expressed as the realization that the space $\mathcal{C}_{\eta'\to\eta}$, together with the projection from $\mathcal{C}_{\eta'}$ into $\mathcal{C}_{\eta'\to\eta}$, can be completely specified by identifying a maximal set of variables in $\mathcal{C}_{\eta'}$ which are not acted upon by the fluxes retained in the label $\eta$. On the other hand, the projection from $\mathcal{C}_{\eta'}$ into $\mathcal{C}_\eta$ is obtained by writing down the edges in $\eta$ as compositions of edges in $\eta'$. Thus, the first step toward the determination of the factorization map is to state precisely how the edges and faces of the label $\eta$ lie within $\eta'$. For this, we will classify the edges of $\eta'$ into various categories depending on whether they belong to some face and/or are part of some edge of $\eta$. Note that no edge in $\gamma(\eta')$ can be a subedge of two different edges in $\gamma(\eta)$, nor can it belong to two different faces in $\mathcal{F}(\eta)$, nor can a subedge of an edge $e \in \gamma(\eta)$ belongs to any face in $\mathcal{F}(\eta)$ but $\chi(e)$. So we are left with the 4 options listed in prop. 3.4 and depicted in fig. 3.1.

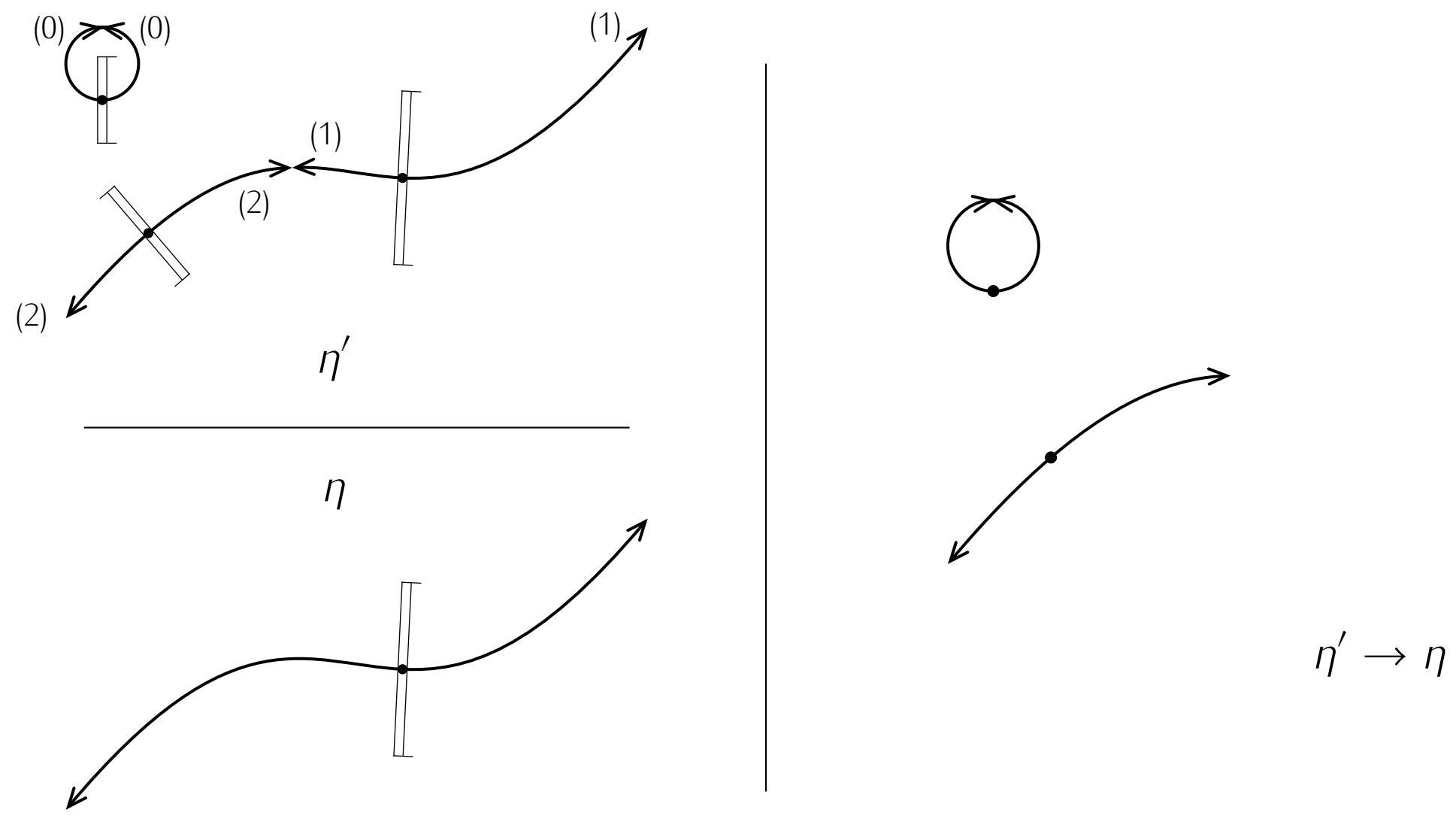

Figure 3.2 – Complementary variables related to edges of type '0' or '2', edges of type '1' do not demand any extra complementary variable

Clearly, the group variables corresponding to edges of $\eta'$ of type '0' or '2' (those that do not belong to any face among $\mathcal{F}(\eta)$) qualify as complementary variables (they are not acted upon by the fluxes retained in $\eta$). Also, knowing the holonomy along some edge of $\eta$, as well as the holonomies along all edges of $\eta'$ that compose it *except* the first one, the holonomy along this first part (which is of type '1') can be reconstructed. Hence, the group variables corresponding to edges of type '1' can be safely droped when extracting $\mathcal{C}_{\eta' \to \eta}$ (fig. 3.2).

Dealing with the group variables corresponding to edges of type '3' is slightly more subtle. Such an edge $e' \in \gamma(\eta')$ belongs to some face $F \in \mathcal{F}(\eta)$ without being the initial part of the conjugate edge $e \in \gamma(\eta)$. To build a group variable invariant under the flux corresponding to $F$, we will compose the holonomy along $e^{-1}$ (which ends in $F$) followed by the holonomy along $e'$ (which starts in $F$). In this way the action of the flux through $F$ cancel out, while the variable in $\mathcal{C}_{\eta'}$ that corresponds to the considered edge $e'$ can still be reconstructed from the variables in $\mathcal{C}_{\eta' \to \eta}$ and $\mathcal{C}_{\eta}$ (fig. 3.3). Note that we could equally well take the composition of the holonomy along $e_1^{-1}$ followed by the holonomy along $e'$, with $e_1$ the first part of $e$ (in its decomposition into edges of $\eta'$). These two alternatives only differ by a function of the group variables attached to the remaining parts of $e$, which already belongs to $\mathcal{C}_{\eta' \to \eta}$ (these remaining parts being edges of type '2' as outlined above), so it is nothing but a change of coordinates on $\mathcal{C}_{\eta' \to \eta}$, and therefore of no consequences for the construction.

**Proposition 3.4** Let $\eta \preccurlyeq \eta' \in \mathcal{L}$. We define:

1. $H^{(0)}_{\eta' \to \eta} := \{e' \in \gamma(\eta') \mid \forall e \in \gamma(\eta),\ \ r(e') \not\subset r(e) \quad \& \quad e' \notin \chi_\eta(e)\}$ ;

and, for any $e \in \gamma(\eta)$ :

2. $H^{(1)}_{\eta' \to \eta, e} := \{e' \in \gamma(\eta') \mid r(e') \subset r(e) \quad \& \quad e' \in \chi_\eta(e)\}$ ;

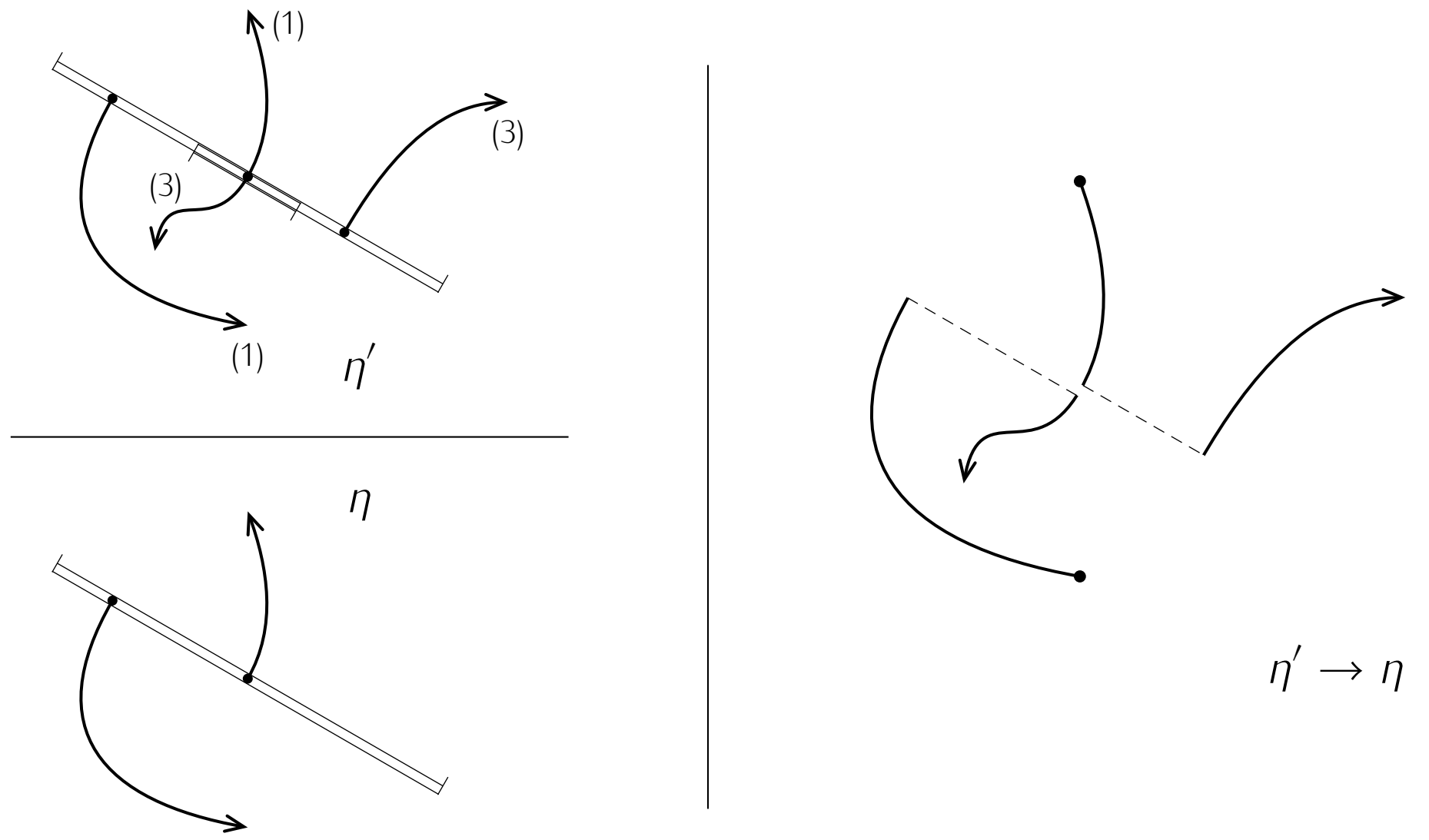

Figure 3.3 – Complementary variables related to edges of type '3'

3. $H^{(2)}_{\eta'\to\eta,e} := \{e' \in \gamma(\eta') \mid r(e') \subset r(e) \quad \& \quad e' \notin \chi_\eta(e)\}$ ;

4. $H^{(3)}_{\eta'\to\eta,e} := \{e' \in \gamma(\eta') \mid r(e') \not\subset r(e) \quad \& \quad e' \in \chi_\eta(e)\}$ ;

Then,

$$\left\{H^{(0)}_{\eta'\to\eta}\right\} \cup \bigcup_{e\in\gamma(\eta)} \left\{H^{(1)}_{\eta'\to\eta,e},\, H^{(2)}_{\eta'\to\eta,e},\, H^{(3)}_{\eta'\to\eta,e}\right\}$$

is a partition of $\gamma(\eta')$.

Additionally, we define:

$$\forall \kappa \in \{1,2,3\}\,,\; H^{(\kappa)}_{\eta'\to\eta} := \bigcup_{e\in\gamma(\eta)} H^{(\kappa)}_{\eta'\to\eta,e}\,.$$

**Proof** For any $e \in \gamma(\eta)$, $\left\{H^{(1)}_{\eta'\to\eta,e},\, H^{(2)}_{\eta'\to\eta,e},\, H^{(3)}_{\eta'\to\eta,e}\right\}$ is a partition of:

$$H^{(4)}_{\eta'\to\eta,e} := H^{(1)}_{\eta'\to\eta,e} \cup H^{(2)}_{\eta'\to\eta,e} \cup H^{(3)}_{\eta'\to\eta,e} = \{e' \in \gamma(\eta') \mid r(e') \subset r(e) \;\text{ or }\; e' \in \chi_\eta(e)\}\,.$$

Since we have:

$$H^{(0)}_{\eta'\to\eta} = \left\{e' \in \gamma(\eta') \,\middle|\, \forall e \in \gamma(\eta),\;\; e' \notin H^{(4)}_{\eta'\to\eta,e}\right\},$$

there only remains to prove that the $H^{(4)}_{\eta'\to\eta,e}$ for $e \in \gamma(\eta)$ are mutually disjoint.

Let $e \in \gamma(\eta)$ and let $e' \in \gamma(\eta')$ such that $r(e') \subset r(e)$. There exists $p \in r(e') \setminus \{b(e'),\, f(e')\}$. From prop. 2.2, we get $p \notin \{b(e),\, f(e)\}$, thus $\forall \widetilde{e} \in \gamma(\eta) \setminus \{e\}$, $p \notin r(\widetilde{e})$, for $\gamma(\eta)$ is a graph. Therefore, $\forall \widetilde{e} \in \gamma(\eta) \setminus \{e\}$, $r(e') \not\subset r(\widetilde{e})$. Moreover, $e \in \chi_\eta(e)$ and for any $\widetilde{e} \in \gamma(\eta) \setminus \{e\}$, $\chi_\eta(\widetilde{e}) \neq \chi_\eta(e)$, so using the auxiliary result at the beginning of the proof of lemma 2.19, $e' \notin \chi_\eta(\widetilde{e})$. Hence, we have proved:

$$\forall e \neq \widetilde{e} \in \gamma(\eta),\, \forall e' \in \gamma(\eta'),\, \Big(r(e') \subset r(e) \;\Rightarrow\; e' \notin H^{(4)}_{\eta'\to\eta,\widetilde{e}}\Big)\,.$$

Now let $e \in \gamma(\eta)$ and let $e' \in \gamma(\eta')$ such that $e' \in \chi_\eta(e)$. Then, from the previous point, we get $\forall \widetilde{e} \in \gamma(\eta) \setminus \{e\}$, $r(e') \not\subset r(\widetilde{e})$. And, since the elements of $\mathfrak{F}(\eta)$ are disjoints and $\chi_\eta$ is bijective, we also have $\forall \widetilde{e} \in \gamma(\eta) \setminus \{e\}$, $r(e') \notin \chi_\eta(\widetilde{e})$. Therefore, we obtain the desired result:

$$\forall e \neq \widetilde{e} \in \gamma(\eta),\ H^{(4)}_{\eta'\to\eta,e} \cap H^{(4)}_{\eta'\to\eta,\widetilde{e}} = \varnothing\,.$$

□

**Proposition 3.5** Let $\eta \preccurlyeq \eta' \in \mathcal{L}$ and $e \in \gamma(\eta)$. We have $\{e\} \preccurlyeq \gamma(\eta')$, and making use of prop. 3.2, we define:

$$n_{\eta'\to\eta,e} := n_{\gamma(\eta')\to e}\,,\quad a_{\eta'\to\eta,e} := a_{\gamma(\eta')\to e} \quad \& \quad \epsilon_{\eta'\to\eta,e} := \epsilon_{\gamma(\eta')\to e}\,.$$

We then have:

$$\epsilon_{\eta'\to\eta,e}(1) = +1\,,\quad H^{(1)}_{\eta'\to\eta,e} = \{a_{\eta'\to\eta,e}(1)\} \quad \& \quad H^{(2)}_{\eta'\to\eta,e} = \{a_{\eta'\to\eta,e}(k) \mid k > 1\}\,, \tag{3.5.1}$$

therefore $n_{\eta'\to\eta,e} = \#H^{(2)}_{\eta'\to\eta,e} + 1$ and $a_{\eta'\to\eta,e}$ induces a bijection $\{1,\ldots,n_{\eta'\to\eta,e}\} \to H^{(1)}_{\eta'\to\eta,e} \cup H^{(2)}_{\eta'\to\eta,e}$.

Also, for any $F \in \mathfrak{F}(\eta)$, we have, using the notations of prop. 3.3 with $\overline{F} = F^\perp \circ F$, $\lambda(\eta') \in \mathcal{L}_{\text{profls}/\overline{F}}$ and:

$$H_{\lambda(\eta')\to\overline{F}} = H^{(1,3)}_{\eta'\to\eta,F} := \left\{F' \in \mathfrak{F}(\eta') \;\middle|\; \chi^{-1}_{\eta'}(F') \in F\right\} = \left\{\chi_{\eta'}(e') \;\middle|\; e' \in H^{(1)}_{\eta'\to\eta,\chi^{-1}_\eta(F)} \cup H^{(3)}_{\eta'\to\eta,\chi^{-1}_\eta(F)}\right\}. \tag{3.5.2}$$

**Proof** Let $e \in \gamma(\eta)$. From eq. (3.2.3) $a_{\eta'\to\eta,e}$ induces a bijection into its image:

$$H_{\gamma(\eta')\to e} = H^{(1)}_{\eta'\to\eta,e} \cup H^{(2)}_{\eta'\to\eta,e}\,,$$

and from eq. (3.2.1) together with props. 2.8.5 and 2.8.6:

$$a_{\eta'\to\eta,e}(1)^{\epsilon_{\eta'\to\eta,e}(1)} \in \chi_\eta(e) \quad \& \quad \forall k > 1,\ a_{\eta'\to\eta,e}(k) \notin \chi_\eta(e).$$

Then, writing $a_{\eta'\to\eta,e}(1)^{\epsilon_{\eta'\to\eta,e}(1)} = e_{[p,p']}$, there exists $p'' \in r\left(e_{[p,p']}\right) \setminus \{p,p'\} \subset r(e) \setminus \{b(e), f(e)\}$ such that $e_{[p,p'']} \in \chi_\eta(e)$. This can only hold if $p = b(e)$, ie. $\epsilon_{\eta'\to\eta,e}(1) = +1$. Thus, we get:

$$a_{\eta'\to\eta,e}(1) \in H^{(1)}_{\eta'\to\eta,e} \quad \& \quad \forall k > 1,\ a_{\eta'\to\eta,e}(k) \in H^{(2)}_{\eta'\to\eta,e}.$$

Let $F \in \mathfrak{F}(\eta)$ and $\overline{F} = F^\perp \circ F$. For any $F' \in H_{\lambda(\eta')\to\overline{F}}$, we have $\chi^{-1}_{\eta'}(F') \in F' \subset \overline{F}$, hence there exist $e_1 \in F$ and $e_2 \in F^\perp$ such that $\chi^{-1}_{\eta'}(F') = e_2 \circ e_1$. But from prop. 2.8.6 together with $F_\urcorner(\eta') \subset F_\urcorner(\eta)$ (as in the proof of prop. 2.13), we have $e_2 \in F_\urcorner(\eta') \subset F_\urcorner(\eta)$, so $\chi^{-1}_{\eta'}(F') \in F$. Thus, we get $H_{\lambda(\eta')\to\overline{F}} \subset H^{(1,3)}_{\eta'\to\eta,F}$. Reciprocally, let $F' \in \mathfrak{F}(\eta')$ such that $\chi^{-1}_{\eta'}(F') \in F \subset \overline{F}$. Then, from eq. (3.3.1), there exist $e_1 \in F''$, for some $F'' \in H_{\lambda(\eta')\to\overline{F}}$, and $e_2 \in \overline{F}^\perp$, such that $\chi^{-1}_{\eta'}(F') = e_2 \circ e_1$. And since $\chi^{-1}_{\eta'}(F') \in F'$, we also have $e_1 \in F'$, therefore $F' = F'' \in H_{\lambda(\eta')\to\overline{F}}$ (for the elements of $\mathfrak{F}(\eta')$ are disjoint from prop. 2.11.1). This proves $H^{(1,3)}_{\eta'\to\eta,F} \subset H_{\lambda(\eta')\to\overline{F}}$, hence $H_{\lambda(\eta')\to\overline{F}} = H^{(1,3)}_{\eta'\to\eta,F}$. □

**Proposition 3.6** Let $\eta \preccurlyeq \eta' \in \mathcal{L}$ and define:

$$\mathcal{C}_{\eta'\to\eta} := \left\{h^{(0)} : H^{(0)}_{\eta'\to\eta} \to G\right\} \times \left\{h^{(2)} : H^{(2)}_{\eta'\to\eta} \to G\right\} \times \left\{h^{(3)} : H^{(3)}_{\eta'\to\eta} \to G\right\}.$$

Like $\mathcal{C}_\eta$, $\mathcal{C}_{\eta'\to\eta}$ is a finite-dimensional smooth manifold.

To any $h_{\eta'} \in \mathcal{C}_{\eta'}$ we associate maps $h_\eta : \gamma(\eta) \to G$, $h^{(0)}_{\eta'\to\eta} : H^{(0)}_{\eta'\to\eta} \to G$, $h^{(2)}_{\eta'\to\eta} : H^{(2)}_{\eta'\to\eta} \to G$, $h^{(3)}_{\eta'\to\eta} : H^{(3)}_{\eta'\to\eta} \to G$ by:

1. $\forall e' \in H^{(0)}_{\eta'\to\eta},\ h^{(0)}_{\eta'\to\eta}(e') := h_{\eta'}(e')$;
2. $\forall e \in \gamma(\eta),\ h_\eta(e) := \left(\prod_{k=2}^{n_{\eta'\to\eta,e}} [h_{\eta'} \circ a_{\eta'\to\eta,e}(k)]^{\epsilon_{\eta'\to\eta,e}(k)}\right) . [h_{\eta'} \circ a_{\eta'\to\eta,e}(1)]$

   with the convention that products of group elements are ordered from right to left:

   $$\forall g_1, \dots, g_n \in G,\ \prod_{k=1}^{n} g_k := g_n . \dots . g_1 ;$$
3. $\forall e' \in H^{(2)}_{\eta'\to\eta},\ h^{(2)}_{\eta'\to\eta}(e') := h_{\eta'}(e')$;
4. $\forall e \in \gamma(\eta),\ \forall e' \in H^{(3)}_{\eta'\to\eta,e},\ h^{(3)}_{\eta'\to\eta}(e') := h_{\eta'}(e') . \left(h_\eta(e)\right)^{-1}$ (with $h_\eta(e)$ from 3.6.2).

Then, the map $\varphi_{\eta'\to\eta} : h_{\eta'} \mapsto h^{(0)}_{\eta'\to\eta}, h^{(2)}_{\eta'\to\eta}, h^{(3)}_{\eta'\to\eta} ; h_\eta$ is a diffeomorphism $\mathcal{C}_{\eta'} \to \mathcal{C}_{\eta'\to\eta} \times \mathcal{C}_\eta$.

**Proof** $\varphi_{\eta'\to\eta}$ is smooth for $G$ is a Lie group. Next, for any $j_\eta \in \mathcal{C}_\eta$ and any $\left(j^{(0)}_{\eta'\to\eta}, j^{(2)}_{\eta'\to\eta}, j^{(3)}_{\eta'\to\eta}\right) \in \mathcal{C}_{\eta'\to\eta}$ we define a map $j_{\eta'}$ by:

5. $\forall e' \in H^{(0)}_{\eta'\to\eta},\ j_{\eta'}(e') := j^{(0)}_{\eta'\to\eta}(e')$;
6. $\forall e \in \gamma(\eta),\ \forall e' \in H^{(1)}_{\eta'\to\eta,e},\ j_{\eta'}(e') := \left(\prod_{k=n_{\eta'\to\eta,e}}^{2} \left[j^{(2)}_{\eta'\to\eta} \circ a_{\eta'\to\eta,e}(k)\right]^{-\epsilon_{\eta'\to\eta,e}(k)}\right) . j_\eta(e)$;
7. $\forall e' \in H^{(2)}_{\eta'\to\eta},\ j_{\eta'}(e') := j^{(2)}_{\eta'\to\eta}(e')$;
8. $\forall e \in \gamma(\eta),\ \forall e' \in H^{(3)}_{\eta'\to\eta,e},\ j_{\eta'}(e') := j^{(3)}_{\eta'\to\eta}(e') . j_\eta(e)$.

Having a partition of $\gamma(\eta')$ (prop. 3.4) ensures that $j_{\eta'}$ is well-defined and, again because $G$ is a Lie group, the map $\widetilde{\varphi}_{\eta'\to\eta} : j^{(0)}_{\eta'\to\eta}, j^{(2)}_{\eta'\to\eta}, j^{(3)}_{\eta'\to\eta} ; j_\eta \mapsto j_{\eta'}$ is a smooth map $\mathcal{C}_{\eta'\to\eta} \times \mathcal{C}_\eta \to \mathcal{C}_{\eta'}$. We can check that $\widetilde{\varphi}_{\eta'\to\eta} \circ \varphi_{\eta'\to\eta} = \mathrm{id}_{\mathcal{C}_{\eta'}}$ and $\varphi_{\eta'\to\eta} \circ \widetilde{\varphi}_{\eta'\to\eta} = \mathrm{id}_{\mathcal{C}_{\eta'\to\eta}\times\mathcal{C}_\eta}$, thus $\varphi_{\eta'\to\eta}$ is a diffeomorphism. □

In order for the previously defined factorization maps to provide a valid factorizing system, they should fulfill the three-spaces consistency condition [17, eq. (2.11.1) and fig. 2.2]: given three labels $\eta \preccurlyeq \eta' \preccurlyeq \eta''$, the variables discarded when going down in one step from $\eta''$ to $\eta$ should be the same as the ones discarded when going, in two successive steps, first from $\eta''$ to $\eta'$ and then from $\eta'$ to $\eta$. In other words, we need a map $\varphi_{\eta''\to\eta'\to\eta}$ identifying $\mathcal{C}_{\eta''\to\eta}$ with $\mathcal{C}_{\eta''\to\eta'} \times \mathcal{C}_{\eta'\to\eta}$, in agreement with the factorization maps $\varphi_{\eta''\to\eta}$, $\varphi_{\eta''\to\eta'}$ and $\varphi_{\eta'\to\eta}$. This will ensure that there is no ambiguity as to how the variables associated to the label $\eta$ are to be extracted from the phase space corresponding to $\eta''$.

To ascertain that this is indeed the case, we have to distinguish the different ways, for an edge $e''$ of $\eta''$, to be positioned with respect to the edges and faces of $\eta'$ and $\eta$: this yields 13 inequivalent possibilities, as depicted in fig. 3.4. Only one of them is compatible with $e''$ being of type '1' for transition $\eta'' \to \eta$, while, in the 12 others case, the group variable attached to $e''$ contributes to the variables of $\mathcal{C}_{\eta''\to\eta}$, in terms of which $\varphi_{\eta''\to\eta'\to\eta}$ then needs to be appropriately specified (see the points 3.7.5 to 3.7.10 of the proof below).

Note that we could have made use of [17, prop. 2.17] to obtain the three-spaces consistency of the factorization maps from a similar condition formulated at the level of the projections between the small phase spaces [17, def. 2.3 and fig. 2.1]: that these projections fulfills such a condition can indeed be read out from their expression (that will be given in prop. 3.8). Yet, using this result would require $G$ to be connected: a restriction that appears quite artificial when the factorization map $\varphi_{\eta'\to\eta}$ in prop. 3.6 has been expressed solely in terms of the group operations (multiplication and inverse). Preferably, the three-spaces consistency can be obtained in full generality directly at the level of the factorization maps: once the correct explicit expression for $\varphi_{\eta''\to\eta'\to\eta}$ has been deduced from the one for $\varphi_{\eta'\to\eta}$, it is a straightforward (albeit rather fastidious) check that [17, eq. (2.11.1)] holds.

**Theorem 3.7** Let $\eta \preccurlyeq \eta' \preccurlyeq \eta'' \in \mathcal{L}$. There exists a diffeomorphism $\varphi_{\eta''\to\eta'\to\eta} : \mathcal{C}_{\eta''\to\eta} \to \mathcal{C}_{\eta''\to\eta'} \times \mathcal{C}_{\eta'\to\eta}$ such that:

$$\left(\varphi_{\eta''\to\eta'\to\eta} \times \mathrm{id}_{\mathcal{C}_\eta}\right) \circ \varphi_{\eta''\to\eta} = \left(\mathrm{id}_{\mathcal{C}_{\eta''\to\eta'}} \times \varphi_{\eta'\to\eta}\right) \circ \varphi_{\eta''\to\eta'} \quad \text{(aka. [17, eq. (2.11.1)])}.$$

This provides a factorizing system of smooth manifolds $(\mathcal{L}, \mathcal{C}, \varphi)^\times$ [17, def. 2.15].

**Proof** Let $\eta \preccurlyeq \eta' \preccurlyeq \eta'' \in \mathcal{L}$. Let $e \in \gamma(\eta)$ and, for any $k \in \{1, \dots, n_{\eta'\to\eta,e} + 1\}$, define:

$$s^{(k)}_{\eta''\to\eta'\to\eta,e} := \sum_{r=1}^{k-1} n_{\eta''\to\eta', a_{\eta'\to\eta,e}(r)}\,.$$

Using prop. 2.3 together with the uniqueness of $a_{\eta''\to\eta,e}$ (prop. 3.2), we then get $n_{\eta''\to\eta,e} = s^{(n_{\eta'\to\eta,e}+1)}_{\eta''\to\eta'\to\eta,e}$ and, for any $k \in \{1, \dots, n_{\eta'\to\eta,e}\}$ and any $l \in \left\{s^{(k)}_{\eta''\to\eta'\to\eta,e} + 1,\ \dots,\ s^{(k+1)}_{\eta''\to\eta'\to\eta,e}\right\}$:

$$a_{\eta''\to\eta,e}(l) = \begin{cases} a_{\eta''\to\eta', a_{\eta'\to\eta,e}(k)}\left(l - s^{(k)}_{\eta''\to\eta'\to\eta,e}\right) & \text{if } k = 1 \text{ or } \epsilon_{\eta'\to\eta,e}(k) = +1 \\ a_{\eta''\to\eta', a_{\eta'\to\eta,e}(k)}\left(s^{(k+1)}_{\eta''\to\eta'\to\eta,e} + 1 - l\right) & \text{otherwise} \end{cases},$$

$$\epsilon_{\eta''\to\eta,e}(l) = \begin{cases} \epsilon_{\eta''\to\eta', a_{\eta'\to\eta,e}(k)}\left(l - s^{(k)}_{\eta''\to\eta'\to\eta,e}\right) & \text{if } k = 1 \text{ or } \epsilon_{\eta'\to\eta,e}(k) = +1 \\ -\epsilon_{\eta''\to\eta', a_{\eta'\to\eta,e}(k)}\left(s^{(k+1)}_{\eta''\to\eta'\to\eta,e} + 1 - l\right) & \text{otherwise} \end{cases}.$$

Next, for any $F \in \mathcal{F}(\eta)$, we have:

$$H^{(1,3)}_{\eta''\to\eta,F} = \bigcup_{F' \in H^{(1,3)}_{\eta'\to\eta,F}} H^{(1,3)}_{\eta''\to\eta',F'}$$

(the proof is similar to the one for the second part of prop. 3.5).

In particular, using eqs. (3.5.1) and (3.5.2), we then get, for any $e \in \gamma(\eta)$:

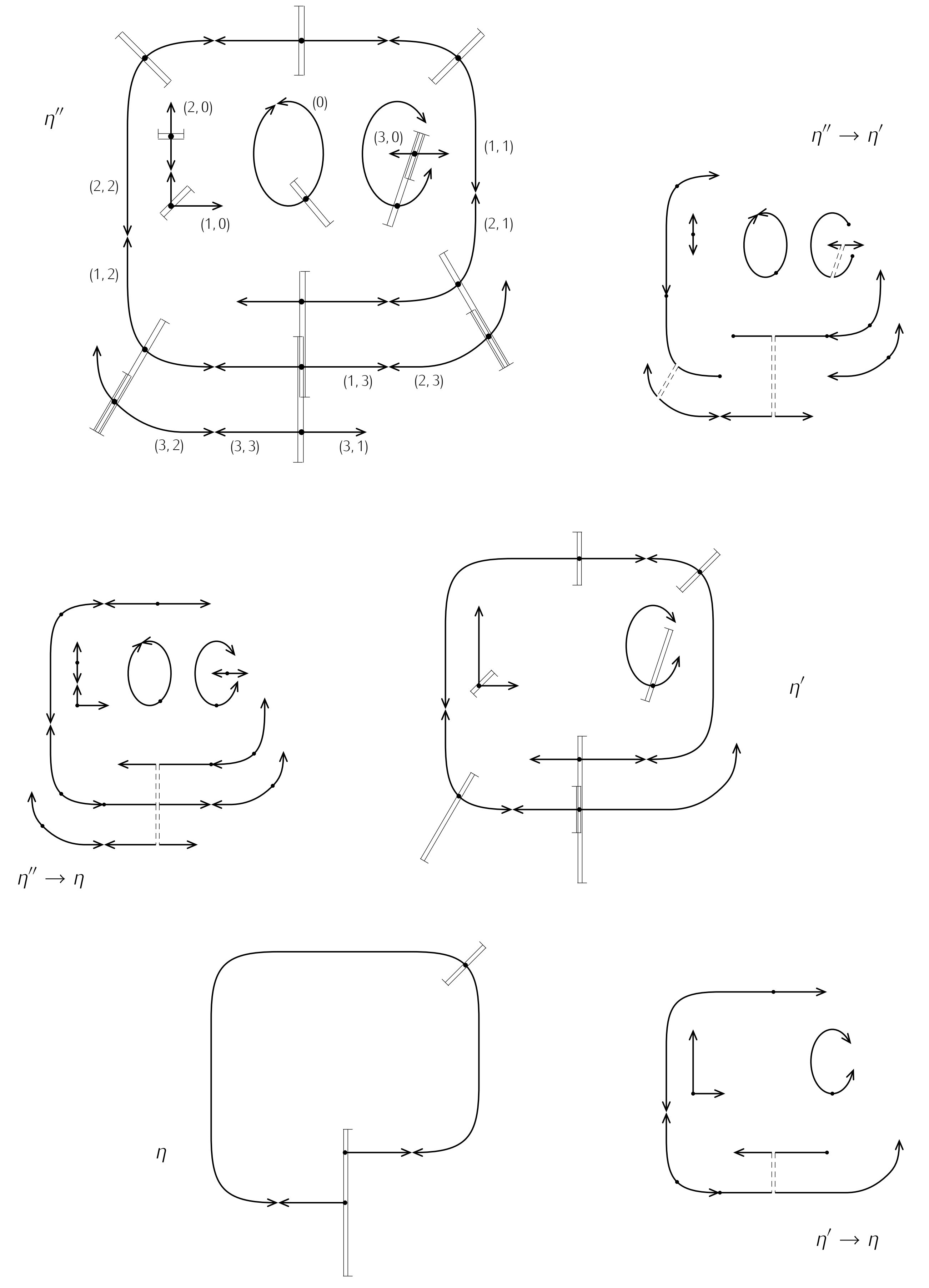

Figure 3.4 – Three-spaces consistency: in the illustration of $\eta''$, a tag $^{(\kappa',\kappa)}$ denotes an edge belonging to $H^{(\kappa')}_{\eta''\to\eta',e}$ for some $e$ in $H^{(\kappa)}_{\eta'\to\eta}$ and a tag $^{(0)}$ denotes an edge belonging to $H^{(0)}_{\eta''\to\eta'}$ (for better readability we have tagged only one edge of each type)

1. $H^{(0)}_{\eta''\to\eta} = H^{(0)}_{\eta''\to\eta'} \cup \bigcup_{e'\in H^{(0)}_{\eta'\to\eta}} H^{(4)}_{\eta''\to\eta',e'} \cup \bigcup_{e'\in H^{(2)}_{\eta'\to\eta}} H^{(3)}_{\eta''\to\eta',e'} \cup \bigcup_{e'\in H^{(3)}_{\eta'\to\eta}} H^{(2)}_{\eta''\to\eta',e'}$ ;

2. $H^{(1)}_{\eta''\to\eta,e} = \bigcup_{e'\in H^{(1)}_{\eta'\to\eta,e}} H^{(1)}_{\eta''\to\eta',e'}$ ;

3. $H^{(2)}_{\eta''\to\eta,e} = \bigcup_{e'\in H^{(1)}_{\eta'\to\eta,e}} H^{(2)}_{\eta''\to\eta',e'} \cup \bigcup_{e'\in H^{(2)}_{\eta'\to\eta,e}} \left(H^{(1)}_{\eta''\to\eta',e'} \cup H^{(2)}_{\eta''\to\eta',e'}\right)$ ;

4. $H^{(3)}_{\eta''\to\eta,e} = \bigcup_{e'\in H^{(1)}_{\eta'\to\eta,e}} H^{(3)}_{\eta''\to\eta',e'} \cup \bigcup_{e'\in H^{(3)}_{\eta'\to\eta,e}} \left(H^{(1)}_{\eta''\to\eta',e'} \cup H^{(3)}_{\eta''\to\eta',e'}\right)$ ;

Now, for any $\left(h^{(0)}_{\eta''\to\eta}, h^{(2)}_{\eta''\to\eta}, h^{(3)}_{\eta''\to\eta}\right) \in \mathcal{C}_{\eta''\to\eta}$, we define:

5. $\forall e' \in H^{(0)}_{\eta'\to\eta}\,,\;\; j^{(0)}_{\eta'\to\eta}(e') := \left(\prod_{k=2}^{n_{\eta''\to\eta',e'}} \left[h^{(0)}_{\eta''\to\eta} \circ a_{\eta''\to\eta',e'}(k)\right]^{\epsilon_{\eta''\to\eta',e'}(k)}\right) . \left[h^{(0)}_{\eta''\to\eta} \circ a_{\eta''\to\eta',e'}(1)\right]$

   (well-defined since $H^{(1)}_{\eta''\to\eta',e'} \cup H^{(2)}_{\eta''\to\eta',e'} \subset H^{(0)}_{\eta''\to\eta}$);

6. $\forall e' \in H^{(2)}_{\eta'\to\eta}\,,\;\; j^{(2)}_{\eta'\to\eta}(e') := \left(\prod_{k=2}^{n_{\eta''\to\eta',e'}} \left[h^{(2)}_{\eta''\to\eta} \circ a_{\eta''\to\eta',e'}(k)\right]^{\epsilon_{\eta''\to\eta',e'}(k)}\right) . \left[h^{(2)}_{\eta''\to\eta} \circ a_{\eta''\to\eta',e'}(1)\right]$

   (well-defined since $H^{(1)}_{\eta''\to\eta',e'} \cup H^{(2)}_{\eta''\to\eta',e'} \subset H^{(2)}_{\eta''\to\eta}$);

7. $\forall e' \in H^{(3)}_{\eta'\to\eta}\,,\;\; j^{(3)}_{\eta'\to\eta}(e') := \left(\prod_{k=2}^{n_{\eta''\to\eta',e'}} \left[h^{(0)}_{\eta''\to\eta} \circ a_{\eta''\to\eta',e'}(k)\right]^{\epsilon_{\eta''\to\eta',e'}(k)}\right) . \left[h^{(3)}_{\eta''\to\eta} \circ a_{\eta''\to\eta',e'}(1)\right]$

   (well-defined since $H^{(1)}_{\eta''\to\eta',e'} \subset H^{(3)}_{\eta''\to\eta}$ and $H^{(2)}_{\eta''\to\eta',e'} \cup H^{(0)}_{\eta''\to\eta',e'} \subset H^{(2)}_{\eta''\to\eta}$);

8. $\forall e'' \in H^{(0)}_{\eta''\to\eta'}\,,\;\; j^{(0)}_{\eta''\to\eta'}(e'') := h^{(0)}_{\eta''\to\eta}(e'')$

   (well-defined since $H^{(0)}_{\eta''\to\eta'} \subset H^{(0)}_{\eta''\to\eta}$);

9. $\forall e'' \in H^{(2)}_{\eta''\to\eta'}\,,\;\; j^{(2)}_{\eta''\to\eta'}(e'') := \begin{cases} h^{(0)}_{\eta''\to\eta}(e'') & \text{if } e'' \in H^{(0)}_{\eta''\to\eta} \\ h^{(2)}_{\eta''\to\eta}(e'') & \text{if } e'' \in H^{(2)}_{\eta''\to\eta} \end{cases}$

   (well-defined since $H^{(2)}_{\eta''\to\eta'} \subset H^{(0)}_{\eta''\to\eta} \cup H^{(2)}_{\eta''\to\eta}$);

10. $\forall e' \in \gamma(\eta'),\ \forall e'' \in H^{(3)}_{\eta''\to\eta',e'}$,

$$j^{(3)}_{\eta''\to\eta'}(e'') := \begin{cases} h^{(0)}_{\eta''\to\eta}(e'') \,.\, \left(j^{(0)}_{\eta'\to\eta}(e')\right)^{-1} & \text{if } e' \in H^{(0)}_{\eta''\to\eta} \\ h^{(0)}_{\eta''\to\eta}(e'') \,.\, \left(j^{(2)}_{\eta'\to\eta}(e')\right)^{-1} & \text{if } e' \in H^{(2)}_{\eta''\to\eta} \\ h^{(3)}_{\eta''\to\eta}(e'') \,.\, \prod_{k=2}^{n_{\eta'\to\eta,e}} \left(j^{(2)}_{\eta'\to\eta} \circ a_{\eta'\to\eta,e}(k)\right)^{\epsilon_{\eta'\to\eta,e}(k)} & \text{if } e' \in H^{(1)}_{\eta''\to\eta,e} \text{ (with } e \in \gamma(\eta)\text{)} \\ h^{(3)}_{\eta''\to\eta}(e'') \,.\, \left(j^{(3)}_{\eta'\to\eta}(e')\right)^{-1} & \text{if } e' \in H^{(3)}_{\eta''\to\eta} \end{cases}$$

   (using $j^{(0)}_{\eta'\to\eta}$, $j^{(2)}_{\eta'\to\eta}$ and $j^{(3)}_{\eta'\to\eta}$ from 3.7.5 to 3.7.7; well-defined since $H^{(3)}_{\eta''\to\eta',e'} \subset H^{(0)}_{\eta''\to\eta}$ in the first two cases, and $H^{(3)}_{\eta''\to\eta',e'} \subset H^{(3)}_{\eta''\to\eta}$ in the last two cases).

Then, the map $\varphi_{\eta''\to\eta'\to\eta} : h^{(0)}_{\eta''\to\eta}, h^{(2)}_{\eta''\to\eta}, h^{(3)}_{\eta''\to\eta} \mapsto j^{(0)}_{\eta''\to\eta'}, j^{(2)}_{\eta''\to\eta'}, j^{(3)}_{\eta''\to\eta'} ; j^{(0)}_{\eta'\to\eta}, j^{(2)}_{\eta'\to\eta}, j^{(3)}_{\eta'\to\eta}$ is smooth $\mathcal{C}_{\eta''\to\eta} \to \mathcal{C}_{\eta''\to\eta'} \times \mathcal{C}_{\eta'\to\eta}$.

Let $h_{\eta''} \in \mathcal{C}_{\eta''}$, and define:

11. $\left(h^{(0)}_{\eta''\to\eta'}, h^{(2)}_{\eta''\to\eta'}, h^{(3)}_{\eta''\to\eta'} ; h_{\eta'}\right) := \varphi_{\eta''\to\eta'}\left(h_{\eta''}\right) \in \mathcal{C}_{\eta''\to\eta'} \times \mathcal{C}_{\eta'}$;

12. $\left(h^{(0)}_{\eta'\to\eta}, h^{(2)}_{\eta'\to\eta}, h^{(3)}_{\eta'\to\eta} ; h_{\eta}\right) := \varphi_{\eta'\to\eta}\left(h_{\eta'}\right) \in \mathcal{C}_{\eta'\to\eta} \times \mathcal{C}_{\eta}$;

13. $\left(h^{(0)}_{\eta''\to\eta}, h^{(2)}_{\eta''\to\eta}, h^{(3)}_{\eta''\to\eta} ; j_{\eta}\right) := \varphi_{\eta''\to\eta}\left(h_{\eta''}\right) \in \mathcal{C}_{\eta''\to\eta} \times \mathcal{C}_{\eta}$;

14. $\left(j^{(0)}_{\eta''\to\eta'}, j^{(2)}_{\eta''\to\eta'}, j^{(3)}_{\eta''\to\eta'} ; j^{(0)}_{\eta'\to\eta}, j^{(2)}_{\eta'\to\eta}, j^{(3)}_{\eta'\to\eta}\right) := \varphi_{\eta''\to\eta'\to\eta'}\left(h^{(0)}_{\eta''\to\eta}, h^{(2)}_{\eta''\to\eta}, h^{(3)}_{\eta''\to\eta}\right)$.

Using the definitions of $\varphi_{\eta'\to\eta}$ (from prop. 3.6) and $\varphi_{\eta''\to\eta'\to\eta'}$, we can check that:

$$\left(j^{(0)}_{\eta''\to\eta'}, j^{(2)}_{\eta''\to\eta'}, j^{(3)}_{\eta''\to\eta'} ; j^{(0)}_{\eta'\to\eta}, j^{(2)}_{\eta'\to\eta}, j^{(3)}_{\eta'\to\eta} ; j_{\eta}\right) =$$
$$= \left(h^{(0)}_{\eta''\to\eta'}, h^{(2)}_{\eta''\to\eta'}, h^{(3)}_{\eta''\to\eta'} ; h^{(0)}_{\eta'\to\eta}, h^{(2)}_{\eta'\to\eta}, h^{(3)}_{\eta'\to\eta} ; h_{\eta}\right).$$

In other words, [17, eq. (2.11.1)] is fulfilled. But this also ensures that $\varphi_{\eta''\to\eta'\to\eta}$ is a diffeomorphism for $\varphi_{\eta''\to\eta'}$, $\varphi_{\eta'\to\eta}$, and $\varphi_{\eta''\to\eta}$ are diffeomorphisms (prop. 3.6). Together with the directedness of $\mathcal{L}$ proved in theorem 2.16, this yields the desired factorizing system [17, def. 2.15]. □

Finally, we wrap up the classical side of the construction by writing down the symplectic manifold $\mathcal{M}_\eta$ attached to a label $\eta$, the projection from $\mathcal{M}_{\eta'}$ into $\mathcal{M}_\eta$ (as prescribed by the factorization $\varphi_{\eta'\to\eta}$, for $\eta \preccurlyeq \eta'$), and the expression of the holonomy and flux observables over $\mathcal{M}_\eta$. These explicit formulas validate a posteriori the intuitive arguments that were repeatedly asserted above (on account of the underlying physical interpretation of the labels), the aim of all previous definitions being indeed to eventually arrive at props. 3.8 and 3.9 below. Note that, as announced at the beginning of the present section, the projections $\mathcal{M}_{\eta'} \to \mathcal{M}_\eta$ are of the form that were considered in [18, theorem 3.2] (making $\mathcal{C}_\eta$ into a Lie group via pointwise multiplication and inverse).

**Proposition 3.8** To any $\eta \in \mathcal{L}$ we associate the symplectic manifold $\mathcal{M}_\eta := T^*(\mathcal{C}_\eta)$ (with its canonical symplectic structure as a cotangent bundle), which we identify with $\mathcal{C}_\eta \times \mathcal{P}_\eta$ (def. 3.1) via:

$$\begin{array}{rccl} L_\eta : & \mathcal{M}_\eta & \to & \mathcal{C}_\eta \times \mathcal{P}_\eta \\ & h, p & \mapsto & h, \left(F \mapsto p \circ \left[T_1\left(t \mapsto R^{(F)}_{\eta,t} h\right)\right]\right) \end{array},$$

the map $R^{(F)}_{\eta,t} h \in \mathcal{C}_\eta$ being defined for $F \in \mathcal{F}(\eta)$, $t \in G$, and $h \in \mathcal{C}_\eta$ by:

$$\forall e \in \gamma(\eta),\ R^{(F)}_{\eta,t} h(e) := \begin{cases} h(e)\,.\,t & \text{if } \chi_\eta(e) = F \\ h(e) & \text{else} \end{cases}.$$

For any $\eta \preccurlyeq \eta' \in \mathcal{L}$, we define $\pi_{\eta'\to\eta} : \mathcal{M}_{\eta'} \to \mathcal{M}_\eta$ as $\pi_{\eta'\to\eta} = \widetilde{s}_{\eta'\to\eta} \circ \widetilde{\varphi}_{\eta'\to\eta}$, where $\widetilde{s}_{\eta'\to\eta} : T^*(\mathcal{C}_{\eta'\to\eta} \times \mathcal{C}_\eta) \approx T^*(\mathcal{C}_{\eta'\to\eta}) \times T^*(\mathcal{C}_\eta) \to T^*(\mathcal{C}_\eta)$ is the projection on the second Cartesian factor and $\widetilde{\varphi}_{\eta'\to\eta} : T^*(\mathcal{C}_{\eta'}) \to T^*(\mathcal{C}_{\eta'\to\eta} \times \mathcal{C}_\eta)$ is the cotangent lift of $\varphi_{\eta'\to\eta}$. Then, $(\mathcal{L}, \mathcal{M}, \pi)^\downarrow$ is a projective system of phase spaces [17, props. 2.16 and 2.13].

Moreover, we have:

$$L_\eta \circ \pi_{\eta'\to\eta} \circ L_{\eta'}^{-1} \;:\; \begin{array}{rcl} \mathcal{C}_{\eta'} \times \mathcal{P}_{\eta'} & \to & \mathcal{C}_\eta \times \mathcal{P}_\eta \\ h_{\eta'}\,,\, P_{\eta'} & \mapsto & h_\eta\,,\, P_\eta \end{array},$$

where $h_\eta$ and $P_\eta$ are given in terms of $h_{\eta'}$ and $P_{\eta'}$ by:

$$\forall e \in \gamma(\eta),\ h_\eta(e) = \left( \prod_{k=2}^{n_{\eta'\to\eta,e}} \left[h_{\eta'} \circ a_{\eta'\to\eta,e}(k)\right]^{\epsilon_{\eta'\to\eta,e}(k)} \right) . \left[h_{\eta'} \circ a_{\eta'\to\eta,e}(1)\right], \tag{3.8.1}$$

$$\text{and} \quad \forall F \in \mathcal{F}(\eta),\ P_\eta(F) = \sum_{F' \in H^{(1,3)}_{\eta'\to\eta,F}} P_{\eta'}(F')\,. \tag{3.8.2}$$

**Proof** For any $\eta$, $L_\eta$ is a diffeomorphism $\mathcal{M}_\eta \to \mathcal{C}_\eta \times \mathcal{P}_\eta$ by definition of $\chi_\eta$ (def. 2.14).

Let $(h_{\eta'}\,,\, p_{\eta'}) \in \mathcal{M}_{\eta'}$ and $(h_{\eta'}\,,\, P_{\eta'}) = L_{\eta'}(h_{\eta'}\,,\, p_{\eta'})$. We have:

$$\widetilde{\varphi}_{\eta'\to\eta}\left(h_{\eta'}\,;\, p_{\eta'}\right) = \left(h^{(0)}_{\eta'\to\eta}\,,\, h^{(2)}_{\eta'\to\eta}\,,\, h^{(3)}_{\eta'\to\eta}\,,\, h_\eta\,;\, p_{\eta'} \circ \left[T_{\varphi_{\eta'\to\eta}(h_{\eta'})}\varphi^{-1}_{\eta'\to\eta}\right]\right),$$

with $h^{(0)}_{\eta'\to\eta}\,,\, h^{(2)}_{\eta'\to\eta}\,,\, h^{(3)}_{\eta'\to\eta}\,,\, h_\eta$ constructed from $h_{\eta'}$ as in prop. 3.6 (in particular, $h_\eta$ is given by eq. (3.8.1)). Thus:

$$\pi_{\eta'\to\eta}\left(h_{\eta'}\,;\, p_{\eta'}\right) = \left(h_\eta\,;\, p_{\eta'} \circ \left[T_{h_\eta}\varphi^{-1}_{\eta'\to\eta}\left(h^{(0)}_{\eta'\to\eta}\,,\, h^{(2)}_{\eta'\to\eta}\,,\, h^{(3)}_{\eta'\to\eta}\,,\, \cdot\,\right)\right]\right),$$

and finally $L_\eta \circ \pi_{\eta'\to\eta}\left(h_{\eta'}\,;\, p_{\eta'}\right) = \left(h_\eta\,;\, P_\eta\right)$, where $P_\eta$ is given in terms of $h_{\eta'}\,,\, p_{\eta'}$ by:

$$\forall F \in \mathcal{F}(\eta),\ P_\eta(F) = p_{\eta'} \circ \left[T_1\, t \mapsto \varphi^{-1}_{\eta'\to\eta}\left(h^{(0)}_{\eta'\to\eta}\,,\, h^{(2)}_{\eta'\to\eta}\,,\, h^{(3)}_{\eta'\to\eta}\,,\, R^{(F)}_{\eta,t} h_\eta\right)\right].$$

Now, for any $F \in \mathcal{F}(\eta)$, $t \in G$ and $e' \in \gamma(\eta')$, we get, using the explicit expression for $\varphi^{-1}_{\eta'\to\eta}$ from the proof of prop. 3.6:

$$\varphi^{-1}_{\eta'\to\eta}\left(h^{(0)}_{\eta'\to\eta}\,,\, h^{(2)}_{\eta'\to\eta}\,,\, h^{(3)}_{\eta'\to\eta}\,,\, R^{(F)}_{\eta,t} h_\eta\right)(e') = \begin{cases} h_{\eta'}(e')\,.\,t & \text{if } \chi_{\eta'}(e') \in H^{(1,3)}_{\eta'\to\eta,F} \\ h_{\eta'}(e') & \text{else} \end{cases}.$$

Therefore, $\forall F \in \mathcal{F}(\eta),\ P_\eta(F) = \displaystyle\sum_{F' \in H^{(1,3)}_{\eta'\to\eta,F}} P_{\eta'}(F')\,.$ □

**Proposition 3.9** Let $e \in \mathcal{L}_{\text{edges}}$ and let $\eta \in \mathcal{L}$ such that $\gamma(\eta) \in \mathcal{L}_{\text{graphs}/e}$ (prop. 3.2). For any $\delta \in C^\infty(G, \mathbb{R})$ (with $C^\infty(G, \mathbb{R})$ the set of smooth functions $G \to \mathbb{R}$), we define $\mathrm{h}^{(e,\delta)}_\eta \in C^\infty(\mathcal{M}_\eta\,,\, \mathbb{R})$ by:

$$\forall (h, p) \in \mathcal{M}_\eta\,,\quad \mathrm{h}^{(e,\delta)}_\eta(h, p) := \delta\left( \prod_{k=1}^{n_{\gamma(\eta)\to e}} \left[h \circ a_{\gamma(\eta)\to e}(k)\right]^{\epsilon_{\gamma(\eta)\to e}(k)} \right).$$

Then, for any $\eta, \eta' \in \mathcal{L}$ such that $\gamma(\eta)$, $\gamma(\eta') \in \mathcal{L}_{\text{graphs}/e}$, we have:

$$\forall \delta \in C^\infty(G, \mathbb{R}),\ \mathrm{h}^{(e,\delta)}_\eta \sim \mathrm{h}^{(e,\delta)}_{\eta'}\,,$$

with the equivalence relation of [17, eq. (2.4.1)]. Moreover $\{\eta \in \mathcal{L} \mid \gamma(\eta) \in \mathcal{L}_{\text{graphs}/e}\} \neq \varnothing$, thus we can associate to any $\delta \in C^\infty(G, \mathbb{R})$ a well-defined observable $\mathsf{h}^{(e,\delta)} \in \mathcal{O}^{\downarrow}_{(\mathcal{L},\mathcal{M},\pi)}$ [17, def. 2.4].

Let $\overline{F} \in \mathcal{L}_{\text{faces}}$ and let $\eta \in \mathcal{L}$ such that $\lambda(\eta) \in \mathcal{L}_{\text{profls}/\overline{F}}$ (prop. 3.3). For any $u \in \mathfrak{g}$, we define $\mathsf{P}^{(\overline{F},u)}_\eta \in C^\infty(\mathcal{M}_\eta\,, \mathbb{R})$ by:

$$\forall (h,p) \in \mathcal{M}_\eta\,,\ \ \mathsf{P}^{(\overline{F},u)}_\eta(h,p) := \sum_{F' \in H_{\lambda(\eta)\to\overline{F}}} p \circ \left[ T_1 \left( t \mapsto R^{(F')}_{\eta,t} h \right) \right](u)\,.$$

Then, for any $\eta, \eta' \in \mathcal{L}$ such that $\lambda(\eta)$, $\lambda(\eta') \in \mathcal{L}_{\text{profls}/\overline{F}}$, we have:

$$\forall u \in \mathfrak{g},\ \mathsf{P}^{(\overline{F},u)}_\eta \sim \mathsf{P}^{(\overline{F},u)}_{\eta'}\,.$$

And since $\left\{ \eta \in \mathcal{L} \;\middle|\; \lambda(\eta) \in \mathcal{L}_{\text{profls}/\overline{F}} \right\} \neq \varnothing$, we can associate to any $u \in \mathfrak{g}$ a well-defined observable $\mathsf{P}^{(\overline{F},u)} \in \mathcal{O}^{\downarrow}_{(\mathcal{L},\mathcal{M},\pi)}$.

**Proof** Let $e \in \mathcal{L}_{\text{edges}}$. By chaining lemma 2.17 to 2.20, there exists $(\gamma, \lambda) \in \mathcal{L}$ such that $\{e\} \preccurlyeq \gamma$ (and $[\varnothing]_{\text{profl}} \preccurlyeq \lambda$ where $[\,\cdot\,]_{\text{profl}}$ denotes the equivalence class in $\mathcal{L}_{\text{profls}}$), hence $\{\eta \in \mathcal{L} \mid \gamma(\eta) \in \mathcal{L}_{\text{graphs}/e}\} \neq \varnothing$. Let $\eta$ such that $\gamma(\eta) \in \mathcal{L}_{\text{graphs}/e}$ and let $\eta' \succcurlyeq \eta$. Then, $\gamma(\eta') \in \mathcal{L}_{\text{graphs}/e}$. Moreover, we can express $a_{\gamma(\eta')\to e}$ in terms of $a_{\gamma(\eta)\to e}$ and of the $a_{\eta'\to\eta,e'}$ for $e'$ in the image of $a_{\gamma(\eta)\to e}$ (like we did above in the proof of theorem 3.7). Together with eq. (3.8.1), we readily check that:

$$\forall \delta \in C^\infty(G, \mathbb{R}),\ \mathsf{h}^{(e,\delta)}_\eta \circ \pi_{\eta'\to\eta} = \mathsf{h}^{(e,\delta)}_{\eta'}\,.$$

Finally, for any $\eta, \eta'$ such that $\gamma(\eta), \gamma(\eta') \in \mathcal{L}_{\text{graphs}/e}$, there exists $\eta'' \succcurlyeq \eta, \eta'$, hence:

$$\forall \delta \in C^\infty(G, \mathbb{R}),\ \mathsf{h}^{(e,\delta)}_\eta \circ \pi_{\eta''\to\eta} = \mathsf{h}^{(e,\delta)}_{\eta''} = \mathsf{h}^{(e,\delta)}_{\eta'} \circ \pi_{\eta''\to\eta'}\,,$$

ie. for any $\delta \in C^\infty(G, \mathbb{R})$, $\mathsf{h}^{(e,\delta)}_\eta \sim \mathsf{h}^{(e,\delta)}_{\eta'}$.

Let $\lambda \in \mathcal{L}_{\text{profls}}$, $F \in \mathcal{F}(\lambda)$ and $\overline{F} = F^\perp \circ F$. Again, there exists $(\gamma', \lambda') \in \mathcal{L}$ such that $\lambda \preccurlyeq \lambda'$ (and $\varnothing \preccurlyeq \gamma'$), hence $\left\{ \eta \in \mathcal{L} \;\middle|\; \lambda(\eta) \in \mathcal{L}_{\text{profls}/\overline{F}} \right\} \neq \varnothing$. Let $\eta$ such that $\lambda(\eta) \in \mathcal{L}_{\text{profls}/\overline{F}}$ and let $\eta' \succcurlyeq \eta$. Then, $\lambda(\eta') \in \mathcal{L}_{\text{profls}/\overline{F}}$ and we have (through a reasoning similar to the proof for the second part of prop. 3.5):

$$H_{\lambda(\eta')\to\overline{F}} = \bigcup_{F' \in H_{\lambda(\eta)\to\overline{F}}} H^{(1,3)}_{\eta'\to\eta,F'}\,. \tag{3.9.1}$$

Combining this with eq. (3.8.2), we get:

$$\forall u \in \mathfrak{g},\ \mathsf{P}^{(\overline{F},u)}_\eta \circ \pi_{\eta'\to\eta} = \mathsf{P}^{(\overline{F},u)}_{\eta'}\,.$$

Like above, this ensures that for any $\eta, \eta'$ such that $\lambda(\eta), \lambda(\eta') \in \mathcal{L}_{\text{profls}/\overline{F}}$ and for any $u \in \mathfrak{g}$, $\mathsf{P}^{(\overline{F},u)}_\eta \sim \mathsf{P}^{(\overline{F},u)}_{\eta'}$. □

Quantizing the classical formalism set up above is then a straightforward application of the prescriptions for quantization in the position representation that were detailed in [18, props. 3.3 and 3.4].

**Proposition 3.10** Let $\mu$ be a right-invariant Haar measure on $G$. For $\eta \in \mathcal{L}$, we can, using the natural identification $\mathcal{C}_\eta \approx G^{\#\gamma(\eta)}$, define a measure $\mu_\eta \approx \mu^{\#\gamma(\eta)}$ on $\mathcal{C}_\eta$ (actually the identification $\mathcal{C}_\eta \approx G^{\#\gamma(\eta)}$ is not unique, since it depends on an ordering of the edges in $\gamma(\eta)$; yet the measure $\mu_\eta$ is independent of this choice).
Then, there exists a family of smooth measures $\mu_{\eta'\to\eta}$ on each $\mathcal{C}_{\eta'\to\eta}$ for $\eta \preccurlyeq \eta' \in \mathcal{L}$, such that $(\mathcal{L}, (\mathcal{C}, \mu), \varphi)^\times$ is a factorizing system of measured manifolds [18, def. 3.1]. Defining:

1. $\forall \eta \in \mathcal{L}$, $\mathcal{H}_\eta := L_2(\mathcal{C}_\eta, d\mu_\eta)$ ;
2. $\forall \eta \preccurlyeq \eta' \in \mathcal{L}$, $\mathcal{H}_{\eta'\to\eta} := L_2(\mathcal{C}_{\eta'\to\eta}, d\mu_{\eta'\to\eta})$ and:

$$\begin{array}{rcll} \Phi_{\eta'\to\eta} : & \mathcal{H}_{\eta'} & \to & \mathcal{H}_{\eta'\to\eta} \otimes \mathcal{H}_\eta \approx L_2(\mathcal{C}_{\eta'\to\eta} \times \mathcal{C}_\eta, d\mu_{\eta'\to\eta} \times d\mu_\eta) \\ & \psi & \mapsto & \psi \circ \varphi^{-1}_{\eta'\to\eta} \end{array};$$

there exists a family of Hilbert spaces isomorphisms $\left(\Phi_{\eta''\to\eta'\to\eta}\right)_{\eta \preccurlyeq \eta' \preccurlyeq \eta''}$ such that $(\mathcal{L}, \mathcal{H}, \Phi)^\otimes$ is a projective system of quantum state spaces [18, def. 2.1].

**Proof** The right-invariant Haar measure $\mu$ comes from a smooth, right-invariant volume form on $G$. Hence, for any $\eta$ there exists a smooth, right-invariant volume form $\omega_\eta$ on $\mathcal{C}_\eta$ such that $\mu_\eta$ is the measure associated to $\omega_\eta$ ($\omega_\eta$ is not unique, as there is a freedom in the orientation of $\mathcal{C}_\eta$; however, it is sufficient here to just pick an $\omega_\eta$ for each $\mathcal{C}_\eta$). In particular, $\mu_\eta$ is therefore a smooth measure on $\mathcal{C}_\eta$. Defining:

$$\begin{array}{rcll} R_{\eta,h} : & \mathcal{C}_\eta & \to & \mathcal{C}_\eta \\ & j & \mapsto & (e \mapsto j(e) \,.\, h(e)) \end{array},$$

the right-invariance of $\omega_\eta$ can be written as:

$$\forall h \in \mathcal{C}_\eta,\ R^*_{\eta,h}\, \omega_\eta = \omega_\eta \,. \tag{3.10.1}$$

Let $\eta \preccurlyeq \eta' \in \mathcal{L}$, $d_\eta := \dim \mathcal{C}_\eta$, $d_{\eta'} := \dim \mathcal{C}_{\eta'}$ and $d_{\eta'\to\eta} := d_{\eta'} - d_\eta$. We choose a basis $u_1, \dots, u_{d_\eta}$ in $T_{\mathbf{1}}(\mathcal{C}_\eta)$ (where $\mathbf{1} \in \mathcal{C}_\eta$ is the map $e \mapsto \mathbf{1} \in G$) and we define a smooth volume form $\omega_{\eta'\to\eta}$ on $\mathcal{C}_{\eta'\to\eta}$ by:

$$\forall q \in \mathcal{C}_{\eta'\to\eta},\ \forall w_1, \dots, w_{d_{\eta'\to\eta}} \in T_q(\mathcal{C}_{\eta'\to\eta}),$$

$$\omega_{\eta'\to\eta,q}(w_1, \dots, w_{d_{\eta'\to\eta}}) := \frac{\left[\varphi^{-1,*}_{\eta'\to\eta}\, \omega_{\eta'}\right]_{(q,\mathbf{1})} \Big((w_1, 0), \dots, (w_{d_{\eta'\to\eta}}, 0), (0, u_1), \dots, (0, u_{d_\eta})\Big)}{\omega_{\eta,\mathbf{1}}(u_1, \dots, u_{d_\eta})}\,.$$

Now, for any $h \in \mathcal{C}_\eta$, we define $\widetilde{h} \in \mathcal{C}_{\eta'}$ by:

$$\forall e \in \gamma(\eta),\ \forall e' \in H^{(1)}_{\eta'\to\eta,e} \cup H^{(3)}_{\eta'\to\eta,e},\ \widetilde{h}(e') = h(e)\,;$$

$$\text{and} \quad \forall e' \in H^{(0)}_{\eta'\to\eta} \cup H^{(2)}_{\eta'\to\eta},\ \widetilde{h}(e') = \mathbf{1}\,.$$

Using the explicit expression for $\varphi^{-1}_{\eta'\to\eta}$ from the proof of prop. 3.6, we can check that:

$$\forall h \in \mathcal{C}_\eta,\ R_{\eta',\widetilde{h}} \circ \varphi^{-1}_{\eta'\to\eta} = \varphi^{-1}_{\eta'\to\eta} \circ \left(\mathrm{id}_{\mathcal{C}_{\eta'\to\eta}} \times R_{\eta,h}\right).$$

Applying eq. (3.10.1), we thus get, for any $h \in \mathcal{C}_\eta$ :

$$\omega_{\eta,\mathbf{1}}(u_1, \ldots, u_{d_\eta}) = \omega_{\eta,h}(U_{1,h}, \ldots, U_{d_\eta,h})$$

and $\forall q \in \mathcal{C}_{\eta' \to \eta}$, $\forall w_1, \ldots, w_{d_{\eta' \to \eta}} \in T_q(\mathcal{C}_{\eta' \to \eta})$,

$$\left[\varphi^{-1,*}_{\eta' \to \eta}\, \omega_{\eta'}\right]_{(q,\mathbf{1})} \Big((w_1, 0), \ldots, (w_{d_{\eta' \to \eta}}, 0), (0, u_1), \ldots, (0, u_{d_\eta})\Big)$$

$$= \left[\varphi^{-1,*}_{\eta' \to \eta}\, \omega_{\eta'}\right]_{(q,h)} \Big((w_1, 0), \ldots, (w_{d_{\eta' \to \eta}}, 0), (0, U_{1,h}), \ldots, (0, U_{d_\eta,h})\Big),$$

where $\forall i \in \{1, \ldots, d_\eta\}$, $U_{i,h} := [T_{\mathbf{1}} R_{\eta,h}](u_i) \in T_h(\mathcal{C}_\eta)$. And since $U_{1,h}, \ldots, U_{d_\eta,h}$ is a basis of $T_h(\mathcal{C}_\eta)$, we get:

$$\varphi^{-1,*}_{\eta' \to \eta}\, \omega_{\eta'} = \omega_{\eta' \to \eta} \times \omega_\eta .$$

Therefore, defining, for any $\eta \preccurlyeq \eta'$, $\mu_{\eta' \to \eta}$ to be the smooth measure on $\mathcal{C}_{\eta' \to \eta}$ associated to the volume form $\omega_{\eta' \to \eta}$, we have:

$$\forall \eta \preccurlyeq \eta' \in \mathcal{L},\ \varphi^{-1,*}_{\eta' \to \eta}\, \mu_{\eta'} = \mu_{\eta' \to \eta} \times \mu_\eta ,$$

and, using the 3-spaces consistency condition [17, eq. (2.11.1)] in the factorizing system $(\mathcal{L}, \mathcal{C}, \varphi)^\times$, this also implies:

$$\forall \eta \preccurlyeq \eta' \preccurlyeq \eta'' \in \mathcal{L},\ \varphi^{-1,*}_{\eta'' \to \eta' \to \eta}\, \mu_{\eta'' \to \eta} = \mu_{\eta'' \to \eta'} \times \mu_{\eta' \to \eta} .$$

Thus, $(\mathcal{L}, (\mathcal{C}, \mu), \varphi)^\times$ is a factorizing system of measured manifolds, from which a projective system of quantum state spaces $(\mathcal{L}, \mathcal{H}, \Phi)^\otimes$ can be constructed as described in [18, prop. 3.3]. □

**Proposition 3.11** Let $e \in \mathcal{L}_{\text{edges}}$ and let $\eta \in \mathcal{L}$ such that $\gamma(\eta) \in \mathcal{L}_{\text{graphs}/e}$. For any $\delta \in C^\infty(G, \mathbb{R})$, $\mathrm{h}^{(e,\delta)}_\eta$ fulfills the quantization condition [18, eq. (3.4.1)] and can be quantized as a densely defined, essentially self-adjoint operator $\widehat{\mathrm{h}^{(e,\delta)}_\eta}$ on $\mathcal{H}_\eta$ (with dense domain $\mathcal{D}_\eta$), given by:

$$\forall \psi \in \mathcal{D}_\eta,\ \forall h \in \mathcal{C}_\eta,\ \big(\widehat{\mathrm{h}^{(e,\delta)}_\eta}\psi\big)(h) := \delta\left(\prod_{k=1}^{n_{\gamma(\eta)\to e}} \left[h \circ a_{\gamma(\eta)\to e}(k)\right]^{\epsilon_{\gamma(\eta)\to e}(k)}\right) \psi(h) . \tag{3.11.1}$$

Moreover, we have for any $\eta, \eta' \in \mathcal{L}$ such that $\gamma(\eta), \gamma(\eta') \in \mathcal{L}_{\text{graphs}/e}$ :

$$\forall \delta \in C^\infty(G, \mathbb{R}),\ \widehat{\mathrm{h}^{(e,\delta)}_\eta} \sim \widehat{\mathrm{h}^{(e,\delta)}_{\eta'}},$$

with the equivalence relation of [18, eq. (2.3.2)], thus we can associate to any $e \in \mathcal{L}_{\text{edges}}$ and any $\delta \in C^\infty(G, \mathbb{R})$ a well-defined observable $\widehat{\mathrm{h}^{(e,\delta)}} \in \mathcal{O}^\otimes_{(\mathcal{L},\mathcal{H},\Phi)}$ [18, prop. 2.5].

Let $\overline{F} \in \mathcal{L}_{\text{faces}}$ and let $\eta \in \mathcal{L}$ such that $\lambda(\eta) \in \mathcal{L}_{\text{profls}/\overline{F}}$. For any $u \in \mathfrak{g}$, $\mathrm{P}^{(\overline{F},u)}_\eta$ fulfills the quantization condition and can be quantized as a densely defined, essentially self-adjoint operator $\widehat{\mathrm{P}^{(\overline{F},u)}_\eta}$ on $\mathcal{H}_\eta$, given by:

$$\forall \psi \in \mathcal{D}_\eta,\ \forall h \in \mathcal{C}_\eta,\ \big(\widehat{\mathrm{P}^{(\overline{F},u)}_\eta}\psi\big)(h) := i\left[T_{\mathbf{1}}\, t \mapsto \psi\left(R^{(\overline{F})}_{\eta,t} h\right)\right](u), \tag{3.11.2}$$

the map $R^{(\overline{F})}_{\eta,t} h \in \mathcal{C}_\eta$ being defined for $t \in G$ and $h \in \mathcal{C}_\eta$ by:

$$\forall e \in \gamma(\eta),\ R^{(\overline{F})}_{\eta,t} h(e) := \begin{cases} h(e)\,.\,t & \text{if } e \in \overline{F} \\ h(e) & \text{else} \end{cases} .$$

Moreover, we have for any $\eta, \eta' \in \mathcal{L}$ such that $\lambda(\eta), \lambda(\eta') \in \mathcal{L}_{\text{profls}/\overline{F}}$ :

$$\forall u \in \mathfrak{g},\ \widehat{\mathrm{P}^{(\overline{F},u)}_{\eta}} \sim \widehat{\mathrm{P}^{(\overline{F},u)}_{\eta'}},$$

thus we can associate to any $\overline{F} \in \mathcal{L}_{\text{faces}}$ and any $u \in \mathfrak{g}$ a well-defined observable $\widehat{\mathrm{P}^{(\overline{F},u)}} \in \mathcal{O}^{\otimes}_{(\mathcal{L},\mathcal{H},\Phi)}$ .

**Proof** *Hamiltonian vector field projected on* $\mathcal{C}_\eta$. Let $\eta \in \mathcal{L}$. Using left-translated exponential coordinates around a point $h \in \mathcal{C}_\eta$ (cf. the proof of [18, theorem 3.2]), we can show that the symplectic structure $\Omega_\eta$ on $\mathcal{M}_\eta$ is given by:

$$\forall h \in \mathcal{C}_\eta\,,\ \forall P \in \mathcal{P}_\eta\,,\ \ \forall v, v' \in T_h(\mathcal{C}_\eta),\ \forall w, w' \in \mathcal{P}_\eta\,,\ \ \left(L^{-1,*}_\eta \Omega_\eta\right)_{h,P} \left((v, w), (v', w')\right) =$$

$$= \sum_{e \in \gamma(\eta)} w'(\chi_\eta(e)) \circ \ell^{(e)}_{\eta,h^{-1}}(v) - w(\chi_\eta(e)) \circ \ell^{(e)}_{\eta,h^{-1}}(v') + P(\chi_\eta(e)) \left(\left[\ell^{(e)}_{\eta,h^{-1}}(v),\ \ell^{(e)}_{\eta,h^{-1}}(v')\right]_{\mathfrak{g}}\right),$$

where $L_\eta : \mathcal{M}_\eta \to \mathcal{C}_\eta \times \mathcal{P}_\eta$ was defined in prop. 3.8 and for any $e \in \gamma(\eta)$:

$$\forall h \in \mathcal{C}_\eta\,,\ \ell^{(e)}_{\eta,h^{-1}} := \left[T_h \ \ j \mapsto h(e)^{-1}\,.\,j(e)\right] : T_h\mathcal{C}_\eta \to \mathfrak{g}\,.$$

Let $\mathrm{r} \in C^\infty(\mathcal{C}_\eta\,,\,\mathbb{R})$. For any $(h, P) \in \mathcal{C}_\eta \times \mathcal{P}_\eta$ and any $w \in \mathcal{P}_\eta$, we have:

$$\left[d\mathrm{r} \circ L^{-1}_\eta\right]_{(h,P)} (0, w) =$$

$$= \Omega_{\eta, L^{-1}_\eta(h,P)} \left(X_{\mathrm{r}, L^{-1}_\eta(h,P)}\,,\ \left[T_{(h,P)} L^{-1}_\eta\right](0, w)\right)$$

$$= \left(L^{-1,*}_\eta \Omega_\eta\right)_{h,P} \left(\left[T_{L^{-1}_\eta(h,P)} L_\eta\right]\left(X_{\mathrm{r}, L^{-1}_\eta(h,P)}\right)\ ,\ (0, w)\right)$$

$$= \sum_{e \in \gamma(\eta)} w\left(\chi_\eta(e)\right) \circ \ell^{(e)}_{\eta,h^{-1}} \left(\left[T_{L^{-1}_\eta(h,P)} \gamma_\eta\right]\left(X_{\mathrm{r}, L^{-1}_\eta(h,P)}\right)\right), \tag{3.11.3}$$

where $X_{\mathrm{r}}$ is the Hamiltonian vector field of r and $\gamma_\eta : \mathcal{M}_\eta \to \mathcal{C}_\eta$ is the bundle projection in $\mathcal{M}_\eta = T^*(\mathcal{C}_\eta)$.

Now, for any $h, j \in \mathcal{C}_\eta$ and any $e \in \gamma(\eta)$, we have:

$$\left(t \mapsto R^{(\chi_\eta(e))}{}_{\eta,t} h\right) \circ \left(j' \mapsto h(e)^{-1}\,.\,j'(e)\right)(j) = R^{(\chi_\eta(e))}{}_{\eta, h(e)^{-1}\,.\,j(e)} h : e' \mapsto \begin{cases} j(e') & \text{if } e' = e \\ h(e') & \text{else} \end{cases} .$$

Therefore, we get for any $h \in \mathcal{C}_\eta$ :

$$\sum_{e \in \gamma(\eta)} \left[T_1 \ \ t \mapsto R^{(\chi_\eta(e))}{}_{\eta,t} h\right] \circ \ell^{(e)}_{\eta,h^{-1}} = \mathrm{id}_{T_h(\mathcal{C}_\eta)}\,.$$

For any $h \in \mathcal{C}_\eta$ and any $\upsilon \in T^*_h(\mathcal{C}_\eta)$, we define $w^{(h)}_\upsilon \in \mathcal{P}_\eta$ by:

$$\forall F \in \mathcal{F}(\eta),\ w^{(h)}_{\upsilon}(F) := \upsilon \circ \left[T_1 \ \ t \mapsto R^{(F)}_{\eta,t} h\right],$$

so that eq. (3.11.3) becomes:

$$\forall (h, P) \in \mathcal{C}_\eta \times \mathcal{P}_\eta\,,\ \upsilon \circ \left[T_{L_\eta^{-1}(h,P)} \gamma_\eta\right] \left(X_{\mathrm{r}, L_\eta^{-1}(h,P)}\right) = \left[d\mathrm{r} \circ L_\eta^{-1}\right]_{(h,P)} (0, w^{(h)}_{\upsilon})\,.$$

Thus, r fulfills the quantization condition [18, eq. (3.4.1)] if and only if there exists a smooth vector field $\overline{X}_{\mathrm{r}}$ on $\mathcal{C}_\eta$ such that:

$$\forall (h, P) \in \mathcal{C}_\eta \times \mathcal{P}_\eta\,,\ \forall \upsilon \in T^*_h(\mathcal{C}_\eta),\ \left[d\mathrm{r} \circ L_\eta^{-1}\right]_{(h,P)} (0, w^{(h)}_{\upsilon}) = \upsilon\left(\overline{X}_{\mathrm{r},h}\right).$$

If this is the case, we can then define $\widehat{\mathrm{r}}$ as a densely defined operator on $\mathcal{H}_\eta$ by:

$$\forall \psi \in \mathcal{D}_\eta\,,\ \forall h \in \mathcal{C}_\eta\,,\ \left(\widehat{\mathrm{r}}\,\psi\right)(h) := \left(\mathrm{r}(h,0) + \frac{i}{2}\,\mathrm{div}_{\mu_\eta} \overline{X}_{\mathrm{r}}(h)\right)\ \psi(h) + i\ \left[d\mathrm{r} \circ L_\eta^{-1}\right]_{(h,P)} (0, w^{(h)}_{d\psi_h})\,,$$

where $\mathrm{div}_{\mu_\eta} \overline{X}_{\mathrm{r}} \in C^\infty(\mathcal{C}_\eta)$ is defined by $\mathfrak{L}_{\overline{X}_{\mathrm{r}}} \mu_\eta = \left(\mathrm{div}_{\mu_\eta} \overline{X}_{\mathrm{r}}\right)\ \mu_\eta$ [18, def. A.12]. Taking $\mathcal{D}_\eta$ as the set of smooth, compactly supported functions on $\mathcal{C}_\eta$, Stokes' Theorem ensures that $\widehat{\mathrm{r}}$ is a symmetric operator.

*Holonomies.* Let $e \in \mathcal{L}_{\mathrm{edges}}$ and let $\eta \in \mathcal{L}$ such that $\gamma(\eta) \in \mathcal{L}_{\mathrm{graphs}/e}$. For any $\delta \in C^\infty(G, \mathbb{R})$, we have:

$$\forall (h, P) \in \mathcal{C}_\eta \times \mathrm{P}_\eta\,,\ \mathrm{h}^{(e,\delta)}_\eta \circ L_\eta^{-1}(h, P) = \delta\left(\prod_{k=1}^{n_{\gamma(\eta)\to e}} \left[h \circ a_{\gamma(\eta)\to e}(k)\right]^{\epsilon_{\gamma(\eta)\to e}(k)}\right),$$

hence $\mathrm{h}^{(e,\delta)}_\eta$ fulfills the quantization condition with $\overline{X}_{\mathrm{h}^{(e,\delta)}_\eta} := 0$. This yields the expression in eq. (3.11.1) for $\widehat{\mathrm{h}^{(e,\delta)}_\eta}$. Next, since $\delta$ is real-valued, we get, using bump functions:

$$\psi \in \mathrm{Ker}\left(\widehat{\mathrm{h}^{(e,\delta)}_\eta}^{\dagger} \pm i\right) \ \Leftrightarrow\ \forall \varphi \in \mathcal{D}_\eta\,,\ \int_{\mathcal{C}_\eta} d\mu_\eta\ \ \mathrm{h}^{(e,\delta)}_\eta\, \varphi^*\, \psi = \mp i \int_{\mathcal{C}_\eta} d\mu_\eta\ \ \varphi^*\, \psi$$

$$\Leftrightarrow\ \mathrm{h}^{(e,\delta)}_\eta\, \psi = \mp i\, \psi \ \text{ a.e. in } \left(\mathcal{C}_\eta\,,\ \mu_\eta\right) \ \Leftrightarrow\ \psi = 0 \ \text{ a.e. in } \left(\mathcal{C}_\eta\,,\ \mu_\eta\right)\,,$$

so the symmetric operator $\widehat{\mathrm{h}^{(e,\delta)}_\eta}$ is essentially self-adjoint [26, theorem VIII.3]. Finally, for any $\eta, \eta' \in \mathcal{L}$ such that $\gamma(\eta), \gamma(\eta') \in \mathcal{L}_{\mathrm{graphs}/e}$, we have $\mathrm{h}^{(e,\delta)}_\eta \sim \mathrm{h}^{(e,\delta)}_{\eta'}$ (prop. 3.9), so [18, prop. 3.4] ensures that $\widehat{\mathrm{h}^{(e,\delta)}_\eta} \sim \widehat{\mathrm{h}^{(e,\delta)}_{\eta'}}$, and therefore that an observable $\widehat{\mathrm{h}^{(e,\delta)}}$ can be consistently defined on $\mathcal{S}^\otimes_{(\mathcal{H},\mathcal{L},\Phi)}$.

*Fluxes.* Let $\overline{F} \in \mathcal{L}_{\mathrm{faces}}$ and let $\eta \in \mathcal{L}$ such that $\lambda(\eta) \in \mathcal{L}_{\mathrm{profls}/\overline{F}}$. For any $u \in \mathfrak{g}$, we have:

$$\forall (h, P) \in \mathcal{C}_\eta \times \mathrm{P}_\eta\,,\ \mathrm{P}^{(\overline{F},u)}_\eta \circ L_\eta^{-1}(h, P) = \sum_{F' \in H_{\lambda(\eta)\to\overline{F}}} P(F')(u)\,,$$

hence $\mathrm{P}^{(\overline{F},u)}_\eta$ fulfills the quantization condition with:

$$\forall h \in \mathcal{C}_\eta\,,\ \overline{X}_{\mathrm{P}^{(\overline{F},u)}_\eta, h} := \sum_{F' \in H_{\lambda(\eta)\to\overline{F}}} \left[T_1 \ \ t \mapsto R^{(F')}_{\eta,t} h\right](u)\,.$$

Now, through a reasoning similar to the proof for the second part of prop. 3.5, we have:

$$\left\{ e \in \gamma(\eta) \;\middle|\; e \in \overline{F} \right\} = \left\{ e \in \gamma(\eta) \;\middle|\; \exists F' \in H_{\lambda(\eta) \to \overline{F}} \,/\, \chi_\eta(e) = F' \right\}, \tag{3.11.4}$$

therefore:

$$\forall h \in \mathcal{C}_\eta \,,\; \overline{X}_{\mathrm{P}_\eta^{(\overline{F},u)},h} = \left[ T_1 \;\; t \mapsto R_{\eta,t}^{(\overline{F})} h \right](u)\,,$$

which yields the expression in eq. (3.11.2) for $\widehat{\mathrm{P}_\eta^{(\overline{F},u)}}$ since the flow $\tau \mapsto R_{\eta,e^{\tau u}}^{(\overline{F})}$ of $\overline{X}_{\mathrm{P}_\eta^{(\overline{F},u)}}$ preserves the right-invariant measure $\mu_\eta$ on $\mathcal{C}_\eta$. Moreover, by the dominated convergence theorem, we have:

$$\psi \in \operatorname{Ker}\left( \widehat{\mathrm{P}_\eta^{(\overline{F},u)}}^{\dagger} \pm i \right)$$

$$\Leftrightarrow \quad \forall \varphi \in \mathcal{D}_\eta \,,\, -i \int_{\mathcal{C}_\eta} d\mu_\eta(h) \;\; \left[ T_1 t \mapsto \varphi^*\left( R_{\eta,t}^{(\overline{F})} h \right) \right](u)\, \psi(h) = \mp i \int_{\mathcal{C}_\eta} d\mu_\eta(h) \;\; \varphi^*(h)\, \psi(h)$$

$$\Leftrightarrow \quad \forall \varphi \in \mathcal{D}_\eta \,,\, \left.\frac{d}{d\tau}\right|_{\tau=0} \int_{\mathcal{C}_\eta} d\mu_\eta(h) \;\; \varphi^*\left( R_{\eta,e^{\tau u}}^{(\overline{F})} h \right) \psi(h) = \pm \int_{\mathcal{C}_\eta} d\mu_\eta(h) \;\; \varphi^*(h)\, \psi(h)$$

$$\Leftrightarrow \quad \forall \varphi \in \mathcal{D}_\eta \,,\, \forall \tau \in \mathbb{R},\, \int_{\mathcal{C}_\eta} d\mu_\eta(h) \;\; \varphi^*\left( R_{\eta,e^{\tau u}}^{(\overline{F})} h \right) \psi(h) = e^{\pm\tau} \int_{\mathcal{C}_\eta} d\mu_\eta(h) \;\; \varphi^*(h)\, \psi(h)\,,$$

thus, using again bump functions, together with the invariance of the measure under right translations:

$$\psi \in \operatorname{Ker}\left( \widehat{\mathrm{P}_\eta^{(\overline{F},u)}}^{\dagger} \pm i \right) \quad \Leftrightarrow \quad \forall \tau \in \mathbb{R},\, \psi \circ \left( h \mapsto R_{\eta,e^{-\tau u}}^{(\overline{F})} h \right) = e^{\pm\tau}\, \psi \;\text{ a.e. in }\; \left( \mathcal{C}_\eta \,,\, \mu_\eta \right)$$

$$\Leftrightarrow \quad \forall \tau \in \mathbb{R},\, \|\psi\|_\eta = e^{\pm\tau}\, \|\psi\|_\eta \quad \Leftrightarrow \quad \psi = 0 \;\text{ a.e. in }\; \left( \mathcal{C}_\eta \,,\, \mu_\eta \right),$$

hence $\widehat{\mathrm{P}_\eta^{(\overline{F},u)}}$ is essentially self-adjoint. Finally, like above, prop. 3.9 together with [18, prop. 3.4] ensures that the $\widehat{\mathrm{P}_\eta^{(\overline{F},u)}}$, defined for each $\eta \in \mathcal{L}$ such that $\lambda(\eta) \in \mathcal{L}_{\mathrm{profls}/\overline{F}}$, can be consistently assembled into an observable $\widehat{\mathrm{P}^{(\overline{F},u)}}$ on $\mathcal{S}^{\otimes}_{(\mathcal{H},\mathcal{L},\Phi)}$. □

## 3.2 Relation to the Ashtekar-Lewandowski Hilbert space

The projective state space set up in the previous subsection regrettably cannot be displayed as the rendering [17, def. 2.6] of a continuous classical theory of connections: if we were to define projection maps from the infinite dimensional phase space of such a theory [31, section IV.33] into the various small phase spaces $\mathcal{M}_\eta$, these maps would not be smooth, and in fact they would not even be surjective. Indeed, the holonomy, resp. flux, variables are obtained by smearing the connection, resp. its conjugate 'electric field', along singular geometrical objects (respectively 1- and ($d-1$)-dimensional, in contrast to a smearing by a smooth function on the $d$-dimensional manifold $\Sigma$). Moreover, these non-smooth flux variables from the continuous theory actually combine both faces of a given surface: the 'one-sided' flux variables we introduced above have no equivalent in terms of appropriate smearings of the electric field, while the fluxes attached to submanifolds of dimension less than $d-2$ (arising at the intersection of surfaces) should vanish according to the continuous theory. On the other hand the inclusion of these additional, seemingly non-physical observables is forced upon us by the regularization of the Poisson brackets [31, section II.6.4]: in other words, they are the price for trying to somewhat restore a notion of compatibility of the projection map with the symplectic structure (similar to what was expressed by [17, def. 2.1] in the case of smooth projections). We refer to the LQG literature for further discussion of these issues (see eg. [29, section 6]).

Instead of exploring the relation between the formalism just constructed and the continuous classical theory of interest, we will take the Ashtekar-Lewandowski Hilbert space $\mathcal{H}_{\text{AL}}$ [2, 22] as a starting point, and investigate in which sense the quantum projective state space from prop. 3.10 can be understood as a reasonable extension of the space of states defined over $\mathcal{H}_{\text{AL}}$. This analysis has to to be carried out in the case of a compact group $G$, since this is a prerequisite for $\mathcal{H}_{\text{AL}}$ to exist: from the perspective we are adopting here, the compactness ensures that the measures $\mu_{\eta'\to\eta}$ will be normalizable, which in turn allows to pick out a natural 'reference state' $\zeta_{\eta'\to\eta}$ in each $\mathcal{H}_{\eta'\to\eta}$, and thus to identify $\mathcal{H}_\eta$ with the vector subspace $\left\{\zeta_{\eta'\to\eta}\right\}\otimes\mathcal{H}_\eta$ in $\mathcal{H}_{\eta'}\approx\mathcal{H}_{\eta'\to\eta}\otimes\mathcal{H}_\eta$. This provides an inductive system of Hilbert spaces, whose limit will reproduce $\mathcal{H}_{\text{AL}}$.

As stressed at the beginning of subsection 2.1, the limit of an inductive system is not affected if one restricts its label set to some cofinal part. This is the reason why we have so far only considered graphs with fully analytic edges. Still, the use of graphs with semianalytic edges [31, sections II.6 and IV.20] is favored in LQG, for, although they yields the same Hilbert space, they present it in a form more convenient for writing the action of semianalytic diffeomorphisms (which, unlike fully analytic ones, can be local). Therefore, we briefly sketch below how to switch back to the semianalytic class.

In this subsection the gauge group $G$ is assumed to be *compact*, and the measure $\mu$ (introduced in prop. 3.10) is taken to be the normalized Haar measure on $G$.

**Definition 3.12** Let $k\in\{1,2,\ldots,\infty\}$. We define the set $\mathcal{L}^{(k)}_{\text{edges}}$ of ($k$)-edges like in def. 2.1, by requiring encharted ($k$)-edges to be $C^k$-diffeomorphisms instead of analytic ones. In analogy to props. 2.2 and 2.3, we define the range $r(e)$, the beginning and ending points $b(e)$ and $f(e)$, ($k$)-subedges $e_{[p,p']}$ (for $p\neq p'\in r(e)$), the reversed ($k$)-edge $e^{-1}$, and the order $<_{(e)}$ on the range of a ($k$)-edge $e$, as well as the composition of ($k$)-composable ($k$)-edges. These have the same properties as in the analytic case, since the proofs of props. 2.2 and 2.3 did not made use of the analyticity.

An analytic encharted edge is also an encharted ($k$)-edge, and two analytic encharted edge are equivalent in $\breve{\mathcal{L}}_{\text{edges}}$ iff they are equivalent in $\breve{\mathcal{L}}^{(k)}_{\text{edges}}$. Thus, we have a natural injection of $\mathcal{L}_{\text{edges}}$ into $\mathcal{L}^{(k)}_{\text{edges}}$. In the following, we will always identify $\mathcal{L}_{\text{edges}}$ with the image of this injection and write $\mathcal{L}_{\text{edges}}\subset\mathcal{L}^{(k)}_{\text{edges}}$.

**Proposition 3.13** We define AL-graphs as finite sets of (1)-edges $\gamma \subset \mathcal{L}^{(1)}_{\text{edges}}$ such that:

1. $\forall e \in \gamma,\ \exists e_1, \ldots, e_n \in \mathcal{L}_{\text{edges}} \subset \mathcal{L}^{(1)}_{\text{edges}}$, (1)-composable $/ e = e_n \circ \ldots \circ e_1$ ;
2. $\forall e \neq e' \in \gamma,\ r(e) \cap r(e') \subset \{b(e),\, f(e)\} \cap \{b(e'),\, f(e')\}$.

We denote the set of AL-graphs by $\mathcal{L}_{\text{AL}}$ and we equip it with the preorder $\preccurlyeq$ defined as in def. 2.4.

Then, $\mathcal{L}_{\text{graphs}}\,, \preccurlyeq$ is a cofinal subset of $\mathcal{L}_{\text{AL}}\,, \preccurlyeq$, so in particular $\mathcal{L}_{\text{AL}}\,, \preccurlyeq$ is a directed preordered set.

**Proof** By construction $\mathcal{L}_{\text{graphs}} \subset \mathcal{L}_{\text{AL}}$, and the order $\preccurlyeq$ between two elements of $\mathcal{L}_{\text{graphs}}$ coincides with their order as elements of $\mathcal{L}_{\text{AL}}$ (indeed, if an analytic edge is the composition of (1)-composable analytic edges, then, by definition, these edges are composable in $\mathcal{L}_{\text{edges}}$).

Next, let $\gamma \in \mathcal{L}_{\text{AL}}$ and for any $e \in \gamma$, choose $\gamma_e = \{e_1\,, \ldots,\, e_n\} \subset \mathcal{L}_{\text{edges}}$ such that $e = e_n \circ \ldots \circ e_1$. We have:

$$\forall e \in \gamma,\ \forall \widetilde{e} \neq \widetilde{e}' \in \gamma_e,\ r(\widetilde{e}) \cap r(\widetilde{e}') \subset \{b(\widetilde{e}),\, f(\widetilde{e})\} \cap \{b(\widetilde{e}'),\, f(\widetilde{e}')\}\,,$$

$$\begin{aligned} \text{and} \quad \forall e \neq e' \in \gamma,\ \forall \widetilde{e} \in \gamma_e\,,\ \forall \widetilde{e}' \in \gamma_{e'}\,,& \\ r(\widetilde{e}) \cap r(\widetilde{e}') &\subset r(\widetilde{e}) \cap r(\widetilde{e}') \cap r(e) \cap r(e') \\ &\subset r(\widetilde{e}) \cap \{b(e),\, f(e)\} \cap r(\widetilde{e}') \cap \{b(e'),\, f(e')\} \\ &\subset \{b(\widetilde{e}),\, f(\widetilde{e})\} \cap \{b(\widetilde{e}'),\, f(\widetilde{e}')\}\,, \end{aligned}$$

by definition of the composition of edges. Therefore $\gamma' := \bigcup_{e \in \gamma} \gamma_e \in \mathcal{L}_{\text{edges}}$, and we have $\gamma \preccurlyeq \gamma'$. □

We recall here the classical construction underlying the Ashtekar-Lewandowski Hilbert space, namely the composition of a projective limit of configuration spaces. The projection maps involved here match exactly the ones induced by the projections between phase spaces considered in prop. 3.8 (modulo the straightforward extension to semianalytic graphs). However, the momentum variables do not come into play in this context, since we are not setting up an actual projective limit of symplectic manifolds, so we can directly use graphs as labels, without having to decorate them with faces.

**Proposition 3.14** Let $\gamma \preccurlyeq \gamma' \in \mathcal{L}_{\text{AL}}$. Then, for any $e \in \gamma$, there exists a *unique* map $a_{\gamma' \to \gamma, e}$ : $\{1, \ldots,\, n_{\gamma' \to \gamma, e}\} \to \gamma'$ (with $n_{\gamma' \to \gamma, e} \geqslant 1$) such that:

$$e = a_{\gamma' \to \gamma, e}(n_{\gamma' \to \gamma, e})^{\epsilon_{\gamma' \to \gamma, e}(n_{\gamma' \to \gamma, e})} \circ \ldots \circ a_{\gamma' \to \gamma, e}(1)^{\epsilon_{\gamma' \to \gamma, e}(1)}\,,$$

where, for any $k \in \{1, \ldots,\, n_{\gamma' \to \gamma, e}\}$, $\epsilon_{\gamma' \to \gamma, e}(k)$ is defined from $a_{\gamma' \to \gamma, e}$ as in eq. (3.2.2).

**Proof** The proof works exactly as in the analytic case (prop. 3.2). □

**Proposition 3.15** For any $\gamma \in \mathcal{L}_{\text{AL}}$ we define the finite-dimensional smooth configuration space $\mathcal{C}_\gamma := \{h : \gamma \to G\}$ (like in def. 3.1). And for any $\gamma \preccurlyeq \gamma' \in \mathcal{L}_{\text{AL}}$, we define the map $\pi_{\gamma' \to \gamma} : \mathcal{C}_{\gamma'} \to \mathcal{C}_\gamma$ by:

$$\forall h' \in \mathcal{C}_{\gamma'},\ \forall e \in \gamma,\ \pi_{\gamma'\to\gamma}(h')(e) := \left( \prod_{k=1}^{n_{\gamma'\to\gamma,e}} \left[h' \circ a_{\gamma'\to\gamma,e}(k)\right]^{\epsilon_{\gamma'\to\gamma,e}(k)} \right).$$

$\pi_{\gamma'\to\gamma}$ is smooth, for $G$ is a Lie group, and it is moreover surjective.

In addition, for any $\gamma \preccurlyeq \gamma' \preccurlyeq \gamma'' \in \mathcal{L}_{\text{AL}}$, we have:

$$\pi_{\gamma'\to\gamma} \circ \pi_{\gamma''\to\gamma'} = \pi_{\gamma''\to\gamma}. \tag{3.15.1}$$

**Proof** Let $\gamma \preccurlyeq \gamma'$ and let $h \in \mathcal{C}_\gamma$. For any $e \neq e' \in \gamma$, we have:

$$\forall \widetilde{e} \in a_{\gamma'\to\gamma,e} \left\langle \{1, \ldots, n_{\gamma'\to\gamma,e}\} \right\rangle,\ \forall \widetilde{e}' \in a_{\gamma'\to\gamma,e'} \left\langle \{1, \ldots, n_{\gamma'\to\gamma,e'}\} \right\rangle,$$
$$r(\widetilde{e}) \cap r(\widetilde{e}') \subset \{b(\widetilde{e}), f(\widetilde{e})\} \cap \{b(\widetilde{e}'), f(\widetilde{e}')\},$$

as in the proof of prop. 3.13, hence $a_{\gamma'\to\gamma,e} \left\langle \{1, \ldots, n_{\gamma'\to\gamma,e}\} \right\rangle \cap a_{\gamma'\to\gamma,e'} \left\langle \{1, \ldots, n_{\gamma'\to\gamma,e'}\} \right\rangle = \varnothing$. Moreover, $a_{\gamma'\to\gamma,e}$ is injective (by definition of the composition of edges), hence the map $h' \in \mathcal{C}_{\gamma'}$ given by:

$$\forall e \in \gamma,\ h' \circ a_{\gamma'\to\gamma,e}(1) = h(e)^{\epsilon_{\gamma'\to\gamma,e}(1)},$$

$$\forall e' \in \gamma' / e' \notin \{a_{\gamma'\to\gamma,e}(1) | e \in \gamma\},\ h'(e') = \mathbf{1},$$

is well-defined and is such that $\pi_{\gamma'\to\gamma}(h') = h$. Thus, $\pi_{\gamma'\to\gamma}$ is surjective.
Finally, eq. (3.15.1) follows from the uniqueness of $a_{\gamma''\to\gamma,e}$ as in the proof of theorem 3.7. □

As mentioned earlier, there are many projections between the phase spaces $T^*(\mathcal{C}_{\gamma'})$ and $T^*(\mathcal{C}_\gamma)$ (with $\gamma \preccurlyeq \gamma'$) that can reproduce a given projection between the configuration spaces $\mathcal{C}_{\gamma'}$ and $\mathcal{C}_\gamma$, or, equivalently, many ways of choosing in $\mathcal{C}_{\gamma'}$ a set of variables complementary to the ones coming from $\mathcal{C}_\gamma$. In the proof of prop. 3.16, we choose, for each pair of graphs $\gamma \preccurlyeq \gamma'$, a factorization of $\mathcal{C}_{\gamma'}$ as $\mathcal{C}_{\gamma'-\gamma} \times \mathcal{C}_\gamma$ consistent with the projection $\pi_{\gamma'\to\gamma}$ defined in prop. 3.15. The injective maps $\tau_{\gamma'\leftarrow\gamma}$ between the Hilbert spaces $\mathcal{H}_\gamma$ that serve as building blocks for $\mathcal{H}_{\text{AL}}$ can then be understood as arising from the corresponding factorization $\mathcal{H}_{\gamma'} \approx \mathcal{H}_{\gamma'-\gamma} \otimes \mathcal{H}_\gamma$, via the selection of a 'reference state' $\zeta_{\gamma'-\gamma}$ in $\mathcal{H}_{\gamma'-\gamma}$. Because this reference state is taken as the constant function $\zeta_{\gamma'-\gamma} \equiv 1$ on $\mathcal{C}_{\gamma'-\gamma}$, the thus obtained identification of $\mathcal{H}_\gamma$ with a vector subspace of $\mathcal{H}_{\gamma'}$ in the end does not depend on which particular factorization of $\mathcal{C}_{\gamma'}$ has been used, but solely on the projection $\pi_{\gamma'\to\gamma}$ from $\mathcal{C}_{\gamma'}$ into $\mathcal{C}_\gamma$. Also, as announced earlier, the need for a *compact* gauge group $G$ manifests itself in this approach as a condition for $\zeta_{\gamma'-\gamma}$ to be a normalizable element of $\mathcal{H}_{\gamma'-\gamma}$ (otherwise the map $\tau_{\gamma'\leftarrow\gamma}$ would not be well-defined).

Of course, it is not really necessary for assembling the inductive limit $\mathcal{H}_{\text{AL}}$ to ever introduce such factorization maps (a more standard path being to directly check that $\pi_{\gamma'\to\gamma,*}\,\mu_{\gamma'} = \mu_\gamma$). Still, it is worth looking closer at the particular family of factorizations elected below. The projections between phase spaces to which they give rise turn out to be precisely the ones that were considered in [29, def. 3.8] and although they do not fulfill the three-spaces consistency needed for a projective system (expressed in [17, def. 2.3], or, at the level of the factorization maps, in [17, eq. (2.11.1)]), this can be quickly fixed: by somewhat tightening the ordering among graphs (requiring, in addition to eq. (2.4.1), that the first part of an edge $e \in \gamma$, in its decomposition into edges of $\gamma'$, should be oriented like $e$ itself, ie. that $\epsilon_1 = +1$), we can, without voiding the directedness of $\mathcal{L}_{\text{AL}}$, rescue this pivotal consistency condition.

So why did we go through the intricacy of carefully setting up a label set with edges and surfaces if an apparently valid projective system could readily have been built over $\mathcal{L}_{\text{AL}}$? If we examine carefully the structure of observables that would arise from such a projective structure (and in particular, if we compare it to the one obtained in the previous subsection), we realize that its momentum variables are fluxes carried by single edge germs (defining the germ of an edge $e$ as the equivalence class consisting of all edges sharing an initial subedge with $e$).

This sheds light on how a projective limit of *phase* spaces can at all be constructed using labels that seems to only know about *configuration* variables. What makes it possible, is the availability of a preferred pairing of conjugate variables, binding each configuration variable to its own particular momentum variable (eg. the holonomy along an edge with the flux carried by its germ). This pairing is such that whenever an edge $e$ belongs to some graph $\gamma$, its companion flux does not act on any other edge of $\gamma$ (indeed, edges in a graph cannot share a common subedge). In this way, any graph, as it selects specific holonomy variables, selects at the same time their conjugate flux variables (and the slight sharpening of the ordering described above ensures that if $\gamma' \succcurlyeq \gamma$ the fluxes thus attached to $\gamma$ are also in $\gamma'$).

A similar mechanism underlies the projective quantum state space built in [23]. In this work, holonomies along analytic loops were used as a complete set of *independent* configuration variables, which again provides an a priori pairing of conjugate variables, since the selected configuration variables can be thought as coordinates and mapped to their dual differential operator.

In both cases, the resulting factorizing systems leads to a theory whose basic momentum variables have no equivalent in the continuum classical theory. The trouble is then that fluxes associated to non-degenerate surfaces (aka. $(d-1)$-dimensional ones) cannot be represented on the corresponding quantum projective state space. If we however insist on using regular holonomies and fluxes as our primary variables (so as to implement an algebra of observables that separates the points in the continuum classical theory), then there is no way of choosing beforehand a canonical pairing that would, as described above, automatically provides a suitable set of canonically conjugate variables for any arbitrary graph $\gamma$.

**Proposition 3.16** For any $\gamma \in \mathcal{L}_{\text{AL}}$, we define the measure $\mu_\gamma \approx \mu^{\#\gamma}$ on $\mathcal{C}_\gamma \approx G^{\#\gamma}$ (as in prop. 3.10) and the Hilbert space $\mathcal{H}_\gamma := L_2(\mathcal{C}_\gamma\,,\, d\mu_\gamma)$. Next, for any $\gamma \preccurlyeq \gamma' \in \mathcal{L}_{\text{AL}}$, we define the map $\tau_{\gamma'\leftarrow\gamma}$ by:

$$\begin{array}{rccc} \tau_{\gamma'\leftarrow\gamma} : & \mathcal{H}_\gamma & \to & \mathcal{H}_{\gamma'} \\ & \psi & \mapsto & \psi \circ \pi_{\gamma'\to\gamma} \end{array}.$$

$\tau_{\gamma'\leftarrow\gamma}$ is an isometry and, for any $\gamma \preccurlyeq \gamma' \preccurlyeq \gamma'' \in \mathcal{L}_{\text{AL}}$:

$$\tau_{\gamma''\leftarrow\gamma'} \circ \tau_{\gamma'\leftarrow\gamma} = \tau_{\gamma''\leftarrow\gamma}\,. \tag{3.16.1}$$

We define the Ashtekar-Lewandowski Hilbert space $\mathcal{H}_{\text{AL}}$ as the (norm completion of) the inductive limit of the system $\left(\left(\mathcal{H}_\gamma\right)_{\gamma\in\mathcal{L}_{\text{AL}}}, \left(\tau_{\gamma'\leftarrow\gamma}\right)_{\gamma\preccurlyeq\gamma'}\right)$. For $\gamma \in \mathcal{L}_{\text{AL}}$, we will denote by $\tau_{\text{AL}\leftarrow\gamma}$ the natural isometric injection of $\mathcal{H}_\gamma$ in $\mathcal{H}_{\text{AL}}$.

**Proof** Let $\gamma \preccurlyeq \gamma' \in \mathcal{L}_{\text{AL}}$. We define:

$$\gamma' - \gamma := \gamma' \setminus \{a_{\gamma'\to\gamma,e}(1) \mid e \in \gamma\}\,,$$

as well as the finite dimensional smooth manifold $\mathcal{C}_{\gamma'-\gamma} := \left\{\widetilde{h} : (\gamma' - \gamma) \to G\right\}$. The smooth map $\varphi_{\gamma'-\gamma}$ given by:

$$\begin{array}{rcl} \varphi_{\gamma'-\gamma} \ : \ \mathcal{C}_{\gamma'} & \to & \mathcal{C}_{\gamma'-\gamma} \times \mathcal{C}_\gamma \\ h' & \mapsto & h'\big|_{\gamma'-\gamma} \, , \, \pi_{\gamma'\to\gamma}(h') \end{array} ,$$

is a diffeomorphism (as can be checked by expressing its inverse like in the proof of prop. 3.6). Moreover, defining for any $h \in \mathcal{C}_\gamma$ the maps $T^{(\gamma)}_{\gamma',h} : \mathcal{C}_{\gamma'} \to \mathcal{C}_{\gamma'}$ and $R_{\gamma,h} : \mathcal{C}_\gamma \to \mathcal{C}_\gamma$ by:

$$\forall j' \in \mathcal{C}_{\gamma'} , \, \forall e' \in \gamma',$$

$$\left[T^{(\gamma)}_{\gamma',h}(j')\right](e') = \begin{cases} j'(e') \, . \, h(e) & \text{if } \exists e \in \gamma \,/\, e' = a_{\gamma'\to\gamma,e}(1) \ \ \& \ \ \epsilon_{\gamma'\to\gamma,e} = +1 \\ h(e)^{-1} \, . \, j'(e') & \text{if } \exists e \in \gamma \,/\, e' = a_{\gamma'\to\gamma,e}(1) \ \ \& \ \ \epsilon_{\gamma'\to\gamma,e} = -1 \\ j'(e') & \text{else} \end{cases} ,$$

$$\text{and} \quad \forall j \in \mathcal{C}_\gamma , \, \forall e \in \gamma, \, [R_{\gamma,h}(j)](e) = j(e) \, . \, h(e) ,$$

we have $\varphi_{\gamma'-\gamma} \circ T^{(\gamma)}_{\gamma',h} = \left(\mathrm{id}_{\mathcal{C}_{\gamma'-\gamma}} \times R_{\gamma,h}\right) \circ \varphi_{\gamma'-\gamma}$. Let $\omega_\gamma$, resp. $\omega_{\gamma'}$, be a right invariant volume form on $\mathcal{C}_\gamma$, resp. $\mathcal{C}_{\gamma'}$, such that $\mu_\gamma$, resp. $\mu_{\gamma'}$, is the corresponding measure. Because the Haar measure on a compact group is left-invariant as well as right-invariant, there exists a smooth map $\varepsilon : \mathcal{C}_\gamma \to \{+1, -1\}$ such that, for any $h \in \mathcal{C}_\gamma$, $T^{(\gamma),*}_{\gamma',h} \omega_{\gamma'} = \varepsilon(h)\, \omega_{\gamma'}$. Then, we can, like in the proof of prop. 3.10, construct a smooth volume form $\omega_{\gamma'-\gamma}$ on $\mathcal{C}_{\gamma'-\gamma}$ such that $\left[\varphi^{-1,*}_{\gamma'-\gamma}\, \omega_{\gamma'}\right] = \omega_{\gamma'-\gamma} \times \left(\varepsilon\omega_\gamma\right)$. Therefore, there exists a smooth measure $\mu_{\gamma'-\gamma}$ on $\mathcal{C}_{\gamma'-\gamma}$ such that $\varphi_{\gamma'-\gamma,*}\, \mu_{\gamma'} = \mu_{\gamma'-\gamma} \times \mu_\gamma$. And from $\mu_\gamma\left(\mathcal{C}_\gamma\right) = 1 = \mu_{\gamma'}\left(\mathcal{C}_{\gamma'}\right)$, we get $\mu_{\gamma'-\gamma}\left(\mathcal{C}_{\gamma'-\gamma}\right) = 1$.

Thus, for any measurable function $\zeta : \mathcal{C}_\gamma \to \mathbb{R}_+$, we have, by Fubini's theorem:

$$\int_{\mathcal{C}_\gamma} d\mu_\gamma(h)\, \zeta(h) = \int_{\mathcal{C}_{\gamma'}} d\mu_{\gamma'}(h')\, \zeta \circ \pi_{\gamma'-\gamma}(h') ,$$

so that $\tau_{\gamma'\leftarrow\gamma}$ is well-defined as a map $\mathcal{H}_\gamma \to \mathcal{H}_{\gamma'}$ and is an isometry. Finally, eq. (3.16.1) follows from eq. (3.15.1).

*Note.* Denoting by $\gamma'_o$ the AL-graph $\gamma' - \gamma \subset \gamma'$, we have $\mu_{\gamma'-\gamma} = \mu_{\gamma'_o}$. Indeed, for any measurable function $\zeta : \mathcal{C}_{\gamma'-\gamma} \to \mathbb{R}_+$:

$$\begin{aligned} \int_{\mathcal{C}_{\gamma'-\gamma}} d\mu_{\gamma'-\gamma}(j) \ \ \zeta(j) &= \int_{\mathcal{C}_{\gamma'-\gamma}\times\mathcal{C}_\gamma} d\mu_{\gamma'-\gamma}(j)\, d\mu_\gamma(h) \ \ \zeta(j) \quad (\text{for } \mu_\gamma(\mathcal{C}_\gamma) = 1) \\ &= \int_{\mathcal{C}_{\gamma'}} d\mu_{\gamma'}(h') \ \ \zeta\left(h'\big|_{\gamma_o}\right) \\ &= \int_{\mathcal{C}_{\gamma'_o}\times\mathcal{C}_{\gamma'_1}} d\mu_{\gamma'_o}(h'_o)\, d\mu_{\gamma'_1}(h'_1) \ \ \zeta(h'_o) \end{aligned}$$

(with $\gamma'_1 := \gamma' \setminus \gamma'_o \in \mathcal{L}_{\text{AL}}$; $\gamma' = \gamma'_o \cup \gamma'_1$ implies $\left(\mathcal{C}_{\gamma'}, d\mu_{\gamma'}\right) \approx \left(\mathcal{C}_{\gamma'_o}, d\mu_{\gamma'_o}\right) \times \left(\mathcal{C}_{\gamma'_1}, d\mu_{\gamma'_1}\right)$)

$$= \int_{\mathcal{C}_{\gamma'_o}} d\mu_{\gamma'_o}(h'_o) \ \ \zeta(h'_o) \quad (\text{for } \mu_{\gamma'_1}(\mathcal{C}_{\gamma'_1}) = 1).$$

□

Finally, we also recall the definition of the holonomy and flux operators on $\mathcal{H}_{\text{AL}}$ [3, 1, 31]. Indeed, if we want to investigate the relation between the Ashtekar-Lewandowski construction and the just developed projective formalism, it is not enough to produce a map $\sigma$ between the state spaces: we should also check that $\sigma$ is dual to the map $\alpha$ that transports the observables according to their physical interpretation. Actually, the second part of prop. 3.19 shows that the map $\sigma$ is *uniquely* specified as soon as we require it to intertwine the implementation of holonomies and exponentiated fluxes on both sides.

**Proposition 3.17** Let $e \in \mathcal{L}_{\text{edges}}$ and define:

$$\mathcal{L}_{\text{AL}/e} := \{\gamma \in \mathcal{L}_{\text{AL}} \mid \{e\} \preccurlyeq \gamma\}.$$

For any $\gamma \in \mathcal{L}_{\text{AL}/e}$ and any $\delta \in C^\infty(G, \mathbb{R})$, we have on $\mathcal{H}_\gamma$ a densely defined, essentially self-adjoint operator $\widehat{\mathrm{h}^{(e,\delta)}_\gamma}$ on $\mathcal{H}_\gamma$ (with dense domain $\mathcal{D}_\gamma$) given by:

$$\forall \psi \in \mathcal{D}_\gamma\,,\ \forall h \in \mathcal{C}_\gamma\,,\ \left(\widehat{\mathrm{h}^{(e,\delta)}_\gamma}\psi\right)(h) := \delta\left(\prod_{k=1}^{n_{\gamma\to\{e\},e}} \left[h \circ a_{\gamma\to\{e\},e}(k)\right]^{\epsilon_{\gamma\to\{e\},e}(k)}\right)\psi(h)\,.$$

Moreover, for any $\gamma' \succcurlyeq \gamma$, we have $\gamma' \in \mathcal{L}_{\text{AL}/e}$ and:

$$\forall \psi \in \mathcal{D}_\gamma\,,\ \tau_{\gamma'\leftarrow\gamma}(\psi) \in \mathcal{D}_{\gamma'} \quad \& \quad \widehat{\mathrm{h}^{(e,\delta)}_{\gamma'}} \circ \tau_{\gamma'\leftarrow\gamma}(\psi) = \tau_{\gamma'\leftarrow\gamma} \circ \widehat{\mathrm{h}^{(e,\delta)}_{\gamma}}(\psi)\,.$$

Thus, the family $\left(\widehat{\mathrm{h}^{(e,\delta)}_\gamma}\right)_{\gamma\in\mathcal{L}_{\text{AL}/e}}$ provides a densely defined, essentially self-adjoint operator $\widehat{\mathrm{h}^{(e,\delta)}_{\text{AL}}}$ on $\mathcal{H}_{\text{AL}}$.

**Proof** Let $\gamma \in \mathcal{L}_{\text{AL}/e}$ and $\delta \in C^\infty(G, \mathbb{R})$. Taking $\mathcal{D}_\gamma = C^\infty(\mathcal{C}_\gamma, \mathbb{C}) \subset \mathcal{H}_\gamma$ (this matches the compactly supported smooth functions used in prop. 3.11 since $\mathcal{C}_\gamma$ is compact), $\widehat{\mathrm{h}^{(e,\delta)}_\gamma}$ is well-defined and essentially self-adjoint (actually, it is a bounded operator). Moreover, we have:

$$\forall h \in \mathcal{C}_\gamma\,,\ \delta\left(\prod_{k=1}^{n_{\gamma\to\{e\},e}} \left[h \circ a_{\gamma\to\{e\},e}(k)\right]^{\epsilon_{\gamma\to\{e\},e}(k)}\right) = \delta\left(\left[\pi_{\gamma\to\{e\}}(h)\right](e)\right). \tag{3.17.1}$$

Now let $\gamma' \succcurlyeq \gamma$. By transitivity of $\succcurlyeq$ on $\mathcal{L}_{\text{AL}}$, $\gamma' \in \mathcal{L}_{\text{AL}/e}$. For any $\psi \in \mathcal{D}_\gamma$, $\tau_{\gamma'\leftarrow\gamma}(\psi) \in \mathcal{D}_{\gamma'}$ (for $\pi_{\gamma'\to\gamma}$ is smooth) and, using eq. (3.15.1):

$$\forall h' \in \mathcal{C}_{\gamma'}\,,\ \left[\widehat{\mathrm{h}^{(e,\delta)}_{\gamma'}} \circ \tau_{\gamma'\leftarrow\gamma}(\psi)\right](h') = \delta\left(\left[\pi_{\gamma'\to\{e\}}(h')\right](e)\right)\,[\psi\circ\pi_{\gamma'\to\gamma}](h') = \left[\tau_{\gamma'\leftarrow\gamma}\circ\widehat{\mathrm{h}^{(e,\delta)}_{\gamma}}(\psi)\right](h')\,.$$

$\mathcal{L}_{\text{AL}/e}$ is a cofinal part of $\mathcal{L}_{\text{AL}}$ (for $\mathcal{L}_{\text{AL}}$ is directed), and this allows us to construct a symmetric operator $\widehat{\mathrm{h}^{(e,\delta)}_{\text{AL}}}$ on the vector subspace $\mathcal{D}_{\text{AL}} \subset \mathcal{H}_{\text{AL}}$, defined as the inductive limit of vector spaces $\left(\mathcal{D}_\gamma\right)_{\gamma\in\mathcal{L}_{\text{AL}}}$, $\left(\tau_{\gamma'\leftarrow\gamma}|_{\mathcal{D}_\gamma\to\mathcal{D}_{\gamma'}}\right)_{\gamma\preccurlyeq\gamma'}$ (without any completion). $\mathcal{D}_{\text{AL}}$ is dense in $\mathcal{H}_{\text{AL}}$ and $\widehat{\mathrm{h}^{(e,\delta)}_{\text{AL}}}$ is bounded, hence essentially self-adjoint. □

**Proposition 3.18** Let $\overline{F} \in \mathcal{L}_{\text{faces}}$ (prop. 3.3) and define:

$$\mathcal{L}_{\text{AL}/\overline{F}} := \left\{\gamma \in \mathcal{L}_{\text{AL}} \,\middle|\, \forall e \in \gamma,\ \forall p \neq p' \in r(e),\ \left(e_{[p,p']} \in \overline{F} \Rightarrow p \in \{b(e), f(e)\}\right)\right\}.$$

Let $\gamma \in \mathcal{L}_{\text{AL}/\overline{F}}$ and define, for any $h \in \mathcal{C}_\gamma$ and any $t \in G$, the map $T^{(\overline{F})}_{\gamma,t} h \in \mathcal{C}_\gamma$ by:

$$\forall e \in \gamma, \quad T^{(\overline{F})}_{\gamma,t} h(e) := \begin{cases} h(e)\,.\,t & \text{if } c(e,\,\overline{F}) = \{b(e)\} \\ t^{-1}\,.\,h(e) & \text{if } c(e,\,\overline{F}) = \{f(e)\} \\ t^{-1}\,.\,h(e)\,.\,t & \text{if } c(e,\,\overline{F}) = \{b(e),\,f(e)\} \\ h(e) & \text{if } c(e,\,\overline{F}) = \varnothing \end{cases},$$

where $c(e,\,\overline{F}) := \left\{p \in r(e) \,\middle|\, \exists p' \neq p \,/\, e_{[p,p']} \in \overline{F}\right\} \subset \{b(e),\,f(e)\}$. Then, for any $u \in \mathfrak{g}$, we have a densely defined, essentially self-adjoint operator $\widehat{\mathrm{P}^{(\overline{F},u)}_\gamma}$ on $\mathcal{H}_\gamma$ (with dense domain $\mathcal{D}_\gamma$) given by:

$$\forall \psi \in \mathcal{D}_\gamma\,,\ \forall h \in \mathcal{C}_\gamma\,,\ \left(\widehat{\mathrm{P}^{(\overline{F},u)}_\gamma} \psi\right)(h) := i\left[T_1\, t \mapsto \psi\left(T^{(\overline{F})}_{\gamma,t} h\right)\right](u)\,,$$

and, for any $t \in G$, we have a unitary operator $\widehat{\mathrm{T}^{(\overline{F},t)}_\gamma}$ on $\mathcal{H}_\gamma$ given by:

$$\forall \psi \in \mathcal{H}_\gamma\,,\ \forall h \in \mathcal{C}_\gamma\,,\ \left(\widehat{\mathrm{T}^{(\overline{F},t)}_\gamma} \psi\right)(h) := \psi\left(T^{(\overline{F})}_{\gamma,t} h\right).$$

Moreover, $\mathcal{L}_{\text{AL}/\overline{F}}$ is a cofinal part of $\mathcal{L}_{\text{AL}}$ and for any $\gamma \preccurlyeq \gamma' \in \mathcal{L}_{\text{AL}/\overline{F}}$ :

$$\forall \psi \in \mathcal{D}_\gamma\,,\ \tau_{\gamma' \leftarrow \gamma}(\psi) \in \mathcal{D}_{\gamma'} \quad \& \quad \widehat{\mathrm{P}^{(\overline{F},u)}_{\gamma'}} \circ \tau_{\gamma' \leftarrow \gamma}(\psi) = \tau_{\gamma' \leftarrow \gamma} \circ \widehat{\mathrm{P}^{(\overline{F},u)}_\gamma}(\psi)\,,$$

$$\text{and} \quad \forall \psi \in \mathcal{H}_\gamma\,,\ \widehat{\mathrm{T}^{(\overline{F},t)}_{\gamma'}} \circ \tau_{\gamma' \leftarrow \gamma}(\psi) = \tau_{\gamma' \leftarrow \gamma} \circ \widehat{\mathrm{T}^{(\overline{F},t)}_\gamma}(\psi)\,.$$

Thus, the family $\left(\widehat{\mathrm{P}^{(\overline{F},u)}_\gamma}\right)_{\gamma \in \mathcal{L}_{\text{AL}/\overline{F}}}$ provides a densely defined, essentially self-adjoint operator $\widehat{\mathrm{P}^{(\overline{F},u)}_{\text{AL}}}$ on $\mathcal{H}_{\text{AL}}$, while the family $\left(\widehat{\mathrm{T}^{(\overline{F},t)}_\gamma}\right)_{\gamma \in \mathcal{L}_{\text{AL}/\overline{F}}}$ provides a unitary operator $\widehat{\mathrm{T}^{(\overline{F},t)}_{\text{AL}}}$.

**Proof** Let $\gamma \in \mathcal{L}_{\text{AL}/\overline{F}}$ and $u \in \mathfrak{g}$. Taking as dense domain $\mathcal{D}_\gamma = C^\infty(\mathcal{C}_\gamma\,,\,\mathbb{C}) \subset \mathcal{H}_\gamma$, $\widehat{\mathrm{P}^{(\overline{F},u)}_\gamma}$ is well-defined, and, using the invariance of the measure $\mu_\gamma$ under left- and right-translations, we can check that it is a symmetric operator with $\text{Ker}\left(\widehat{\mathrm{P}^{(\overline{F},u)}_\gamma}^{\dagger} \pm i\right) = \{0\}$ (like in the proof of prop. 3.11), therefore it is an essentially self-adjoint operator. The left- and right-invariance of the measure also ensures that, for any $t \in G$, $T^{(\overline{F})}_{\gamma,t,*}\,\mu_\gamma = \mu_\gamma$, so $\widehat{\mathrm{T}^{(\overline{F},t)}_\gamma}$ is a well-defined unitary operator on $\mathcal{H}_\gamma$.

Now, let $\gamma \in \mathcal{L}_{\text{AL}}$ and let $\lambda \in \mathcal{L}_{\text{profls}}$ such that there exists $F \in \mathcal{F}(\lambda)$ with $\overline{F} = F^\perp \circ F$. Since $\mathcal{L}_{\text{graphs}}$ is cofinal in $\mathcal{L}_{\text{AL}}$ (prop. 3.13), there exists $\gamma' \in \mathcal{L}_{\text{graphs}}$ with $\gamma' \succcurlyeq \gamma$, and by subsection 2.2, there exists $(\gamma'', \lambda'') \in \mathcal{L}$ with $\gamma'' \succcurlyeq \gamma'$ and $\lambda'' \succcurlyeq \lambda$. Next, let $e'' \in \gamma''$ and suppose that there exists $p \neq p' \in r(e'')$ such that $e''_{[p,p']} \in \overline{F}$. Then, using $\overline{F} = F^\perp \circ F$ together with $\lambda'' \succcurlyeq \lambda$, there exists $p'' \in r\left(e''_{[p,p']}\right) \setminus \{p,\,p'\} \subset r(e'') \setminus \{b(e''),\,f(e'')\}$ such that $e''_{[p,p'']} \in F''$ for some $F'' \in \mathcal{F}(\lambda'')$. But since $e'' \in \chi_{(\gamma'',\lambda'')}(e'')$, this can only be the case if $p = b(e)$ (props. 2.8.5 and 2.8.6). Therefore $\gamma'' \in \mathcal{L}_{\text{AL}/\overline{F}}$. Thus, $\mathcal{L}_{\text{AL}/\overline{F}}$ is cofinal in $\mathcal{L}_{\text{AL}}$.

Let $e \in \mathcal{L}^{(1)}_{\text{edges}}$ such that $c(e,\overline{F}) \subset \{b(e), f(e)\}$ and let $e_1\,,\,\dots\,,\,e_n \in \mathcal{L}^{(1)}_{\text{edges}}\,,\ \epsilon_1\,,\,\dots\,,\,\epsilon_n \in \{\pm 1\}$

such that $e = e_n^{\epsilon_n} \circ \ldots \circ e_1^{\epsilon_1}$. Then, using:

$$\forall \widetilde{e} \in \overline{F},\ \forall p \in r(\widetilde{e}) \setminus \{b(e)\}\,,\ \widetilde{e}_{[b(e),p]} \in \overline{F}\,,$$

we have:

$$b\left(e_i^{\epsilon_i}\right) \in c(e_i, \overline{F}) \Leftrightarrow \left(i = 1 \quad \& \quad b(e) \in c(e, \overline{F})\right),$$

$$f\left(e_i^{\epsilon_i}\right) \in c(e_i, \overline{F}) \Leftrightarrow \left(i = n \quad \& \quad f(e) \in c(e, \overline{F})\right).$$

Thus, for any $\gamma \preccurlyeq \gamma' \in \mathcal{L}_{\text{AL}/\overline{F}}$, we can check that:

$$\forall h' \in \mathcal{C}_{\gamma'}\,,\ \forall t \in G,\quad \pi_{\gamma' \to \gamma}\left(T^{(\overline{F})}_{\gamma',t} h'\right) = T^{(\overline{F})}_{\gamma,t}\left(\pi_{\gamma' \to \gamma}(h')\right). \tag{3.18.1}$$

Hence, for any $t \in G$, the unitary operators $\widehat{\mathrm{T}^{(\overline{F},t)}_{\gamma}}$ defined on each $\mathcal{H}_\gamma$ for $\gamma \in \mathcal{L}_{\text{AL}/\overline{F}}$ can be assembled into a unitary operator $\widehat{\mathrm{T}^{(\overline{F},t)}_{\text{AL}}}$ on $\mathcal{H}_{\text{AL}}$, while, for any $u \in \mathrm{g}$, a densely defined, symmetric operator $\widehat{\mathrm{P}^{(\overline{F},u)}_{\text{AL}}}$ can be constructed on the dense subspace $\mathcal{D}_{\text{AL}} \subset \mathcal{H}_{\text{AL}}$, defined as the inductive limit of vector spaces $\left(\mathcal{D}_\gamma\right)_{\gamma \in \mathcal{L}_{\text{AL}}}$, $\left(\tau_{\gamma' \leftarrow \gamma}|_{\mathcal{D}_\gamma \to \mathcal{D}_{\gamma'}}\right)_{\gamma \preccurlyeq \gamma'}$.

Finally, let $\psi \in \text{Ker}\left(\widehat{\mathrm{P}^{(\overline{F},u)}_{\text{AL}}}^{\dagger} \pm i\right)$ and define, for $\gamma \in \mathcal{L}_{\text{AL}/\overline{F}}$, $\psi_\gamma \in \mathcal{H}_\gamma$ such that $\tau_{\text{AL} \leftarrow \gamma}\left(\psi_\gamma\right)$ is the orthogonal projection of $\psi$ on the closed vector subspace $\tau_{\text{AL} \leftarrow \gamma}\left\langle \mathcal{H}_\gamma \right\rangle$. Then, $\psi_\gamma \in \text{Ker}\left(\widehat{\mathrm{P}^{(\overline{F},u)}_{\gamma}}^{\dagger} \pm i\right) = \{0\}$, so:

$$\psi \in \left(\bigcup_{\gamma \in \mathcal{L}_{\text{AL}/\overline{F}}} \tau_{\text{AL} \leftarrow \gamma}\left\langle \mathcal{H}_\gamma \right\rangle\right)^{\perp} = \left(\bigcup_{\gamma \in \mathcal{L}_{\text{AL}}} \tau_{\text{AL} \leftarrow \gamma}\left\langle \mathcal{H}_\gamma \right\rangle\right)^{\perp} = \{0\}\,.$$

Hence, $\widehat{\mathrm{P}^{(\overline{F},u)}_{\text{AL}}}$ is essentially self-adjoint. □

**Proposition 3.19** Let $\overline{F} \in \mathcal{L}_{\text{faces}}$ and $t \in G$. For any $\eta \in \mathcal{L}$ such that $\lambda(\eta) \in \mathcal{L}_{\text{profls}/\overline{F}}$, we define a unitary operator $\widehat{\mathrm{T}^{(\overline{F},t)}_{\eta}}$ on $\mathcal{H}_\eta$ by:

$$\forall \psi \in \mathcal{H}_\eta\,,\ \forall h \in \mathcal{C}_\eta\,,\ \left(\widehat{\mathrm{T}^{(\overline{F},t)}_{\eta}} \psi\right)(h) := \psi\left(R^{(\overline{F})}_{\eta,t} h\right),$$

where $R^{(\overline{F})}_{\eta,t} : \mathcal{C}_\eta \to \mathcal{C}_\eta$ was defined in prop. 3.11. Then, for any $\eta, \eta' \in \mathcal{L}$ such that $\lambda(\eta), \lambda(\eta') \in \mathcal{L}_{\text{profls}/\overline{F}}$:

$$\widehat{\mathrm{T}^{(\overline{F},t)}_{\eta}} \sim \widehat{\mathrm{T}^{(\overline{F},t)}_{\eta'}}\,,$$

so this provides an element $\widehat{\mathrm{T}^{(\overline{F},t)}} \in \mathcal{A}^{\otimes}_{(\mathcal{L},\mathcal{H},\Phi)}$ [18, def. 2.3].

Let $\eta \in \mathcal{L}$ and denote by $\mathcal{I}_\eta$ the algebra of bounded operators on $\mathcal{H}_\eta$ generated by:

$$\left\{\widehat{\mathrm{h}^{(e,\delta)}_{\eta}} \,\middle|\, \delta \in C^\infty(G, \mathbb{R}) \quad \& \quad e \in \gamma(\eta)\right\} \cup \left\{\widehat{\mathrm{T}^{(F^{\perp} \circ F,t)}_{\eta}} \,\middle|\, t \in G \quad \& \quad F \in \mathcal{F}(\eta)\right\}.$$

If $\rho$, $\rho'$ are (self-adjoint) positive semi-definite, traceclass operators on $\mathcal{H}_\eta$ such that:

$$\forall A \in \mathfrak{I}_\eta\,,\ \mathrm{Tr}_{\mathcal{H}_\eta}\, \rho\, A = \mathrm{Tr}_{\mathcal{H}_\eta}\, \rho' A\,,$$

then $\rho = \rho'$.

**Proof** *Definition of* $\widehat{T^{(\overline{F},t)}}$. For any $\eta \in \mathcal{L}$ such that $\lambda(\eta) \in \mathcal{L}_{\mathrm{profls}/\overline{F}}$, we have $R^{(\overline{F})}_{\eta,t,*}\mu_\eta = \mu_\eta$ (for $\mu$ is invariant under right-translations) hence $\widehat{\mathrm{T}^{(\overline{F},t)}_\eta}$ is a well-defined and unitary operator on $\mathcal{H}_\eta$.

Next, for any $\eta$ such that $\lambda(\eta) \in \mathcal{L}_{\mathrm{profls}/\overline{F}}$ and any $\eta' \succcurlyeq \eta$, we have $\lambda(\eta') \in \mathcal{L}_{\mathrm{profls}/\overline{F}}$. Moreover, using eqs. (3.11.4), (3.9.1) and (3.5.2), we can check that:

$$\left(\mathrm{id}_{\mathcal{C}_{\eta'\to\eta}} \times R^{(\overline{F})}_{\eta,t}\right) \circ \varphi_{\eta'\to\eta} = \varphi_{\eta'\to\eta} \circ R^{(\overline{F})}_{\eta',t}\,.$$

Thus, we get $\Phi_{\eta'\to\eta} \circ \widehat{\mathrm{T}^{(\overline{F},t)}_{\eta'}} = \left(\mathrm{id}_{\mathcal{H}_{\eta'\to\eta}} \times \widehat{\mathrm{T}^{(\overline{F},t)}_{\eta}}\right) \circ \Phi_{\eta'\to\eta}$. Therefore, the directedness of $\mathcal{L}$ ensures that, for any $\eta$, $\eta' \in \mathcal{L}$ such that $\lambda(\eta), \lambda(\eta') \in \mathcal{L}_{\mathrm{profls}/\overline{F}}$, $\widehat{\mathrm{T}^{(\overline{F},t)}_\eta} \sim \widehat{\mathrm{T}^{(\overline{F},t)}_{\eta'}}$ (with the equivalence relation defined in [18, eq. (2.3.2)]), so we can define an element $\widehat{\mathrm{T}^{(\overline{F},t)}} \in \mathcal{A}^{\otimes}_{(\mathcal{L},\mathcal{H},\Phi)}$.

*Definition of* $\widehat{h^{(\Delta)}_\eta}$ *and* $\widehat{T^{(j)}_\eta}$. Let $\eta \in \mathcal{L}$ and $\Delta : \gamma(\eta) \to C^\infty(G, \mathbb{R})$. Then we can define an element $\widehat{\mathrm{h}^{(\Delta)}_\eta} \in \mathfrak{I}_\eta$ by:

$$\widehat{\mathrm{h}^{(\Delta)}_\eta} := \prod_{e\in\gamma(\eta)} \widehat{\mathrm{h}^{(e,\Delta(e))}}_\eta\,.$$

The right-hand side does not depend on the ordering of the product and we have:

$$\forall \psi \in \mathcal{H}_\eta\,,\ \forall h \in \mathcal{C}_\eta\,,\ \left(\widehat{\mathrm{h}^{(\Delta)}_\eta}\psi\right)(h) = \left(\prod_{e\in\gamma(\eta)} [\Delta(e)\circ h](e)\right)\psi(h)\,.$$

Next, for any $F \in \mathcal{F}(\eta)$ and any $t \in G$, we have $\lambda(\eta) \in \mathcal{L}_{\mathrm{profls}/F^\perp\circ F}$ and, with the help of prop. 2.8:

$$\forall h \in \mathcal{C}_\eta\,,\ \forall e \in \gamma(\eta)\,,\ \left[R^{(F^\perp\circ F)}_{\eta,t}\, h\right](e) := \begin{cases} h(e)\,.\,t & \text{if } \chi_\eta(e) = F \\ h(e) & \text{else} \end{cases}\,.$$

Hence, for any $j \in \mathcal{C}_\eta$ we can define a unitary operator $\widehat{\mathrm{T}^{(j)}_\eta} \in \mathfrak{I}_\eta$ by:

$$\widehat{\mathrm{T}^{(j)}_\eta} := \prod_{F\in\mathcal{F}(\eta)} \widehat{\mathrm{T}^{(F^\perp\circ F,\, j\circ\chi_\eta^{-1}(F))}_\eta}\,,$$

and we have (the product ordering being again irrelevant):

$$\forall \psi \in \mathcal{H}_\eta\,,\ \forall h \in \mathcal{C}_\eta\,,\ \left(\widehat{\mathrm{T}^{(j)}_\eta}\psi\right)(h) = \psi(h\,.\,j)\,,$$

where, for any $h$, $h' \in \mathcal{C}_\eta$, $h\,.\,h'$ denotes their pointwise multiplication as maps $\gamma(\eta) \to G$.

*Characterization of* $\rho \in \overline{\mathcal{S}}_\eta$ *by its evaluations over* $\mathfrak{I}_\eta$. Let $\rho, \rho'$ be non-negative traceclass operators on $\mathcal{H}_\eta$ and $\varphi, \varphi' \in \mathcal{H}_\eta$. Let $\epsilon > 0$ and define $\epsilon_o := \frac{\epsilon}{10}$, as well as:

$$\epsilon_1 := \min\left(\frac{\epsilon_o}{1 + \operatorname{Tr}_{\mathcal{H}_\eta}\rho}, \frac{\epsilon_o}{1 + \operatorname{Tr}_{\mathcal{H}_\eta}\rho'}, 1\right) > 0.$$

The $\mathbb{C}$-vector space generated by $C^\infty(G, \mathbb{R})$ is dense in $L_2(G, d\mu)$ ($C^\infty(G, \mathbb{R}) \subset L_2(G, d\mu)$ for $G$ is compact), hence the $\mathbb{C}$-vector space generated by $\{\otimes_{e\in\gamma(\eta)}\Delta(e) \mid \Delta : \gamma(\eta) \to C^\infty(G, \mathbb{R})\}$ is dense in $\bigotimes_{e\in\gamma(\eta)} L_2(G, d\mu) \approx \mathcal{H}_\eta$ (using the isometric identification provided by Fubini's theorem). Thus, there exist a finite family of maps $\left(\Delta_l : \gamma(\eta) \to C^\infty(G, \mathbb{R})\right)_{1\leqslant l\leqslant L}$ and a finite family of complex numbers $(\mu_l)_{1\leqslant l\leqslant L}$ such that:

$$\left\| \varphi - \sum_{l=1}^{L} \mu_l \left[\otimes_{\gamma(\eta)}\Delta_l\right] \right\|_{\mathcal{H}_\eta} < \frac{\epsilon_1}{1 + \|\varphi'\|_{\mathcal{H}_\eta}} ,$$

where, for any $\Delta : \gamma(\eta) \to C^\infty(G, \mathbb{R})$, the vector $\otimes_{\gamma(\eta)}\Delta \in \mathcal{H}_\eta$ is defined by:

$$\forall h \in \mathcal{C}_\eta, \left[\otimes_{\gamma(\eta)}\Delta\right](h) := \prod_{e\in\gamma(\eta)} \left[\Delta(e) \circ h\right](e).$$

Similarly, there exist $\left(\Delta'_{l'} : \gamma(\eta) \to C^\infty(G, \mathbb{R})\right)_{1\leqslant l'\leqslant L'}$ and $\left(\mu'_{l'}\right)_{1\leqslant l'\leqslant L'}$ ($0 \leqslant L, L' < \infty$) such that:

$$\left\| \varphi' - \sum_{l'=1}^{L'} \mu'_{l'} \left[\otimes_{\gamma(\eta)}\Delta'_{l'}\right] \right\|_{\mathcal{H}_\eta} < \frac{\epsilon_1}{1 + \|\varphi\|_{\mathcal{H}_\eta}} .$$

Thus, we have:

$$\left| \langle \varphi' \mid \rho \mid \varphi\rangle - \sum_{l=1}^{L}\sum_{l'=1}^{L'} {\mu'_{l'}}^{*} \mu_l \left\langle \otimes_{\gamma(\eta)}\Delta'_{l'} \mid \rho \mid \otimes_{\gamma(\eta)}\Delta_l \right\rangle \right|$$
$$< \frac{\epsilon_1 \|\varphi\|_{\mathcal{H}_\eta} \operatorname{Tr}_{\mathcal{H}_\eta}\rho}{1 + \|\varphi\|_{\mathcal{H}_\eta}} + \frac{\|\varphi'\|_{\mathcal{H}_\eta} \epsilon_1 \operatorname{Tr}_{\mathcal{H}_\eta}\rho}{1 + \|\varphi'\|_{\mathcal{H}_\eta}} + \frac{\epsilon_1^2 \operatorname{Tr}_{\mathcal{H}_\eta}\rho}{\left(1 + \|\varphi\|_{\mathcal{H}_\eta}\right)\left(1 + \|\varphi'\|_{\mathcal{H}_\eta}\right)} \leqslant 2\,\epsilon_o , \qquad (3.19.1)$$

and similarly for $\rho'$. We define:

$$\epsilon_2 := \frac{\epsilon_o}{1 + \sum_{l=1}^{L}\sum_{l'=1}^{L'} \left(|\mu_l|\, |\mu'_{l'}| \prod_{e\in\gamma(\eta)} \|\Delta_l(e)\|_\infty \, \|\Delta'_{l'}(e)\|_\infty\right)} > 0,$$

where, for any $\delta \in C^\infty(G, \mathbb{R})$, $\|\delta\|_\infty := \sup_{t\in G} |\delta(t)|$ ($< \infty$ for $G$ is compact). Note that, for any $\Delta : \gamma(\eta) \to C^\infty(G, \mathbb{R})$, we have the bound:

$$\left\| \widehat{\mathrm{h}_\eta^{(\Delta)}} \right\|_{\mathcal{A}_\eta} \leqslant \prod_{e\in\gamma(\eta)} \|\Delta(e)\|_\infty ,$$

where $\|\cdot\|_{\mathcal{A}_\eta}$ denotes the operator norm on the algebra $\mathcal{A}_\eta$ of bounded operators over $\mathcal{H}_\eta$ [18, def. 2.3], as well as:

$$\|\otimes_{\gamma(\eta)}\Delta\|_{\mathcal{H}_\eta} \leqslant \prod_{e\in\gamma(\eta)} \|\Delta(e)\|_\infty$$

(using the fact that $\mu_\eta$ is normalized).

Now, from the spectral theorem, together with the non-negative and traceclass conditions, there exist an orthonormal family $\left(\widetilde{\psi}_k\right)_{1\leqslant k\leqslant \widetilde{K}}$ in $\mathcal{H}_\eta$ (with $0 \leqslant \widetilde{K} \leqslant \infty$) and a family of strictly positive reals $(p_k)_{1\leqslant k\leqslant \widetilde{K}}$ such that:

$$\rho = \sum_{k=1}^{\widetilde{K}} p_k \left| \widetilde{\psi}_k \right\rangle \left\langle \widetilde{\psi}_k \right| \quad \& \quad \sum_{k=1}^{\widetilde{K}} p_k =: \mathrm{Tr}_{\mathcal{H}_\eta} \rho < \infty .$$

Let $K < \infty$ such that $\left|\mathrm{Tr}_{\mathcal{H}_\eta} \rho - \sum_{k=1}^{K} p_k\right| < \frac{\epsilon_2}{4}$. Then, for any $k \leqslant K$, there exists $\psi_k \in C^\infty(\mathcal{C}_\eta\,,\,\mathbb{C}) \subset \mathcal{H}_\eta$ such that:

$$\left\|\widetilde{\psi}_k - \psi_k\right\|_{\mathcal{H}_\eta} < \min\left(1,\, \frac{\epsilon_2}{4\,\mathrm{Tr}_{\mathcal{H}_\eta} \rho + 1}\right).$$

Thus, we have:

$$\left|\sum_{l=1}^{L}\sum_{l'=1}^{L'} {\mu'_{l'}}^* \mu_l \,\langle \otimes_{\gamma(\eta)}\Delta'_{l'} \mid \rho \mid \otimes_{\gamma(\eta)}\Delta\rangle - \sum_{k=1}^{K}\sum_{l=1}^{L}\sum_{l'=1}^{L'} p_k\, {\mu'_{l'}}^* \mu_l \,\langle \otimes_{\gamma(\eta)}\Delta'_{l'} \mid \psi_k\rangle \,\langle \psi_k \mid \otimes_{\gamma(\eta)}\Delta\rangle\right|$$
$$< \sum_{l=1}^{L}\sum_{l'=1}^{L'} |\mu'_{l'}|\,|\mu_l|\,\|\otimes_{\gamma(\eta)}\Delta'_{l'}\|_{\mathcal{H}_\eta}\,\|\otimes_{\gamma(\eta)}\Delta\|_{\mathcal{H}_\eta}\left[\frac{\epsilon_2}{4} + \sum_{k=1}^{K} p_k \frac{3\epsilon_2}{4\,\mathrm{Tr}_{\mathcal{H}_\eta} \rho + 1}\right] \leqslant \epsilon_o\,, \tag{3.19.2}$$

and, for any $j \in \mathcal{C}_\eta$:

$$\left|\sum_{l=1}^{L}\sum_{l'=1}^{L'} {\mu'_{l'}}^* \mu_l \,\mathrm{Tr}_{\mathcal{H}_\eta}\left(\rho\, \widehat{\mathrm{h}_\eta^{(\Delta_l)}}\,\widehat{\mathrm{T}_\eta^{(j)}}\,\widehat{\mathrm{h}_\eta^{(\Delta'_{l'})}}\right) - \sum_{k=1}^{K}\sum_{l=1}^{L}\sum_{l'=1}^{L'} p_k\, {\mu'_{l'}}^* \mu_l \left\langle \psi_k \,\middle|\, \widehat{\mathrm{h}_\eta^{(\Delta_l)}}\,\widehat{\mathrm{T}_\eta^{(j)}}\,\widehat{\mathrm{h}_\eta^{(\Delta'_{l'})}} \,\middle|\, \psi_k\right\rangle\right|$$
$$< \sum_{l=1}^{L}\sum_{l'=1}^{L'} |\mu'_{l'}|\,|\mu_l| \left\|\widehat{\mathrm{h}_\eta^{(\Delta_l)}}\,\widehat{\mathrm{T}_\eta^{(j)}}\,\widehat{\mathrm{h}_\eta^{(\Delta'_{l'})}}\right\|_{\mathcal{A}_\eta}\left[\frac{\epsilon_2}{4} + \sum_{k=1}^{K} p_k \frac{3\,\epsilon_2}{4\,\mathrm{Tr}_{\mathcal{H}_\eta} \rho + 1}\right] \leqslant \epsilon_o\,. \tag{3.19.3}$$

Similarly, there exist a finite family $\left(\psi'_k\right)_{1\leqslant k\leqslant K'}$ of functions in $C^\infty(\mathcal{C}_\eta\,,\,\mathbb{C})$ (with $0 \leqslant K' < \infty$) and a finite family $\left(p'_k\right)_{1\leqslant k\leqslant K'}$ of strictly positive reals such that the equivalents of eqs. (3.19.2) and (3.19.3) are fulfilled for $\rho'$. We define:

$$\epsilon_3 := \frac{\epsilon_o}{1 + \sum_{k=1}^{K}\sum_{l=1}^{L}\sum_{l'=1}^{L'} p_k\,|\mu_l|\,|\mu'_{l'}|\,\|\psi_k\|_\infty\,\|\otimes_{\gamma(\eta)}\Delta_l\|_\infty} > 0\,,$$

where, for any $\zeta \in C^\infty(\mathcal{C}_\eta\,,\,\mathbb{C})$, $\|\zeta\|_\infty := \sup_{j\in\mathcal{C}_\eta} |\zeta(j)|$. Similarly, we define $\epsilon'_3$ using $\left(\psi'_k\right)_{1\leqslant k\leqslant K'}$ and $\left(p'_k\right)_{1\leqslant k\leqslant K'}$.

Let $l' \leqslant L'$ and $k \leqslant K$. The function $\zeta_{l',k}$, defined on $\mathcal{C}_\eta \times \mathcal{C}_\eta$ by:

$$\forall h \in \mathcal{C}_\eta\,,\, \forall j \in \mathcal{C}_\eta\,,\, \zeta_{l',k}(h,j) := [\otimes_{\gamma(\eta)}\Delta'_{l'}](h.j)\,\psi_k(h.j)\,,$$

is smooth by construction. Hence, for any $h \in \mathcal{C}_\eta$, there exists an open, measurable neighborhood $V_{k,l'}^{(h)}$ of $\mathbf{1}$ in $\mathcal{C}_\eta$ and an open, measurable neighborhood $W_{k,l'}^{(h)}$ of $h$ in $\mathcal{C}_\eta$ such that:

$$\forall h' \in W^{(h)}_{k,l'}\,,\ \forall j \in V^{(h)}_{k,l'}\,,\ |\zeta_{l',k}(h',j) - \zeta_{l',k}(h,\mathbf{1})| < \frac{\epsilon_3}{2}\,,$$

and therefore:

$$\forall h' \in W^{(h)}_{k,l'}\,,\ \forall j \in V^{(h)}_{k,l'}\,,\ |\zeta_{l',k}(h',j) - \zeta_{l',k}(h',\mathbf{1})| < \epsilon_3\,.$$

Now, since $\mathfrak{C}_\eta$ is compact, there exists a finite subset $H_{k,l'}$ of $\mathfrak{C}_\eta$ such that $\mathfrak{C}_\eta \subset \bigcup_{h\in H_{k,l'}} W^{(h)}_{k,l'}$, so, defining $V_{k,l'} := \bigcap_{h\in H_{k,l'}} V^{(h)}_{k,l'}$, $V_{k,l'}$ is an open, measurable neighborhood of $\mathbf{1}$ in $\mathfrak{C}_\eta$ and we have:

$$\forall h' \in \mathfrak{C}_\eta\,,\ \forall j \in V_{k,l'}\,,\ |\zeta_{l',k}(h',j) - \zeta_{l',k}(h',\mathbf{1})| < \epsilon_3\,.$$

Next, defining $V := \bigcap_{k=1}^{K}\bigcap_{l'=1}^{L'} V_{k,l'}$, we get, for any measurable subset $\widetilde{V} \subset V$, any $k \leqslant K$ and any $l' \leqslant L'$:

$$\forall h \in \mathfrak{C}_\eta\,,\ \left|\left(\int_{\widetilde{V}} d\mu_\eta(j)\ [\otimes_{\gamma(\eta)}\Delta'_{l'}](h.j)\,\psi_k(h.j)\right) - \left(\mu(\widetilde{V})\ [\otimes_{\gamma(\eta)}\Delta'_{l'}](h)\,\psi_k(h)\right)\right| < \mu(\widetilde{V})\,\epsilon_3\,. \quad (3.19.4)$$

Similarly, we have an open, measurable neighborhood $V'$ of $\mathbf{1}$ in $\mathfrak{C}_\eta$ such that, for any measurable subset $\widetilde{V} \subset V'$, any $k' \leqslant K'$ and any $l' \leqslant L'$, the equivalent of eq. (3.19.4) is fulfilled for $\psi'_{k'}$ (with respect to $\epsilon'_3$ instead of $\epsilon_3$). We define $V_o := V \cap V'$.

For any $j \in \mathfrak{C}_\eta$, $j\,.\,V_o$ is an open, measurable neighborhood of $j$ in $\mathfrak{C}_\eta$, hence, from the compacity of $\mathfrak{C}_\eta$, there exist $j_1, \ldots, j_M$ in $\mathfrak{C}_\eta$ $(1 \leqslant M < \infty)$ such that $\mathfrak{C}_\eta \subset \bigcup_{m=1}^{M} j_m\,.\,V_o$. For $m \leqslant M$ we define:

$$V_m := (j_m\,.\,V_o) \cap \left(\mathfrak{C}_\eta \setminus \bigcup_{n=1}^{m-1} j_n\,.\,V_o\right).$$

Thus, $(V_m)_{1\leqslant m\leqslant M}$ is a partition of $\mathfrak{C}_\eta$ into finitely many measurable parts. Moreover, we have, for any $m \leqslant M$:

$$j_m^{-1}\,.\,V_m \subset V_o\,,$$

so that, for any $k \leqslant K$ and any $l' \leqslant L'$, eq. (3.19.4) yields:

$$\forall h \in \mathfrak{C}_\eta\,,\ \left|\langle \otimes_{\gamma(\eta)}\Delta'_{l'} \mid \psi_k\rangle - \sum_{m=1}^{M} \mu(V_m)\left[\widehat{\mathrm{T}^{(j_m)}_\eta}\,\widehat{\mathrm{h}^{(\Delta'_{l'})}_\eta}\,\psi_k\right](h)\right| =$$

$$= \left|\left(\int_{\mathfrak{C}_\eta} d\mu_\eta(j)\ [\otimes_{\gamma(\eta)}\Delta'_{l'}](h.j)\,\psi_k(h.j)\right) - \sum_{m=1}^{M}\left(\mu(V_m)\ [\otimes_{\gamma(\eta)}\Delta'_{l'}](h.j_m)\,\psi_k(h.j_m)\right)\right|$$

$$\leqslant \sum_{m=1}^{M}\left|\left(\int_{j_m^{-1}\,.\,V_m} d\mu_\eta(j)\ [\otimes_{\gamma(\eta)}\Delta'_{l'}](h.j_m\,.\,j)\,\psi_k(h.j_m\,.\,j)\right) - \left(\mu(j_m^{-1}\,.\,V_m)\ [\otimes_{\gamma(\eta)}\Delta'_{l'}](h.j_m)\,\psi_k(h.j_m)\right)\right|$$

$$< \epsilon_3\,.$$

Therefore, for any $k \leqslant K$, any $l \leqslant L$ and any $l' \leqslant L'$, we get:

$$\left|\langle \psi_k \mid \otimes_{\gamma(\eta)}\Delta_l\rangle\,\langle \otimes_{\gamma(\eta)}\Delta'_{l'} \mid \psi_k\rangle - \sum_{m=1}^{M}\mu(V_m)\left\langle \psi_k \middle| \widehat{\mathrm{h}^{(\Delta_l)}_\eta}\,\widehat{\mathrm{T}^{(j_m)}_\eta}\,\widehat{\mathrm{h}^{(\Delta'_{l'})}_\eta} \middle| \psi_k\right\rangle\right|$$

$$\leqslant \left| \int_{\mathcal{C}_\eta} d\mu(h)\, \psi_k^*(h)\, [\otimes_{\gamma(\eta)}\Delta_l](h) \left( \langle \otimes_{\gamma(\eta)}\Delta'_{l'} \mid \psi_k \rangle - \sum_{m=1}^{M} \mu(V_m) \left[ \widehat{\mathsf{T}_\eta^{(j_m)}}\, \widehat{\mathsf{h}_\eta^{(\Delta'_{l'})}}\, \psi_k \right](h) \right) \right|$$

$$< \epsilon_3\, \|\psi_k\|_\infty\, \|\otimes_{\gamma(\eta)}\Delta_l\|_\infty .$$

Now, using eqs. (3.19.1) to (3.19.3) together with the definition of $\epsilon_3$, this implies:

$$\left| \langle \varphi' \mid \rho \mid \varphi \rangle - \sum_{l=1}^{L} \sum_{l'=1}^{L'} \sum_{m=1}^{M} {\mu'_{l'}}^{*}\, \mu_l\, \mu(V_m)\, \mathrm{Tr}_{\mathcal{H}_\eta} \left( \rho\, \widehat{\mathsf{h}_\eta^{(\Delta_l)}}\, \widehat{\mathsf{T}_\eta^{(j_m)}}\, \widehat{\mathsf{h}_\eta^{(\Delta'_{l'})}} \right) \right| < 5\, \epsilon_o ,$$

and the same holds for $\rho'$.

Finally, if $\rho$, $\rho'$ are such that:

$$\forall A \in \mathfrak{I}_\eta ,\ \mathrm{Tr}_{\mathcal{H}_\eta} \rho A = \mathrm{Tr}_{\mathcal{H}_\eta} \rho' A ,$$

we thus have:

$$\forall \varphi, \varphi' \in \mathcal{H}_\eta ,\ \forall \epsilon > 0,\ |\langle \varphi' \mid \rho \mid \varphi \rangle - \langle \varphi' \mid \rho' \mid \varphi \rangle| < \epsilon ,$$

and therefore $\rho = \rho'$. □

We can now formulate the relation between the inductive construction just reviewed and the projective construction from subsection 3.1, by displaying how an arbitrary state over $\mathcal{H}_{\mathrm{AL}}$ can be unambiguously identified with a projective family of density matrices over the Hilbert spaces $\mathcal{H}_\eta$. Note that, as stressed above, we are not merely stating that there exists *some* injective map $\sigma$ between the state spaces (which would only be an assertion about the respective cardinalities, with very little physical content): we also make sure that the mapping of the states considered here intertwines the evaluation of observables in agreement with their physical interpretation.

By using [18, prop. 3.5] (which is itself a straightforward application of the more general result in [18, theorem 2.9]), we first obtain a map from the state space of an inductive limit of Hilbert spaces built over the label set $\mathcal{L}$. As previously announced, the insensitivity of the injections with respect to the faces in each label then allows to collapse this inductive system into a simpler one, built over a subset of $\mathcal{L}_{\text{graphs}}$ (namely those analytic graphs that are the underlying graph of some label). Finally, since this set of graphs is cofinal in $\mathcal{L}_{\mathrm{AL}}$ (as follows from $\mathcal{L}$ being cofinal in $\mathcal{L}_{\text{graphs}} \times \mathcal{L}_{\text{profls}}$ and $\mathcal{L}_{\text{graphs}}$ in $\mathcal{L}_{\mathrm{AL}}$), the corresponding inductive limit can be identified with $\mathcal{H}_{\mathrm{AL}}$.

**Theorem 3.20** There exist maps $\sigma : \overline{\mathcal{S}}_{\mathrm{AL}} \to \overline{\mathcal{S}}^{\otimes}_{(\mathcal{L},\mathcal{H},\Phi)}$ and $\alpha : \overline{\mathcal{A}}^{\otimes}_{(\mathcal{L},\mathcal{H},\Phi)} \to \mathcal{A}_{\mathrm{AL}}$ (where $\overline{\mathcal{S}}_{\mathrm{AL}}$ is the space of self-adjoint, positive semi-definite, traceclass operators over $\mathcal{H}_{\mathrm{AL}}$, $\mathcal{A}_{\mathrm{AL}}$ is the space of bounded operators on $\mathcal{H}_{\mathrm{AL}}$, and $\overline{\mathcal{S}}^{\otimes}_{(\mathcal{L},\mathcal{H},\Phi)}$ and $\overline{\mathcal{A}}^{\otimes}_{(\mathcal{L},\mathcal{H},\Phi)}$ were defined respectively in [18, def. 2.2 and prop. 2.4]) such that:

1. $\alpha$ is a $C^*$-algebra morphism;
2. for any $e \in \mathcal{L}_{\text{edges}}$ and any $\delta \in C^\infty(G, \mathbb{R})$, $\alpha\big(\widehat{\mathsf{h}^{(e,\delta)}}\big) = \widehat{\mathsf{h}^{(e,\delta)}_{\mathrm{AL}}}$, while for any $\overline{F} \in \mathcal{L}_{\text{faces}}$ and any $t \in G$, $\alpha\big(\widehat{\mathsf{T}^{(\overline{F},t)}}\big) = \widehat{\mathsf{T}^{(\overline{F},t)}_{\mathrm{AL}}}$;
3. for any $\rho \in \overline{\mathcal{S}}_{\mathrm{AL}}$ and any $A \in \overline{\mathcal{A}}^{\otimes}_{(\mathcal{L},\mathcal{H},\Phi)}$, $\mathrm{Tr}_{\mathcal{H}_{\mathrm{AL}}} (\rho\, \alpha(A)) = \mathrm{Tr}\, (\sigma(\rho)\, A)$;

4. $\sigma$ is an injective map;

5. $\sigma\left\langle \mathcal{S}_{\text{AL}}\right\rangle = \left\{ \left(\rho_\eta\right)_{\eta\in\mathcal{L}} \,\middle|\, \sup_{\eta\in\mathcal{L}} \inf_{\eta'\succcurlyeq\eta} \mathrm{Tr}_{\mathcal{H}_{\eta'}}\, \rho_{\eta'}\, \Theta_{\eta'|\eta} = 1 \right\}$, where $\mathcal{S}_{\text{AL}}$ is the space of density matrices over $\mathcal{H}_{\text{AL}}$ and the bounded operator $\Theta_{\eta'|\eta}$ is defined on $\mathcal{H}_{\eta'}$ by:

$$\forall \psi \in \mathcal{H}_{\eta'}\,,\ \forall j \in \mathcal{C}_{\eta'\to\eta}\,,\ \forall h \in \mathcal{C}_\eta\,,\ \left[\Theta_{\eta'|\eta}\,\psi\right] \circ \varphi^{-1}_{\eta'\to\eta}(j,\,h) := \int_{\mathcal{C}_{\eta'\to\eta}} d\mu_{\eta'\to\eta}(\widetilde{j})\ \ \psi \circ \varphi^{-1}_{\eta'\to\eta}(\widetilde{j},\,h)\,.$$

**Proof** *Auxiliary inductive limit of Hilbert spaces.* For any $\eta \preccurlyeq \eta' \in \mathcal{L}$, we define the map $\tau_{\eta'\leftarrow\eta} : \mathcal{H}_\eta \to \mathcal{H}_{\eta'}$ by:

$$\forall \psi \in \mathcal{H}_\eta\,,\ \forall h_{\eta'} \in \mathcal{C}_{\eta'}\,,$$
$$\left[\tau_{\eta'\leftarrow\eta}(\psi)\right](h_{\eta'}) := \psi\left(e \mapsto \left[\prod_{k=2}^{n_{\eta'\to\eta,e}} \left[h_{\eta'} \circ a_{\eta'\to\eta,e}(k)\right]^{\epsilon_{\eta'\to\eta,e}(k)}\right].\left[h_{\eta'} \circ a_{\eta'\to\eta,e}(1)\right]\right).$$

As was shown in [18, prop. 3.5], $\tau_{\eta'\leftarrow\eta}$ is well-defined, and $\left(\mathcal{L},\ \left(\mathcal{H}_\eta\right)_{\eta\in\mathcal{L}},\ \left(\tau_{\eta'\leftarrow\eta}\right)_{\eta\preccurlyeq\eta'}\right)$ is an inductive system of Hilbert spaces whose limit we denote by $\mathcal{H}_{\widetilde{\text{AL}}}$. For any $\eta \in \mathcal{L}$, we call $\tau_{\widetilde{\text{AL}}\leftarrow\eta}$ the natural injection of $\mathcal{H}_\eta$ into $\mathcal{H}_{\widetilde{\text{AL}}}$. Also by [18, prop. 3.5], there exist maps $\widetilde{\sigma} : \overline{\mathcal{S}}_{\widetilde{\text{AL}}} \to \overline{\mathcal{S}}^{\otimes}_{(\mathcal{L},\mathcal{H},\Phi)}$ and $\widetilde{\alpha} : \overline{\mathcal{A}}^{\otimes}_{(\mathcal{L},\mathcal{H},\Phi)} \to \mathcal{A}_{\widetilde{\text{AL}}}$ (where $\overline{\mathcal{S}}_{\widetilde{\text{AL}}}$, resp. $\mathcal{A}_{\widetilde{\text{AL}}}$, denote the space of non-negative traceclass operators, resp. of bounded operators, over $\mathcal{H}_{\widetilde{\text{AL}}}$) satisfying:

$$\forall \rho \in \overline{\mathcal{S}}_{\widetilde{\text{AL}}}\,,\ \forall A \in \overline{\mathcal{A}}^{\otimes}_{(\mathcal{L},\mathcal{H},\Phi)}\,,\ \mathrm{Tr}_{\mathcal{H}_{\widetilde{\text{AL}}}}\big(\rho\, \widetilde{\alpha}(A)\big) = \mathrm{Tr}\big(\widetilde{\sigma}(\rho)\, A\big)\,.$$

Moreover, $\widetilde{\sigma}$ is injective and:

$$\widetilde{\sigma}\left\langle \mathcal{S}_{\widetilde{\text{AL}}}\right\rangle = \left\{ \left(\rho_\eta\right)_{\eta\in\mathcal{L}} \,\middle|\, \sup_{\eta\in\mathcal{L}} \inf_{\eta'\succcurlyeq\eta} \int_{\mathcal{C}_{\eta'\to\eta}\times\mathcal{C}_{\eta'\to\eta}} d^{(2)}\mu_{\eta'\to\eta}(j,\,j') \int_{\mathcal{C}_\eta} d\mu_\eta(h)\ \ \rho_{\eta'}\left(\varphi^{-1}_{\eta'\to\eta}(j,h),\, \varphi^{-1}_{\eta'\to\eta}(j',h)\right) = 1 \right\},$$

where $\mathcal{S}_{\widetilde{\text{AL}}}$ is the space of density matrices over $\mathcal{H}_{\widetilde{\text{AL}}}$ and, for any density matrix $\rho_\eta$ over $\mathcal{H}_\eta$, $\rho_\eta(\,\cdot\,,\,\cdot\,)$ denotes the integral kernel of $\rho_\eta$.

To further specify $\widetilde{\alpha}$, we now fetch its explicit definition from the proof of [18, theorem 2.9]. First, we can define, for any $\eta \in \mathcal{L}$, an Hilbert space $\mathcal{H}_{\widetilde{\text{AL}}\to\eta}$, and an Hilbert space isomorphism $\Phi_{\widetilde{\text{AL}}\to\eta} : \mathcal{H}_{\widetilde{\text{AL}}} \to \mathcal{H}_{\widetilde{\text{AL}}\to\eta} \otimes \mathcal{H}_\eta$. $\mathcal{H}_{\widetilde{\text{AL}}\to\eta}$ is given as an inductive limit and we have, for any $\kappa \succcurlyeq \eta \in \mathcal{L}$, a natural isometric injection $\tau_{\widetilde{\text{AL}}\leftarrow\kappa\to\eta} : \mathcal{H}_{\kappa\to\eta} \to \mathcal{H}_{\widetilde{\text{AL}}\to\eta}$ satisfying:

$$\left(\tau_{\widetilde{\text{AL}}\leftarrow\kappa\to\eta} \otimes \mathrm{id}_{\mathcal{H}_\eta}\right) \circ \Phi_{\kappa\to\eta} = \Phi_{\widetilde{\text{AL}}\to\eta} \circ \tau_{\widetilde{\text{AL}}\leftarrow\kappa}\,.$$

Then, $\widetilde{\alpha}$ is the $C^*$-algebra morphism $\overline{\mathcal{A}}^{\otimes}_{(\mathcal{L},\mathcal{H},\Phi)} \to \mathcal{A}_{\widetilde{\text{AL}}}$ such that, for any $\eta \in \mathcal{H}_\eta$ and any bounded operator $A_\eta$ on $\mathcal{H}_\eta$:

$$\widetilde{\alpha}\left([A_\eta]_\sim\right) = \Phi^{-1}_{\widetilde{\text{AL}}\to\eta} \circ \left(\mathrm{id}_{\mathcal{H}_{\widetilde{\text{AL}}\to\eta}} \otimes A_\eta\right) \circ \Phi_{\widetilde{\text{AL}}\to\eta}\,,$$

where $[A_\eta]_\sim$ denotes the equivalence class of $A_\eta$ in $\mathcal{A}^{\otimes}_{(\mathcal{L},\mathcal{H},\Phi)}$ [18, eq. (2.3.2)]. In particular, for any $\kappa \succcurlyeq \eta$, we have:

$$\widetilde{\alpha}\left([A_\eta]_\sim\right) \circ \tau_{\widetilde{\text{AL}}\leftarrow\kappa} = \tau_{\widetilde{\text{AL}}\leftarrow\kappa} \circ \Phi^{-1}_{\kappa\to\eta} \circ \left(\mathrm{id}_{\mathcal{H}_{\kappa\to\eta}} \otimes A_\eta\right) \circ \Phi_{\kappa\to\eta}\,.$$

*Identification of* $\mathcal{H}_{\widetilde{\text{AL}}}$ *with* $\mathcal{H}_{\text{AL}}$. For any $\eta \in \mathcal{L}$, we define $\Gamma_{\eta\to\gamma(\eta)} : \mathcal{H}_\eta \to \mathcal{H}_{\gamma(\eta)}$ as the identity map on $\mathcal{H}_\eta = L_2(\mathcal{C}_\eta\,, d\mu_\eta) = L_2(\mathcal{C}_{\gamma(\eta)}\,, d\mu_{\gamma(\eta)}) = \mathcal{H}_{\gamma(\eta)}$. Then, for any $\eta \preccurlyeq \eta' \in \mathcal{L}$, we have $\gamma(\eta) \preccurlyeq \gamma(\eta')$ and (from props. 3.2 and 3.14):

$$\Gamma_{\eta'\to\gamma(\eta')} \circ \tau_{\eta'\leftarrow\eta} = \tau_{\gamma(\eta')\leftarrow\gamma(\eta)} \circ \Gamma_{\eta\to\gamma(\eta)}\,.$$

Thus, there exists an isometric injection $\Gamma_{\widetilde{\text{AL}}\to\text{AL}} : \mathcal{H}_{\widetilde{\text{AL}}} \to \mathcal{H}_{\text{AL}}$ satisfying:

$$\forall \eta \in \mathcal{L},\, \Gamma_{\widetilde{\text{AL}}\to\text{AL}} \circ \tau_{\widetilde{\text{AL}}\leftarrow\eta} = \tau_{\text{AL}\leftarrow\gamma(\eta)} \circ \Gamma_{\eta\to\gamma(\eta)}\,.$$

Moreover, $\left\{\gamma \in \mathcal{L}_{\text{AL}} \,\middle|\, \exists \eta \in \mathcal{L} \,/\, \gamma(\eta) = \gamma\right\}$ is cofinal in $\mathcal{L}_{\text{AL}}$ (from subsection 2.2 and prop. 3.13), hence:

$$\bigcup_{\eta\in\mathcal{L}} \Gamma_{\widetilde{\text{AL}}\to\text{AL}} \circ \tau_{\widetilde{\text{AL}}\leftarrow\eta} \left\langle \mathcal{H}_\eta \right\rangle = \bigcup_{\gamma\in\mathcal{L}_{\text{AL}}} \tau_{\text{AL}\leftarrow\gamma} \left\langle \mathcal{H}_\gamma \right\rangle,$$

is dense in $\mathcal{H}_{\text{AL}}$, and therefore $\Gamma_{\widetilde{\text{AL}}\to\text{AL}}$ is an Hilbert space isomorphism.

Now, we define, for any $\rho \in \overline{\mathcal{S}}_{\text{AL}}$:

$$\sigma(\rho) := \widetilde{\sigma}\left(\Gamma^{-1}_{\widetilde{\text{AL}}\to\text{AL}} \circ \rho \circ \Gamma_{\widetilde{\text{AL}}\to\text{AL}}\right),$$

and for any $A \in \overline{\mathcal{A}}^{\otimes}_{(\mathcal{L},\mathcal{H},\Phi)}$:

$$\alpha(A) := \Gamma_{\widetilde{\text{AL}}\to\text{AL}} \circ \widetilde{\alpha}(A) \circ \Gamma^{-1}_{\widetilde{\text{AL}}\to\text{AL}}\,.$$

The points 3.20.1, 3.20.3 and 3.20.4 follow from the corresponding properties of $\widetilde{\sigma}$ and $\widetilde{\alpha}$. Moreover, we have, for any $\eta \in \mathcal{L}$ and any bounded operator $A_\eta$ on $\mathcal{H}_\eta$:

$$\alpha\left([A_\eta]_\sim\right) \circ \tau_{\text{AL}\leftarrow\gamma(\eta)} = \tau_{\text{AL}\leftarrow\gamma(\eta)} \circ \Gamma_{\eta\to\gamma(\eta)} \circ A_\eta \circ \Gamma^{-1}_{\eta\to\gamma(\eta)}\,. \tag{3.20.1}$$

Let $e \in \mathcal{L}_{\text{edges}}$, $\delta \in C^\infty(G, \mathbb{R})$ and $\eta \in \mathcal{L}$ such that $\gamma(\eta) \in \mathcal{L}_{\text{graphs}/e}$. Then, $\gamma(\eta) \in \mathcal{L}_{\text{AL}/e}$ and, using again props. 3.2 and 3.14, we can check that:

$$\Gamma_{\eta\to\gamma(\eta)} \circ \widehat{\mathrm{h}^{(e,\delta)}_{\eta}} \circ \Gamma^{-1}_{\eta\to\gamma(\eta)} = \widehat{\mathrm{h}^{(e,\delta)}_{\gamma(\eta)}}\,,$$

hence:

$$\alpha\left(\widehat{\mathrm{h}^{(e,\delta)}}\right) \circ \tau_{\text{AL}\leftarrow\gamma(\eta)} = \tau_{\text{AL}\leftarrow\gamma(\eta)} \circ \widehat{\mathrm{h}^{(e,\delta)}_{\gamma(\eta)}} = \widehat{\mathrm{h}^{(e,\delta)}_{\text{AL}}} \circ \tau_{\text{AL}\leftarrow\gamma(\eta)}\,.$$

Now, $\alpha$ is a $C^*$-algebra isomorphism, so in particular an isometry, and $\left\{\gamma(\eta) \,\middle|\, \eta \in \mathcal{L} \,/\, \gamma(\eta) \in \mathcal{L}_{\text{graphs}/e}\right\}$ is cofinal in $\mathcal{L}_{\text{AL}}$ so that:

$$\bigcup_{\eta\in\mathcal{L}\,/\,\gamma(\eta)\in\mathcal{L}_{\text{graphs}/e}} \tau_{\text{AL}\leftarrow\gamma(\eta)} \left\langle \mathcal{H}_{\gamma(\eta)} \right\rangle = \bigcup_{\gamma\in\mathcal{L}_{\text{AL}}} \tau_{\text{AL}\leftarrow\gamma} \left\langle \mathcal{H}_\gamma \right\rangle$$

is dense in $\mathcal{H}_{\text{AL}}$. Therefore, $\alpha\left(\widehat{\mathrm{h}^{(e,\delta)}}\right) = \widehat{\mathrm{h}^{(e,\delta)}_{\text{AL}}}$.

Next, let $\overline{F} \in \mathcal{L}_{\text{faces}}$, $t \in G$ and $\eta \in \mathcal{L}$ such that $\lambda(\eta) \in \mathcal{L}_{\text{profls}/\overline{F}}$. From eq. (3.3.1) and prop. 2.8, we get:

$$\forall e \in \gamma(\eta),\, \forall p \neq p' \in r(e),\, \left(e_{[p,p']} \in \overline{F} \Leftrightarrow \left[p = b(e) \;\;\&\;\; \chi_\eta(e) \in H_{\lambda(\eta)\to\overline{F}}\right]\right).$$

Hence, $\gamma(\eta) \in \mathcal{L}_{\text{AL}/\overline{F}}$ and, using eq. (3.11.4):

$$\forall h \in \mathcal{C}_\eta = \mathcal{C}_{\gamma(\eta)}\,,\ T^{(\overline{F})}_{\gamma(\eta),t} h = R^{(\overline{F})}_{\eta,t} h\,.$$

Thus, we get:

$$\Gamma_{\eta\to\gamma(\eta)} \circ \widehat{\mathrm{T}^{(\overline{F},t)}_{\eta}} \circ \Gamma^{-1}_{\eta\to\gamma(\eta)} = \widehat{\mathrm{T}^{(\overline{F},t)}_{\gamma(\eta)}}\,.$$

Like above, this ensures that $\alpha\left(\widehat{\mathrm{T}^{(\overline{F},t)}}\right) = \widehat{\mathrm{T}^{(\overline{F},t)}_{\mathrm{AL}}}$, for $\left\{\gamma(\eta) \;\middle|\; \eta \in \mathcal{L} \,/\, \lambda(\eta) \in \mathcal{L}_{\mathrm{profls}/\overline{F}}\right\}$ is cofinal in $\mathcal{L}_{\mathrm{AL}}$.

Finally, for any $\eta \preccurlyeq \eta' \in \mathcal{L}$, we define $\zeta_{\eta'\to\eta} : \mathcal{C}_{\eta'\to\eta} \to \mathbb{C}$ by:

$$\forall j \in \mathcal{C}_{\eta'\to\eta}\,,\ \zeta_{\eta'\to\eta}(j) = 1\,.$$

Observe that $\|\zeta_{\eta'\to\eta}\|_{\mathcal{H}_{\eta'\to\eta}} = \mu_{\eta'\to\eta}\left(\mathcal{C}_{\eta'\to\eta}\right) = \mu_{\eta'}\left(\mathcal{C}_{\eta'}\right) / \mu_\eta\left(\mathcal{C}_\eta\right) = 1$. Hence, the operator $\Theta_{\eta'|\eta}$ is well-defined and bounded on $\mathcal{H}_{\eta'}$, for it is given by:

$$\Theta_{\eta'|\eta} = \Phi^{-1}_{\eta'\to\eta} \circ \left(|\zeta_{\eta'\to\eta}\,\rangle\langle\, \zeta_{\eta'\to\eta}| \otimes \mathrm{id}_{\mathcal{H}_\eta}\right) \circ \Phi_{\eta'\to\eta}\,. \tag{3.20.2}$$

And, since for any non-negative traceclass operator $\rho_{\eta'}$ on $\mathcal{H}_{\eta'}$ we have:

$$\int_{\mathcal{C}_{\eta'\to\eta}\times\mathcal{C}_{\eta'\to\eta}} d^{(2)}\mu_{\eta'\to\eta}(j,j') \int_{\mathcal{C}_\eta\times\mathcal{C}_\eta} d^{(2)}\mu_\eta(h,h')\,\delta_{\mu_\eta}(h,h')\,\rho_{\eta'}\left(\varphi^{-1}_{\eta'\to\eta}(j,h),\ \varphi^{-1}_{\eta'\to\eta}(j',h')\right) =$$
$$= \sum_{\psi\in\mathrm{ONB}_\eta} \left\langle \Phi^{-1}_{\eta'\to\eta}(\zeta_{\eta'\to\eta}\otimes\psi) \;\middle|\; \rho_{\eta'} \;\middle|\; \Phi^{-1}_{\eta'\to\eta}(\zeta_{\eta'\to\eta}\otimes\psi)\right\rangle,$$

(with $\mathrm{ONB}_\eta$ some orthonormal basis of $\mathcal{H}_\eta$), we get:

$$\sigma\left\langle \mathcal{S}_{\mathrm{AL}}\right\rangle = \widetilde{\sigma}\left\langle \mathcal{S}_{\widetilde{\mathrm{AL}}}\right\rangle = \left\{ \left(\rho_\eta\right)_{\eta\in\mathcal{L}} \;\middle|\; \sup_{\eta\in\mathcal{L}} \inf_{\eta'\succcurlyeq\eta} \mathrm{Tr}_{\mathcal{H}_{\eta'}} \rho_{\eta'}\,\Theta_{\eta'|\eta} = 1 \right\}.$$

□

The injective map $\sigma$ allows to identify the space $\mathcal{S}_{\mathrm{AL}}$ of states over $\mathcal{H}_{\mathrm{AL}}$ with a certain subset in the space $\mathcal{S}^{\otimes}_{(\mathcal{L},\mathcal{H},\Phi)}$ of all projective states. The results below suggests that $\mathcal{S}^{\otimes}_{(\mathcal{L},\mathcal{H},\Phi)}$ can even be thought as a closure of $\mathcal{S}_{\mathrm{AL}}$. Indeed, $\mathcal{S}^{\otimes}_{(\mathcal{L},\mathcal{H},\Phi)}$ is complete, in the sense that a net of projective states admits a limit as soon as it converges over each $\mathcal{H}_\eta$, and $\sigma\left\langle \mathcal{S}_{\mathrm{AL}}\right\rangle$ is dense in the same sense, namely the restrictions of its elements over each $\mathcal{H}_\eta$ fill a dense subset of the associated space $\mathcal{S}_\eta$ of density matrices (in fact, they fill all $\mathcal{S}_\eta$). This is somewhat reminiscent of the Fell's theorem [8], but, while the Fell's theorem tells us that $\mathcal{S}_{\mathrm{AL}}$ is dense in the space of all states [11, part III, def. 2.2.8] over the $C^*$-algebra $\overline{\mathcal{A}}^{\otimes}_{(\mathcal{L},\mathcal{H},\Phi)}$ [18, prop. 2.4], with respect to the weakest topology that makes the evaluation maps $\rho \mapsto \mathrm{Tr}\,\rho A$ continuous, we show here that $\mathcal{S}_{\mathrm{AL}}$ is dense in the (presumably smaller) set of all projective states with respect to a much stronger topology.

**Proposition 3.21** For any $\eta \in \mathcal{L}$, we define on $\mathcal{S}^{\otimes}_{(\mathcal{L},\mathcal{H},\Phi)}$ [18, def. 2.2] the semimetric (aka. possibly degenerate metric, see [7, section IX.10]) $d^{(\eta)}$ by:

$$\forall \left(\rho_{\eta'}\right)_{\eta'\in\mathcal{L}}\,,\ \left(\rho'_{\eta'}\right)_{\eta'\in\mathcal{L}} \in \mathcal{S}^{\otimes}_{(\mathcal{L},\mathcal{H},\Phi)}\,,\ d^{(\eta)}\left[\left(\rho_{\eta'}\right)_{\eta'\in\mathcal{L}}\,,\ \left(\rho'_{\eta'}\right)_{\eta'\in\mathcal{L}}\right] := \left\|\rho_\eta - \rho'_\eta\right\|_1,$$

where $\|\cdot\|_1$ denotes the trace norm on the space of traceclass operators over $\mathcal{H}_\eta$ (see [18, lemma 2.10]

and/or [27]).

The family of semimetrics $\left(d^{(\eta)}\right)_{\eta\in\mathcal{L}}$ equips $\mathcal{S}^{\otimes}_{(\mathcal{L},\mathcal{H},\Phi)}$ with the structure of a *complete* uniform space [7, sections IX.11 and XIV.9], and $\sigma\left\langle \mathcal{S}_{\text{AL}}\right\rangle$ is dense in the induced topology.

**Proof** For any $\eta$, the space $\mathcal{S}_\eta$ of density matrices over $\mathcal{H}_\eta$ equipped with the metric $d^{(\eta)}_\eta$ defined by:

$$\forall \rho_\eta\,,\, \rho_{\eta'} \in \mathcal{S}_\eta\,,\quad d^{(\eta)}_\eta[\rho_\eta\,,\,\rho_{\eta'}] := \left\|\rho_\eta - \rho'_\eta\right\|_1\,,$$

is complete (for the traceclass operators over $\mathcal{H}_\eta$ form, with respect to the trace norm, a Banach space [27] in which $\mathcal{S}_\eta$ is a closed subset). Hence, $\prod_{\eta\in\mathcal{L}}\mathcal{S}_\eta$ has a structure of complete and Hausdorff uniform space as a Cartesian product of complete metric spaces [7, theorem XIV.9.4].

Let $\eta \preccurlyeq \eta' \in \mathcal{L}$. For any self-adjoint traceclass operator $\delta$ on $\mathcal{H}_{\eta'}$, we have, writing $\delta = \delta^{(+)} - \delta^{(-)}$ with $\delta^{(+)}$ and $(-\delta^{(-)})$ respectively the positive and negative parts of $\delta$:

$$\begin{aligned}\left\|\mathrm{Tr}_{\eta'\to\eta}\,\delta\right\|_1 &\leqslant \left\|\mathrm{Tr}_{\eta'\to\eta}\,\delta^{(+)}\right\|_1 + \left\|\mathrm{Tr}_{\eta'\to\eta}\,\delta^{(-)}\right\|_1\\ &= \mathrm{Tr}_{\mathcal{H}_\eta}\,\mathrm{Tr}_{\eta'\to\eta}\,\delta^{(+)} + \mathrm{Tr}_{\mathcal{H}_\eta}\,\mathrm{Tr}_{\eta'\to\eta}\,\delta^{(-)}\\ &= \mathrm{Tr}_{\mathcal{H}_{\eta'}}\,\delta^{(+)} + \mathrm{Tr}_{\mathcal{H}_{\eta'}}\,\delta^{(-)} = \left\|\delta\right\|_1.\end{aligned}$$

Thus, $\mathrm{Tr}_{\eta'\to\eta}$ is a continuous map between the metric spaces $\left(\mathcal{S}_{\eta'}\,,\, d^{(\eta')}_{\eta'}\right)$ and $\left(\mathcal{S}_\eta\,,\, d^{(\eta)}_\eta\right)$, so its graph is closed in their Cartesian product. The projective limit $\mathcal{S}^{\otimes}_{(\mathcal{L},\mathcal{H},\Phi)}$ is therefore a closed subset of $\prod_{\eta\in\mathcal{L}}\mathcal{S}_\eta$, hence inherits a structure of complete and Hausdorff uniform space, which is precisely the one induced by the family of semimetrics $\left(d^{(\eta)}\right)_{\eta\in\mathcal{L}}$.

Let $\rho = \left(\rho_\eta\right)_{\eta\in\mathcal{L}} \in \mathcal{S}^{\otimes}_{(\mathcal{L},\mathcal{H},\Phi)}$. For any $\eta \in \mathcal{L}$, we define $\rho^{(\eta)} \in \mathcal{S}_{\text{AL}}$ by:

$$\rho^{(\eta)} := \tau_{\text{AL}\leftarrow\gamma(\eta)} \circ \Gamma_{\eta\to\gamma(\eta)} \circ \rho_\eta \circ \Gamma^{-1}_{\eta\to\gamma(\eta)} \circ \tau^{+}_{\text{AL}\leftarrow\gamma(\eta)}\,.$$

From theorem 3.20.3 and eq. (3.20.1), we have:

$$\forall A_\eta \in \mathcal{A}_\eta\,,\ \mathrm{Tr}_{\mathcal{H}_\eta}\left(\left[\sigma(\rho^{(\eta)})\right]_\eta A_\eta\right) = \mathrm{Tr}_{\mathcal{H}_{\text{AL}}}\left(\rho^{(\eta)}\,\alpha\left([A_\eta]_\sim\right)\right) = \mathrm{Tr}_{\mathcal{H}_\eta}\left(\rho_\eta A_\eta\right),$$

hence $\left[\sigma\left(\rho^{(\eta)}\right)\right]_\eta = \rho_\eta$. Thus, the net $\left(\sigma\left(\rho^{(\kappa)}\right)\right)_{\kappa\in\mathcal{L}}$ converges to $\rho$ with respect to the family of semimetrics $\left(d^{(\eta)}\right)_{\eta\in\mathcal{L}}$ (like in the proof of [17, prop. 2.7]). Therefore, $\sigma\left\langle\mathcal{S}_{\text{AL}}\right\rangle$ is dense in $\mathcal{S}^{\otimes}_{(\mathcal{L},\mathcal{H},\Phi)}$ for the corresponding topology. □

Finally, we want to check that the projective quantum state space $\mathcal{S}^{\otimes}_{(,\mathcal{H},\Phi)}$ is not a mere rewriting of the space of density matrices over $\mathcal{H}_{\text{AL}}$, ie. that $\sigma$, while being injective, is *not* surjective, and yields a *strict* embedding of $\mathcal{S}_{\text{AL}}$ in $\mathcal{S}^{\otimes}_{(,\mathcal{H},\Phi)}$. To exhibit a projective state that cannot be realized as a density matrix on $\mathcal{H}_{\text{AL}}$, we will, in line with the previous result, consider a sequence of states over $\mathcal{H}_{\text{AL}}$, which, although it does not converge in $\mathcal{S}_{\text{AL}}$, does converge in $\mathcal{S}^{\otimes}_{(\mathcal{L},\mathcal{H},\Phi)}$.

To this intend, we cut some analytic edge $e$ into infinitely many pieces (with an accumulation point at one extremity of the edge) and we denote by $\psi^{(n)}$ the state in $\mathcal{H}_{\text{AL}}$ that assigns to the first $n$ pieces a given, non-trivial, sample wavefunction $\chi$. It can then be shown that the sequence $\left[\sigma\left(\left|\psi^{(n)}\right\rangle\left\langle\psi^{(n)}\right|\right)\right]_\eta$ (as a sequence in $n$, with $\eta$ held fixed) is eventually constant: indeed the evaluations on $\psi^{(n)}$ of the observables covered by the label $\eta$ freeze as soon as $n$ exceeds a certain ($\eta$-dependent) threshold $n_\eta$ (fig. 3.5), and we have proved in prop. 3.19 that these evaluations completely specify a state over $\mathcal{H}_\eta$.

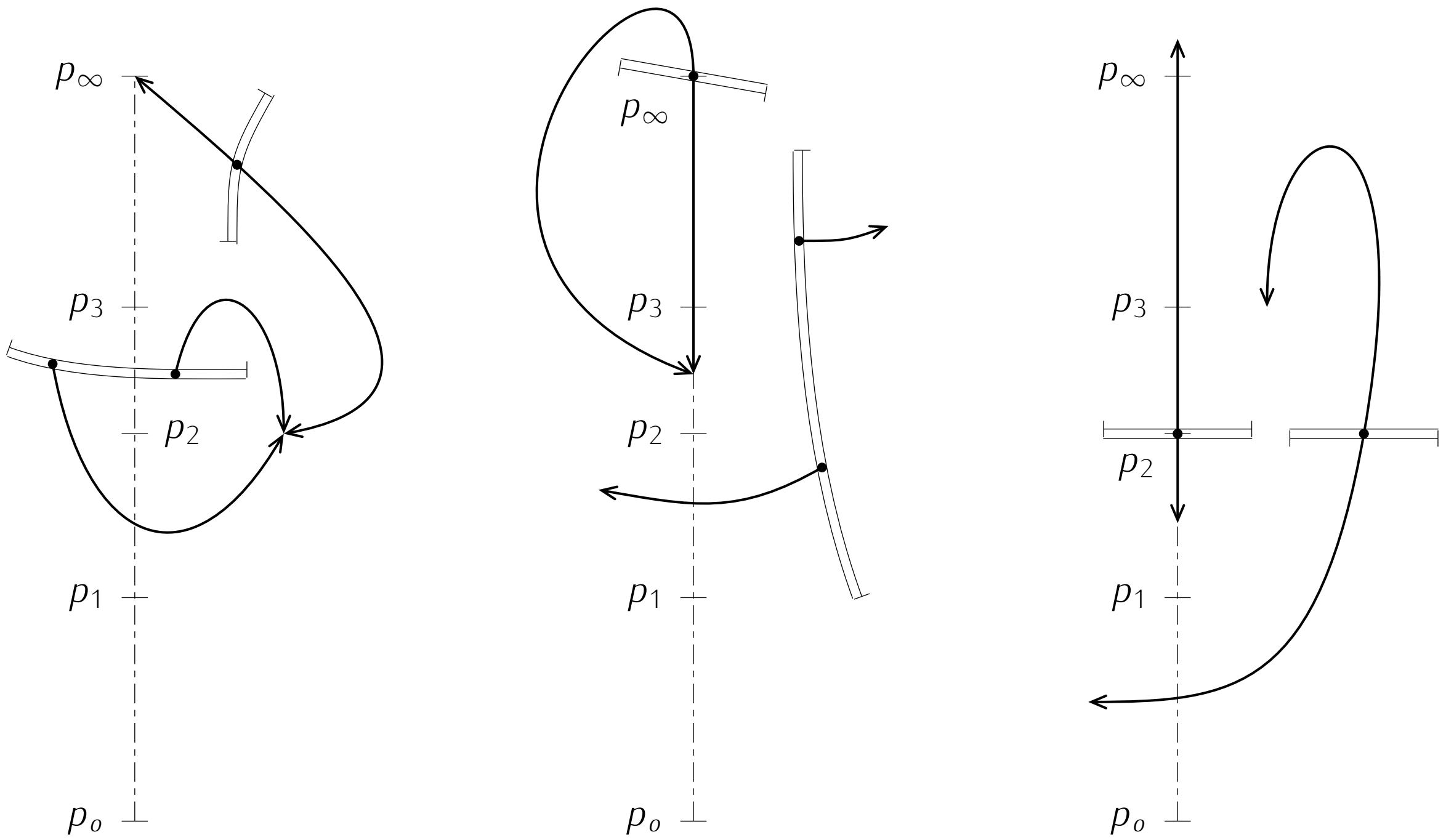

Figure 3.5 – A few labels having 3 as the smallest possible choice for $n_\eta$ (the dashed line indicates the base edge $e$)

Next, we need to confirm that the thus constructed projective state is not in the image of $\sigma$. Let us look closer at the characterization of this image given in theorem 3.20.5. The orthogonal projection $\Theta_{\eta'|\eta}$ selects in $\mathcal{H}_{\eta'}$ those states that do not depend on the complementary variables $\eta' \to \eta$. Letting the *upper* label $\eta'$ become infinitely fine corresponds, in $\mathcal{H}_{\text{AL}}$, to the orthogonal projection on the fixed-graph subspace $\mathcal{H}_{\gamma(\eta)}$. Thus, if we now let the *lower* label $\eta$ get finer and finer, we will recover the state we started from, provided it was a state on $\mathcal{H}_{\text{AL}}$ to begin with. By contrast, for a state that is *not* realizable on $\mathcal{H}_{\text{AL}}$, chopping those parts of the state that depart from the Ashtekar-Lewandowski vacuum over the $\eta' \to \eta$ degrees of freedom and taking first the net limit on $\eta'$, we loose a significant part of the state, no matter how fine $\eta$: such states are not just excitations around the AL-vacuum, but differ from it all the way down to infinitely fine labels.

Now, for any label $\eta$ and any integer $M$, there exists a set of labels $\eta' \succcurlyeq \eta$, which is cofinal in $\mathcal{L}$ and such that, for each $\eta'$ in this set, the holonomies along $M$ distinct pieces of $e$ can be discovered among the complementary variables $\eta' \to \eta$. Since the state we are considering here attributes to each such piece a distribution $\chi$ distinct from the uniform one, its agreement with the AL-vacuum, as far as the $\eta' \to \eta$ degrees of freedom are concerned, can thus be bounded by an exponentially decreasing function of $M$. Taking first the limit on finer and finer $\eta'$, we can then let $M$ goes to infinity, so that the overlap, over the degrees of freedom beyond $\eta$, between this state and the AL-vacuum is actually zero. If it would be a state in $\mathcal{S}_{\text{AL}}$ its projection on any finite-graph subspace $\mathcal{H}_{\gamma(\eta)}$ should therefore vanish, in contradiction with the state having unit trace.

**Proposition 3.22** The map $\sigma$ is *not* surjective.

**Proof** *A family of states in* $\mathcal{H}_{\text{AL}}$. Let $e \in \mathcal{L}_{\text{edges}}$ and let $\check{e} : U \to V$ be a representative of $e$. We define $p_o = b(e)$, $p_\infty = f(e)$ and, for any $n \geqslant 1$:

$$p_n = \check{e}\left(\frac{n}{n+1}\,,\,0\right).$$

For any $n \geqslant 1$, we define:

$$\gamma^{(n)} := \left\{e_{[p_{k-1}\,,\,p_k]} \;\middle|\; 1 \leqslant k \leqslant n\right\} \in \mathcal{L}_{\text{AL}}\,.$$

Next, we choose $\chi \in L_2(G, d\mu)$ such that $\|\chi\| = 1$ and $\int_G d\mu(g)\ |\chi(g)| < 1$ (for example, we can take $\chi := 1/\sqrt{v}\,\mathbb{1}_V$ where $\mathbb{1}_V$ is the indicator function of a measurable region of G with $0 < v := \mu(V) < 1$). For any $n \geqslant 1$, we define $\psi^{(n)}_{\gamma^{(n)}} \in \mathcal{H}_{\gamma^{(n)}}$ by:

$$\forall h \in \mathcal{C}_{\gamma^{(n)}}\,,\ \psi^{(n)}_{\gamma^{(n)}}(h) := \prod_{k=1}^{n} [\chi \circ h]\left(e_{[p_{k-1}\,,\,p_k]}\right),$$

and $\psi^{(n)} := \tau_{\text{AL}\leftarrow\gamma^{(n)}}\left(\psi^{(n)}_{\gamma^{(n)}}\right) \in \mathcal{H}_{\text{AL}}$. We have $\left\|\psi^{(n)}\right\|_{\mathcal{H}_{\text{AL}}} = \left\|\psi^{(n)}_{\gamma^{(n)}}\right\|_{\mathcal{H}_{\gamma^{(n)}}} = 1$.

*Evaluation of the* $\eta$*-observables is* $(n)$*-eventually-constant.* Let $\eta \in \mathcal{L}$. From prop. 2.11.4, there exist $p \in r(e) \setminus \{f(e)\}$, $\epsilon = \pm 1$ and $F \in \mathcal{F}(\eta) \cup \left\{F_{\daleth}(\eta)\right\}$ such that $e^{\epsilon}_{[p,f(e)]} \in F$. Let $p' \in r\left(e_{[p,f(e)]}\right) \setminus \{p,\, f(e)\}$. Then, $p' <_{(e)} f(e)$ and, from prop. 2.8:

$$e_{[f(e),p']} \in F_{\daleth}(\eta) \quad \text{or} \quad e_{[f(e),p']} \in F \in \mathcal{F}(\eta)\,.$$

Moreover, for any $e' \in \gamma(\eta)$, applying lemma 2.6 to $e^{-1}$ and $e'$ yields:

$$\exists q \in r(e) \setminus \{f(e)\} \;/\; \left(r\left(e_{[f(e),q]}\right) \subset r(e') \quad \text{or} \quad r\left(e_{[f(e),q]}\right) \cap r(e') \subset \{f(e)\}\right).$$

Thus, there exists $q' <_{(e)} f(e)$ such that:

$$\forall e' \in \gamma(\eta),\ \left(r\left(e_{[f(e),q']}\right) \subset r(e') \quad \text{or} \quad r\left(e_{[f(e),q']}\right) \cap r(e') \subset \{f(e)\}\right).$$

Therefore, there exists $n_\eta \geqslant 1$ such that (fig. 3.5):

$$r\left(e_{[f(e),p_{n_\eta}]}\right) \cap \left(\bigcup_{e' \in \gamma(\eta)} r(e')\right) \subset \{f(e)\} \quad \text{or} \quad \exists\, e' \in \gamma(\eta) \,/\, r\left(e_{[f(e),p_{n_\eta}]}\right) \subset r(e'),$$

$$\text{and: } e_{[f(e),p_{n_\eta}]} \in F_{\daleth}(\eta) \quad \text{or} \quad \exists\, F \in \mathcal{F}(\eta) \,/\, e_{[f(e),p_{n_\eta}]} \in F\,.$$

Let $m > n \geqslant n_\eta$. We have:

$$r\left(e_{[p_\infty,p_n]}\right) \cap \left(\bigcup_{e' \in \gamma(\eta)} r(e')\right) \subset \{p_\infty\} \quad \text{or} \quad \exists\, e' \in \gamma(\eta) \,/\, r\left(e_{[p_\infty,p_n]}\right) \subset r(e'), \tag{3.22.1}$$

$$\text{and: } e_{[p_\infty,p_n]} \in F_{\daleth}(\eta) \quad \text{or} \quad \exists\, F \in \mathcal{F}(\eta) \,/\, e_{[p_\infty,p_n]} \in F\,. \tag{3.22.2}$$

Let $\gamma := \gamma^{(m)} \cup \left\{e_{[p_\infty,p_m]}\right\} \in \mathcal{L}_{\text{AL}}$. For any $e' \in \gamma(\eta)$, $\mathcal{L}_{\text{AL}/e'}$ is cofinal in $\mathcal{L}_{\text{AL}}$ and, for any $F \in \mathcal{F}(\eta)$,

$\mathcal{L}_{\text{AL}/F^\perp \circ F}$ is cofinal in $\mathcal{L}_{\text{AL}}$, hence there exists $\gamma' \in \mathcal{L}_{\text{AL}}$, with $\gamma \preccurlyeq \gamma'$, such that:

$$\forall e' \in \gamma(\eta),\ \gamma' \in \mathcal{L}_{\text{AL}/e'} \quad \& \quad \forall F \in \mathcal{F}(\eta),\ \gamma' \in \mathcal{L}_{\text{AL}/F^\perp \circ F}\,.$$

Next, let $\widetilde{\gamma} := \gamma^{(n)} \cup \left\{e_{[p_\infty,p_n]}\right\} \in \mathcal{L}_{\text{AL}}$. We have $\widetilde{\gamma} \preccurlyeq \gamma$ for:

$$\gamma^{(n)} \subset \gamma^{(m)} \quad \& \quad e_{[p_\infty,p_n]} = e^{-1}_{[p_n,p_{n+1}]} \circ \ldots \circ e^{-1}_{[p_{m-1},p_m]} \circ e_{[p_\infty,p_m]}\,.$$

Thus, we can define $\varphi_{\gamma'-(m)-(n)} := s_{\gamma'-(m)-(n)} \circ \left(\text{id}_{\mathcal{C}_{\gamma'-\gamma}} \times \varphi_{\gamma-\widetilde{\gamma}}\right) \circ \varphi_{\gamma'-\gamma}$, where $\varphi_{\gamma_2-\gamma_1} : \mathcal{C}_{\gamma_2} \to \mathcal{C}_{\gamma_2-\gamma_1} \times \mathcal{C}_{\gamma_1}$ has been defined for any $\gamma_1 \preccurlyeq \gamma_2 \in \mathcal{L}_{\text{AL}}$ in the proof of prop. 3.16 and $s_{\gamma'-(m)-(n)}$ is given by:

$$\begin{array}{rccl} s_{\gamma'-(m)-(n)} : & \mathcal{C}_{\gamma'-\gamma} \times \mathcal{C}_{\gamma-\widetilde{\gamma}} \times \mathcal{C}_{\widetilde{\gamma}} & \to & \mathcal{C}_{\gamma'-(m)-(n)} \times \mathcal{C}_{(m)-(n)} \\ & j', h, j & \mapsto & (j', j), h \end{array},$$

with $\mathcal{C}_{\gamma'-(m)-(n)} := \mathcal{C}_{\gamma'-\gamma} \times \mathcal{C}_{\widetilde{\gamma}}$ and $\mathcal{C}_{(m)-(n)} := \mathcal{C}_{\gamma-\widetilde{\gamma}}$. Using the definition of $\varphi_{\gamma_2-\gamma_1}$ for $\gamma_1 \preccurlyeq \gamma_2$ together with eq. (3.15.1), we get:

$$\forall h' \in \mathcal{C}_{\gamma'},\ \varphi_{\gamma'-(m)-(n)}(h') = \left(h'\big|_{\gamma'-\gamma},\ \pi_{\gamma'\to\widetilde{\gamma}}(h')\right),\ \pi_{\gamma'\to\gamma}(h')\big|_{\gamma^{(m)-(n)}}, \tag{3.22.3}$$

where $\gamma^{(m)-(n)} := \gamma - \widetilde{\gamma} = \left\{e_{[p_{k-1}\,,\,p_k]} \;\middle|\; n+1 \leqslant k \leqslant m\right\}$.

We define $\mathcal{H}_{\gamma'-(m)-(n)} := L_2\left(\mathcal{C}_{\gamma'-\gamma}\,,\, d\mu_{\gamma'-\gamma}\right) \otimes L_2\left(\mathcal{C}_{\widetilde{\gamma}}\,,\, d\mu_{\widetilde{\gamma}}\right)$ and $\mathcal{H}_{(m)-(n)} := L_2\left(\mathcal{C}_{\gamma-\widetilde{\gamma}}\,,\, d\mu_{\gamma-\widetilde{\gamma}}\right)$, so that $\varphi_{\gamma'-(m)-(n)}$ provides a unitary map $\Phi_{\gamma'-(m)-(n)}$:

$$\begin{array}{rccl} \Phi_{\gamma'-(m)-(n)} : & \mathcal{H}_{\gamma'} & \to & \mathcal{H}_{\gamma'-(m)-(n)} \otimes \mathcal{H}_{(m)-(n)} \\ & \psi & \mapsto & \psi \circ \varphi^{-1}_{\gamma'-(m)-(n)} \end{array}.$$

Since $\gamma^{(n)}$, $\gamma^{(m)} \preccurlyeq \gamma'$, we have:

$$\psi^{(n)} = \tau_{\text{AL}\leftarrow\gamma'} \circ \tau_{\gamma'\leftarrow\gamma^{(n)}}\left(\psi^{(n)}_{\gamma^{(n)}}\right) \quad \& \quad \psi^{(n)} = \tau_{\text{AL}\leftarrow\gamma'} \circ \tau_{\gamma'\leftarrow\gamma^{(m)}}\left(\psi^{(m)}_{\gamma^{(m)}}\right),$$

and eq. (3.22.3) yields, for any $(j', j) \in \mathcal{C}_{\gamma'-(m)-(n)}$ and any $h \in \mathcal{C}_{(m)-(n)}$:

$$\begin{aligned} \Phi_{\gamma'-(m)-(n)} \circ \tau_{\gamma'\leftarrow\gamma^{(n)}}\left(\psi^{(n)}_{\gamma^{(n)}}\right)((j', j),\ h) &= \psi^{(n)}_{\gamma^{(n)}} \circ \pi_{\gamma'\to\gamma^{(n)}} \circ \varphi^{-1}_{\gamma'-(m)-(n)}((j', j),\ h) \\ &= \psi^{(n)}_{\gamma^{(n)}} \circ \pi_{\widetilde{\gamma}\to\gamma^{(n)}}(j) = \psi^{(n)}_{\gamma^{(n)}}\left(j\big|_{\gamma^{(n)}}\right) \\ &= \prod_{k=1}^{n} [\chi \circ j]\left(e_{[p_{k-1}\,,\,p_k]}\right), \end{aligned}$$

as well as:

$$\begin{aligned} \Phi_{\gamma'-(m)-(n)} \circ \tau_{\gamma'\leftarrow\gamma^{(m)}}\left(\psi^{(m)}_{\gamma^{(m)}}\right)((j', j),\ h) &= \prod_{k=1}^{m} \left[\chi \circ \left(\pi_{\gamma'\to\gamma^{(m)}} \circ \varphi^{-1}_{\gamma'-(m)-(n)}((j', j),\ h)\right)\right]\left(e_{[p_{k-1}\,,\,p_k]}\right) \\ &= \prod_{k=1}^{n} [\chi \circ j]\left(e_{[p_{k-1}\,,\,p_k]}\right) \prod_{k=n+1}^{m} [\chi \circ h]\left(e_{[p_{k-1}\,,\,p_k]}\right). \end{aligned}$$

Thus there exist $\psi_{\gamma'-(m)-(n)} \in \mathcal{H}_{\gamma'-(m)-(n)}$ and $\zeta_{(m)-(n)}$, $\chi_{(m)-(n)} \in \mathcal{H}_{(m)-(n)}$ such that:

$$\psi^{(n)} = \tau_{\text{AL}\leftarrow\gamma'} \circ \Phi^{-1}_{\gamma'-(m)-(n)}\left(\psi_{\gamma'-(m)-(n)} \otimes \zeta_{(m)-(n)}\right),$$

$$\& \quad \psi^{(m)} = \tau_{\mathrm{AL}\leftarrow\gamma'} \circ \Phi^{-1}_{\gamma'-(m)-(n)} \left(\psi_{\gamma'-(m)-(n)} \otimes \chi_{(m)-(n)}\right). \tag{3.22.4}$$

In addition, $\left\|\zeta_{(m)-(n)}\right\|_{\mathcal{H}_{(m)-(n)}} = \left\|\chi_{(m)-(n)}\right\|_{\mathcal{H}_{(m)-(n)}}$, for $\left\|\psi^{(n)}\right\|_{\mathcal{H}_{\mathrm{AL}}} = 1 = \left\|\psi^{(m)}\right\|_{\mathcal{H}_{\mathrm{AL}}}$.

Now, let $\gamma'' := \left\{e'' \in \gamma' \;\middle|\; r(e'') \subset r\left(e_{[p_o\,,\,p_n]}\right)\right\} \cup \left\{e_{[p_\infty\,,\,p_n]}\right\} \in \mathcal{L}_{\mathrm{AL}}$. We have $\widetilde{\gamma} \preccurlyeq \gamma'' \preccurlyeq \gamma'$, hence:

$$\forall h' \in \mathcal{C}_{\gamma'}\,,\ \varphi_{\gamma''-\widetilde{\gamma}} \circ \pi_{\gamma'\to\gamma''}(h') = \pi_{\gamma'\to(\gamma''-\widetilde{\gamma})}(h')\,,\ \pi_{\gamma'\to\widetilde{\gamma}}(h')\,,$$

and, since $(\gamma'' - \widetilde{\gamma}) \subset (\gamma' - \gamma)$, we get, for any $(j', j) \in \mathcal{C}_{\gamma'-(m)-(n)}$ and any $h \in \mathcal{C}_{(m)-(n)}$:

$$\pi_{\gamma'\to\gamma''} \circ \varphi^{-1}_{\gamma'-(m)-(n)}((j', j),\, h) = \varphi^{-1}_{\gamma''-\widetilde{\gamma}}\left(j'|_{\gamma''-\widetilde{\gamma}}\,,\, j\right). \tag{3.22.5}$$

Next, for any $e' \in \gamma(\eta)$, we define:

$$\gamma'_{(e')} = \{e'' \in \gamma' \mid r(e'') \subset r(e') \ \ \&\ \ r(e'') \not\subset r(e)\} \ \ \cup\ \ \{e'' \in \gamma'' \mid r(e'') \subset r(e')\} \in \mathcal{L}_{\mathrm{AL}}.$$

$\gamma'_{(e')} \preccurlyeq \gamma'$ and, from eq. (3.22.1), $\gamma'_{(e')} \in \mathcal{L}_{\mathrm{AL}/e'}$, so $\pi_{\gamma'\to\{e'\}} = \pi_{\gamma'_{(e')}\to\{e'\}} \circ \pi_{\gamma'\to\gamma'_{(e')}}$. Moreover, since $\gamma'_{(e')} \subset (\gamma' - \gamma) \cup \gamma''$, we have, for any $(j', j) \in \mathcal{C}_{\gamma'-(m)-(n)}$ and any $h \in \mathcal{C}_{(m)-(n)}$:

$$\forall e'' \in \gamma'_{(e')}\,,\ \left[\pi_{\gamma'\to\gamma'_{(e')}} \circ \varphi^{-1}_{\gamma'-(m)-(n)}((j', j),\, h)\right](e'') =$$

$$= \begin{cases} \left[\pi_{\gamma'\to(\gamma'-\gamma)} \circ \varphi^{-1}_{\gamma'-(m)-(n)}((j', j),\, h)\right](e'') = j'(e'') & \text{if } e'' \in (\gamma' - \gamma) \\ \left[\pi_{\gamma'\to\gamma''} \circ \varphi^{-1}_{\gamma'-(m)-(n)}((j', j),\, h)\right](e'') \quad = \left[\varphi^{-1}_{\gamma''-\widetilde{\gamma}}\left(j'|_{\gamma''-\widetilde{\gamma}}\,,\, j\right)\right](e'') & \text{if } e'' \in \gamma'' \end{cases}.$$

Thus, $\pi_{\gamma'\to\gamma'_{(e')}} \circ \varphi^{-1}_{\gamma'-(m)-(n)}((j', j),\, h)$ does not depend on $h$. Therefore, using eq. (3.17.1), there exists, for any $\delta \in C^\infty(G,\, \mathbb{R})$, an operator $A \in \mathcal{A}_{\gamma'-(m)-(n)}$ (with $\mathcal{A}_{\gamma'-(m)-(n)}$ the algebra of bounded operators on $\mathcal{H}_{\gamma'-(m)-(n)}$) such that:

$$\widehat{\mathrm{h}^{(e',\delta)}_{\gamma'}} = \Phi^{-1}_{\gamma'-(m)-(n)} \circ \left[A \otimes \mathrm{id}_{\mathcal{H}_{(m)-(n)}}\right] \circ \Phi_{\gamma'-(m)-(n)}.$$

Let $F \in \mathcal{F}(\eta)$. Since $\widetilde{\gamma} \preccurlyeq \gamma'' \preccurlyeq \gamma'$, eq. (3.22.3) implies:

$$\forall h' \in \mathcal{C}_{\gamma'}\,,\ \varphi_{\gamma'-(m)-(n)}(h') = \left(\pi_{\gamma'\to(\gamma'-\gamma)}(h'),\ \pi_{\gamma''\to\widetilde{\gamma}} \circ \pi_{\gamma'\to\gamma''}(h')\right),\ \pi_{\gamma'\to\gamma^{(m)-(n)}}(h').$$

But since $\gamma' \in \mathcal{L}_{\mathrm{AL}/F^\perp\circ F}$ and $e_{[p_\infty,p_n]} \in F'$ with $F' \in \mathcal{F}(\eta) \cup \left\{F_{\urcorner}(\eta)\right\}$, we have $(\gamma' - \gamma)$, $\gamma''$ & $\gamma^{(m)-(n)} \in \mathcal{L}_{\mathrm{AL}/F^\perp\circ F}$ and, moreover:

$$\forall e' \in \gamma^{(m)-(n)},\ c(e',\, F^\perp \circ F) = \varnothing\,,$$

so that $\forall t \in G$, $\forall h \in \mathcal{C}_{(m)-(n)}$, $T^{(F^\perp\circ F)}_{\gamma^{(m)-(n)},t} h = h$. Thus, using eq. (3.18.1), we get, for any $t \in G$ and any $h' \in \mathcal{C}_{\gamma'}$:

$$\varphi_{\gamma'-(m)-(n)}\left(T^{(F^\perp\circ F)}_{\gamma',t} h'\right) = \left(T^{(F^\perp\circ F)}_{(\gamma'-\gamma),t} \circ \pi_{\gamma'\to(\gamma'-\gamma)}(h'),\ \pi_{\gamma''\to\widetilde{\gamma}} \circ T^{(F^\perp\circ F)}_{\gamma'',t} \circ \pi_{\gamma'\to\gamma''}(h')\right),\ \pi_{\gamma'\to\gamma^{(m)-(n)}}(h').$$

Hence, eq. (3.22.5) yields, for any $(j', j) \in \mathcal{C}_{\gamma'-(m)-(n)}$ and any $h \in \mathcal{C}_{(m)-(n)}$:

$$T^{(F^\perp\circ F)}_{\gamma',t}\left(\varphi^{-1}_{\gamma'-(m)-(n)}\left((j', j),\, h\right)\right) = \varphi^{-1}_{\gamma'-(m)-(n)}\left(T^{(F^\perp\circ F)}_{(\gamma'-(m)-(n)),t}(j', j),\, h\right),$$

where $T^{(F^\perp\circ F)}_{(\gamma'-(m)-(n)),t}(j', j) := \left(T^{(F^\perp\circ F)}_{(\gamma'-\gamma),t} j',\ \pi_{\gamma''\to\widetilde{\gamma}} \circ T^{(F^\perp\circ F)}_{\gamma'',t} \circ \varphi^{-1}_{\gamma''\to\widetilde{\gamma}}\left(j'|_{\gamma''-\widetilde{\gamma}}\,,\, j\right)\right).$

Therefore, there exists, for any $t \in G$, an operator $A \in \mathcal{A}_{\gamma'-(m)-(n)}$ such that:

$$\widehat{\mathrm{T}^{(F^\perp \circ F, t)}_{\gamma'}} = \Phi^{-1}_{\gamma'-(m)-(n)} \circ \left[A \otimes \mathrm{id}_{\mathcal{H}_{(m)-(n)}}\right] \circ \Phi_{\gamma'-(m)-(n)} .$$

So we have proved that for any $A_{\mathrm{AL}}$ in:

$$\left\{ \widehat{\mathrm{h}^{(e',\delta)}_{\mathrm{AL}}} \,\middle|\, \delta \in C^\infty(G, \mathbb{R}) \;\;\&\;\; e' \in \gamma(\eta) \right\} \cup \left\{ \widehat{\mathrm{T}^{(F^\perp \circ F, t)}_{\mathrm{AL}}} \,\middle|\, t \in G \;\;\&\;\; F \in \mathcal{F}(\eta) \right\},$$

there exists $A \in \mathcal{A}_{\gamma'-(m)-(n)}$ such that:

$$A_{\mathrm{AL}} \circ \tau_{\mathrm{AL}\leftarrow\gamma'} = \tau_{\mathrm{AL}\leftarrow\gamma'} \circ \Phi^{-1}_{\gamma'-(m)-(n)} \circ \left[A \otimes \mathrm{id}_{\mathcal{H}_{(m)-(n)}}\right] \circ \Phi_{\gamma'-(m)-(n)} .$$

Theorem 3.20.1 and 3.20.2 then implies that for any $A_\eta \in \mathfrak{I}_\eta$, there exists $A \in \mathcal{A}_{\gamma'-(m)-(n)}$ such that:

$$\alpha\left([A_\eta]_\sim\right) \circ \tau_{\mathrm{AL}\leftarrow\gamma'} = \tau_{\mathrm{AL}\leftarrow\gamma'} \circ \Phi^{-1}_{\gamma'-(m)-(n)} \circ \left[A \otimes \mathrm{id}_{\mathcal{H}_{(m)-(n)}}\right] \circ \Phi_{\gamma'-(m)-(n)} ,$$

hence, using the expression for $\psi^{(n)}$ and $\psi^{(m)}$ from eq. (3.22.4):

$$\forall A_\eta \in \mathfrak{I}_\eta ,\ \left\langle \psi^{(n)}, \alpha\left([A_\eta]_\sim\right) \psi^{(n)} \right\rangle_{\mathcal{H}_{\mathrm{AL}}} = \left\langle \psi^{(m)}, \alpha\left([A_\eta]_\sim\right) \psi^{(m)} \right\rangle_{\mathcal{H}_{\mathrm{AL}}} .$$

*Constructing a projective state from the $\psi^{(n)}$.* For each $\eta \in \mathcal{L}$, we choose $n_\eta$ as above, and we define:

$$\rho_\eta := \left[\sigma\Big( \left|\psi^{(n_\eta)} \right\rangle \left\langle \psi^{(n_\eta)}\right| \Big)\right]_\eta .$$

Let $\eta \preccurlyeq \eta' \in \mathcal{L}$ and let $m \geqslant n_\eta , n_{\eta'}$. From the previous point, together with theorem 3.20.3, we have, for any $A_\eta \in \mathfrak{I}_\eta$:

$$\begin{aligned}\mathrm{Tr}_{\mathcal{H}_\eta} \rho_\eta A_\eta &= \left\langle \psi^{(n_\eta)} \,\middle|\, \alpha\left([A_\eta]_\sim\right) \psi^{(n_\eta)} \right\rangle_{\mathcal{H}_{\mathrm{AL}}} \\ &= \left\langle \psi^{(m)} \,\middle|\, \alpha\left([A_\eta]_\sim\right) \psi^{(m)} \right\rangle_{\mathcal{H}_{\mathrm{AL}}} = \mathrm{Tr}_{\mathcal{H}_\eta} \left[\sigma\Big( \left|\psi^{(m)} \right\rangle \left\langle \psi^{(m)}\right| \Big)\right]_\eta A_\eta .\end{aligned}$$

Hence, the second part of prop. 3.19 implies:

$$\rho_\eta = \left[\sigma\Big( \left|\psi^{(m)} \right\rangle \left\langle \psi^{(m)}\right| \Big)\right]_\eta ,$$

and similarly for $\rho_{\eta'}$. But $\sigma\big( \left| \psi^{(m)} \right\rangle\!\left\langle \psi^{(m)} \right| \big) \in \overline{\mathcal{S}}^{\otimes}_{(\mathcal{L},\mathcal{H},\Phi)}$ [18, def. 2.2], so we get:

$$\mathrm{Tr}_{\eta'\to\eta} \rho_{\eta'} = \mathrm{Tr}_{\eta'\to\eta} \left[\sigma\Big( \left|\psi^{(m)} \right\rangle \left\langle \psi^{(m)}\right| \Big)\right]_{\eta'} = \left[\sigma\Big( \left|\psi^{(m)} \right\rangle \left\langle \psi^{(m)}\right| \Big)\right]_\eta = \rho_\eta ,$$

and therefore $\rho := \left(\rho_\eta\right)_{\eta\in\mathcal{L}} \in \overline{\mathcal{S}}^{\otimes}_{(\mathcal{L},\mathcal{H},\Phi)}$.

*$\rho$ is not in the image of $\sigma$.* Let $\eta \in \mathcal{L}$ and $n = n_\eta \geqslant 1$. Let $M \geqslant 1$ be an *odd* integer and $m = n + M$. By definition of $n_\eta$ we have, for any $l$ such that $0 \leqslant 2l \leqslant M - 1$:

$$p_{n+2l+1} \notin \bigcup_{e'\in\gamma_\sharp} r(e') ,$$

where $\gamma_\sharp := \gamma(\eta) \setminus \left\{e' \in \gamma(\eta) \,\middle|\, r(e_{[p_\infty,p_n]}) \subset r(e')\right\}$. If there exists $e' \in \gamma(\eta)$ such that $r(e_{[p_\infty,p_n]}) \subset$

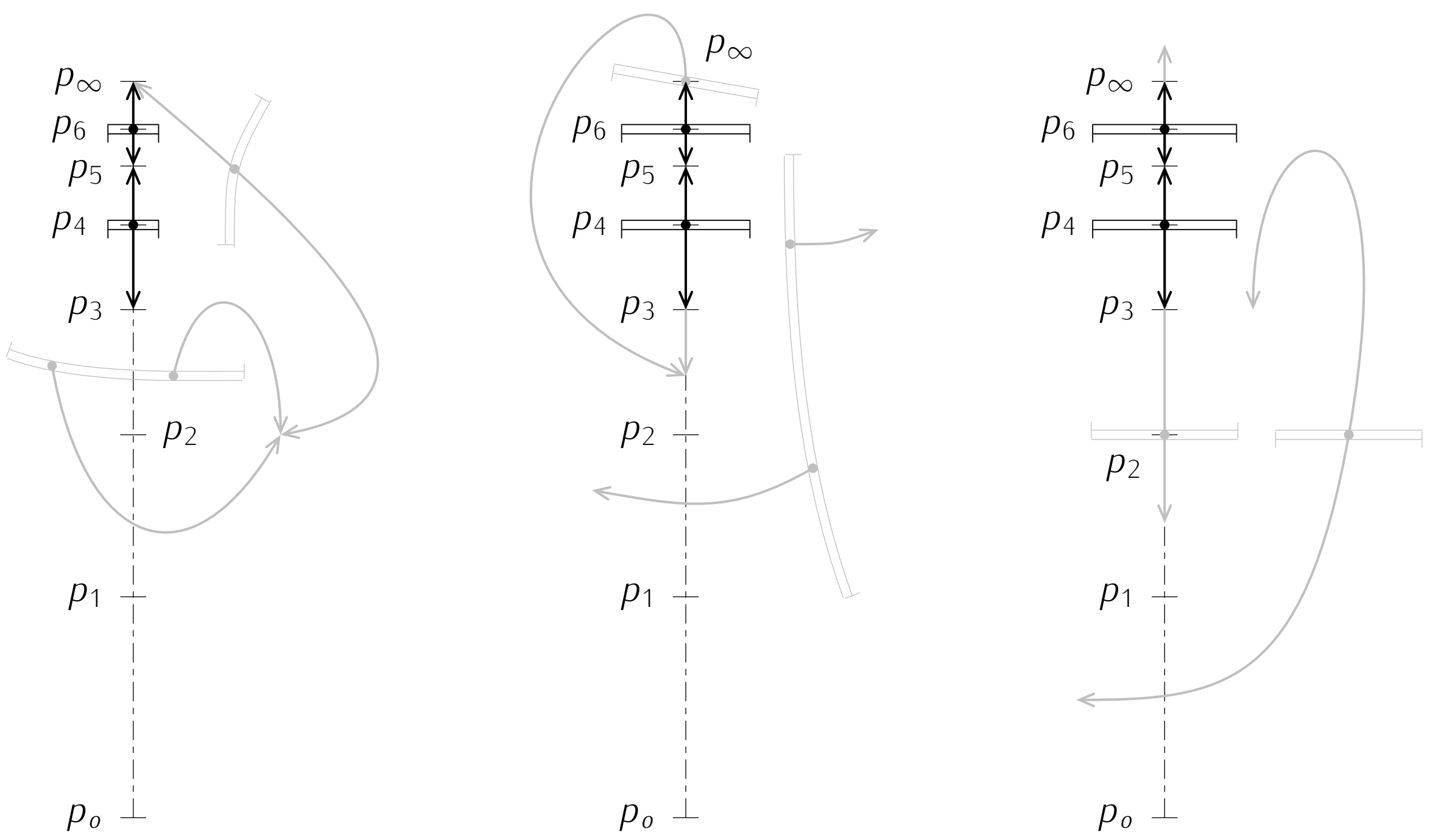

Figure 3.6 – Construction of the auxiliary label $\widetilde{\eta}$ with $M = 3$ for the labels of fig. 3.5 (recalled in light gray)

$r(e')$ (note that there can be as much one such $e'$ for $\gamma(\eta)$ is a graph), we define $\gamma_\flat \in \mathcal{L}_{\text{graph}}$ by:

$$\gamma_\flat = \begin{cases} \left\{e'_{[b(e'),p]},\ e'_{[q,f(e')]}\right\} & \text{if } b(e') <_{(e')} p <_{(e')} q <_{(e')} f(e') \\ \left\{e'_{[b(e'),p]}\right\} & \text{if } b(e') <_{(e')} p <_{(e')} q = f(e') \\ \left\{e'_{[q,f(e')]}\right\} & \text{if } b(e') = p <_{(e')} q <_{(e')} f(e') \\ \varnothing & \text{if } b(e') = p <_{(e')} q = f(e') \end{cases} \quad \text{with } \{p,q\} = \{p_\infty, p_n\}\,,$$

otherwise we define $\gamma_\flat = \varnothing$. By construction, we have $\gamma \setminus \gamma_\sharp \preccurlyeq \gamma_\flat \cup \left\{e_{[p_\infty,p_n]}\right\}$ and, for any $l$ such that $0 \leqslant 2l \leqslant M-1$, $p_{n+2l+1} \notin \bigcup_{e'' \in \gamma_\flat} r(e'')$. Since $\bigcup_{e' \in \gamma_\sharp \cup \gamma_\flat} r(e')$ is compact, $\breve{e}^{-1}\left\langle V \setminus \bigcup_{e' \in \gamma_\sharp \cup \gamma_\flat} r(e') \right\rangle \subset U$ is an open neighborhood of $\left\{ \frac{n+2l+1}{n+2l+2} \,\middle|\, 0 \leqslant 2l \leqslant M-1 \right\} \times \{0\}^{d-1}$ in $\mathbb{R}^d$ so there exists $R > 0$ such that, for any $l$ in $\{0, \dots, (M-1)/2\}$:

$$\left\{\frac{n+2l+1}{n+2l+2}\right\} \times B_R^{(d-1)} \subset U \quad \& \quad \breve{e}\left\langle \left\{\frac{n+2l+1}{n+2l+2}\right\} \times B_R^{(d-1)} \right\rangle \cap \bigcup_{e' \in \gamma_\sharp \cup \gamma_\flat} r(e') = \varnothing$$

(where $B_R^{(d-1)}$ is the closed ball of radius $R$ and center 0 in $\mathbb{R}^{d-1}$). Thus, this allows us to construct a label $\widetilde{\eta} \in \mathcal{L}$ such that (fig. 3.6):

$$\gamma(\widetilde{\eta}) = \widetilde{\gamma} := \left\{e_{[p_m,p_{m-1}]}\,,\ e_{[p_m,p_\infty]}\right\} \ \cup \bigcup_{0 \leqslant 2l < M-1} \left\{e_{[p_{n+2l+1},p_{n+2l}]}\,,\ e_{[p_{n+2l+1},p_{n+2l+2}]}\right\}$$

$$\& \quad \forall e' \in \gamma_\sharp \cup \gamma_\flat \cup \left\{e_{[p_o,p_n]}\right\}\,,\ e' \in F_{\daleth}(\widetilde{\eta}).$$

We have $e_{[p_\infty,p_n]} = e_{[p_{n+1},p_n]} \circ e^{-1}_{[p_{n+1},p_{n+2}]} \circ \ldots \circ e_{[p_m,p_{m-1}]} \circ e^{-1}_{[p_m,p_\infty]}$, hence $\left\{e_{[p_\infty,p_n]}\right\} \preccurlyeq \widetilde{\gamma}$ and:

$$\widetilde{\gamma}_\sharp := \widetilde{\gamma} - \left\{e_{[p_\infty,p_n]}\right\} = \widetilde{\gamma} \setminus \left\{e_{[p_m,p_\infty]}\right\}.$$

Note that $\gamma^{(m)-(n)} = \left\{e_{[p_{k-1},p_k]} \;\middle|\; n+1 \leqslant k \leqslant m\right\} \preccurlyeq \widetilde{\gamma}_\sharp$ with $\widetilde{\gamma}_\sharp - \gamma^{(m)-(n)} = \varnothing$, so that $\pi_{\widetilde{\gamma}_\sharp \to \gamma^{(m)-(n)}}$ is a diffeomorphism $\mathcal{C}_{\widetilde{\gamma}_\sharp} \to \mathcal{C}_{\gamma^{(m)-(n)}} = \mathcal{C}_{(m)-(n)}$.

Next, let $\eta' \succcurlyeq \eta, \widetilde{\eta}$. Since $\left\{e_{[p_m,p_{m-1}]}\right\} \preccurlyeq \widetilde{\gamma} \preccurlyeq \gamma(\eta')$, there exists $e' \in \gamma(\eta')$ such that $r(e') \subset r\left(e_{[p_m,p_{m-1}]}\right)$, so $n' := n_{\eta'}$ has to be bigger than $m$. Also, since $\mathcal{L}_{\text{graphs}}$ is directed, there exists $\gamma' \succcurlyeq \gamma(\eta')$, $\gamma^{(n')}$. So, applying the construction from subsection 2.2 to $(\gamma', \lambda(\eta')) \in \mathcal{L}_{\text{graphs}} \times \mathcal{L}_{\text{profls}}$, there exists $\eta'' \succcurlyeq \eta'$ with $\gamma(\eta'') \succcurlyeq \gamma^{(n')}$. We define a diffeomorphism $\varphi_{\eta'' \to (m)-(n)} : \mathcal{C}_{\eta''} \to \mathcal{C}_{\eta'' \to (m)-(n)} \times \mathcal{C}_{(m)-(n)}$ by:

$$\varphi_{\eta'' \to (m)-(n)} = \left[\widetilde{j}', (\widetilde{h}, \widetilde{j}) \mapsto (\widetilde{j}', \widetilde{j}), \pi_{\widetilde{\gamma}_\sharp \to \gamma^{(m)-(n)}}(\widetilde{h})\right] \circ \left(\mathrm{id}_{\mathcal{C}_{\eta'' \to \widetilde{\eta}}} \times \varphi_{\widetilde{\gamma} - \left\{e_{[p_\infty,p_n]}\right\}}\right) \circ \varphi_{\eta'' \to \widetilde{\eta}},$$

with $\mathcal{C}_{\eta'' \to (m)-(n)} := \mathcal{C}_{\eta'' \to \widetilde{\eta}} \times \mathcal{C}_{\left\{e_{[p_\infty,p_n]}\right\}}$.

Let $h'' \in \mathcal{C}_{\eta''}$ and define:

$$\widetilde{j}', \widetilde{h}, \widetilde{j} = \left(\mathrm{id}_{\mathcal{C}_{\eta'' \to \widetilde{\eta}}} \times \varphi_{\widetilde{\gamma} - \left\{e_{[p_\infty,p_n]}\right\}}\right) \circ \varphi_{\eta'' \to \widetilde{\eta}}(h''),$$

$$\& \quad j', j, h = \left(\mathrm{id}_{\mathcal{C}_{\eta'' \to \eta'}} \times \varphi_{\eta' \to \eta}\right) \circ \varphi_{\eta'' \to \eta'}(h'').$$

Using [17, eq. (2.11.1)] (which was proved in theorem 3.7), we have:

$$\begin{aligned}
j' &= [j'_\circ, h' \mapsto j'_\circ] \circ \varphi_{\eta'' \to \eta'}(h'') \\
&= [j'_\circ, h' \mapsto j'_\circ] \circ \varphi_{\eta'' \to \eta'} \circ \varphi^{-1}_{\eta'' \to \widetilde{\eta}} \circ \varphi_{\eta'' \to \widetilde{\eta}}(h'') \\
&= [j'_\circ, h' \mapsto j'_\circ] \circ \left(\mathrm{id}_{\mathcal{C}_{\eta'' \to \eta'}} \times \varphi^{-1}_{\eta' \to \widetilde{\eta}}\right) \circ \left(\varphi_{\eta'' \to \eta' \to \widetilde{\eta}} \times \mathrm{id}_{\mathcal{C}_{\widetilde{\eta}}}\right) \circ \varphi_{\eta'' \to \widetilde{\eta}}(h'') \\
&= \left[j'_\circ, \widetilde{j}'', \widetilde{h}' \mapsto j'_\circ\right] \circ \left(\varphi_{\eta'' \to \eta' \to \widetilde{\eta}} \times \mathrm{id}_{\mathcal{C}_{\widetilde{\eta}}}\right) \circ \varphi_{\eta'' \to \widetilde{\eta}}(h'') \\
&= \left[j'_\circ, \widetilde{j}'' \mapsto j'_\circ\right] \circ \left(\varphi_{\eta'' \to \eta' \to \widetilde{\eta}}\right)(\widetilde{j}'),
\end{aligned}$$

and:

$$\begin{aligned}
h &= [j'_\circ, j_\circ, h_\circ \mapsto h_\circ] \circ \left(\varphi_{\eta'' \to \eta' \to \eta} \times \mathrm{id}_{\mathcal{C}_\eta}\right) \circ \varphi_{\eta'' \to \eta}(h'') \\
&= [j'', h_\circ \mapsto h_\circ] \circ \varphi_{\eta'' \to \eta}(h'') \\
&= e' \mapsto \left(\prod_{k=1}^{n_{\eta'' \to \eta, e'}} \left[h'' \circ a_{\eta'' \to \eta, e'}(k)\right]^{\epsilon_{\eta'' \to \eta, e'}(k)}\right) \\
&= \pi_{\gamma(\eta'') \to \gamma(\eta)}(h'').
\end{aligned}$$

Let $e' \in \gamma_\sharp$. By construction, we have, for any $k \in \{1, \ldots, n_{\eta'' \to \eta, e'}\}$, $a_{\eta'' \to \eta, e'}(k) \in F_{\daleth}(\widetilde{\eta})$, as well as $\forall \widetilde{e} \in \widetilde{\gamma}$, $r\left(a_{\eta'' \to \eta, e'}(k)\right) \not\subset r(\widetilde{e})$ (for $r(e') \cap r\left(e_{[p_\infty,p_n]}\right) \subset \{p_\infty, p_n\}$). Hence, $a_{\eta'' \to \eta, e'}(k) \in H^{(0)}_{\eta'' \to \widetilde{\eta}}$. Writing $\widetilde{j}' =: \left(\widetilde{j}'^{(0)}, \widetilde{j}'^{(2)}, \widetilde{j}'^{(3)}\right)$, we thus get:

$$\forall k \in \{1, \ldots, n_{\eta'' \to \eta, e'}\}, \; h'' \circ a_{\eta'' \to \eta, e'}(k) = \widetilde{j}'^{(0)} \circ a_{\eta'' \to \eta, e'}(k).$$

So, for any $e' \in \gamma_\sharp$, $[\pi_{\gamma(\eta'') \to \gamma(\eta)}(h'')](e')$ only depends on $h''$ via $\widetilde{j}'^{(0)}$.

If there exists $e' \in \gamma(\eta)$ such that $r\left(e_{[p_\infty,p_n]}\right) \subset r(e')$, then we have at the same time $\{e'\} \preccurlyeq$ $\gamma(\eta) \preccurlyeq \gamma(\eta')$ and $\left\{e_{[p_\infty,p_n]}\right\} \preccurlyeq \gamma(\widetilde{\eta}) \preccurlyeq \gamma(\eta')$, so $\gamma_\flat \preccurlyeq \gamma(\eta') \preccurlyeq \gamma(\eta'')$ must hold (we can check this by writing both $e'$ and $e_{[p_\infty,p_n]}$ as compositions of edges in $\gamma(\eta')$, and by showing that the edges that appears in the decomposition of $e_{[p_\infty,p_n]}$ should appear in the decomposition of $e'$ as well, for $\gamma(\eta')$ is a graph; then, the remaining edges from the decomposition of $e'$ build up precisely the edges in $\gamma_\flat$). With the same argument as for $\gamma_\sharp$ above, we have for any $e'' \in \gamma_\flat$ and any $k \in \{1, \dots, n_{\gamma(\eta'')\to e''}\}$, $a_{\gamma(\eta'')\to e''}(k) \in H^{(0)}_{\eta''\to\widetilde{\eta}}$, and therefore $h'' \circ a_{\gamma(\eta'')\to e''}(k) = \widetilde{j}'^{(0)} \circ a_{\gamma(\eta'')\to e''}(k)$. On the other hand, we have $\pi_{\gamma(\eta'')\to\left\{e_{[p_\infty,p_n]}\right\}}(h'') = \pi_{\widetilde{\gamma}\to\left\{e_{[p_\infty,p_n]}\right\}} \circ \pi_{\gamma(\eta'')\to\gamma(\widetilde{\eta})}(h'') = \widetilde{j}$. Thus, $\pi_{\gamma(\eta'')\to\gamma_\flat\cup\left\{e_{[p_\infty,p_n]}\right\}}(h'')$ only depends on $h''$ via $\widetilde{j}'^{(0)}$ and $\widetilde{j}$, and the same holds for $[\pi_{\gamma(\eta'')\to\gamma(\eta)}(h'')](e')$ since:

$$[\pi_{\gamma(\eta'')\to\gamma(\eta)}(h'')](e') = \left[\pi_{\gamma(\eta'')\to\{e'\}}(h'')\right](e') = \left[\pi_{\gamma_\flat\cup\left\{e_{[p_\infty,p_n]}\right\}\to\{e'\}} \circ \pi_{\gamma(\eta'')\to\gamma_\flat\cup\left\{e_{[p_\infty,p_n]}\right\}}(h'')\right](e')\,.$$

Hence, we have proved so far that there exist two maps $\theta_{\eta''\to\eta'} : \mathcal{C}_{\eta''\to(m)-(n)} \to \mathcal{C}_{\eta''\to\eta'}$ and $\theta_\eta : \mathcal{C}_{\eta''\to(m)-(n)} \to \mathcal{C}_\eta$ such that:

$$j' = \theta_{\eta''\to\eta'}\left(\widetilde{j}', \widetilde{j}\right) \quad \& \quad h = \theta_\eta\left(\widetilde{j}', \widetilde{j}\right).$$

Now, we define an Hilbert space isomorphism $\Phi_{\eta''\to(m)-(n)} : \mathcal{H}_{\eta''} \to \mathcal{H}_{\eta''\to(m)-(n)} \otimes \mathcal{H}_{(m)-(n)}$ by:

$$\begin{array}{rccl} \Phi_{\eta''\to(m)-(n)} & : \mathcal{H}_{\eta''} & \to & \mathcal{H}_{\eta''\to(m)-(n)} \otimes \mathcal{H}_{(m)-(n)} \\ & \psi & \mapsto & \psi \circ \varphi^{-1}_{\eta''\to(m)-(n)} \end{array},$$

with $\mathcal{H}_{\eta''\to(m)-(n)} := \mathcal{H}_{\eta''\to\widetilde{\eta}} \otimes \mathcal{H}_{\left\{e_{[p_\infty,p_n]}\right\}}$ and $\mathcal{H}_{(m)-(n)} := \mathcal{H}_{\gamma^{(m)-(n)}}$. $\Phi_{\eta''\to(m)-(n)}$ being an Hilbert space isomorphism follows from the note at the end of the proof of prop. 3.16 and from the fact that $\pi_{\widetilde{\gamma}_\sharp\to\gamma^{(m)-(n)},*}\,\mu_{\widetilde{\gamma}_\sharp} = \mu_{\gamma^{(m)-(n)}}$ (for the Haar measure on a compact group is invariant under taking the inverse). Then, we have, for any $\psi_{\eta''\to\eta'} \in \mathcal{H}_{\eta''\to\eta'}$ and any $\psi_\eta \in \mathcal{H}_\eta$:

$$\Phi^{-1}_{\eta''\to\eta'} \circ \left(\mathrm{id}_{\mathcal{H}_{\eta''\to\eta'}} \otimes \Phi^{-1}_{\eta'\to\eta}\right)\left(\psi_{\eta''\to\eta'} \otimes \zeta_{\eta'\to\eta} \otimes \psi_\eta\right) = \Phi^{-1}_{\eta''\to(m)-(n)}\left([\psi_{\eta''\to\eta'} \circ \theta_{\eta''\to\eta'}\, \psi_\eta \circ \theta_\eta] \otimes \zeta_{(m)-(n)}\right),$$

where $\zeta_{\eta'\to\eta} \equiv 1$ and $\zeta_{(m)-(n)} \equiv 1$. Since $\Phi_{\eta''\to(m)-(n)}$, $\Phi_{\eta''\to\eta'}$ and $\Phi_{\eta'\to\eta}$ are Hilbert space isomorphism, $\Phi_{\eta''\to(m)-(n)} \circ \Phi^{-1}_{\eta''\to\eta'} \circ \left(\mathrm{id}_{\mathcal{H}_{\eta''\to\eta'}} \otimes \Phi^{-1}_{\eta'\to\eta}\right)$ thus induces a unitary (injective) map from:

$$\overline{\mathrm{Vect}\left\{\psi_{\eta''\to\eta'} \otimes \zeta_{\eta'\to\eta} \otimes \psi_\eta \,\middle|\, \psi_{\eta''\to\eta'} \in \mathcal{H}_{\eta''\to\eta'}\,,\ \psi_\eta \in \mathcal{H}_\eta\right\}}$$

(where $\overline{\ \cdot\ }$ denotes the completion) into $\mathcal{H}_{\eta''\to(m)-(n)} \otimes \left\{\zeta_{(m)-(n)}\right\}$. Therefore, we get, using the characterization of $\Theta_{\eta'|\eta}$ from the proof of theorem 3.20 (eq. (3.20.2)) together with [18, def. 2.3]:

$$\begin{aligned} &\mathrm{Tr}_{\mathcal{H}_{\eta'}}\, \rho_{\eta'}\, \Theta_{\eta'|\eta} = \\ &\quad = \mathrm{Tr}_{\mathcal{H}_{\eta'}}\left[\sigma\left(\left|\psi^{(n')}\right\rangle\left\langle\psi^{(n')}\right|\right)\right]_{\eta'} \Theta_{\eta'|\eta} \\ &\quad = \mathrm{Tr}_{\mathcal{H}_{\eta''}}\left[\sigma\left(\left|\psi^{(n')}\right\rangle\left\langle\psi^{(n')}\right|\right)\right]_{\eta''} \Phi^{-1}_{\eta''\to\eta'}\left(\mathrm{id}_{\mathcal{H}_{\eta''\to\eta'}} \otimes \Theta_{\eta'|\eta}\right)\Phi_{\eta''\to\eta'} \\ &\quad = \mathrm{Tr}_{\mathcal{H}_{\eta''}}\left[\sigma\left(\left|\psi^{(n')}\right\rangle\left\langle\psi^{(n')}\right|\right)\right]_{\eta''} \Phi^{-1}_{\eta''\to\eta'} \circ \left(\mathrm{id}_{\mathcal{H}_{\eta''\to\eta'}} \otimes \Phi^{-1}_{\eta'\to\eta}\right) \\ &\qquad\qquad \left(\mathrm{id}_{\mathcal{H}_{\eta''\to\eta'}} \otimes |\zeta_{\eta'\to\eta}\rangle\langle\zeta_{\eta'\to\eta}| \otimes \mathrm{id}_{\mathcal{H}_\eta}\right)\left(\mathrm{id}_{\mathcal{H}_{\eta''\to\eta'}} \otimes \Phi_{\eta'\to\eta}\right) \circ \Phi_{\eta''\to\eta'} \end{aligned}$$

$$\leqslant \mathrm{Tr}_{\mathcal{H}_{\eta''}} \left[\sigma\left(\left|\psi^{(n')}\right\rangle\left\langle\psi^{(n')}\right|\right)\right]_{\eta''} \Phi^{-1}_{\eta''\to(m)-(n)} \left(\mathrm{id}_{\mathcal{H}_{\eta''\to(m)-(n)}} \otimes \left|\zeta_{(m)-(n)}\right\rangle\left\langle\zeta_{(m)-(n)}\right|\right) \Phi_{\eta''\to(m)-(n)} \,.$$

This implies, using theorem 3.20.3 and eq. (3.20.1):

$$\mathrm{Tr}_{\mathcal{H}_{\eta'}} \rho_{\eta'} \Theta_{\eta'|\eta} \leqslant \left\langle \Phi_{\eta''\to(m)-(n)} \Gamma^{-1}_{\eta''\to\gamma(\eta'')} \tau_{\gamma(\eta'')\leftarrow\gamma^{(n')}} \psi^{(n')}_{\gamma^{(n')}} \,\middle|\, \left(\mathrm{id}_{\mathcal{H}_{\eta''\to(m)-(n)}} \otimes \left|\zeta_{(m)-(n)}\right\rangle\left\langle\zeta_{(m)-(n)}\right|\right) \,\middle|\, \Phi_{\eta''\to(m)-(n)} \Gamma^{-1}_{\eta''\to\gamma(\eta'')} \tau_{\gamma(\eta'')\leftarrow\gamma^{(n')}} \psi^{(n')}_{\gamma^{(n')}} \right\rangle_{\mathcal{H}_{\eta''\to(m)-(n)}\otimes\mathcal{H}_{(m)-(n)}} . \tag{3.22.6}$$

Let $h'' \in \mathcal{C}_{\eta''}$ and $\left(\left(\widetilde{j}'^{(0)}, \widetilde{j}'^{(2)}, \widetilde{j}'^{(3)}\right), \widetilde{j}; \widetilde{h}\right) := \varphi_{\eta''\to(m)-(n)}(h'')$. Let $\widetilde{\gamma}^{(0)} := H^{(0)}_{\eta''\to\widetilde{\eta}}$ and $\widetilde{\gamma}^{(2)} := H^{(2)}_{\eta''\to\widetilde{\eta}}$. Since $\widetilde{\gamma}^{(0)}, \widetilde{\gamma}^{(2)} \subset \gamma(\eta'') \in \mathcal{L}_{\text{graphs}}$, we have $\widetilde{\gamma}^{(0)}, \widetilde{\gamma}^{(2)} \in \mathcal{L}_{\text{graphs}} \subset \mathcal{L}_{\text{AL}}$ and $\widetilde{\gamma}^{(0)}, \widetilde{\gamma}^{(2)} \preccurlyeq \gamma(\eta'')$. Moreover, the expression for $\varphi_{\eta''\to\widetilde{\eta}}$ (prop. 3.6) together with the uniqueness part of prop. 3.14 yields:

$$\widetilde{j}'^{(0)} = h''|_{\widetilde{\gamma}^{(0)}} = \pi_{\gamma(\eta'')\to\widetilde{\gamma}^{(0)}}(h'')\,,\quad \widetilde{j}'^{(2)} = h''|_{\widetilde{\gamma}^{(2)}} = \pi_{\gamma(\eta'')\to\widetilde{\gamma}^{(2)}}(h'')\,,$$
$$\&\quad \pi^{-1}_{\widetilde{\gamma}_\sharp\to\gamma^{(m)-(n)}}(\widetilde{h}), \widetilde{j} = \varphi_{\widetilde{\gamma}-\left\{e_{[p_\infty,p_n]}\right\}} \circ \pi_{\gamma(\eta'')\to\widetilde{\gamma}}(h'')\,.$$

Using eq. (3.15.1) and the expression for $\varphi_{\gamma_2-\gamma_1}$ from the proof of prop. 3.16, the last relation above becomes:

$$\widetilde{h} = \pi_{\gamma(\eta'')\to\gamma^{(m)-(n)}}(h'') \quad \& \quad \widetilde{j} = \pi_{\gamma(\eta'')\to\left\{e_{[p_\infty,p_n]}\right\}}(h'')\,.$$

Next, since $e_{[p_o,p_n]} \in F_{\daleth}(\widetilde{\eta})$ and $\bigcup_{\widetilde{e}\in\widetilde{\gamma}} r(\widetilde{e}) = r(e_{[p_n,p_\infty]})$, we have, for any $e' \in \gamma^{(n)}$ and any $r \in \left\{1, \dots, n_{\gamma(\eta'')\to\gamma^{(n)},e'}\right\}$:

$$a_{\gamma(\eta'')\to\gamma^{(n)},e'}(r) \in F_{\daleth}(\widetilde{\eta}) \quad \& \quad \forall \widetilde{e} \in \widetilde{\gamma},\ r\left[a_{\gamma(\eta'')\to\gamma^{(n)},e'}(r)\right] \not\subset r(\widetilde{e})\,.$$

Hence $a_{\gamma(\eta'')\to\gamma^{(n)},e'}(r) \in H^{(0)}_{\eta''\to\widetilde{\eta}}$, so that $\gamma^{(n)} \preccurlyeq \widetilde{\gamma}^{(0)}$. Similarly, if $n' \geqslant m+1$, $e_{[p_m,p_\infty]} \in \widetilde{\gamma}$ implies $\left\{e_{[p_{m+1},p_\infty]}\right\} \preccurlyeq \widetilde{\gamma}^{(2)}$ (we have $\left\{e_{[p_{m+1},p_\infty]}\right\} \preccurlyeq \gamma(\eta'')$ for $\left\{e_{[p_m,p_{m+1}]}\right\} \preccurlyeq \gamma^{(n')} \preccurlyeq \gamma(\eta'')$ and $\left\{e_{[p_m,p_\infty]}\right\} \preccurlyeq \widetilde{\gamma} \preccurlyeq \gamma(\eta'')$), as well as $\gamma^{(n')-(m+1)} \preccurlyeq \widetilde{\gamma}^{(2)}$ (if $n'$ happens to be bigger than $m+2$). Then, using repeatedly eq. (3.15.1), we get:

$$\forall k \in \{1, \dots, n\}\,,\ \pi_{\gamma(\eta'')\to\gamma^{(n')}}(h'')\left(e_{[p_{k-1},p_k]}\right) = \pi_{\widetilde{\gamma}^{(0)}\to\gamma^{(n)}}(\widetilde{j}'^{(0)})\left(e_{[p_{k-1},p_k]}\right),$$
$$\forall k \in \{n+1, \dots, m\}\,,\ \pi_{\gamma(\eta'')\to\gamma^{(n')}}(h'')\left(e_{[p_{k-1},p_k]}\right) = \widetilde{h}\left(e_{[p_{k-1},p_k]}\right)$$

(note that $m \leqslant n'$ as underlined earlier),

$$\forall k \in \{m+2, \dots, n'\}\,,\ \pi_{\gamma(\eta'')\to\gamma^{(n')}}(h'')\left(e_{[p_{k-1},p_k]}\right) = \pi_{\widetilde{\gamma}^{(2)}\to\gamma^{(n')-(m+1)}}(\widetilde{j}'^{(2)})\left(e_{[p_{k-1},p_k]}\right)$$

(of course, this only applies if $n' \geqslant m+2$).

If $n' > m$ we also need the evaluation of $\pi_{\gamma(\eta'')\to\gamma^{(n')}}(h'')$ on $e_{[p_m,p_{m+1}]}$. For this, we notice that $\left\{e_{[p_\infty,p_n]}\right\} \preccurlyeq \gamma^{[m,m+1]} \preccurlyeq \gamma(\eta'')$, where $\gamma^{[m,m+1]} := \left\{e_{[p_n,p_m]}, e_{[p_m,p_{m+1}]}, e_{[p_{m+1},p_\infty]}\right\}$. Using the explicit expression for $\pi_{\gamma^{[m,m+1]}\to\left\{e_{[p_\infty,p_n]}\right\}}$ together with $\left\{e_{[p_n,p_m]}\right\} \preccurlyeq \gamma^{(m)-(n)}$ and $\left\{e_{[p_{m+1},p_\infty]}\right\} \preccurlyeq \widetilde{\gamma}^{(2)}$ (and again repeatedly applying eq. (3.15.1)), we get:

$$\widetilde{j}(e_{[p_\infty,p_n]}) = \pi_{\gamma(\eta'')\to\left\{e_{[p_\infty,p_n]}\right\}}(h'')(e_{[p_\infty,p_n]})$$
$$= \pi_{\gamma(\eta'')\to\gamma^{[m,m+1]}}(h'')(e_{[p_n,p_m]})^{-1} \,.\, \pi_{\gamma(\eta'')\to\gamma^{[m,m+1]}}(h'')(e_{[p_m,p_{m+1}]})^{-1} \,.\, \pi_{\gamma(\eta'')\to\gamma^{[m,m+1]}}(h'')(e_{[p_{m+1},p_\infty]})^{-1}$$

$$= \pi_{\gamma^{(m)-(n)} \to \left\{e_{[p_n,p_m]}\right\}}(\widetilde{h})(e_{[p_n,p_m]})^{-1} \,.\, \pi_{\gamma(\eta'') \to \gamma^{(n')}}(h'')(e_{[p_m,p_{m+1}]})^{-1} \,.\, \pi_{\widetilde{\gamma}^{(2)} \to \left\{e_{[p_{m+1},p_\infty]}\right\}}(\widetilde{j}'^{(2)})(e_{[p_{m+1},p_\infty]})^{-1}$$

$$= \left[\prod_{k=n+1}^{m} \widetilde{h}(e_{[p_{k-1},p_k]})\right]^{-1} .\, \pi_{\gamma(\eta'') \to \gamma^{(n')}}(h'')(e_{[p_m,p_{m+1}]})^{-1} \,.\, \pi_{\widetilde{\gamma}^{(2)} \to \left\{e_{[p_{m+1},p_\infty]}\right\}}(\widetilde{j}'^{(2)})(e_{[p_{m+1},p_\infty]})^{-1},$$

so that:

$$\pi_{\gamma(\eta'') \to \gamma^{(n')}}(h'')(e_{[p_m,p_{m+1}]}) = \left[\prod_{k=n+1}^{m} \widetilde{h}(e_{[p_{k-1},p_k]}) \,.\, \widetilde{j}(e_{[p_\infty,p_n]}) \,.\, \pi_{\widetilde{\gamma}^{(2)} \to \left\{e_{[p_{m+1},p_\infty]}\right\}}(\widetilde{j}'^{(2)})(e_{[p_{m+1},p_\infty]})\right]^{-1}.$$

We want to use the thus obtained relations between $\pi_{\gamma(\eta'') \to \gamma^{(n')}}$ and $\varphi_{\eta'' \to (m)-(n)}$ in order to reformulate eq. (3.22.6). We first consider the case $n' = m$. Then, we have:

$$\Phi_{\eta'' \to (m)-(n)} \, \Gamma^{-1}_{\eta'' \to \gamma(\eta'')} \, \tau_{\gamma(\eta'') \leftarrow \gamma^{(n')}} \, \psi^{(n')}_{\gamma^{(n')}} = \psi_{\eta'' \to (m)-(n)} \otimes \chi_{(m)-(n)} \,,$$

where:

$$\forall \left(\widetilde{j}'^{(0)}, \widetilde{j}'^{(2)}, \widetilde{j}'^{(3)}\right), \widetilde{j} \in \mathcal{C}_{\eta'' \to (m)-(n)} \,,$$

$$\psi_{\eta'' \to (m)-(n)} \left(\widetilde{j}'^{(0)}, \widetilde{j}'^{(2)}, \widetilde{j}'^{(3)} \,;\, \widetilde{j}\right) := \prod_{k=1}^{n} \left[\chi \circ \pi_{\widetilde{\gamma}^{(0)} \to \gamma^{(n)}}(\widetilde{j}'^{(0)})\right](e_{[p_{k-1},p_k]})$$

$$\& \quad \forall \widetilde{h} \in \mathcal{C}_{(m)-(n)} \,,\, \chi_{(m)-(n)}\big(\widetilde{h}\big) := \prod_{k=n+1}^{m} \big[\chi \circ \widetilde{h}\big](e_{[p_{k-1},p_k]}) \,.$$

Thus, eq. (3.22.6) becomes:

$$\operatorname{Tr}_{\mathcal{H}_{\eta'}} \rho_{\eta'} \, \Theta_{\eta'|\eta} \leqslant \left\|\psi_{\eta'' \to (m)-(n)}\right\|^2_{\mathcal{H}_{\eta'' \to (m)-(n)}} \left|\left\langle \zeta_{(m)-(n)} \mid \chi_{(m)-(n)} \right\rangle_{\mathcal{H}_{(m)-(n)}}\right|^2 .$$

And, since $\left\|\psi^{(n')}_{\gamma^{(n')}}\right\|_{\mathcal{H}_{\gamma^{(n')}}} = 1$, we get:

$$\operatorname{Tr}_{\mathcal{H}_{\eta'}} \rho_{\eta'} \, \Theta_{\eta'|\eta} \leqslant \frac{\left|\left\langle \zeta_{(m)-(n)} \mid \chi_{(m)-(n)} \right\rangle_{\mathcal{H}_{(m)-(n)}}\right|^2}{\left\|\chi_{(m)-(n)}\right\|^2_{\mathcal{H}_{(m)-(n)}}} = \frac{\left|\int_G d\mu(g)\, \chi(g)\right|^{2M}}{\|\chi\|^{2M}} \leqslant \left(\int_G d\mu(g)\ |\chi(g)|\right)^{2M}.$$

We now consider the case $n' > m$. Here, we get:

$$\forall \big(\widetilde{j}', \widetilde{j}\big) \in \mathcal{C}_{\eta'' \to (m)-(n)} \,,\, \forall \widetilde{h} \in \mathcal{C}_{(m)-(n)} \,,$$

$$\Phi_{\eta'' \to (m)-(n)} \, \Gamma^{-1}_{\eta'' \to \gamma(\eta'')} \, \tau_{\gamma(\eta'') \leftarrow \gamma^{(n')}} \, \psi^{(n')}_{\gamma^{(n')}} \left[\widetilde{j}', \widetilde{j}; \widetilde{h}\right] =$$

$$= \psi_{\eta'' \to \widetilde{\eta}}\big(\widetilde{j}'\big) \ \ \chi_{(m)-(n)}\big(\widetilde{h}\big) \ \ \chi\left[\beta_{\eta'' \to \widetilde{\eta}}\big(\widetilde{j}'\big) \,.\, \widetilde{j}(e_{[p_\infty,p_n]})^{-1} \,.\, \beta_{(m)-(n)}\big(\widetilde{h}\big)\right],$$

where:

$$\forall \left(\widetilde{j}'^{(0)}, \widetilde{j}'^{(2)}, \widetilde{j}'^{(3)}\right) \in \mathcal{C}_{\eta'' \to \widetilde{\eta}} \,,$$

$$\psi_{\eta'' \to \widetilde{\eta}} \left(\widetilde{j}'^{(0)}, \widetilde{j}'^{(2)}, \widetilde{j}'^{(3)}\right)$$

$$:= \left(\prod_{k=1}^{n} \left[\chi \circ \pi_{\widetilde{\gamma}^{(0)} \to \gamma^{(n)}}(\widetilde{j}'^{(0)})\right](e_{[p_{k-1},p_k]})\right) \ \left(\prod_{k=m+2}^{n'} \left[\chi \circ \pi_{\widetilde{\gamma}^{(2)} \to \gamma^{(n')-(m+1)}}(\widetilde{j}'^{(2)})\right](e_{[p_{k-1},p_k]})\right),$$

$$\beta_{\eta''\to\widetilde{\eta}}\left(\widetilde{j}'^{(0)}, \widetilde{j}'^{(2)}, \widetilde{j}'^{(3)}\right) := \left[\pi_{\widetilde{\gamma}^{(2)}\to\left\{e_{[p_{m+1},p_\infty]}\right\}}(\widetilde{j}'^{(2)})(e_{[p_{m+1},p_\infty]})\right]^{-1},$$

$$\& \quad \forall\, \widetilde{h} \in \mathcal{C}_{(m)-(n)}\,,\ \beta_{(m)-(n)}(\widetilde{h}) := \prod_{k=m}^{n+1} \widetilde{h}(e_{[p_{k-1},p_k]})^{-1}\,.$$

Thus, eq. (3.22.6) now reads:

$$\mathrm{Tr}_{\mathcal{H}_{\eta'}}\, \rho_{\eta'}\, \Theta_{\eta'|\eta} \leqslant$$

$$\int_{\mathcal{C}_{\eta''\to\widetilde{\eta}}} d\mu_{\eta''\to\widetilde{\eta}}(\widetilde{j}') \;\; \left|\psi_{\eta''\to\widetilde{\eta}}\left(\widetilde{j}'\right)\right|^2 \int_{\mathcal{C}_{(m)-(n)}\times\mathcal{C}_{(m)-(n)}} d\mu^{(2)}_{(m)-(n)}(\widetilde{h}, \widetilde{h}') \;\; \chi^*_{(m)-(n)}\left(\widetilde{h}\right) \chi_{(m)-(n)}\left(\widetilde{h}'\right)$$

$$\int_G d\mu(t) \;\; \chi^*\left[\beta_{\eta''\to\widetilde{\eta}}\left(\widetilde{j}'\right) . \, t^{-1} . \, \beta_{(m)-(n)}\left(\widetilde{h}\right)\right] \chi\left[\beta_{\eta''\to\widetilde{\eta}}\left(\widetilde{j}'\right) . \, t^{-1} . \, \beta_{(m)-(n)}\left(\widetilde{h}'\right)\right],$$

while the normalization condition $\left\|\psi^{(n')}_{\gamma^{(n')}}\right\|_{\mathcal{H}_{\gamma^{(n')}}} = 1$ yields:

$$1 = \int_{\mathcal{C}_{\eta''\to\widetilde{\eta}}} d\mu_{\eta''\to\widetilde{\eta}}(\widetilde{j}') \;\; \left|\psi_{\eta''\to\widetilde{\eta}}\left(\widetilde{j}'\right)\right|^2 \int_{\mathcal{C}_{(m)-(n)}} d\mu_{(m)-(n)}(\widetilde{h}) \;\; \left|\chi_{(m)-(n)}\left(\widetilde{h}\right)\right|^2 \int_G d\mu(t) \;\; |\chi(t)|^2$$

(the measure $\mu$ being invariant under the transformation $t \mapsto t_1 . \, t^{-1} . \, t_2$ for any $t_1$ , $t_2 \in G$)

$$= \int_{\mathcal{C}_{\eta''\to\widetilde{\eta}}} d\mu_{\eta''\to\widetilde{\eta}}(\widetilde{j}') \;\; \left|\psi_{\eta''\to\widetilde{\eta}}\left(\widetilde{j}'\right)\right|^2 \qquad \text{(since } \|\chi\| = 1\text{)}.$$

Moreover, for any $t_1$ , $t_2$ , $t'_2 \in G$, the Cauchy-Schwarz inequality ensures that:

$$\left|\int_G d\mu(t) \;\; \chi^*\left[t_1 . \, t^{-1} . \, t_2\right] \chi\left[t_1 . \, t^{-1} . \, t'_2\right]\right| \leqslant \|\chi\|^2 = 1\,,$$

so we again get:

$$\mathrm{Tr}_{\mathcal{H}_{\eta'}}\, \rho_{\eta'}\, \Theta_{\eta'|\eta} \leqslant \int_{\mathcal{C}_{\eta''\to\widetilde{\eta}}} d\mu_{\eta''\to\widetilde{\eta}}(\widetilde{j}') \;\; \left|\psi_{\eta''\to\widetilde{\eta}}\left(\widetilde{j}'\right)\right|^2 \int_{\mathcal{C}_{(m)-(n)}\times\mathcal{C}_{(m)-(n)}} d\mu^{(2)}_{(m)-(n)}(\widetilde{h}, \widetilde{h}') \;\; \left|\chi^*_{(m)-(n)}\left(\widetilde{h}\right)\right| \left|\chi_{(m)-(n)}\left(\widetilde{h}'\right)\right|$$

$$= \left(\int_G d\mu(g)\; |\chi(g)|\right)^{2M}.$$

Since $\chi$ has been chosen so that $\int_G d\mu(g)\; |\chi(g)| < 1$, there exists, for any $\epsilon > 0$, an odd integer $M \geqslant 1$ with $\left(\int_G d\mu(g)\; |\chi(g)|\right)^{2M} < \epsilon$. Thus, there exists, for any $\eta \in \mathcal{L}$ and any $\epsilon > 0$, $\eta' \succcurlyeq \eta$ such that:

$$\mathrm{Tr}_{\mathcal{H}_{\eta'}}\, \rho_{\eta'}\, \Theta_{\eta'|\eta} < \epsilon\,.$$

Hence, for any $\eta \in \mathcal{L}$, $\inf_{\eta' \succcurlyeq \eta} \mathrm{Tr}_{\mathcal{H}_{\eta'}}\, \rho_{\eta'}\, \Theta_{\eta'|\eta} = 0$, and, therefore, $\sup_{\eta\in\mathcal{L}} \inf_{\eta' \succcurlyeq \eta} \mathrm{Tr}_{\mathcal{H}_{\eta'}}\, \rho_{\eta'}\, \Theta_{\eta'|\eta} = 0$. On the other hand theorem 3.20.5 implies:

$$\sigma\left\langle \overline{\mathcal{S}}_{\text{AL}}\right\rangle = \left\{\rho' = \left(\rho'_\eta\right)_{\eta\in\mathcal{L}} \;\middle|\; \sup_{\eta\in\mathcal{L}} \inf_{\eta' \succcurlyeq \eta} \mathrm{Tr}_{\mathcal{H}_{\eta'}}\, \rho'_{\eta'}\, \Theta_{\eta'|\eta} = \mathrm{Tr}\, \rho'\right\},$$

and we have $\mathrm{Tr}\, \rho = \mathrm{Tr}_{\mathcal{H}_\eta}\, \rho_\eta = \left\|\psi^{(n_\eta)}\right\|_{\mathcal{H}_{\text{AL}}} = 1$ (for some $\eta \in \mathcal{L}$), so $\rho \notin \sigma\left\langle \overline{\mathcal{S}}_{\text{AL}}\right\rangle$. □

# 4 Outlook

So we were able to show that the construction developed in [24] can be generalized from the linear case to an arbitrary gauge group $G$. The key ingredient is still the same, namely the use of labels defined as collections of edges and surfaces. However, the somewhat involved algebra built by the holonomies and fluxes in the case of a non-Abelian group requires to be more restrictive as to which such collections qualify as labels. The factorization maps, which are the central objects of the formalism, can then be expressed explicitly in terms of the group operations, so that no further restrictions on the Lie group $G$ are needed.

In the case of a *compact* gauge group, which will be the most relevant for us (since we are interested in the application to LQG, with $G = \mathcal{SU}(2)$), we have a clear picture of how this alternative state space relates to the well-established Ashtekar-Lewandowski one. Recall that our motivation for this construction was to extend the latter, to try and cure the difficulties arising in the search for good semi-classical states. The observation of prop. 3.22, confirming that we indeed have gained new states in the process, is in this respect particularly important.

The state we used to prove this result can even be seen as a first step toward the design of satisfactory semi-classical states. Indeed, if we take as 'pattern' $\chi$ a coherent state (eg. a Hall state [12], which is the generalization over a compact Lie group of a Gaussian state), we obtain a projective state yielding a narrow distribution for infinitely many holonomies (namely the ones along the infinitely many pieces of the base edge $e$), while such a state could not exist over $\mathcal{H}_{\text{AL}}$.

Still, this would not yet be a state suitable for the study of the semi-classical limit, where we would need states presenting narrow distributions for a full set of holonomies and fluxes. There remain in fact further obstructions to this endeavor, the understanding and overcoming of which will be the topic of forthcoming work.

Another issue that will have to be addressed thoroughly before the projective formalism can provide a serious alternative to the successful inductive one, is how to solve at least the Gauss and diffeomorphism constraints [1, 3, 31]. In [17, section 3] we proposed a strategy to deal with constraints in the projective context, with the help of a suitably defined regularization scheme (this proposal was developed at the classical level, ie. in the setting of a projective limit of symplectic manifolds, however we also displayed on an example in [19, subsection 3.2] how a similar approach could be implemented at the quantum level).

Note that while the Gauss constraints are well-adapted to the inductive structure underlying the Ashtekar-Lewandowski Hilbert space (ie. they leave the fixed-graph subspaces $\mathcal{H}_\gamma$ invariant, which allows for their straightforward resolution in $\mathcal{H}_{\text{AL}}$) they are *not* adapted (in the sense of [17, subsection 3.1]) to the projective structure we have introduced: while gauge transformations preserve the algebra of holonomies attached to a graph, they do *not* preserve the algebra of fluxes attached to a profile. In fact, fluxes do not at all transform nicely under gauge transformations. A popular method to circumvent this difficulty is to use, instead of the standard fluxes, appropriately 'anchored' ones [29, def. 3.5]: by choosing, for each face, a supporting system of paths, we can parallel transport the electric field at each point of the face back to a common root, thus forming an observable with better transformations properties. Yet, a complete solution based on this device will require some more work, because again one needs to ensure, at the same time, that the labels we are using can be properly associated to an algebra of observables (ie. that the observables assigned to a given label form a subset closed under Poisson brackets), and that they build a directed preordered set (with a preorder that respects the relations between the associated algebras of observables).

Similarly the techniques developed for the resolution of the diffeomorphism constraints in $\mathcal{H}_{\text{AL}}$ [3] cannot be directly imported into the projective formalism because they critically rely on having states made up of discrete excitations. To make progress on this issue, we will have to understand how the input needed for a resolution along the lines of [17, section 3] can be set up in a background independent manner. In a more speculative line, it might also be possible to combine the known regularization scheme for the Hamiltonian constraint [28] with the strategy from [17, section 3] in order to arrive at a constructive description of a space of states solving the quantum dynamics of gravity.

## Acknowledgements

This work has been financially supported by the Université François Rabelais, Tours, France, and by the Friedrich-Alexander-Universität Erlangen-Nürnberg, Germany (via the Bavarian Equal Opportunities Sponsorship – Förderung von Frauen in Forschung und Lehre (FFL) – Promoting Equal Opportunities for Women in Research and Teaching).

This research project has been supported by funds to Emerging Field Project "Quantum Geometry" from the FAU Erlangen-Nuernberg within its Emerging Fields Initiative.

# A References